\RequirePackage{fix-cm}
\documentclass{svjour3}                     
\smartqed  
\usepackage{graphicx}
\usepackage{bm}
\usepackage{here}
\usepackage{tikz}
\usepackage{float,subcaption}
\begin{document}

\title{Toward the detection of spin-vortex-induced loop currents in a single bilayer Bi$_2$Sr$_2$CaCu$_2$O$_{8+\delta}$ thin film and their possible use as qubits: Model calculations for three nano-island architecture  
}

\titlerunning{Spin-vortex-induced loop currents in a single bilayer Bi-2212 thin film}        

\author{Hayato Taya \and Yuto Takatsu \and  Hiroyasu Koizumi
}
\institute{
           H. Taya, Y. Takatsu \at
              Graduate School of Pure and Applied Sciences, University of Tsukuba, Tsukuba, Ibaraki, Japan
               \and
              H. Koizumi\at
              Center for Computational Sciences, University of Tsukuba, Tsukuba, Ibaraki, Japan \\
              Tel.:+81-29-8536403\\
              Fax:+81-29-8536403\\
              \email{koizumi.hiroyasu.fn@u.tsukuba.ac.jp}           
}

\date{Received: date / Accepted: date}

\maketitle

\begin{abstract}
A theory for cuprate superconductivity predicts the existence of nano-sized loop currents called, ``spin-vortex-induced loop currents (SVILCs)''. 
In this wok, we first calculate magnetic fields produced by them in a single bilayer Bi$_2$Sr$_2$CaCu$_2$O$_{8+\delta}$ (Bi-2212) thin film for the purpose of detecting the SVILCs.
The estimated magnitude of the magnetic field at the point 10$a$ ($a$ is the lattice constant of the CuO$_2$ plane) above the surface could be in the order of 100mT; thus, they may be detectable by currently available detection methods.
Next, we investigate the use of them as qubits (the ``SVILC qubits'') in an architecture
composed of three nano-islands of the thin film; and consider the 
use of the detection of the magnetic field generated by the SVILCs as the qubit readout.
 We show there are a number of energy levels suitable for qubit states
 that can be manipulated by external current feeding, and the magnetic field generated by the SVILCs is large enough to be used for the readout.

\end{abstract}

\section{Introduction}

Since the discovery of high temperature superconductivity in 1986 \cite{Muller1986}, 
extensive efforts have been devoted to elucidate its mechanism. However, no widely-accepted theory exists despite almost 40 years of efforts.
Since the cuprate superconductivity is markedly different from the BCS superconductivity, the elucidation of it will require a marked departure from the standard superconductivity theory.
In this respect, it is noteworthy that experimental facts of superconductivity indicate still several serious loose ends to be tied in the standard theory as follows: 1) the standard theory relies on the use of particle number non-conserving formalism although superconductivity occurs in an isolated system where the particle-number is conserved \cite{Peierls1991}; 2) the superconducting carrier mass obtained by the London moment experiment is the free electron mass $m_e$, although the standard theory prediction is the effective mass $m^\ast$ of the normal state \cite{Hirsch2013b}; 3) the reversible superconducting-normal phase transition in a magnetic field cannot be explained by the standard theory \cite{Hirsch2017}; 4) the dissipative quantum phase transition in a Josephson junction system predicted by the standard theory is absent \cite{PhysRevX2021a}; 5) so-called the `quasiparticle poisoning problem' indicates the existence a large amount of excited single electrons in Josephson junction systems, obtaining the observed ratio of their number to the Cooper pair number  $10^{-9} \sim 10^{-5}$ in disagreement with the standard theory predicted value $10^{-52}$\cite{poisoning2023,Serniak2019}.
The existence of the above problems indicate the need for serious revisions of the superconductivity theory.
It is sensible to consider that the theory explains the cuprate superconductivity will also need to resolve the above problems.

One of the present authors has put forward a new theory of superconductivity that encompasses the BCS theory and lifts the disagreements mentioned above  \cite{koizumi2022,koizumi2022b,koizumi2023}.
In this theory, supercurrent is generated by an emergent $U(1)$ gauge field 
arising from the singularities of the many-body wave function that can be included by the Berry phase formalism \cite{Berry}. 
Especially, such a gauge field arises when spin-twisting itinerant motion of electrons is realized; 
in this case, singularities of the wave function exist at the centers of the spin-twisting, and the  emergent gauge field gives rise to persistent loop currents around them.

The cuprate superconductivity may be elucidated by this new theory since the presence of coherence-length-sized spin-vortices and accompanying loop currents ({\em spin-vortex-induced loop currents} or {\em SVILCs}) are highly plausible from the following facts 
: 1) the superconducting transition temperature for the optimally doped cuprates corresponds to the stabilization temperature of the coherence-length-sized loop currents \cite{Kivelson95}, and the experiment using Bi$_2$Sr$_2$CaCu$_2$O$_{8+\delta}$ (Bi-2212) thin films has confirmed that the superconducting transition is the BKT type \cite{D3RA02701E}; 2) theoretical calculations based on the stabilization of the SVILCs yield reasonable superconducting transition temperatures \cite{HKoizumi2015B,Koizumi2017}; 3)
the magnetic excitation spectra observed  may be taken as the evidence for the existence of nano-sized spin-vortices \cite{Neutron,Hidekata2011}; 4) the presence of SVILCs explains the polar Kerr effect measurement \cite{Kerr1}, enhanced Nernst effect measurement \cite{Nernst}, and the neutron scattering measurement \cite{neutron2015}. 
Recently, spin-vortices have been observed in the cupare superconductors \cite{Wang:2023aa};
although the observed spin-textures are different from the predicted spin-vortices, experiments with improved spatial resolution and sensitivity may detect the predicted nano-sized spin-vortices. 

In the present work, we repot calculated magnetic fields produced by the SVILCs for the purpose of 
detecting them.
The confirmation of the existence of the SVILCs will lead to the elucidation of the cuprate superconductivity; and it may also lead to novel quantum device applications of the cuprate
since methods for preparing Bi-2212 thin films are now established \cite{WANG201213,Jiang:2014aa,Jindal:2017aa,adfm.201807379,SHEN202135067,nwac089,KEPPERT2023157822}. Thus, qubits made of the SVILCs may be realized \cite{WAKAURA201655,Wakaura2017,Koizumi:2022aa}. 
In this work, we will explore such a possibility 
by extending our previous work \cite{Koizumi:2022aa}.
In the present work, we will consider the three nano-island architecture.

The effort to realize fully fault-tolerant quantum computers
with more than 100 logical qubits is the next target of the quantum computing technology.
For this purpose, currently available qubits may not be good enough 
by various reasons. For example, superconducting qubits
using aluminum require extremely low operation temperature (about 10 mK); the coupler between qubits is too large. These two problems
make it difficult to assemble a large number of physical
qubits (estimated to be more than 100,000) that are necessary to build over 100 logical qubit quantum computers.
The SVILC qubits we have been proposing might overcome such difficulties since
each SVILC qubit is topologically protected, coherence-length-sized loop current with the size of the 10 nm order including the coupler; the operation temperature might be elevated to the liquid
nitrogen temperature since the cuprates become superconducting at liquid nitrogen temperature in the ambient pressure. 

The organization of the present work is as follows: In Section~\ref{Bog}, the particle number conserving Bogoliubov-de Gennes (PNC-BdG) formalism employed in this work is briefly explained.
This formalism can deal with the emergent gauge field generated by the spin-twisting itinerant motion of electrons \cite{koizumi2019,koizumi2021,koizumi2021b,Koizumi2021c} (see Appendices A and B for additional explanations for the PNC-BdG). 
 In Section~\ref{sec3}, the PNC-BdG equations for the system of a single CuO$_2$ bilayer with one surface-layer and one bulk-layer are explained. In Section~\ref{sec4}, calculated results for SVILCs and their magnetic fields are presented. 
Then, we consider the SVILC qubit in Section~\ref{subss:result:qubit_state}.
The three nano-island architecture is explained and examined in Section~\ref{subss:3Nano-Islands}.
 Lastly, we conclude the present work in Section~\ref{sec6}. 

\section{Superconducting state described by the particle-number conserving Bogoliubov-de Gennes(PNC-BdG) formalism}
  \label{Bog}
The superconducting state in the present theory is defined as a state in which a collection of 
loop currents exist. They are produced by the $\chi$ mode, where $\chi$ mode arises from the Berry connection from many-electron wave functions \cite{koizumi2023}, and plays the similar role that the Nambu-Goldstone (NG) mode does in the standard theory \cite{Nambu1960} (see Appendices A and B for more detailed explanations for this point).
The $\chi$ mode generates the following velocity field
\begin{eqnarray}
{\bf v}_{\chi}
=-{\hbar \over {2m_e}} \nabla \chi 
\label{eq12}
\end{eqnarray}
where $\chi$ is an angular variable with period $2\pi$, $\hbar$ is the reduced Planck constant, and $m_e$ is the free electron mass.

The current generated by it is topologically protected in the following sense:
$\chi$ is an angular variable with period $2\pi$, and characterized by the topological integer, {\em winding number}, 
defined by
\begin{equation}
w_C[\chi]={ 1 \over {2 \pi}} \oint_C \nabla \chi \cdot d{\bf r}
\end{equation}
where $C$ is a loop in the coordinate space.
If it is time-independent 
\begin{eqnarray}
{d \over {dt}} w_C[\chi]=0
\label{eq15}
\end{eqnarray}
and the time-derivative of the velocity field is approximated by the relaxation time approximation 
\begin{eqnarray}
{{d{\bf v}_{\chi}} \over {dt}}=-{1 \over \tau}{\bf v}_{\chi}
\label{eq16}
\end{eqnarray}
where $\tau$ is the relaxation time, the following relation is obtained
\begin{eqnarray}
\tau {d \over {dt}}w_C[\chi] =-w_C[\chi]
\label{eq17}
\end{eqnarray}
This relation with $w_C[\chi] \neq 0$ and ${d \over {dt}} w_C[\chi]=0$ requires
$\tau$ to be $\infty$. This means that the current produced by $\nabla \chi$ is persistent. This persistent loop current is the element of a macroscopic supercurrent; a macroscopic supercurrent is a collection them.

When an electromagnetic field is present, the velocity field in Eq.~(\ref{eq12}) becomes
\begin{eqnarray}
{\bf v}={e \over m_e}\left( {\bf A} -{\hbar \over {2e}}\nabla \chi \right)
\label{eqvelo}
\end{eqnarray}
where $-e$ is the electron charge, and ${\bf A}$ is the electromagnetic vector potential.
The velocity field in Eq.~(\ref{eqvelo}) realizes the Meissner effect, thus, ${\bf v}$ is zero deep inside the
sample. Let us consider a ring-shaped superconductor 
and  a loop $C$  that goes through the inside region where ${\bf v}=0$ is satisfied; then,
the flux quantization is obtained by integrating ${\bf A}$ along $C$, which is given by
\begin{eqnarray}
\oint_C {\bf A} \cdot d{\bf r}={ h \over {2e}}w_C[\chi]
\label{flux-eq}
\end{eqnarray}
This is the experimentally observed flux quantization. 

  \begin{figure}[H]
    \centering
    \includegraphics[width=4cm]{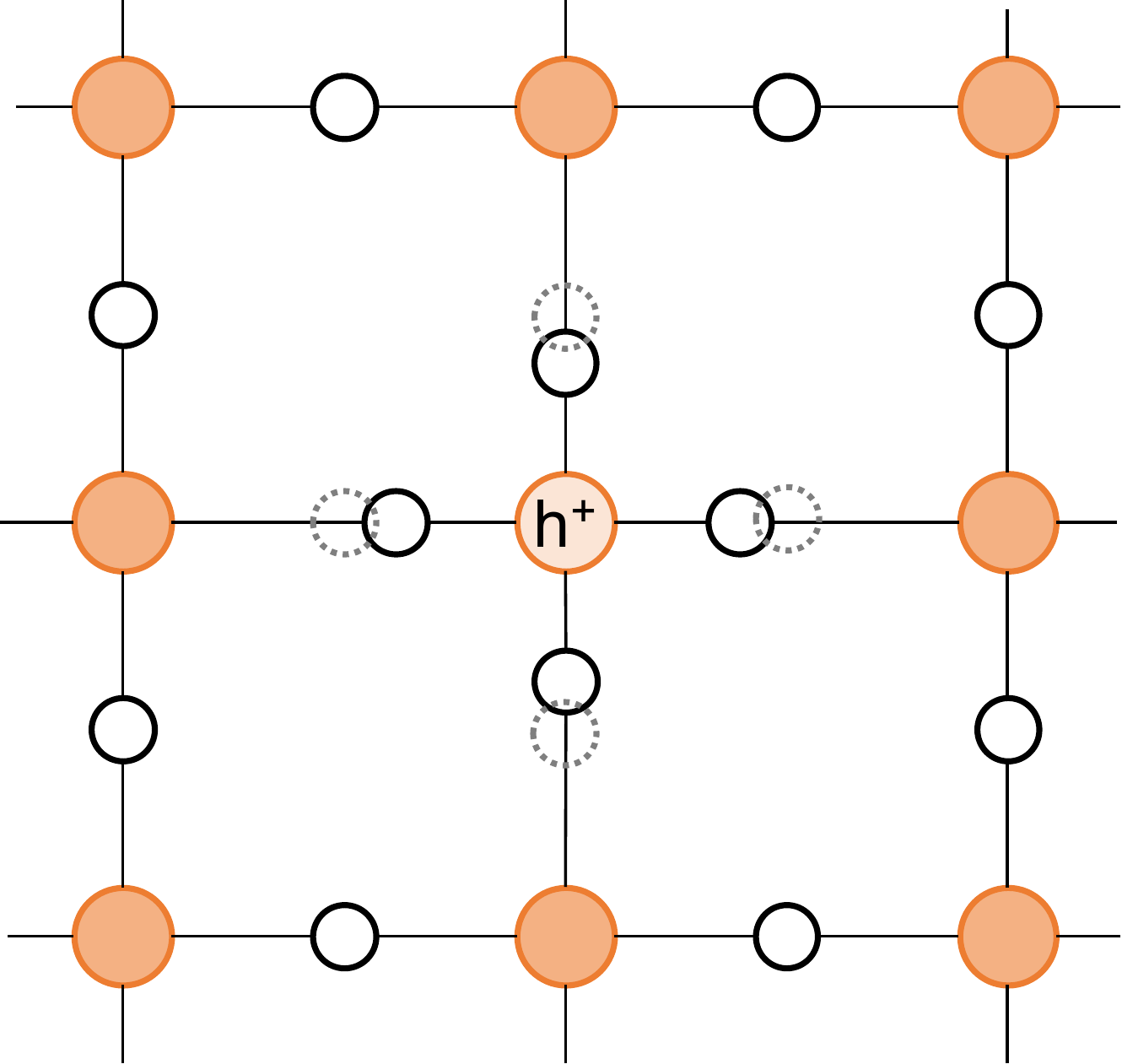}
    \caption{Small polaron formation around a hole occupied site of copper in the CuO$_2$ plane of the bulk of hole-doped cuprate superconductors. Large circles indicate coppers, and small ones oxygens. ``h$^+$'' indicates the hole. Four oxygen atoms around the hole occupied copper atom come closer to the copper atom, and a small lattice polaron is formed.}
  \label{fig:CuO2Plane}
\end{figure}

\begin{figure}[H]
    \centering
    \includegraphics[width=5cm]{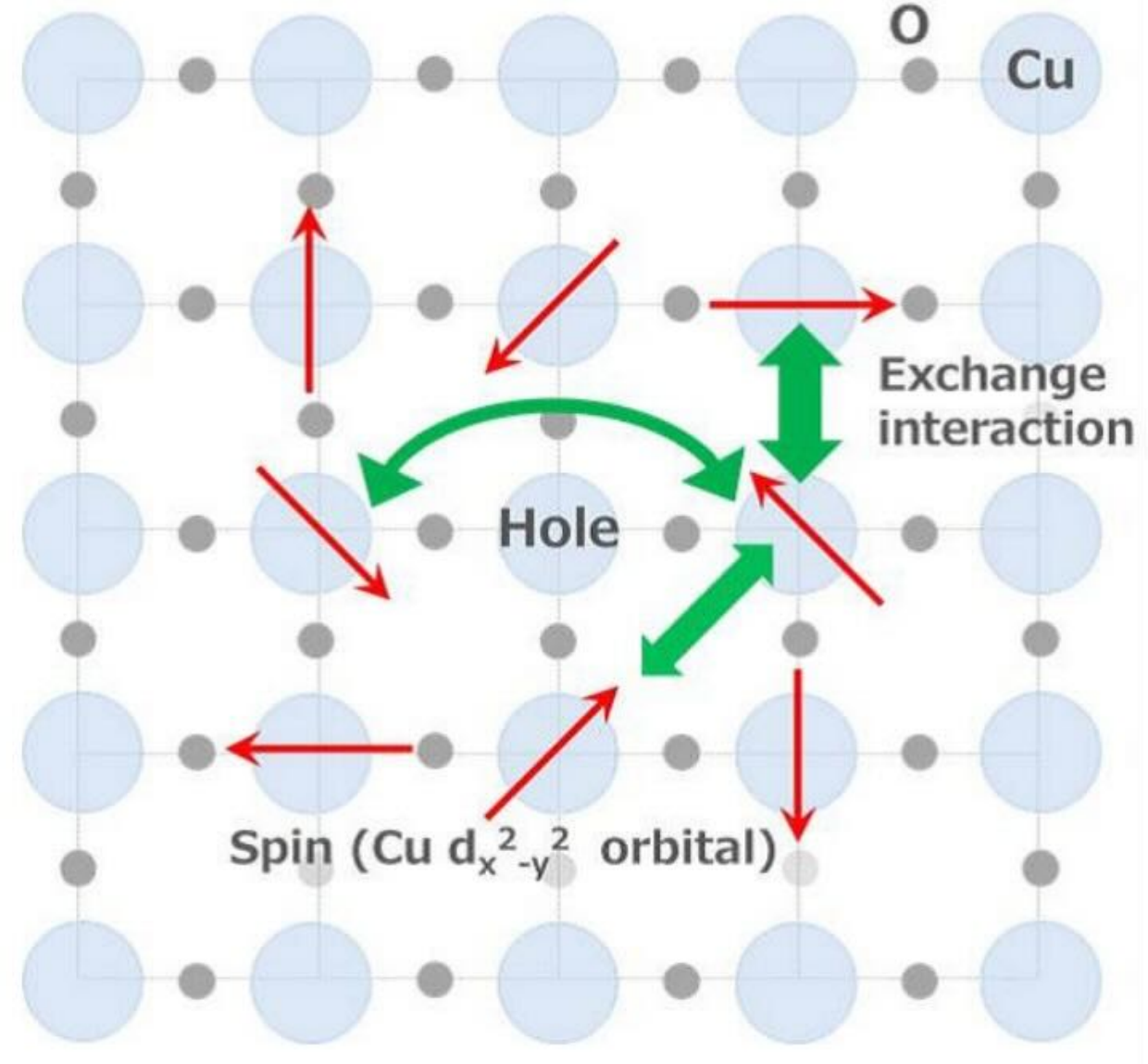}
    \caption{Spin-vortex formation around a hole-doped site of copper in the CuO$_2$ plane of the bulk of hole-doped cuprate superconductors. Double ended arrows indicate antiferromagnetic exchange interactions.
    There are two types of antiferromagnetic exchange interactions; one of them is that acts between pairs of electrons reside in the first nearest neighbor sites, and the other is that acts between pairs of electrons reside across the hole-occupied sites (includes the straight line direction and right angle direction).
 Arrows on copper sites indicate the spin moment of itinerant electrons that perform spin-twisting itinerant motion.}
    \label{fig:spin-vortex}
  \end{figure}

Let us explain how $\chi$ with nonzero winding number is realized in cuprate superconductors. 
The stage of the cuprate superconductivity is the CuO$_2$ plane.
A schematic picture of a doped hole in the bulk of the CuO$_2$ plane is shown in Fig.~\ref{fig:CuO2Plane}.
Oxygen atoms around the hole occupied copper atom come closer to the copper atom, and a small lattice polaron is formed.
Then, a spin-vortex is formed around the small polaron due to the competition between two types of antiferromagnetic exchange interactions, where one of them is that acts between pairs of electrons reside in the first nearest neighbor sites, and the other is that acts between pairs of electrons reside across the hole-occupied 
sites (Fig.~\ref{fig:spin-vortex}).

Spin vortices are characterized by the winding number 
\begin{eqnarray}
 w_C[\xi]={ 1 \over {2 \pi}} \oint_C \nabla \xi \cdot d{\bf r}
 \label{VwindingN}
\end{eqnarray}
where $\xi$ is the angle of spin rotation in the CuO$_2$ plane.
Due to the spinor property of spin functions, spin-twisting itinerant motion along a loop $C$ causes the sign-change of it if $w_C[\xi]$ is odd.
 Since the total wave function is required to be a single-valued function of the coordinates \cite{koizumi2022b}, an extra phase factor appears in the wave function to compensate the sign-change of the spin function. It is attributed to the $U(1)$ phase factor neglected by Dirac \cite{koizumi2023} (see also 
 Appendix~\ref{chi-mode}), and given by $e^{-{i \over 2}\chi}$ in the present work. In other words, $e^{-{i \over 2}\chi}$ arises from the constrain that 
 the total wave function needs to be the single-valued with respect to the electron coordinate
 in the presence of the sign-change caused by the spin-rotation. 
 The condition for the compensation is given by
\begin{eqnarray}
 w_C[\chi]+ w_C[\xi]=\mbox{ even number}
 \label{wcomb-eq}
\end{eqnarray}
This point will be explained in detail, shortly.
The condition in Eq.~(\ref{wcomb-eq}) indicates if $w_C[\xi]$ is odd, $w_C[\chi]$ is also odd, meaning nonzero; consequently, a loop current is generated. This is the {\em spin-vortex-induced loop current (SVILC)} \cite{Koizumi:2022aa,HKoizumi2013}.

\begin{figure}[H]
  \centering
    \includegraphics[width=8cm]{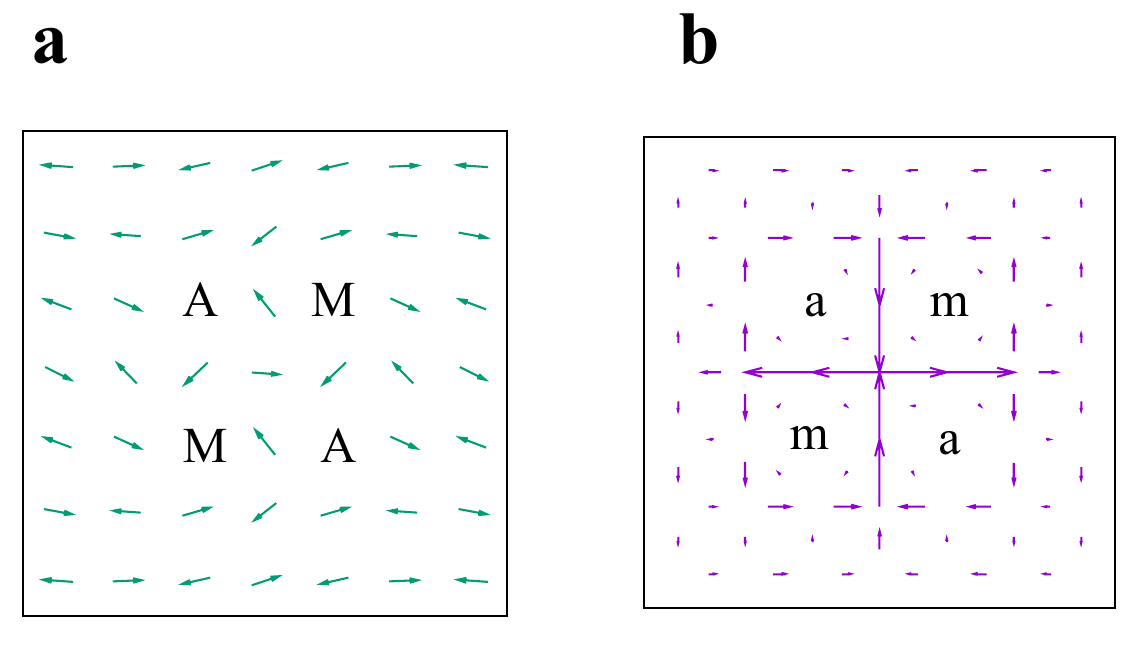}
    \caption{Spin texture and current pattern for  a `spin-vortex-quartet (SVQ)'. {\bf a})  A stable four spin-vortices form an SVQ. `M' and `A' indicate the winding number $+1$ and $-1$ spin-vortices, respectively. Here, the winding number is calculated with removing the background antiferromagnetic contribution.
    {\bf b}) A current generated by SVILCs for the SVQ in {\bf a}. `m' and `a' indicate the winding numbers $+1$ and $-1$ SVILCs, respectively. Other combinations of `m' and `a' are possible if they satisfy the condition in Eq.~(\ref{wcomb-eq}). They are depicted in 
    Fig.~\ref{fig:result:1NI_16_current}.}
    \label{fig:result:oneplane:spin}
  \end{figure}

Examples of spin-vortices and SVILCs are depicted in Fig.~\ref{fig:result:oneplane:spin}.
The spin-vortices are grouped in a `spin-vortex-quartet (SVQ)' in this figure. This is a stable set of spin-vortices with the sum of winding numbers zero. Note that a cluster of even number of spin-vortices with the sum of winding numbers zero is stable. 
In the present work, we assume that spin-vortices exist in the form of SVQs, and all
SVILCs arise from SVQs. It is worth mentioning that if the CuO$_2$ plane is covered by tiling these SVQs,
the hole concentration becomes $x=0.25$, which agrees with the doping level where the superconducting state disappears \cite{HKoizumi2015B,Koizumi2017}; this may indicate that
the superconducting state disappears beyond $x=0.25$ due to the collapse of the spin-vortices.

Now we come back to see how the condition in Eq.~(\ref{wcomb-eq}) arises.
The field operators for the present theory are modified version of those given in Eq.~(\ref{new-fieldO}) of Appendix A. They are given by
 \begin{eqnarray}
\hat{\Psi}_{\uparrow}({\bf r})&=&\sum_{n} e^{-{i \over 2}\hat{\chi} ({\bf r})}\left( \gamma_{{n} } u_{n \uparrow}({\bf r})  -\gamma^{\dagger}_{{n} } v^{\ast}_{n \uparrow}({\bf r}) \right)
\nonumber
\\
\hat{\Psi}_{\downarrow}({\bf r})&=&
\sum_{n} e^{-{i \over 2}\hat{\chi} ({\bf r})} \left( \gamma_{{n} } u_{n \downarrow}({\bf r}) +\gamma^{\dagger}_{{n} } v^{\ast}_{n \downarrow}({\bf r}) \right)
\label{eq1}
\end{eqnarray}
where $u_{n \sigma}({\bf r})$ and $v_{n \sigma}({\bf r})$ are the single-particle basis functions with spin $\sigma$, and $\gamma_{n }$ and $\gamma^\dagger_{n}$ are the particle-number conserving Bogoliubov operators (PNC-BOs) that act in a similar manner as the particle-number nonconserving  Bogoliubov operators as follows: The superconducting ground state, $|{\rm Gnd}(N) \rangle$, is defined by
\begin{eqnarray}
\gamma_{n }|{\rm Gnd}(N) \rangle=0
\label{eq2}
\end{eqnarray}
where $N$ denotes the  total particle-number of the system. The Bogoliubov operators are constructed to diagonalize the effective Hamiltonian for electrons (given later as $H_{\rm eff}$ as in Eq.~(\ref{Heff})) in the following form 
    \begin{eqnarray}
H_{\rm eff}=\sum_{n}'  E_n  \gamma^{\dagger}_{n} \gamma_{n} + E'_{\rm const}
\label{Heff-gamm}
  \end{eqnarray}
where $E'_{\rm const}$ is a constant \cite{Zhu2016}. Only the Bogoliubov energies with $E_n >0$ should be included so that the Bogoliubov operators express excitations from the ground state.

Since the field operator $\hat{\Psi}_{\sigma}({\bf r})$ annihilates an electron with spin $\sigma$ at the spatial position ${\bf r}$, relations between the electron annihilation operators at site $k$ (its coordinate is ${\bf r}_k$), and the PNC-BOs are given by
 \begin{eqnarray}
 c_{ k \uparrow} &=&\sum'_{n} [ u^{n}_{k \uparrow}\gamma_{n}- (v^{n}_{k \uparrow})^{\ast}\gamma_{n}^{\dagger}] e^{ -{i \over 2} \hat{\chi}_k}
 \nonumber
 \\
 c_{ k \downarrow} &=&\sum'_{n} [ u^{n}_{k \downarrow}\gamma_{n}+ (v^{n}_{k \downarrow})^{\ast}\gamma_{n}^{\dagger}] e^{ -{i \over 2} \hat{\chi}_k}
 \label{eq137}
  \end{eqnarray}
where  $c_{k \sigma}$  denotes
  the annihilation operators for electrons with spin $\sigma$ at site $k$; $u^{n}_{k \sigma}$ and  $v^{n}_{k \sigma}$ correspond to $u^{n}_{\sigma}({\bf r}_k)$ and  $v^{n}_{\sigma}({\bf r}_k)$, respectively.  
  
The relation in Eq.~(\ref{Canonical}) of Appendix A for the lattice model is given by
     \begin{eqnarray}
\left[\hat{\chi}_j, \hat{\rho}_k \right]=2i \delta_{jk}
\label{eq29b}
 \end{eqnarray}
 where $j$ and $k$ are site indices. 
From this relation, the following is obtained
 \begin{eqnarray}
 [e^{\pm  {i \over 2}\hat{\chi}_k}, \hat{\rho}_k]=\pm e^{\pm {i \over 2}\hat{\chi}_k}
 \end{eqnarray}
 indicating  $e^{\pm{i \over 2}\hat{\chi}_k}$ are number changing operators for $\rho_k$
 with $e^{{i \over 2}\hat{\chi}_k}$ increasing the number by one, and $e^{-{i \over 2}\hat{\chi}_k}$ decreasing the number by one.
  In the following, we present Hamiltonians using operators, $c_{ k \sigma}$ and $c_{ k \sigma}^\dagger$; however, in numerical computations, expressions using $\gamma_n$, $\gamma_n^\dagger$,
 $e^{ -{i \over 2} \hat{\chi}_k}$, and $ e^{ {i \over 2} \hat{\chi}_k}$ are used. 
 
The Hamiltonian for the bulk CuO$_2$ plane is given by
\begin{eqnarray}
     H_{\rm{EHFS}} 
       &=&
     -t_1\sum_{\langle i, j\rangle_{1},\sigma}[c^\dagger_{i \sigma}c_{j \sigma}+{\rm H.c.}]
     + U \sum_{j} c^\dagger_{j \uparrow} c_{j \uparrow}c^\dagger_{j \downarrow} c_{j \downarrow}+
 J_h \sum_{\left<i, j \right>_h} {\bf S}_{i} \cdot {\bf S}_{j} 
 \nonumber
 \\
 &&+\lambda\sum_{h} \Big[
    c^{\dagger}_{h+y \downarrow} c_{h-x \uparrow}
    -c^{\dagger}_{h+y \uparrow}c_{h-x \downarrow}
    + c^{\dagger}_{h+x \downarrow} c_{h-y \uparrow}
    - c^{\dagger}_{h+x \uparrow} c_{h-y \downarrow} 
    \nonumber
    \\
    &&+ {\rm i}(c^{\dagger}_{h-x\downarrow}c_{h-y\uparrow}
    + c^{\dagger}_{h-x\uparrow}c_{h-y\downarrow})
    + {\rm i}(c^{\dagger}_{h+y\downarrow}c_{h+x\uparrow}
    + c^{\dagger}_{h+y\uparrow}c_{h+x\downarrow})  
    + \rm{h.c.} \Big]
    \nonumber
    \\
    &&-\mu_{\rm bulk} \sum_{j, \sigma} c_{j \sigma}^{\dagger}c_{j \sigma}
    \label{EHFS}
  \end{eqnarray}
Here, `EHFS` stands for `effectively half-filled situation'; this means the situation where the number of sites accessible for the conduction electrons is the same as that of the conduction electrons.  This situation is assumed to be realized in the bulk CuO$_2$ plane due to the formation of the small polaron hole; i.e., the doped holes form small polarons that do not move at low temperatures, thus, the hole occupied sites become inaccessible sites of itinerant electrons.
  The symbol $\langle i,j \rangle_{1}$ indicates the nearest neighbor site-pairs in the CuO$_2$ plane
  with removing the hole occupied sites as inaccessible sites. The parameter $U$ is the on-site Coulomb repulsion parameter, and $t_1$ is the nearest neighbor hopping parameter. The term with $J_h$ describes the superexchange interaction between electrons across 
hole occupied sites. The term with $\lambda$ describes the Rashba spin-orbit interaction around small polarons, where hole occupied sites  are denoted by $h$; $h \pm x$ and $h \pm y$ denote
sites adjacent to $h$ in the $\pm x$ and $\pm y$ directions, respectively.
The parameter $\mu_{\rm bulk}$ is the chemical potential of the bulk layer. The EHFS is achieved by
choosing $\mu_{\rm bulk}=4 t_1=U/2$.
${\bf S}_{j}=({S}^{x}_{j}, S^{y}_{j}, S^{z}_{j})$ are components of the spin operator given by
  \begin{eqnarray}
 {S}^{x}_{j}&=&\frac{1}{2}( c^{\dagger}_{j \uparrow} c_{j \downarrow}+c^{\dagger}_{j \downarrow} c_{j \uparrow})
 \nonumber
 \\
 {S}^{y}_{j}&=&\frac{i}{2}(-c^{\dagger}_{j \uparrow} c_{j \downarrow}+c^{\dagger}_{j \downarrow} c_{j \uparrow})
 \nonumber
 \\
 {S}^{z}_{j}&=&\frac{1}{2}( c^{\dagger}_{j \uparrow} c_{j \uparrow}-c^{\dagger}_{j \downarrow} c_{j \downarrow})
\end{eqnarray}  

The Hamiltonian $H_{\rm{EHFS}}$ is constructed emprically based on the following experimental and theoretical findings:
1) the undoped cuprate is well-described by the Hubbard model with including terms with $t_1$ and $U$ \cite{AndersonBook};
2) experiments indicate the small polaron formation by hole doping \cite{Bianconi}; this fact is taken into account by the removal of hole occupied sites from the electron accessible sites; 3) inelastic neutron scattering reveals 
the magnetic moment lying in the CuO$_2$ plane, and the hourglass-shaped
magnetic excitation spectrum \cite{Neutron}, which can be reproduced by the spin dynamics of spin-vortices in the antiferromagnetic background \cite{Hidekata2011}; the term with $J_h$ which arises from the superexchange interaction across the hole occupied sites (see Fig.~\ref{fig:spin-vortex}) realizes the spin-vortex formation \cite{HKoizumi2013}; 4) nano-sized loop currents exist in the cuprate \cite{Kivelson95,Kerr1}; the stabilization of this loop current is brought about by the term with $\lambda$ which is expected to
arise from the Rashba spin-orbit coupling produced by the electric field around the small polaron hole \cite{HKoizumi2015B,Koizumi2017}. This term favors the spin moment lying in the CuO$_2$ plane.

The Hamiltonian $H_{\rm EHFS}$ is, however, too difficult to solve as it is; thus, we use the mean-field version of it given by   
 \begin{eqnarray}
     H^{\rm{HF}}_{\rm{EHFS}} 
       &=&
     -t_1\sum_{\langle i, j \rangle_{1},\sigma}[c^\dagger_{i \sigma}c_{j \sigma}+{\rm H.c.}]
   \nonumber
   \\
     &&+ U \sum_{j} \Big[(-\frac{2}{3}\langle {S^z_{j}} \rangle + \frac{1}{2})c^\dagger_{j \uparrow} c_{j \uparrow}
     +(\frac{2}{3}\langle{S^z_{j}} \rangle+ \frac{1}{2}) c^\dagger_{{j}\downarrow} c_{j \downarrow} 
     \nonumber
     \\
     &&-\frac{2}{3}(\langle{S^x_{j}}\rangle - {\rm i} \langle{S^y_{j}}\rangle) c^\dagger_{j \uparrow} c_{j \downarrow}
       -\frac{2}{3}(\langle{S^x_{j}} \rangle+ {\rm  i} \langle{S^y_{j}}\rangle) c^\dagger_{j \downarrow} c_{j \uparrow}
     - \frac{2}{3} \langle{\bf S}_{j} \rangle^2 \Big]
     \nonumber
     \\
    &&+ J_h \sum_{\left<i,j\right>_h}
    \Big[\frac{1}{2}(\langle{S^z_{i}} \rangle c^\dagger_{j\uparrow} c_{j\uparrow} + \langle{S^z_{j}}\rangle  c^\dagger_{i \uparrow} c_{i \uparrow})
    - \frac{1}{2}(\langle{S^z_{i} }\rangle c^\dagger_{j \downarrow} c_{j \downarrow} + \langle{S^z_{j}} \rangle c^\dagger_{i \downarrow} c_{i \downarrow})
    \nonumber
     \\
    && + \frac{1}{2}\{(\langle{S^x_{i}}\rangle - {\rm i} \langle{S^y_{i}} \rangle ) c^\dagger_{j \uparrow} c_{j \downarrow} + (\langle{S^x_{j} } \rangle- {\rm i} \langle{S^y_{j}}\rangle) c^\dagger_{i\uparrow} c_{i\downarrow}\}   
    \nonumber
    \\
    &&+\frac{1}{2}\{(\langle{S^x_{i}}\rangle + {\rm i} \langle{S^y_{i}}\rangle) c^\dagger_{j \downarrow} c_{j \uparrow} + (\langle{S^x_{j}}\rangle + {\rm i} \langle{S^y_{j}}\rangle) c^\dagger_{i \downarrow} c_{i \uparrow}\}
    - \langle{\bf S}_{i}\rangle \cdot \langle {\bf S}_{j} \rangle \Big] 
    \nonumber
    \\
    &&+ \lambda\sum_{h} \Big[
    c^{\dagger}_{h+y \downarrow} c_{h-x \uparrow}
    -c^{\dagger}_{h+y \uparrow}c_{h-x \downarrow}
    + c^{\dagger}_{h+x \downarrow} c_{h-y \uparrow}
    - c^{\dagger}_{h+x \uparrow} c_{h-y \downarrow} 
    \nonumber
    \\
    &&+ {\rm i}(c^{\dagger}_{h-x\downarrow}c_{h-y\uparrow}
    + c^{\dagger}_{h-x\uparrow}c_{h-y\downarrow})
    + {\rm i}(c^{\dagger}_{h+y\downarrow}c_{h+x\uparrow}
    + c^{\dagger}_{h+y\uparrow}c_{h+x\downarrow})  
    + {\rm H.c.} \Big]
    \nonumber
    \\
    &&-\mu_{\rm bulk} \sum_{j, \sigma} c_{j \sigma}^{\dagger}c_{j \sigma}
       \label{HF_EHFS}
  \end{eqnarray}
where $\langle \hat{O} \rangle$ represents the expectation value of the operator $\hat{O}$.

Let us examine the appearance of spin-vortices and loop currents that leads to the appearance of the $\chi$ mode and the condition in Eq.~(\ref{wcomb-eq}). Spin-vortices are created by the conduction electrons with spin moments lying in the $xy$ plane (it is CuO$_2$ plane) with small polarons at their centers. The small polaron occupied sites are inaccessible sites of the conduction electrons at low temperatures since small polarons do not move at low enough temperatures. The expectation values of the components of the spin lying in the CuO$_2$ plane are given by
\begin{eqnarray}
 \langle {S}^x_j\rangle +i   \langle { S}^y_{j} \rangle=S_j e^{i \xi_j}, \  \langle S^z_j \rangle=0
 \label{eqSpin}
\end{eqnarray}
where $S_j$ is a real number, and  $\xi_j$ is the polar angle of the spin at the $j$th site; and the winding number of $\xi$ around each inaccessible site is $+1$ or $-1$.

The self-consistent solutions for $u^{n}_{j \sigma}$ and $v^{n}_{j \sigma}$ are obtained by solving a system of equations (Eq.~(\ref{New-BdG-Eq}) in the next section). They are not single-valued when spin-vortices with odd winding numbers exist.
Let us see this point below: The term like $\left(\langle { S}^x_{j} \rangle -i   \langle { S}^y_{j} \rangle \right) c^{\dagger}_{j \uparrow} c_{j \downarrow}$ in Eq.~(\ref{HF_EHFS}) shows $\xi$ dependence as
\begin{eqnarray}
e^{-i\xi_{j}}c^{\dagger}_{j\uparrow} c_{j \downarrow}
\label{eq-xi-1}
\end{eqnarray}
and $\left(\langle { S}^x_{j} \rangle + i   \langle { S}^y_{j} \rangle \right) c^{\dagger}_{j \downarrow} c_{j \uparrow}$
as 
\begin{eqnarray}
e^{i\xi_{j}}c^{\dagger}_{j \downarrow} c_{j \uparrow}
\label{eq-xi-2}
\end{eqnarray}
Since the expectation values should depend on the difference of the phases $\xi$ at nearby sites, we have the following relations,
\begin{eqnarray}
\langle c^{\dagger}_{j \uparrow} c_{j \downarrow} \rangle \sim e^{i\xi_{j}},
\quad 
\langle c^{\dagger}_{j \downarrow} c_{j \uparrow} \rangle \sim e^{-i\xi_{j}},
\end{eqnarray}
to cancel the on-site $\xi_j$ dependency in Eqs.~(\ref{eq-xi-1}) and (\ref{eq-xi-2}).

Taking into account  the relations in Eq.~(\ref{eq137}), $\xi_j$ dependencies
in $u_{j \sigma}^n$ and $v_{j \sigma}^n$ are deduced as
\begin{eqnarray}
u_{j \uparrow}^n, \ v_{j \downarrow}^n \sim  e^{-{i \over 2}\xi_j}, \quad u_{j \downarrow}^n, \ v_{j \uparrow}^n \sim  e^{{i \over 2}\xi_j}, 
\end{eqnarray}
When an excursion of the value of $\xi$ is performed starting from $\xi_j$ around loop $C$ in the coordinate space, $\xi_j$ may become $\xi_j + 2\pi w_C[\xi]$. If $w_C[\xi]$ is odd,
 $e^{\pm{i \over 2}\xi_j}$ becomes  $-e^{\pm{i \over 2}\xi_j}$ after the excursion,
 resulting in the sign-change.
 
The single-valuedness is achieved by the phase factor $e^{-{i \over 2} \chi}$ in Eq.~(\ref{eq137}).
It gives the following dependences
\begin{eqnarray}
u_{j \sigma}^n \sim  e^{{i \over 2}\chi_j}, 
\quad v_{j \sigma }^n \sim  e^{-{i \over 2}\chi_j}
\label{eq-u-chi}
\end{eqnarray}
to cancel $\chi$ dependences of $c_{k \sigma}$.
Thus, overall dependences are
\begin{eqnarray}
u_{j \uparrow}^n \sim  e^{{i \over 2}\chi_j} e^{-{i \over 2}\xi_j}, 
\quad u_{j \downarrow}^n \sim  e^{{i \over 2}\chi_j} e^{{i \over 2}\xi_j}, 
\quad 
v_{j \uparrow}^n \sim  e^{-{i \over 2}\chi_j} e^{{i \over 2}\xi_j}, 
\quad v_{j \downarrow}^n \sim  e^{-{i \over 2}\chi_j} e^{-{i \over 2}\xi_j} 
\nonumber
\\
\label{eq-u-chi2}
\end{eqnarray}
These dependencies indicate that the sign-change is avoided if $\chi_i$ is so chosen to satisfy 
Eq.~(\ref{wcomb-eq}).
Note that although we use the condition $\langle S^z_j \rangle=0$ in Eq.~(\ref{eqSpin}) for the calculation, this condition can be relaxed as far as the twisting rotation only occurs in the $xy$ plane; i.e., $\langle S^z_j \rangle \neq 0$ is allowed if it does not affect the spin-rotation projected on the $xy$ plane.

\section{One surface-layer plus one bulk-layer for a single bilayer model: Spin vortex formation}
  \label{sec3}

Let us explain the model for Bi-2212 thin films used in the present work. In this model, we only retain the sites for copper atoms in CuO$_2$ planes as in our previous work \cite{Koizumi2022aa}. Since the Bi-2212 consists of CuO$_2$ bilayers, we need to modify our previous model to incorporate this fact.

Our model Hamiltonian $H_{\rm eff}$ is given by
  \begin{eqnarray}
H_{\rm eff}=H^{\rm HF}_{\rm b}+H^{\rm HF}_{\rm s}+H^{\rm HF}_{\rm s-b}
\end{eqnarray}
where $H^{\rm HF}_{\rm b}$, $H^{\rm HF}_{\rm s}$, and $H^{\rm HF}_{\rm s-b}$ are
the bulk-layer Hamiltonian, surface-layer Hamiltonian, and surface-bulk interlayer Hamiltonian, respectively.
For  $H^{\rm HF}_{\rm b}$, we use $H^{\rm{HF}}_{\rm{EHFS}}$ in Eq.~(\ref{HF_EHFS}).

The surface CuO$_2$ plane is treated differently from that in the bulk. 
This difference comes from the absence of the small polarons in the surface layer.   Note that the small polaron formation is suppressed in the surface layer due to the absence of the charge reservoir layer that covers the CuO$_2$ layers. Then, we use the following so-called $t$-$J$ model
   \begin{eqnarray}
 H_{t-J}
  &=&-t_1\sum_{\langle i,j \rangle_{1},\sigma}[(1-n_{i, -\sigma})c^\dagger_{i\sigma}c_{j\sigma}(1-n_{j, -\sigma})+{\rm H.c.}]
   \nonumber
   \\
    &&+{{4t_1^2} \over U} \sum_{\langle i,j \rangle_{1}}\left( {\bf S}_i \cdot {\bf S}_j - {1 \over 4} n_i n_j \right) 
    \nonumber
     \\
\label{t-J1}
     \end{eqnarray}
where $n_j$ is the number of electrons at the $j$th site, and  $n_{j, \sigma}$ is the number of electrons with spin $\sigma$ at the $j$th site.
      
We rewrite it in the following form
\begin{eqnarray}
 H_{s}
   &=&-t_1\sum_{\langle i,j \rangle_{1},\sigma}[(1-n_{i, -\sigma})c^\dagger_{i\sigma}c_{j\sigma}(1-n_{j, -\sigma})+{\rm H.c.}]   \nonumber
   \\
   &&-t_2\sum_{\langle i,j \rangle_{2},\sigma}[(1-n_{i, -\sigma})c^\dagger_{i\sigma}c_{j\sigma}(1-n_{j, -\sigma})+{\rm H.c.}]   \nonumber
   \\
        &&-{{2t_1^2} \over U} \sum_{\langle i,j \rangle_{1}}(c_{i \uparrow}^{\dagger}c_{j \downarrow}^{\dagger}
     -c_{i \downarrow}^{\dagger}c_{j \uparrow}^{\dagger})(c_{j \downarrow}c_{i \uparrow}
     -c_{j \uparrow}c_{i \downarrow}) -\mu_{\rm surf} \sum_{j, \sigma} c_{j \sigma}^{\dag}c_{j \sigma}
\label{t-J2}
 \end{eqnarray}
where the second nearest hopping term with hopping parameter $t_2$ and the chemical potential term with $\mu_{\rm surf}$ are added.
    
For the mean field approximation, we first perform the following replacement
     \begin{eqnarray}
    && (c_{i \uparrow}^{\dagger}c_{j \downarrow}^{\dagger}
     -c_{i \downarrow}^{\dagger}c_{j \uparrow}^{\dagger})(c_{j \downarrow}c_{i \uparrow}
     -c_{j \uparrow}c_{i \downarrow})
     \nonumber
     \\
 \rightarrow &&
     (c_{i \uparrow}^{\dagger}c_{j \downarrow}^{\dagger}
     -c_{i \downarrow}^{\dagger}c_{j \uparrow}^{\dagger}) e^{ -{i \over 2} \hat{\chi}_i}e^{ -{i \over 2} \hat{\chi}_j} e^{ {i \over 2} \hat{\chi}_j}e^{ {i \over 2} \hat{\chi}_i}(c_{j \downarrow}c_{i \uparrow}
     -c_{j \uparrow}c_{i \downarrow})
     \end{eqnarray} 
This is due to the fact that the use of the PNC-BOs requires  $c_{j \sigma}$ needs to be multiplied by $e^{ {i \over 2} \hat{\chi}_j}$ as is inferred from Eq.~(\ref{eq137}).
     
Thus, the surface layer Hamiltonian becomes
\begin{eqnarray}
      H^{\rm HF}_{\rm s}
      &=&-t_1\sum_{\langle i,j \rangle_{1},\sigma}[(1-n_{i, -\sigma})c^\dagger_{i\sigma}c_{j\sigma}(1-n_{j, -\sigma})+{\rm H.c.}]   \nonumber
   \\
   &-&t_2\sum_{\langle i,j \rangle_{2},\sigma}[(1-n_{i, -\sigma})c^\dagger_{i\sigma}c_{j\sigma}(1-n_{j, -\sigma})+{\rm H.c.}]   
   \nonumber
   \\
      &+&\sum_{\langle i, j \rangle_1 }\left(\Delta_{i j} e^{ -{i \over 2} \hat{\chi}_j} e^{ -{i \over 2} \hat{\chi}_i} c_{i \uparrow}^{\dag}c_{j \downarrow}^{\dag} + {\rm h.c.} \right)
      -\mu_{\rm surf}  \sum_{j, \sigma} c_{j \sigma}^{\dag}c_{j \sigma}
      \label{H-surf}
  \end{eqnarray}
where the pair potential is given by
    \begin{eqnarray}
  \Delta_{ij}=-{{2t_1^2} \over U} \langle e^{ {i \over 2} \hat{\chi}_j}e^{ {i \over 2} \hat{\chi}_i}(c_{j \downarrow}c_{i \uparrow}
     -c_{j \uparrow}c_{i \downarrow} )\rangle= \Delta_{ji}
     \label{Delta}
  \end{eqnarray}
This pair potential shows the so-called `d-wave' type behavior, and explains the observed
STS and ARPES results as demonstrated in our previous work \cite{Koizumi2022aa}. The value of $ \mu_{\rm surf}$ is so chosen that the average electron densities in the baulk and surface layers are the same.
     
Each CuO$_2$ plane extends in the $xy$ plane. The two layers in a bilayer are connected by the hopping in the $z$-direction. The Hamiltonian for this interlayer connection is
  \begin{eqnarray}
   H^{\rm HF}_{\rm s-b}
   &=&
  -t_{sb}\sum_{\langle k_{s},j_{b} \rangle_{z}, \sigma}\left[\left(1-n_{k_{s}, -\sigma} \right)c^\dagger_{k_{s}\sigma}c_{j_{b}\sigma}\left(1-n_{j_{b}, -\sigma}\right)+{\rm H.c.} \right]
         \nonumber
       \\
  &&+{{2t_{sb}^2} \over U} \sum_{\langle k_{s},j_{b} \rangle_{z}}\Big[ \left( \langle { S}^x_{k_{s}} \rangle -i   \langle { S}^y_{k_{s}} \rangle \right) c^{\dagger}_{j_{b} \uparrow} c_{j_{b} \downarrow}+ \left(\langle { S}^x_{k_{s}} \rangle + i   \langle { S}^y_{k_{s}} \rangle \right) c^{\dagger}_{j_{b} \downarrow} c_{j_{b} \uparrow}
  \nonumber
  \\
  &&+
 \left( \langle { S}^z_{k_{s}} \rangle-{1 \over 2}  n_{k_{s}}  \right)c^{\dagger}_{j_{b} \uparrow} c_{j_{b} \uparrow}- \left( \langle { S}^z_{k_{s}} \rangle+{1 \over 2} n_{k_{s}} \right)c^{\dagger}_{j_{b} \downarrow} c_{j_{b} \downarrow} \Big]
  \nonumber
  \\
    &&+{{2t_{s-b}^2} \over U} \sum_{\langle k_{s},j_{b} \rangle_{z}}\Big[ \left( \langle { S}^x_{j_{b}} \rangle -i   \langle { S}^y_{j_{b}} \rangle \right) c^{\dagger}_{k_{s} \uparrow} c_{k_{s} \downarrow}+ \left(\langle { S}^x_{j_{b}} \rangle + i   \langle { S}^y_{j_{b}} \rangle \right) c^{\dagger}_{k_{s} \downarrow} c_{k_{s} \uparrow}
  \nonumber
  \\
  &&+
 \left( \langle { S}^z_{j_{b}} \rangle-{1 \over 2}  n_{j_{b}} \right)c^{\dagger}_{k_{s} \uparrow} c_{k_{s} \uparrow}- \left( \langle { S}^z_{j_{b}} \rangle+{1 \over 2} n_{j_{b}} \right) 
 c^{\dagger}_{k_{s} \downarrow} c_{k_{s} \downarrow} \Big]
  \nonumber
  \\
  &&-{{4t_{s-b}^2} \over U} \sum_{\langle k_{s},j_{b} \rangle_{z}}\left( \langle {\bf S}_{k_{s}}\rangle \cdot \langle {\bf S}_{j_{b}}\rangle -{1 \over 4} n_{k_{s}} n_{j_{b}}  \right)
  \label{H-interlayer}
  \end{eqnarray}
where $k_{s}$ and $j_{b}$ denote sites in the surface layer and those in the bulk layer, respectively; $\langle k_{s},j_{b} \rangle_{z}$ indicates a pair of sites connected by the $z$-direction hopping.  

Overall, the model Hamiltonian is given in the following form
\begin{eqnarray}
H_{\rm eff}&=&
\sum_{k,j,\sigma, \sigma'} h_{k \sigma, j \sigma'} c_{k \sigma}^{\dagger} c_{j \sigma'}
  + \sum_{k,j} [ \Delta_{kj} e^{ -{i \over 2} \hat{\chi}_j}e^{ -{i \over 2} \hat{\chi}_k}c_{k \uparrow}^{\dagger}c_{j \downarrow}^{\dagger}+ \mbox{H.c.}]
  +E_{\rm const}
  \nonumber
  \\
  \label{Heff}
  \end{eqnarray}
where 
$k,j$ are site indices and $\sigma, \sigma'$ are spin indices; $E_{\rm const}$ is a constant arising from the mean field treatment; $c_{k \sigma}^\dagger$ and  $c_{k \sigma}$  denote
the creation and annihilation operators for electrons with spin $\sigma$ at site $k$, respectively. The elements $h_{k \sigma, j \sigma'}$ can be found in $H^{\rm HF}_b=H^{\rm{HF}}_{\rm{EHFS}}$ in Eq.~(\ref{HF_EHFS}), $H^{\rm HF}_b$ in Eq.~(\ref{H-surf}), and $H^{\rm HF}_{\rm s-b}$ in Eq.~(\ref{H-interlayer}); they are collections of contributions from all the terms that appear as parameters for the products of electron creation and annihilation operators $c_{k \sigma}^{\dagger} c_{j \sigma'}$.

We choose the PNC-BOs to satisfy  the requirement in Eq.~(\ref{eq2}), and also 
$H_{\rm eff}$ is expressed in Eq.~(\ref{Heff-gamm}).
Using Eqs.~(\ref{Heff-gamm}), (\ref{eq137}), and (\ref{Heff}), the following PNC-BdG equations are obtained
    \begin{eqnarray}
 E_{n} u^{n}_{k \uparrow}&=& \sum_{j \sigma'}  e^{ {i \over 2} \hat{\chi}_k} h_{k \uparrow, j \sigma'} e^{- {i \over 2} \hat{\chi}_j} u^{n}_{j \sigma'} + 
 \sum_j \Delta_{kj} v^{n}_{j \downarrow} 
  \nonumber
 \\
  E_{n} u^{n}_{k \downarrow}&=& \sum_{j \sigma'}  e^{ {i \over 2} \hat{\chi}_k} h_{k \downarrow, j \sigma'} e^{- {i \over 2} \hat{\chi}_j} u^{n}_{j \sigma'} + \sum_j \Delta_{jk} v^{n}_{j \uparrow} 
  \nonumber
 \\
  E_{n} v^{n}_{k \uparrow}&=& -\sum_{j } e^{ -{i \over 2} \hat{\chi}_k} h^{\ast}_{k \uparrow, j \uparrow} e^{ {i \over 2} \hat{\chi}_j} v^{n}_{j \uparrow} +\sum_{j}  e^{ -{i \over 2} \hat{\chi}_k} h^{\ast}_{k \uparrow, j \downarrow} e^{ {i \over 2} \hat{\chi}_j} v^{n}_{j \downarrow} + \sum_j \Delta_{kj}^{\ast} u^{n}_{j \downarrow} 
  \nonumber
 \\
  E_{n} v^{n}_{k \downarrow}&=& \sum_{j} e^{ -{i \over 2} \hat{\chi}_k} h^{\ast}_{k \downarrow, j \uparrow} e^{ {i \over 2} \hat{\chi}_j} v^{n}_{j \uparrow} -\sum_{j} e^{ -{i \over 2} \hat{\chi}_k} h^{\ast}_{k \downarrow, j \downarrow} e^{ {i \over 2} \hat{\chi}_j} v^{n}_{j \downarrow} +\sum_j \Delta_{jk}^{\ast} u^{n}_{j \uparrow} 
  \nonumber
  \\
  \label{New-BdG-Eq}
  \end{eqnarray}
  By self-consistently solving the above equations, $E_n, u^{n}_{j \sigma}$, and $v^{n}_{j \sigma}$ are obtained. Note that the self-consistent solution yields the  case with $\chi_k={\rm const.}$; it is the solution for a currentless state.
  
   The parameters for the model Hamiltonian are following: The first nearest neighbor hopping parameter in the CuO$_2$ plane, $t_1$, is taken to be $t_1=0.2$ eV; the second nearest neighbor hopping parameter in the CuO$_2$ plane, $t_2 =-0.12t_1$; the on-site Coulomb repulsion parameter, $U = 8t_1$; the Rashba spin-orbit interaction parameter, $\lambda= 0.02t_1$; and the hopping parameter between layers in a bilayer, $t_{s-b} = 0.25 t_1$. The bulk chemical potential, $\mu_{\rm bulk}$, and surface chemical potential, $\mu_{\rm surf}$,  are taken to be $\mu_{\rm bulk} = 4t_1$ and $\mu_{\rm surf} =-0.1t_1$, respectively.
 The unit of length is the CuO$_2$ plane lattice constant $a \approx 0.4$ nm. The bulk layer is located at $z=0.0$, and the surface layer is at $z=0.8$.
 
The calculation is done with the open boundary condition. The spin-vortices are created by itinerant electrons with spin moments lying in the $xy$ plane, and given in the form of 
Eq.~(\ref{eqSpin}). 
Each spin-vortex is characterized by the winding number given in Eq.~(\ref{VwindingN}), which 
is calculated for a loop $C$ formed by surrounding lattice sites around each center of small polaron.
If the winding number is $+1$, it is called a `meron' and  denoted by ``M'', and  if it is $-1$,
it is called an `antimeron' and denoted by ``A'' in  Fig.~\ref{fig:result:oneplane:spin}. 

The winding numbers for the bulk spin-vortices are input parameters for the self-consistent calculation. The combination of four-vortex unit is found energetically favorable in our previous works, and called the `spin-vortex-quartet (SVQ)'. The input spin winding numbers are so arranged that SVQs are placed periodically.
On the other hand, the spin vortices in the surface layers are obtained without winding number inputs;
they are automatically formed through the interaction from the bulk layers.

    \begin{figure}[H]
    \centering
    \begin{subfigure}{0.49\columnwidth}
      \centering
      \includegraphics[width=6cm]{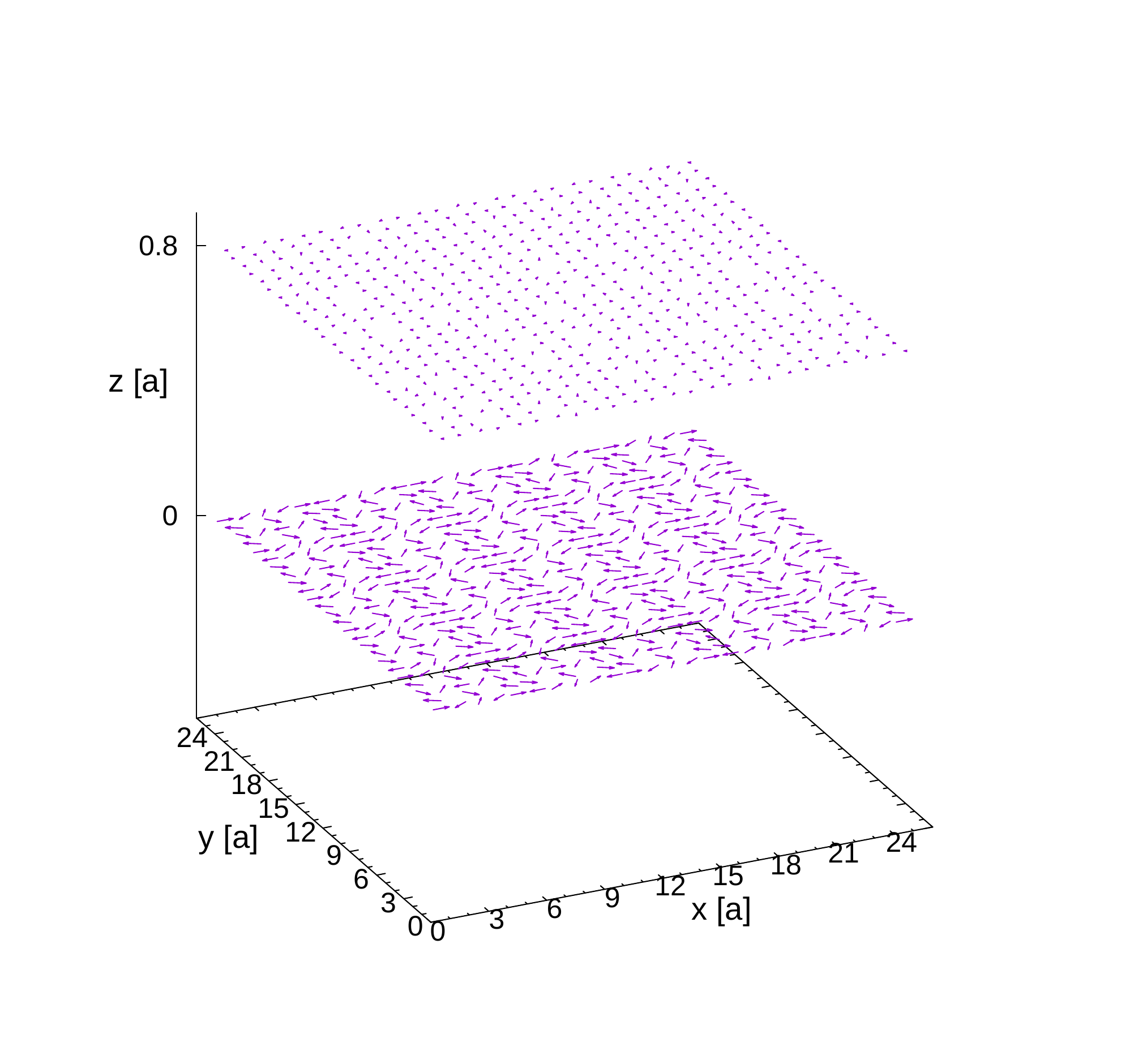}
      \caption{Spin moment distribution.}
      \label{fig:25_spin}
    \end{subfigure}
    \begin{subfigure}{0.49\columnwidth}
      \centering
      \includegraphics[width=6cm]{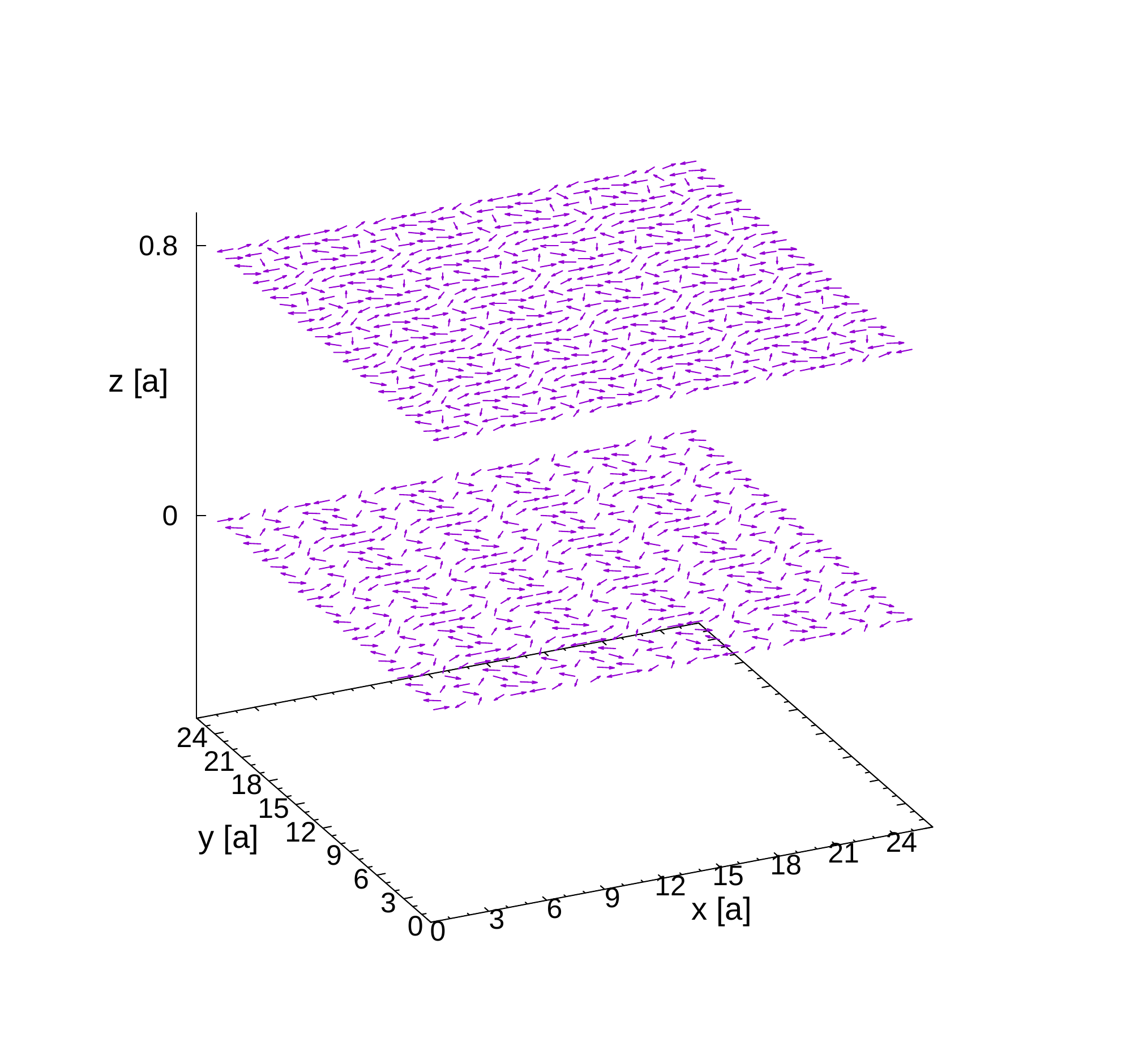}
      \caption{Spin moment distribution with normalized spins in each layer.}
      \label{fig:25_spin_n}
    \end{subfigure}
    \begin{subfigure}{0.49\columnwidth}
      \centering
      \includegraphics[width=6cm]{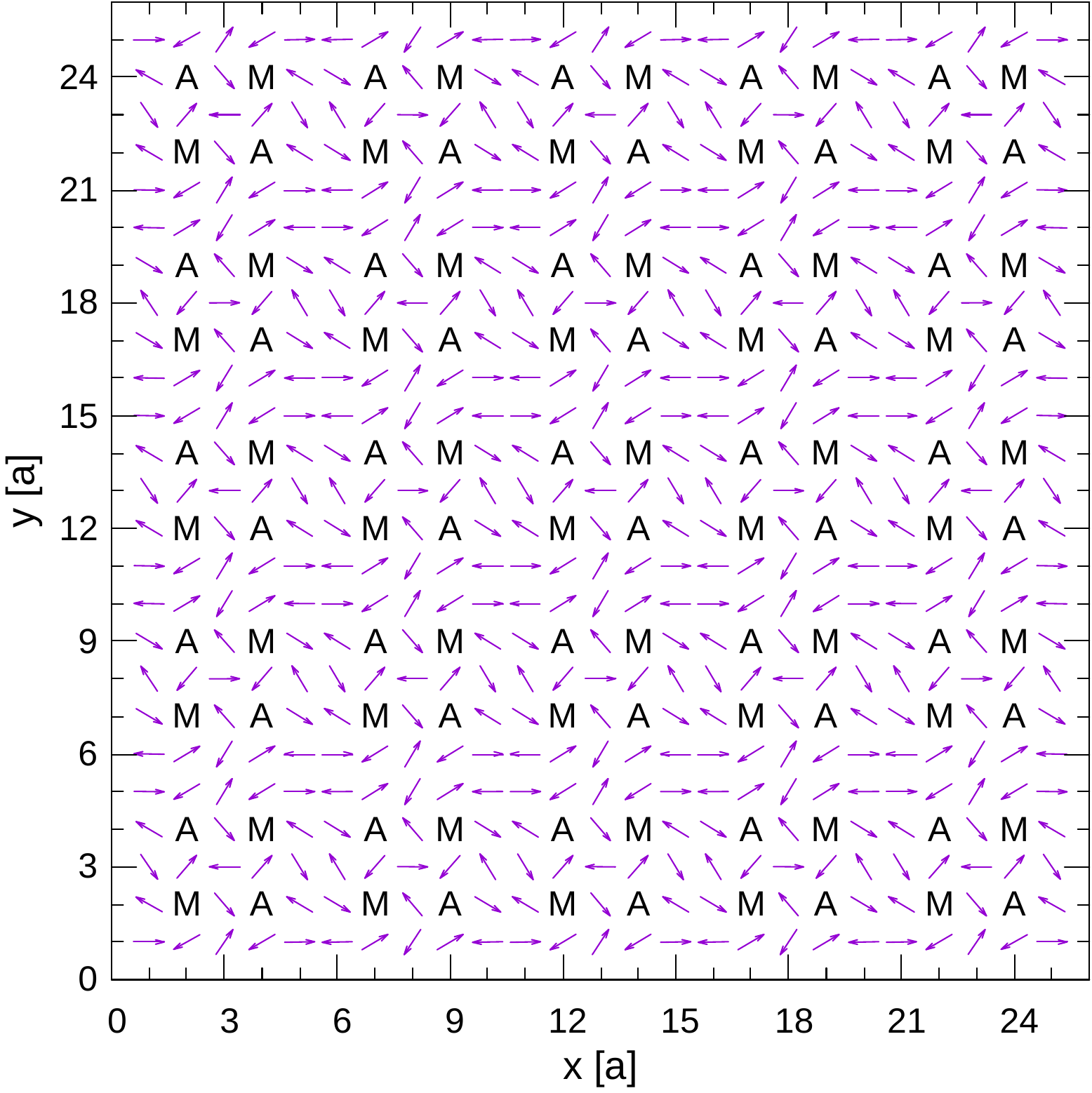}
      \caption{Bulk layer distribution}
      \label{fig:25_spin_bulk}
    \end{subfigure}
    \begin{subfigure}{0.49\columnwidth}
      \centering
      \includegraphics[width=6cm]{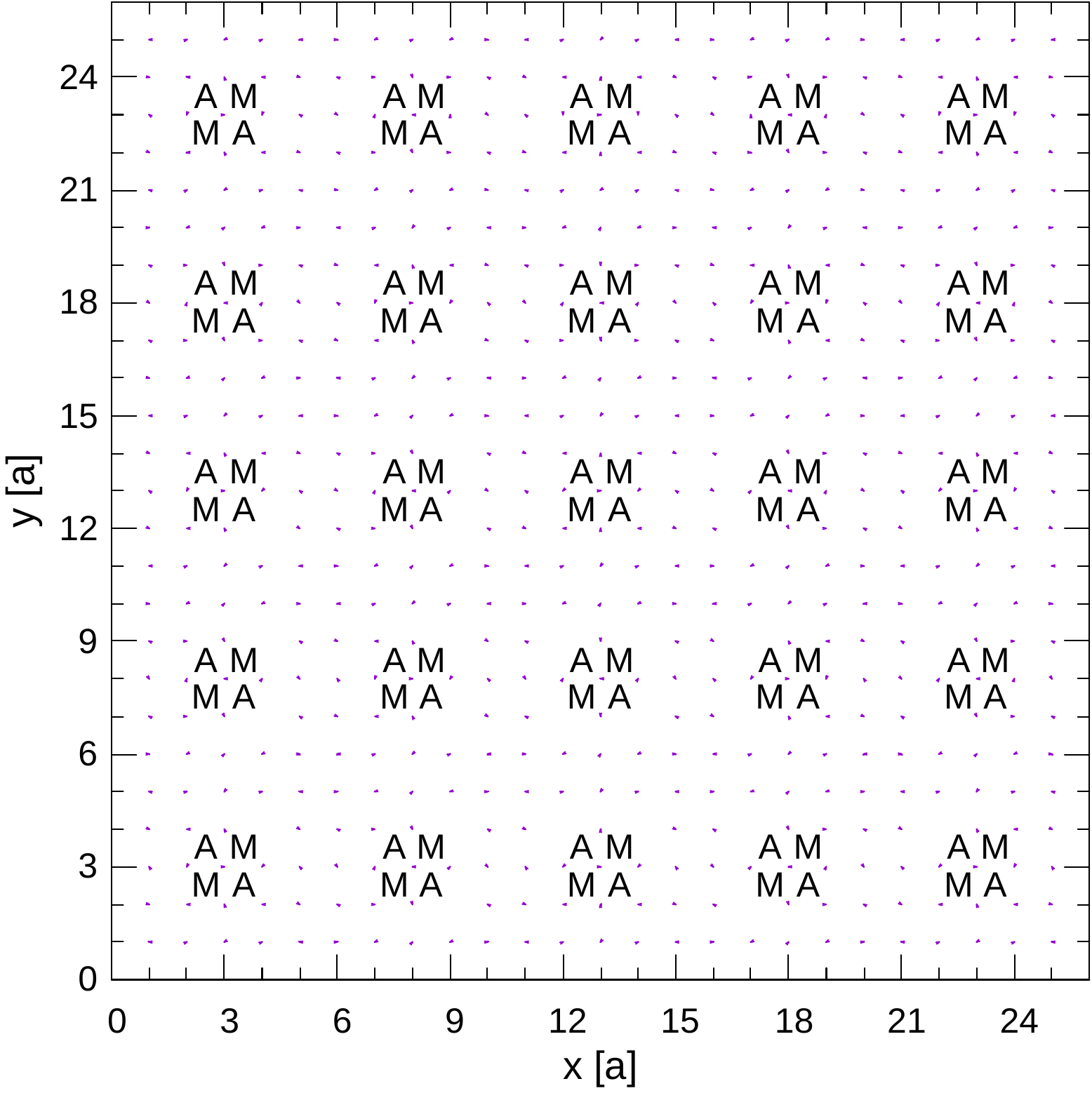}
      \caption{Surface layer distribution}
      \label{fig:25_spin_surface}
    \end{subfigure}
    \caption{Spin moment distribution for the $25 \times 25$ SVQ system.}
    \label{fig:25_spin_vortex}
  \end{figure}
  
In Fig.~\ref{fig:25_spin_vortex}, the result for the lattice with $25 \times 25$ CuO$_2$ planes is depicted.
Spin vortices are formed around small polarons in the bulk layers.
The spin-texture in the bulk layers are those arising from $4 \times 4$ SVQs with their winding numbers 
equal to the input winding numbers.
Each bulk layer is in the state of the `effectively-half-filled-situation', where the number of electrons and that for the sites allowed for electron hopping are the same.
The bulk chemical potential, $\mu_{\rm bulk}$, is chosen to realize this effectively-half-filled-situation.
The surface chemical potential, $\mu_{\rm surf}$, is so chosen that the number of electrons in the surface layer is almost equal to that of the bulk layer.
Spin-vortices are also generated in the surface layers although small polarons are absent.
 
 \section{One surface-layer plus one bulk-layer for a single bilayer model: SVILCs}
  \label{sec4}
  
The self-consistent solutions $u^{n}_{j \sigma}$, and $v^{n}_{j \sigma}$ may not be valid ones when spin-vortices exist since they may be multi-valued functions of the coordinate $j$.
This is the case when spin-vortices are present.
In this case we need to obtain $\chi$ that satisfies the condition in Eq.~(\ref{wcomb-eq}).
We obtain $\chi_j$ by requiring this condition for independent loops $\{ C_1, \cdots, C_{N_{\rm loop}} \}$ formed as the boundaries of plaques of the flattened lattice, where the flattened lattice is a 2D lattice constructed from the original 3D lattice by removing some faces and walls
(Consult Ref.~\cite{koizumi2022b} for the detail).
Here, 
$N_{\rm loop}$ is the total number of plaques of the flattened lattice. Actually, what we need to know is the differences of the phases, $(\chi_i - \chi_j)$'s, between bonds connecting sites. Therefore, the number of unknowns to be evaluated is equal to the number of the bonds, $N_{\rm bond}$.

The winding number requirement provides $N_{\rm loop}$ equations from Eq.~(\ref{wcomb-eq}) with
\begin{eqnarray}
    \label{eq:winding_number_xi}
    w_{C_l}[\xi] = 
    \frac{1}{2\pi} \sum_{i=1}^{N_l} (\xi_{C_l(i+1)} - \xi_{C_l(i)})
    \label{xi-eta}
  \end{eqnarray}
  and
  \begin{eqnarray}
    \label{eq:winding_number_chi}
    w_{C_l}[\chi] = 
    \frac{1}{2\pi} \sum_{i=1}^{N_l} (\chi_{C_l(i+1)} - \chi_{C_l(i)})
  \end{eqnarray}
  where $C_{l(i)}, \ i=1, \cdots N_l$ are sites in the loop $C_l$ and the $N_l$ is the number of sites of it. Actually, we use $\eta$ instead of $\xi$ for the calculation of Eq.~(\ref{xi-eta})
  for numerical convenience, where $\eta$ is the angle that removes the background antiferromagnetic contribution from $\xi$ (see Ref.~\cite{Koizumi2021c} for the detail).
  
The conservation of the local charge requirements give rise to the number of sites (we denote it as $N_{\rm site}$) minus one conditions, where minus one comes from the fact that the calculation is done by conserving the total number of electrons. These requirements are expressed using the current density.
The electric current density is obtained in the following way:
The Berry connection can be viewed as a fictitious magnetic field given by the vector potential
  \begin{eqnarray}
    \label{eq:berry_A_fic}
    \bm{A}^{\rm fic} = -\frac{\hbar}{2e} \nabla \chi
  \end{eqnarray}
as indicated in Eq.~(\ref{eqvelo}).
Thus, the current density ${\bf j}$ is calculated as
  \begin{eqnarray}
    \label{eq:current_dEdA}
    {\bf j} = -\frac{\delta E[{\bf A}^{\rm fic}]}{\delta {\bf A}^{\rm fic}}
    = \frac{2e}{\hbar} \frac{\delta E[\nabla \chi]}{\delta \nabla \chi}
  \end{eqnarray}
where
  $E[{\bf A}^{\rm fic}]$ is the total energy functional depends on ${\bf A}^{\rm fic}$.
We denote $\tau_{k \leftarrow j}$ as the difference of $\chi$ values at sites $k$ and $j$ connected by the bond
  \begin{eqnarray}
    \label{eq:tau_DifferenceOfChi}
    \tau_{k \leftarrow j} = \chi_k - \chi_j
  \end{eqnarray}
Then, the current for the lattice system is given by
  \begin{eqnarray}
    \label{eq:current_lattice}
    J_{k \leftarrow j} = \frac{2e}{\hbar} \frac{\partial E}{\partial \tau_{k \leftarrow j}}
  \end{eqnarray}
  where $J_{k \leftarrow j}$ is the electric current flows from $j$ to $k$.
Thus, the conservation of the charge at $j$ is given by
  \begin{eqnarray}
    \label{eq:current_conservation}
    J_j^{\rm EX} + \sum_i \frac{2e}{\hbar} \frac{\partial E}{\partial \tau_{j \leftarrow i}} = 0
  \end{eqnarray}
where
  $J_j^{\rm EX}$ is the external current fed into $j$. This gives  $N_{\rm site}-1$ equations.
  Then, the solvability condition for all $(\chi_i - \chi_j)$'s is
\begin{eqnarray}
N_{\rm bond}=N_{\rm loop}+N_{\rm site}-1
\label{Euler1}
\end{eqnarray}
This condition is satisfied since it agrees with the Euler's theorem for 2D lattice given by
\begin{eqnarray}
[\mbox{\# edges}]=[\mbox{\# faces}]+[\mbox{\# vertices}-1]
\label{Euler2}
\end{eqnarray}
where `$\mbox{\# A}$' means `the number of $A$'; here `2D' is a broad sense 2D, meaning
that if the bonds are flexible enough, it can be flattened into two-dimensional lattice.

   \begin{figure}[H]
    \centering
    \begin{subfigure}{0.31\columnwidth}
      \centering
      \includegraphics[width=3.5cm]{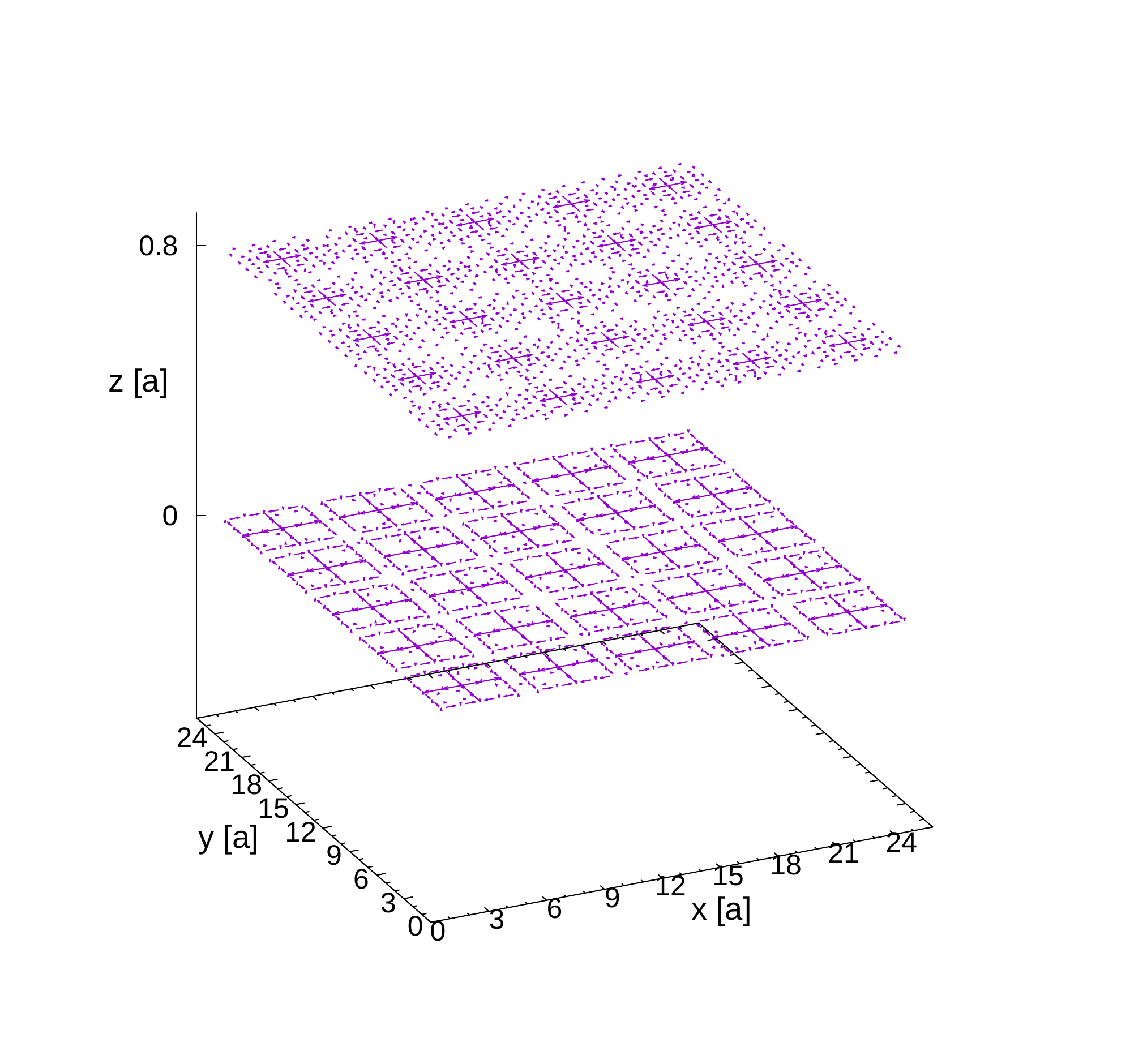}
      \caption{Current distribution.}
      \label{fig:25_current_maam_all}
    \end{subfigure}
    \hspace*{1mm}
    \begin{subfigure}{0.31\columnwidth}
      \centering
      \includegraphics[width=3cm]{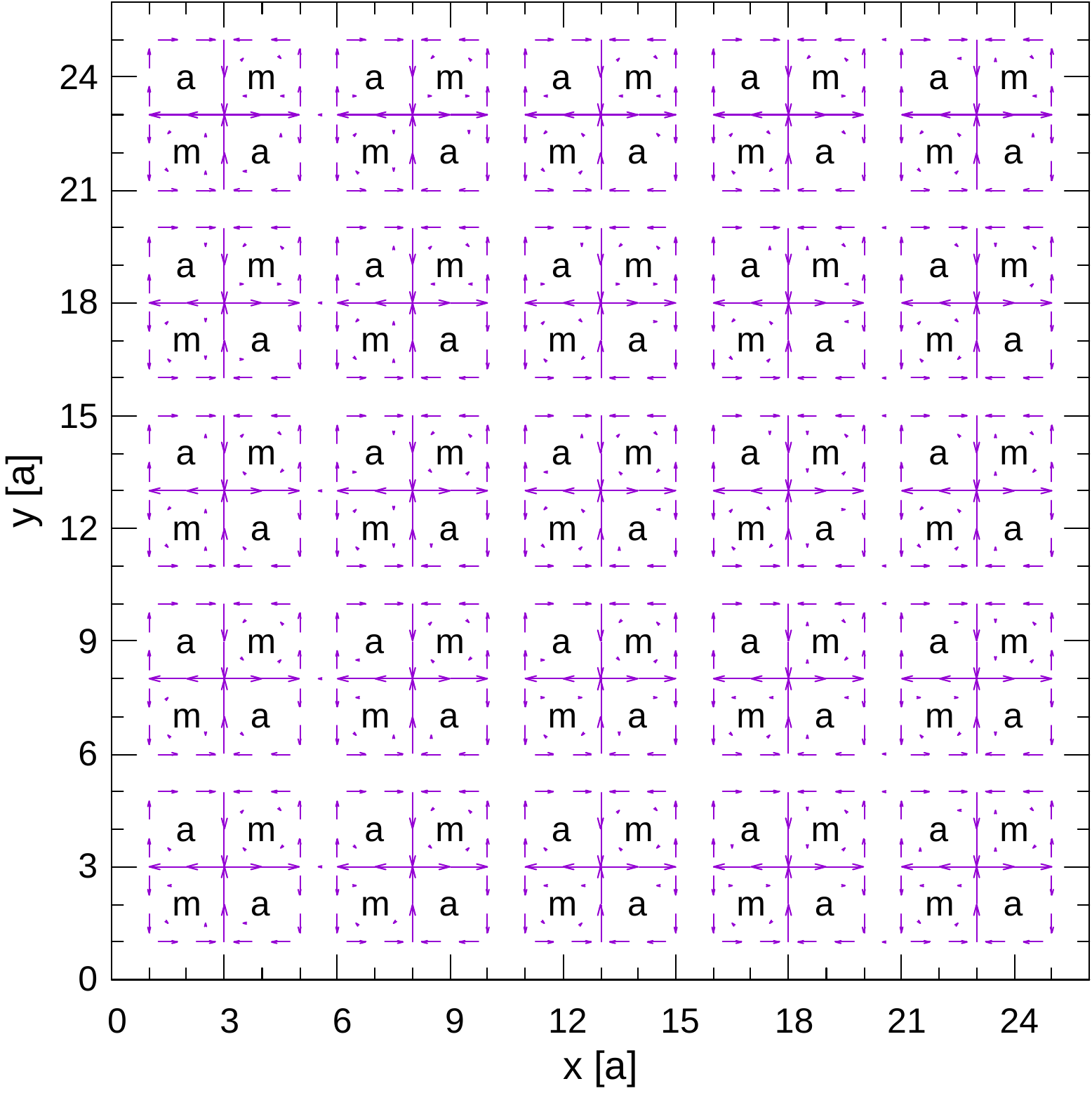}
      \caption{The bulk layer current.}
      \label{fig:25_current_maam_bulk}
    \end{subfigure}
    \hspace*{1mm}
    \begin{subfigure}{0.31\columnwidth}
      \centering
      \includegraphics[width=3cm]{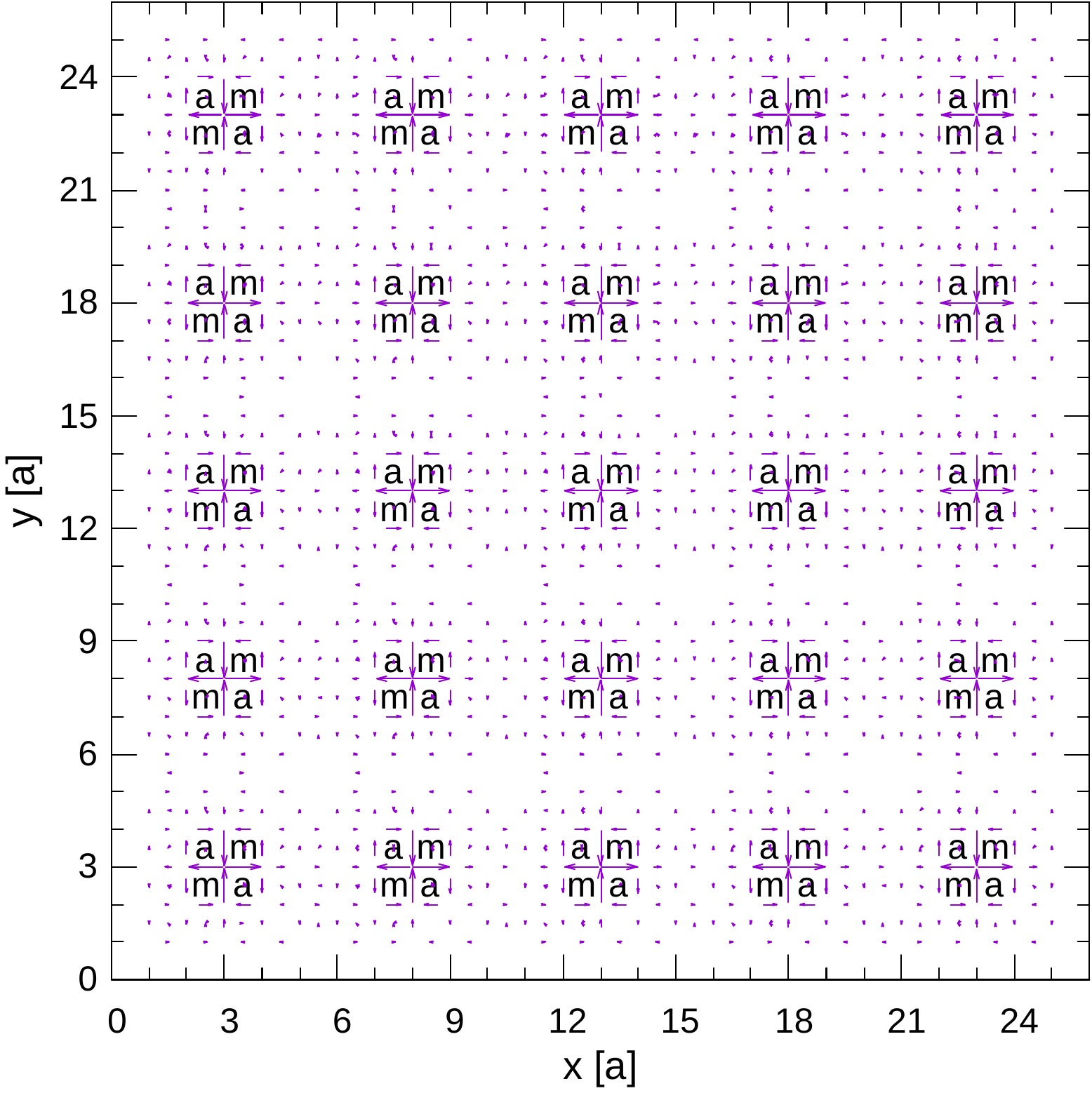}
      \caption{The surface layer current.}
      \label{fig:25_current_maam_surface}
    \end{subfigure}
    \begin{subfigure}{0.31\columnwidth}
      \centering
      \includegraphics[width=3.5cm]{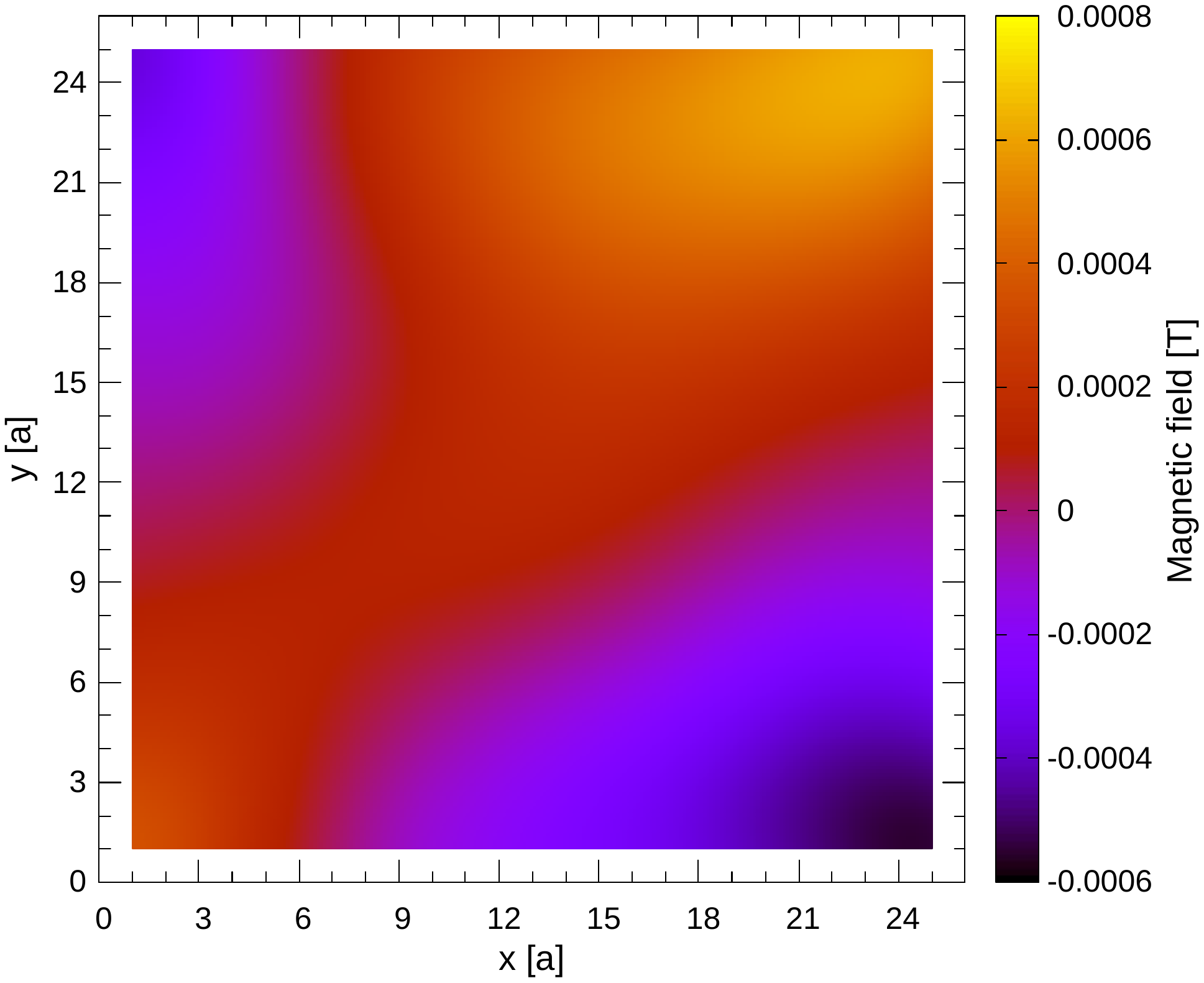}
      \caption{The contour plot of $B_z$ at $z=10$.}
      \label{fig:25_magnetic_maam_all}
    \end{subfigure}
    \hspace*{1mm}
    \begin{subfigure}{0.31\columnwidth}
      \centering
      \includegraphics[width=3.5cm]{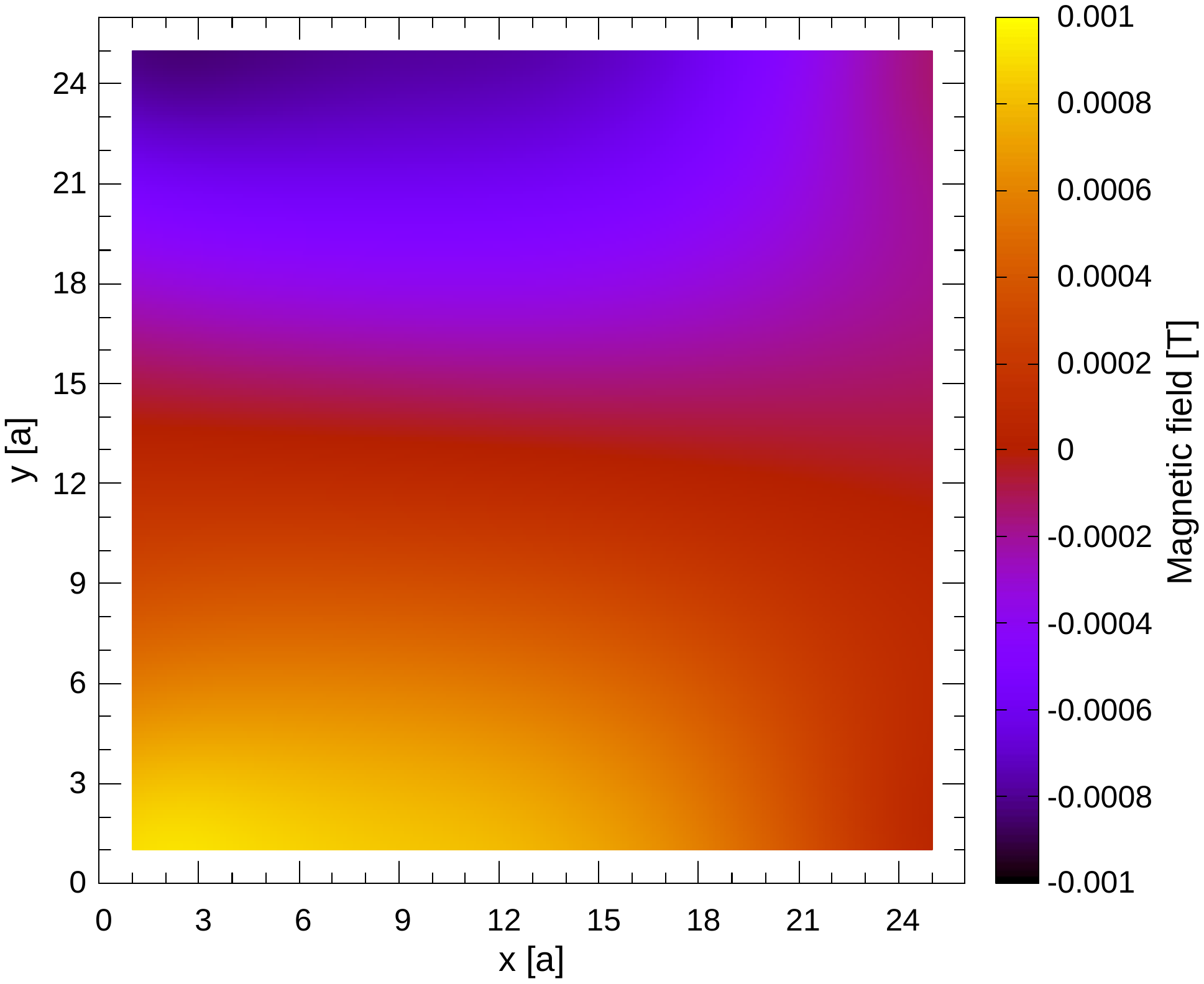}
      \caption{The bulk layer contribution.}
      \label{fig:25_magnetic_maam_bulk}
    \end{subfigure}
    \hspace*{1mm}
    \begin{subfigure}{0.31\columnwidth}
      \centering
      \includegraphics[width=3.5cm]{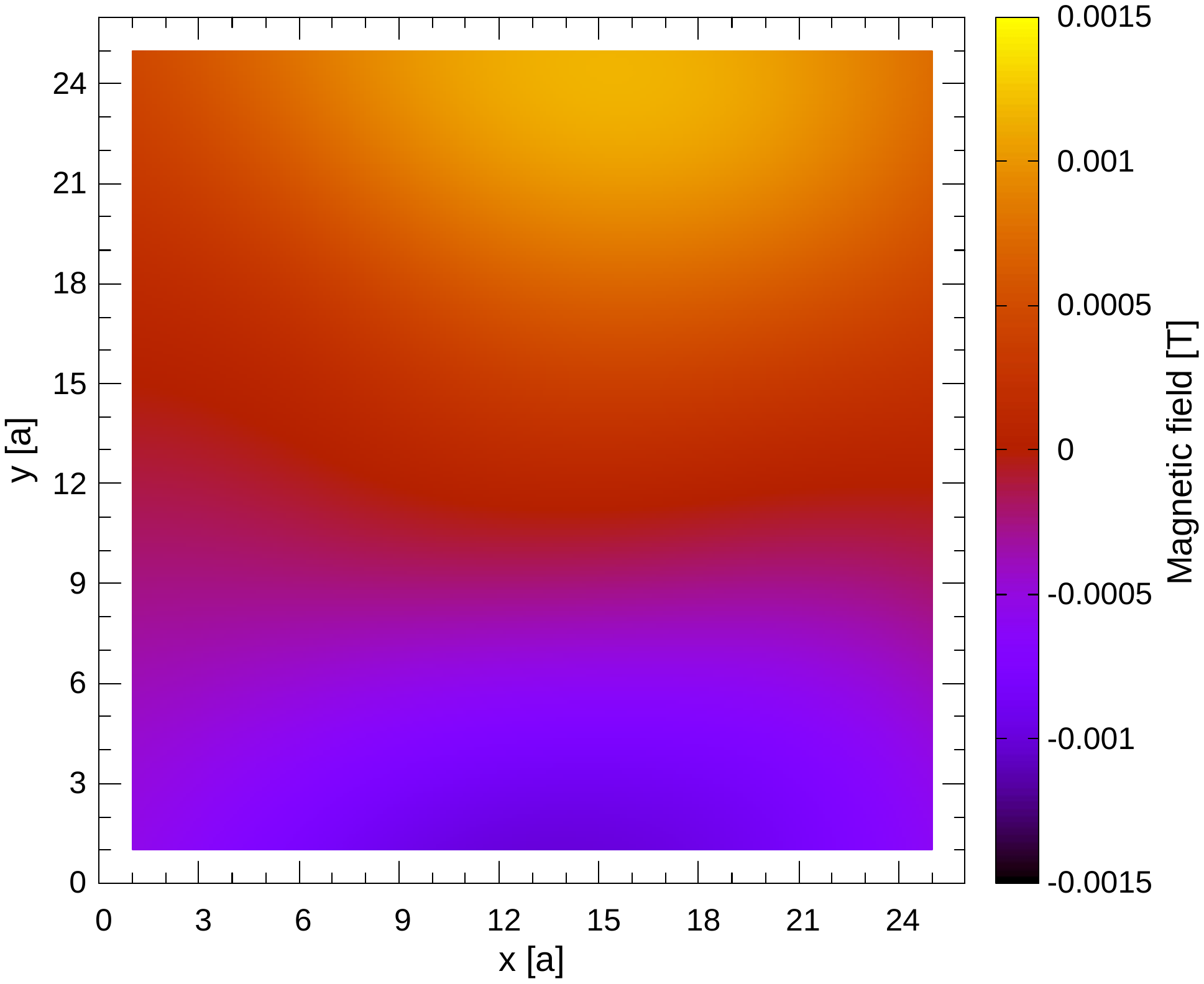}
      \caption{The surface layer contribution.}
      \label{fig:25_magnetic_maam_surface}
    \end{subfigure}
    \caption{The lowest energy state current for the $25 \times 25$ lattice system with
    the current pattern around the central SVQ, `maam' in 
      Fig.~\ref{fig:result:1NI_16_current}, and the contour plot of $B_z$ at $z=10$ generated by it.}
    \label{fig:25_current_maam}
  \end{figure}

In Fig.~\ref{fig:25_current_maam}, a current distribution generated by the SVILCs for the spin texture given in Fig.~\ref{fig:25_spin_vortex} is depicted.
In this current distribution, the  winding numbers of the SVILCs are assumed to be the same as the underlying SVQs winding numbers; from our previous calculation experience in similar systems, we noticed that this case will be the lowest energy state. 
Although the magnitude of the moments is very small, spin-vortices are also formed in the surface layer by the influence of the bulk layer as seen in Fig.~\ref{fig:25_spin_vortex}(d).
Therefore, they also induce SVILCs with the magnitude comparable to those generated in the bulk layers as seen in Fig.~\ref{fig:25_current_maam}(c). 
However, the magnetic field generated is too small to be used as the detection of the SVILCs in this case.

\begin{figure}[H]
    \centering
    \begin{subfigure}{0.31\columnwidth}
      \centering
      \includegraphics[width=3.5cm]{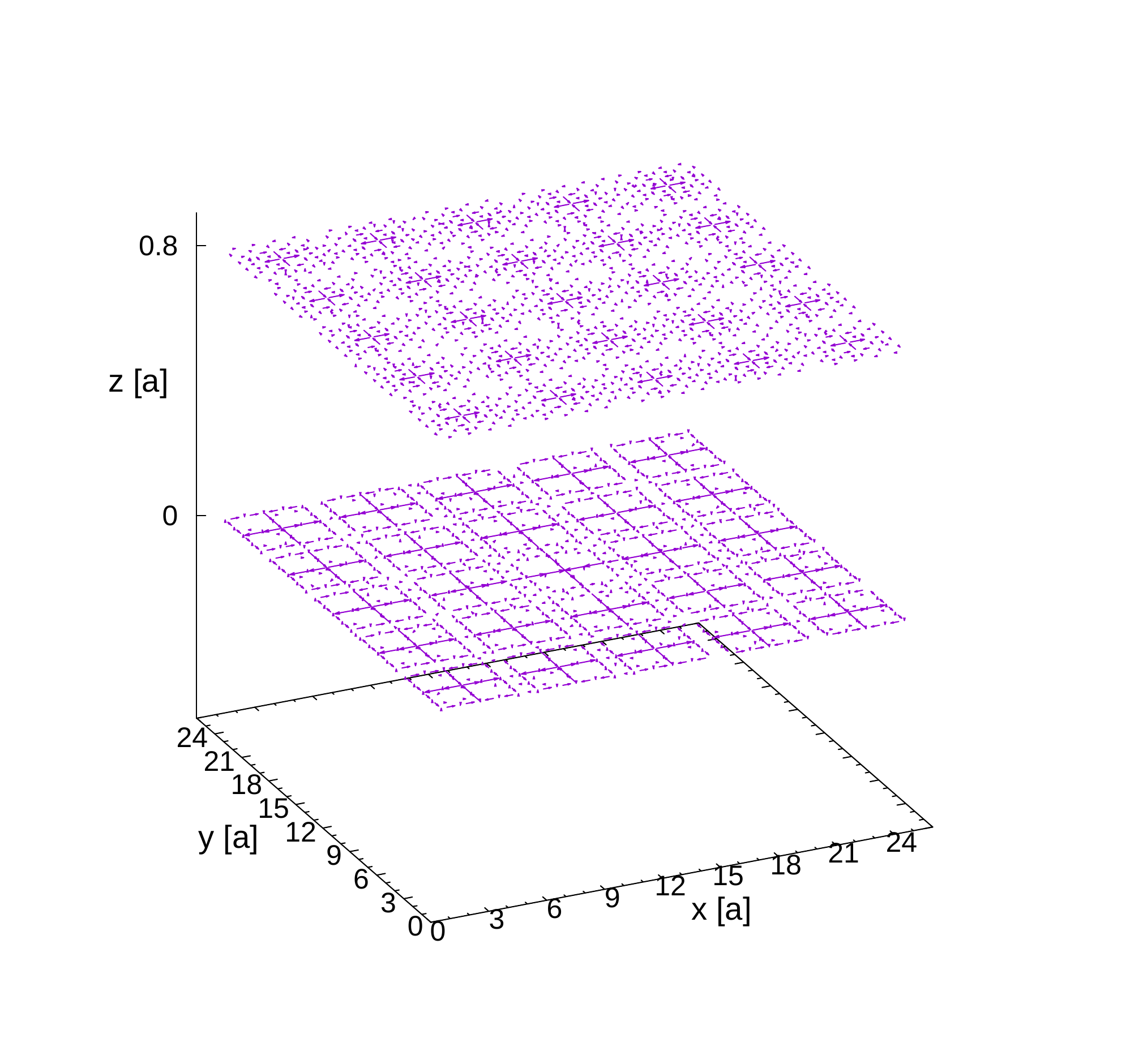}
      \caption{Current distribution.}
      \label{fig:25_current_amma_all}
    \end{subfigure}
    \hspace*{1mm}
    \begin{subfigure}{0.31\columnwidth}
      \centering
      \includegraphics[width=3cm]{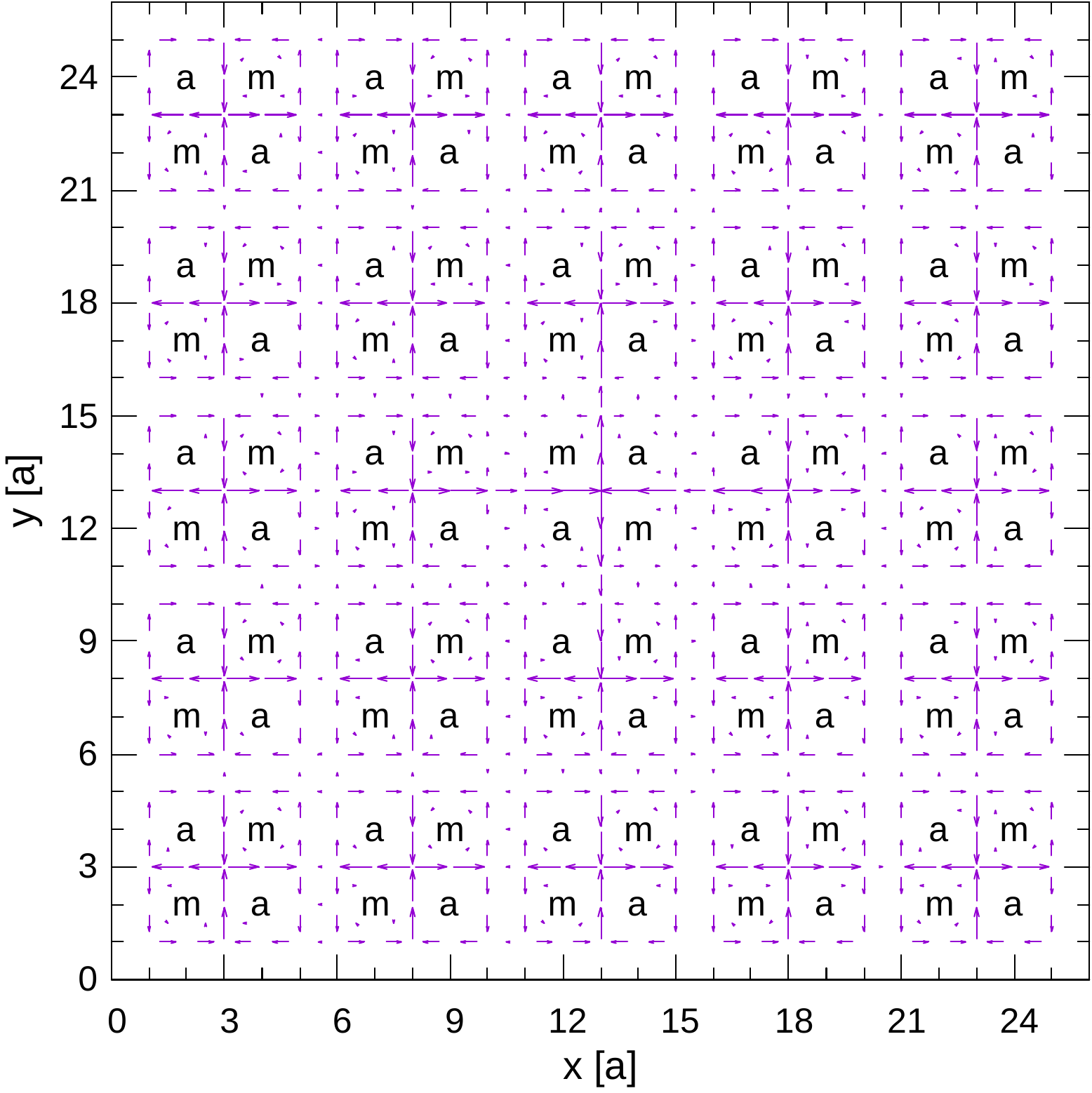}
      \caption{The bulk layer current.}
      \label{fig:25_current_amma_bulk}
    \end{subfigure}
    \hspace*{1mm}
    \begin{subfigure}{0.31\columnwidth}
      \centering
      \includegraphics[width=3cm]{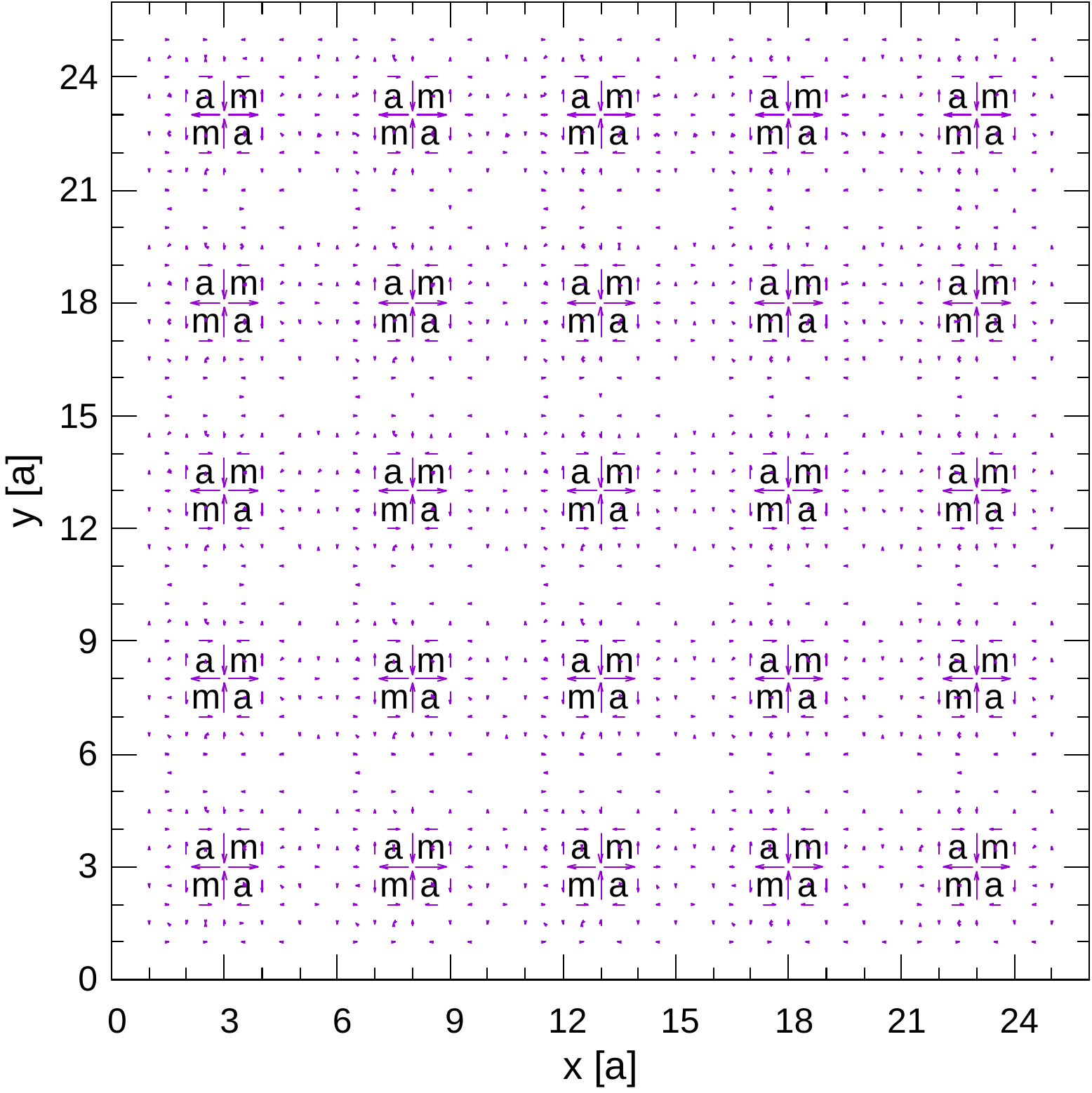}
      \caption{The surface layer current.}
      \label{fig:25_current_amma_surface}
    \end{subfigure}
     \centering
    \begin{subfigure}{0.31\columnwidth}
      \centering
      \includegraphics[width=3.5cm]{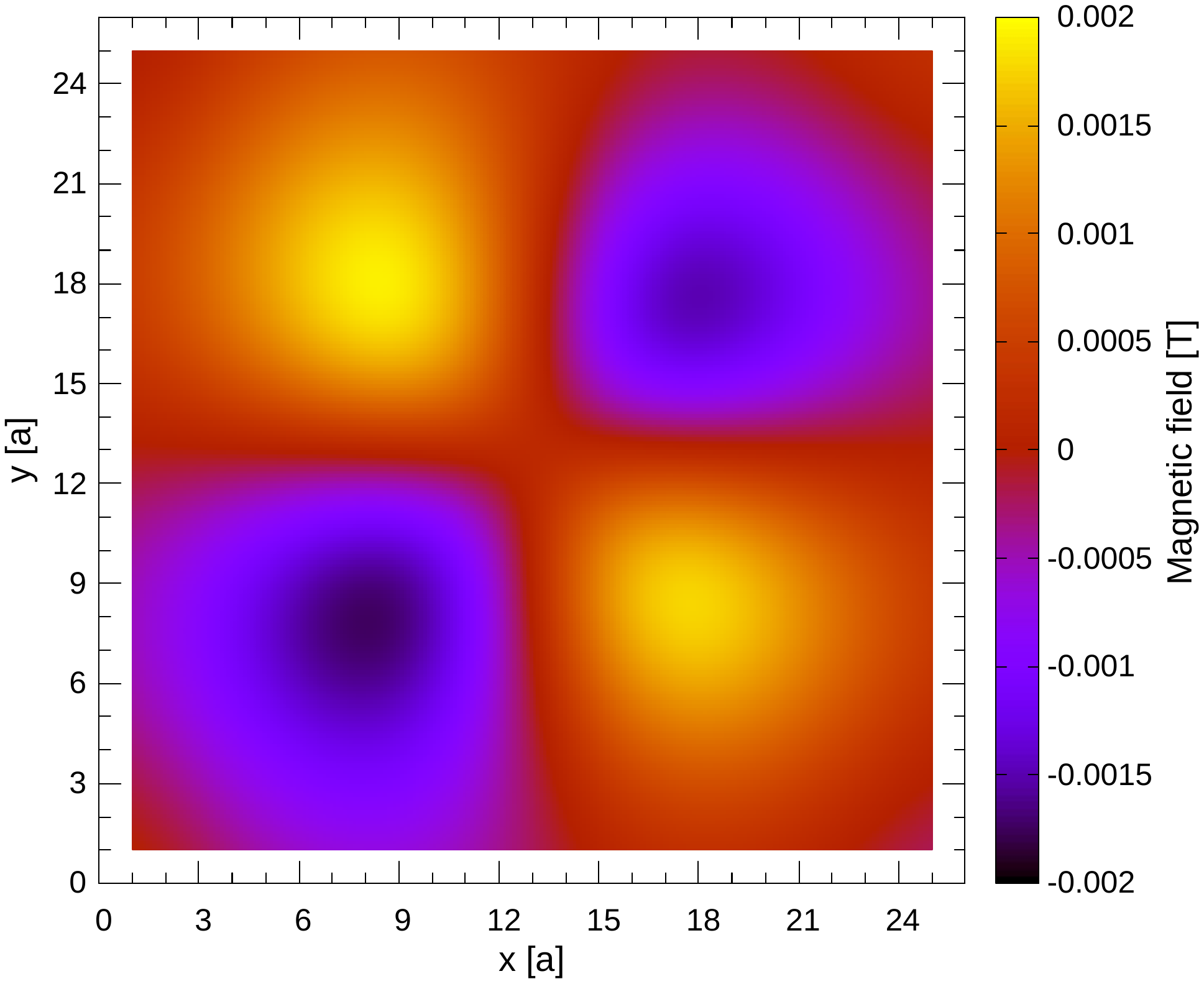}
      \caption{The contour plot of $B_z$ at $z=10$.}
      \label{fig:25_magnetic_amma_all}
    \end{subfigure}
    \hspace*{1mm}
    \begin{subfigure}{0.31\columnwidth}
      \centering
      \includegraphics[width=3.5cm]{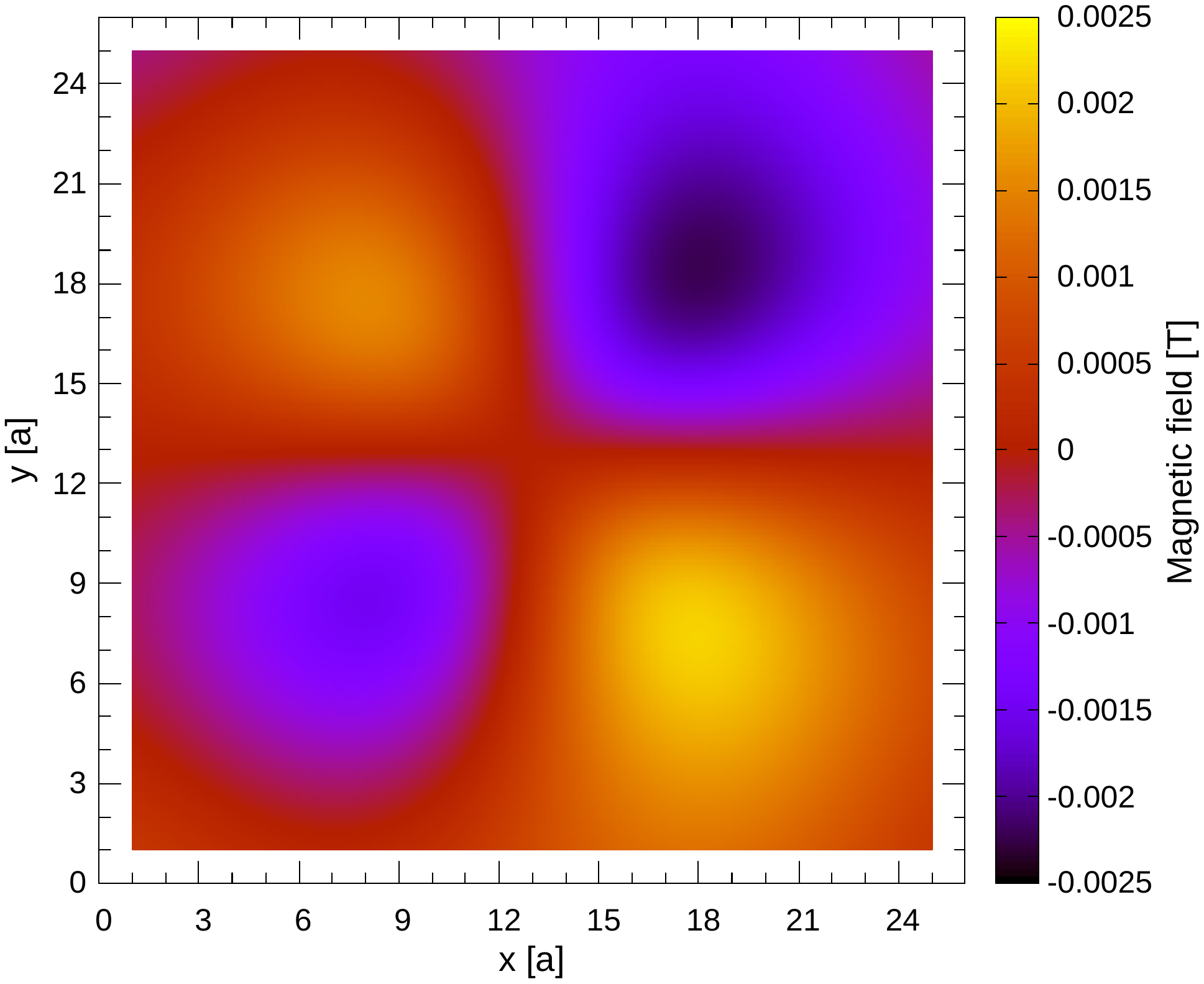}
      \caption{The bulk layer contribution from }
      \label{fig:25_magnetic_amma_bulk}
    \end{subfigure}
    \hspace*{1mm}
    \begin{subfigure}{0.31\columnwidth}
      \centering
      \includegraphics[width=\columnwidth]{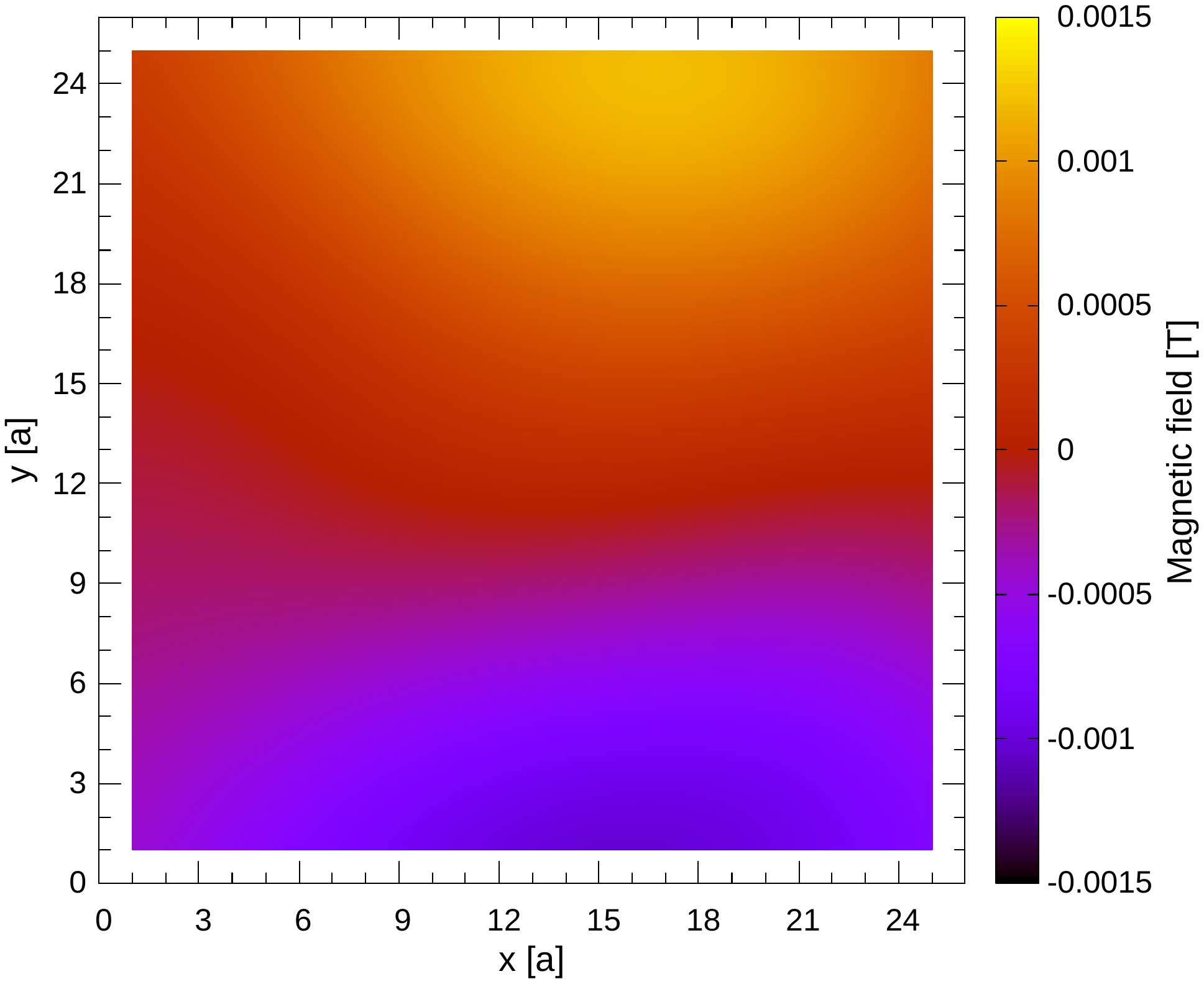}
      \caption{The surface layer contribution.}
      \label{fig:25_magnetic_amma_surface}
    \end{subfigure}
    \caption{The $25 \times 25$ lattice system with the current pattern around the central SVQ,  `amma' in Fig.~\ref{fig:result:1NI_16_current}, and the contour plot of $B_z$ at $z=10$ generated by it.}
    \label{fig:25_current_amma}
  \end{figure}

\begin{figure}[H]
    \centering
    \begin{subfigure}{0.31\columnwidth}
      \centering
      \includegraphics[width=3.5cm]{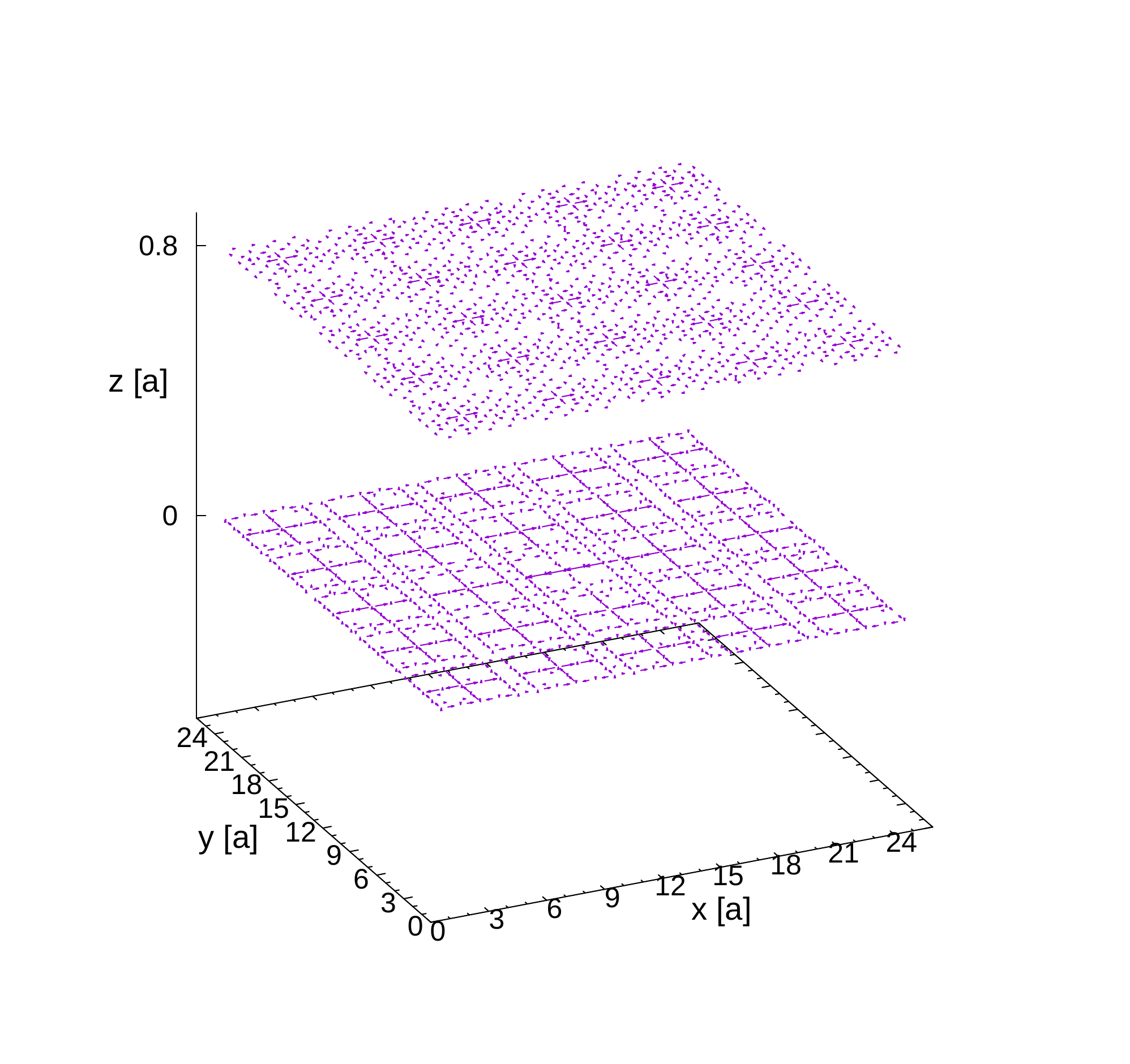}
      \caption{Current distribution.}
      \label{fig:25_current_mama_all}
    \end{subfigure}
    \hspace*{1mm}
    \begin{subfigure}{0.31\columnwidth}
      \centering
      \includegraphics[width=3cm]{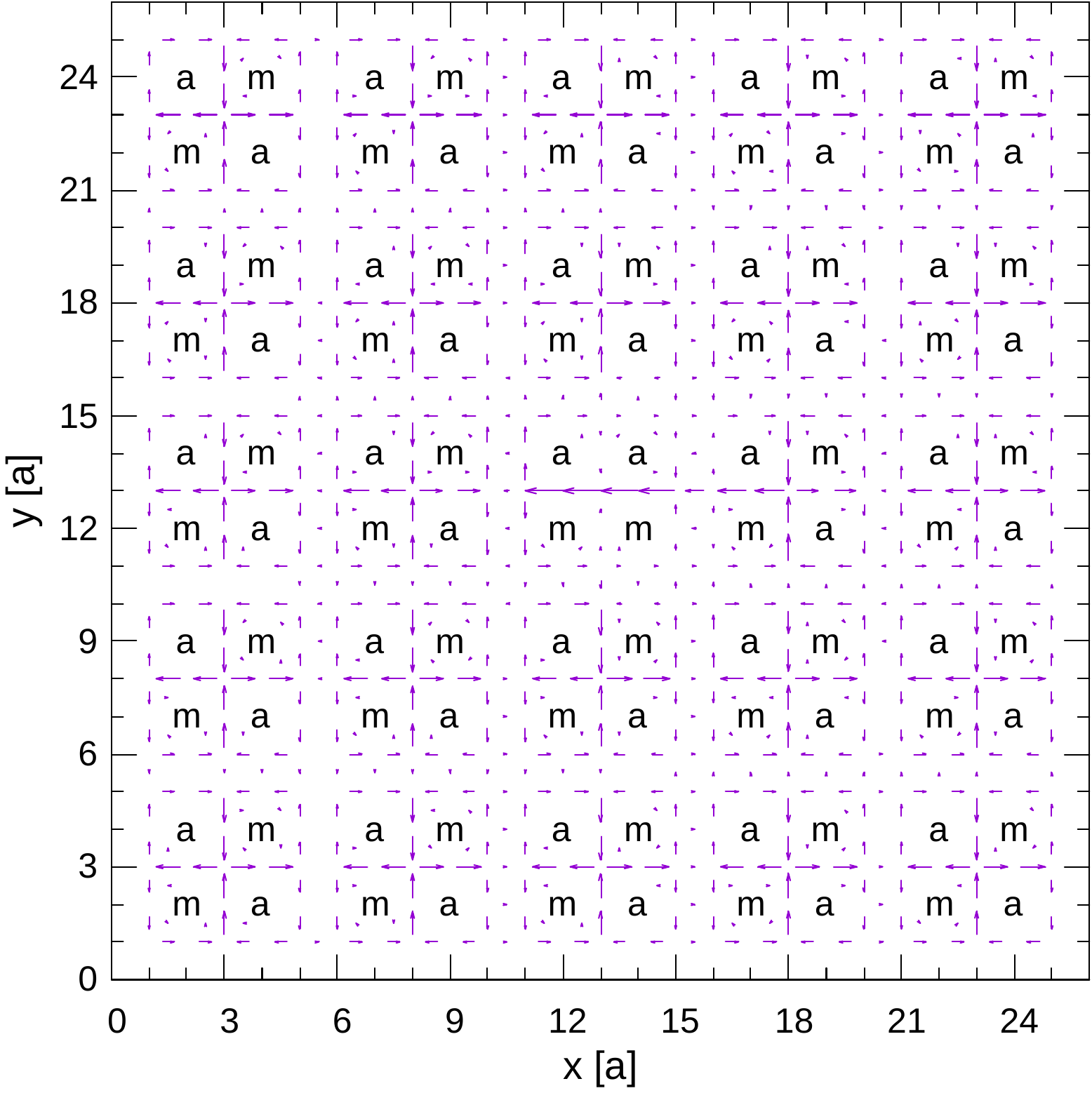}
      \caption{The bulk layer current.}
      \label{fig:25_current_mama_bulk}
    \end{subfigure}
    \hspace*{1mm}
    \begin{subfigure}{0.31\columnwidth}
      \centering
      \includegraphics[width=3cm]{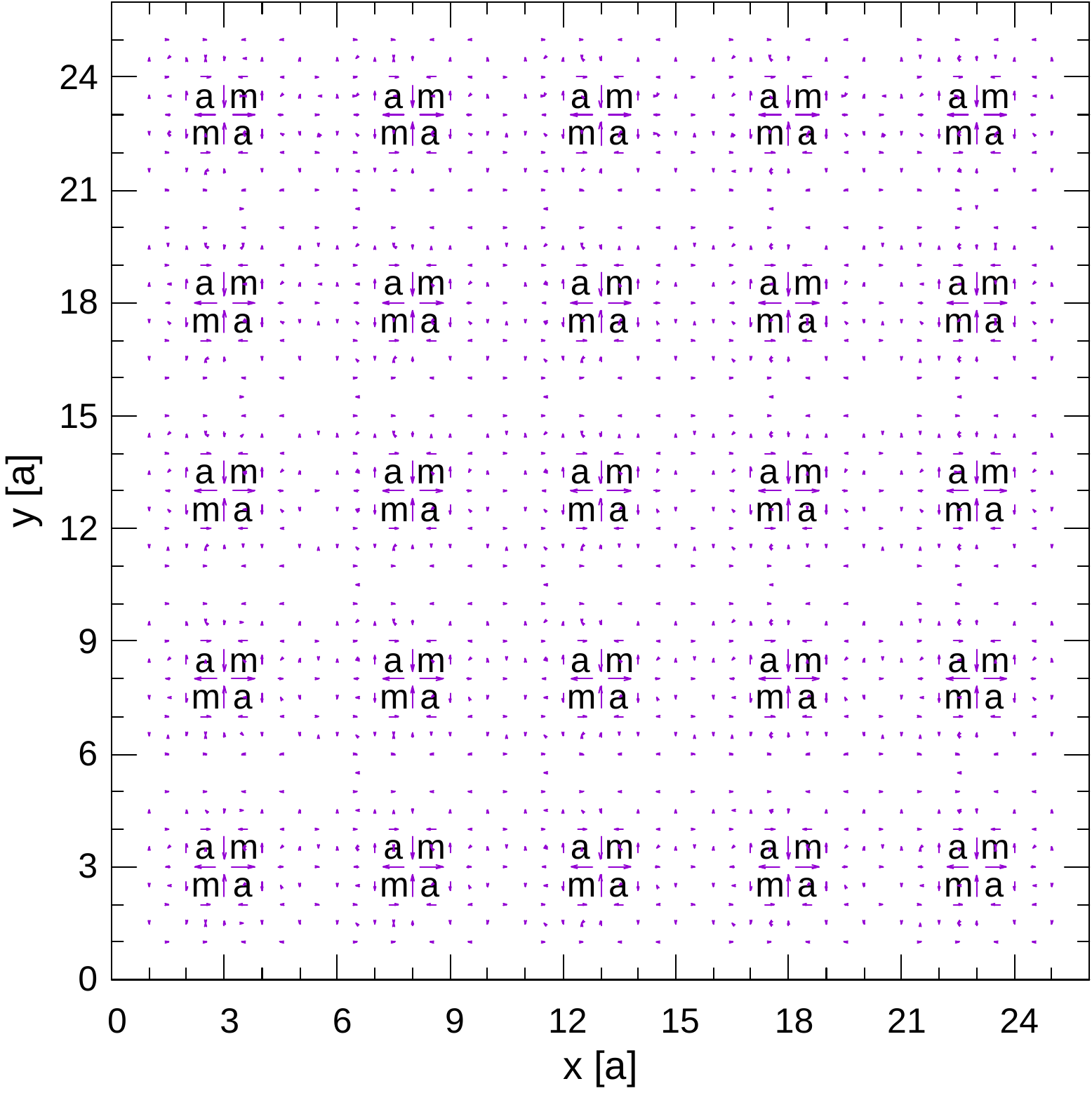}
      \caption{The surface layer current.}
      \label{fig:25_current_mama_surface}
    \end{subfigure}
    \centering
    \begin{subfigure}{0.31\columnwidth}
      \centering
      \includegraphics[width=3.5cm]{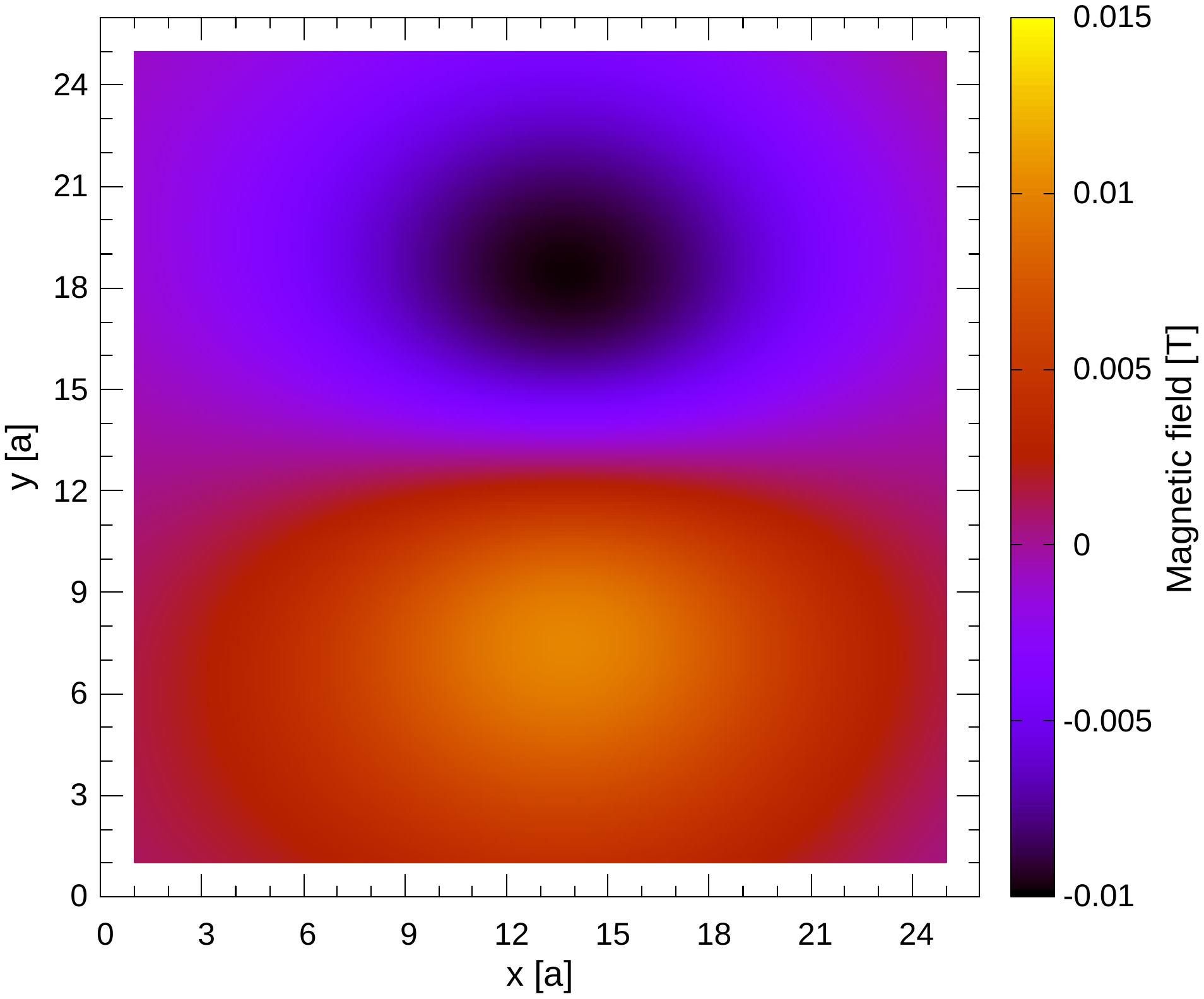}
      \caption{The contour plot of $B_z$ at $z=10$.}
      \label{fig:25_magnetic_mama_all}
    \end{subfigure}
    \hspace*{1mm}
    \begin{subfigure}{0.31\columnwidth}
      \centering
      \includegraphics[width=3.5cm]{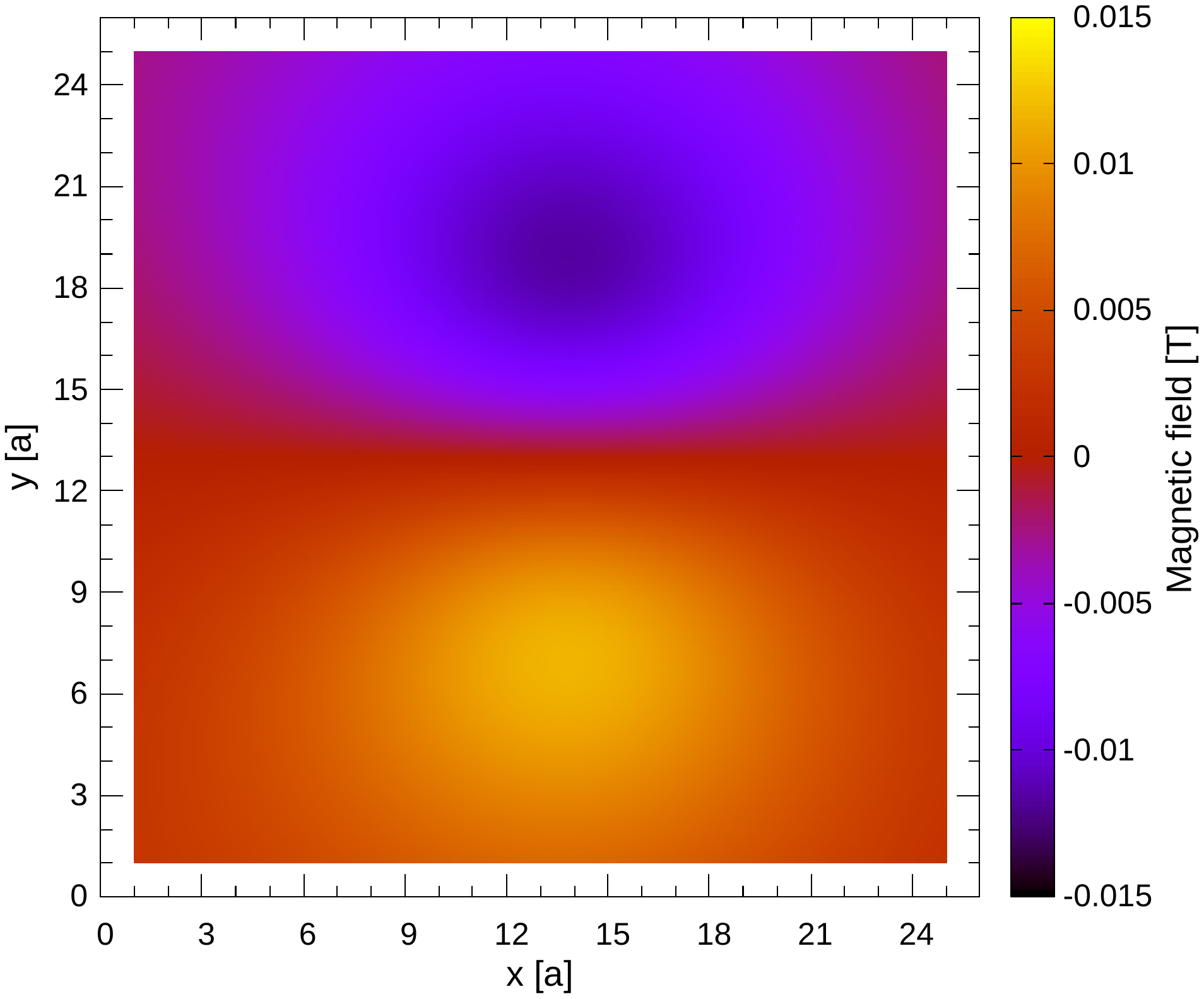}
      \caption{The bulk layer contribution.}
      \label{fig:25_magnetic_mama_bulk}
    \end{subfigure}
    \hspace*{1mm}
    \begin{subfigure}{0.31\columnwidth}
      \centering
      \includegraphics[width=3.5cm]{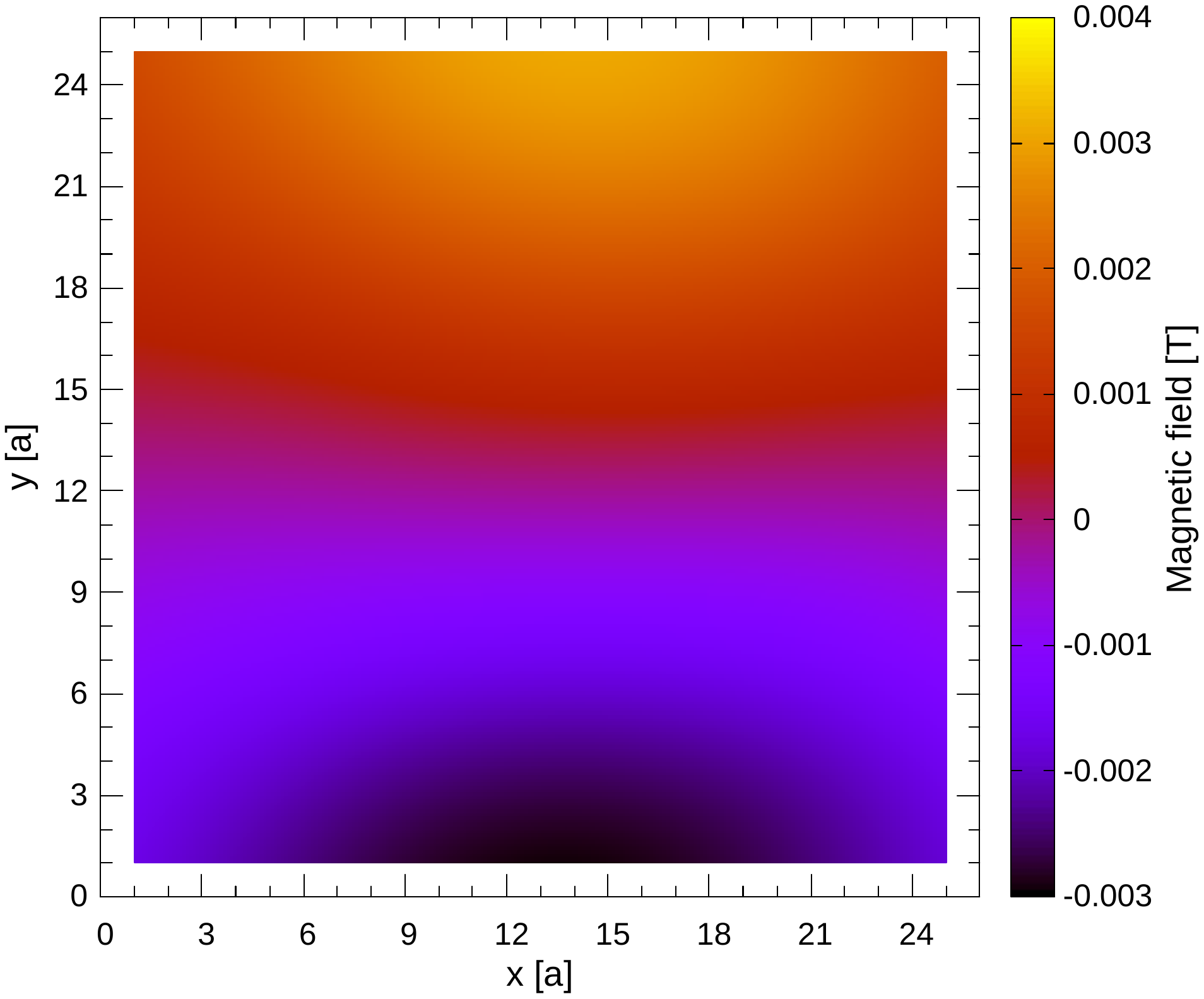}
      \caption{The surface layer contribution.}
      \label{fig:25_magnetic_mama_surface}
    \end{subfigure}
    \caption{
    The $25 \times 25$ lattice system with the current pattern around the central SVQ,  `mama' in Fig.~\ref{fig:result:1NI_16_current}, and the contour plot of $B_z$ at $z=10$ generated by it.
  }
    \label{fig:25_current_mama}
  \end{figure}

\begin{figure}[H]
    \centering
    \begin{subfigure}{0.31\columnwidth}
      \centering
      \includegraphics[width=3.5cm]{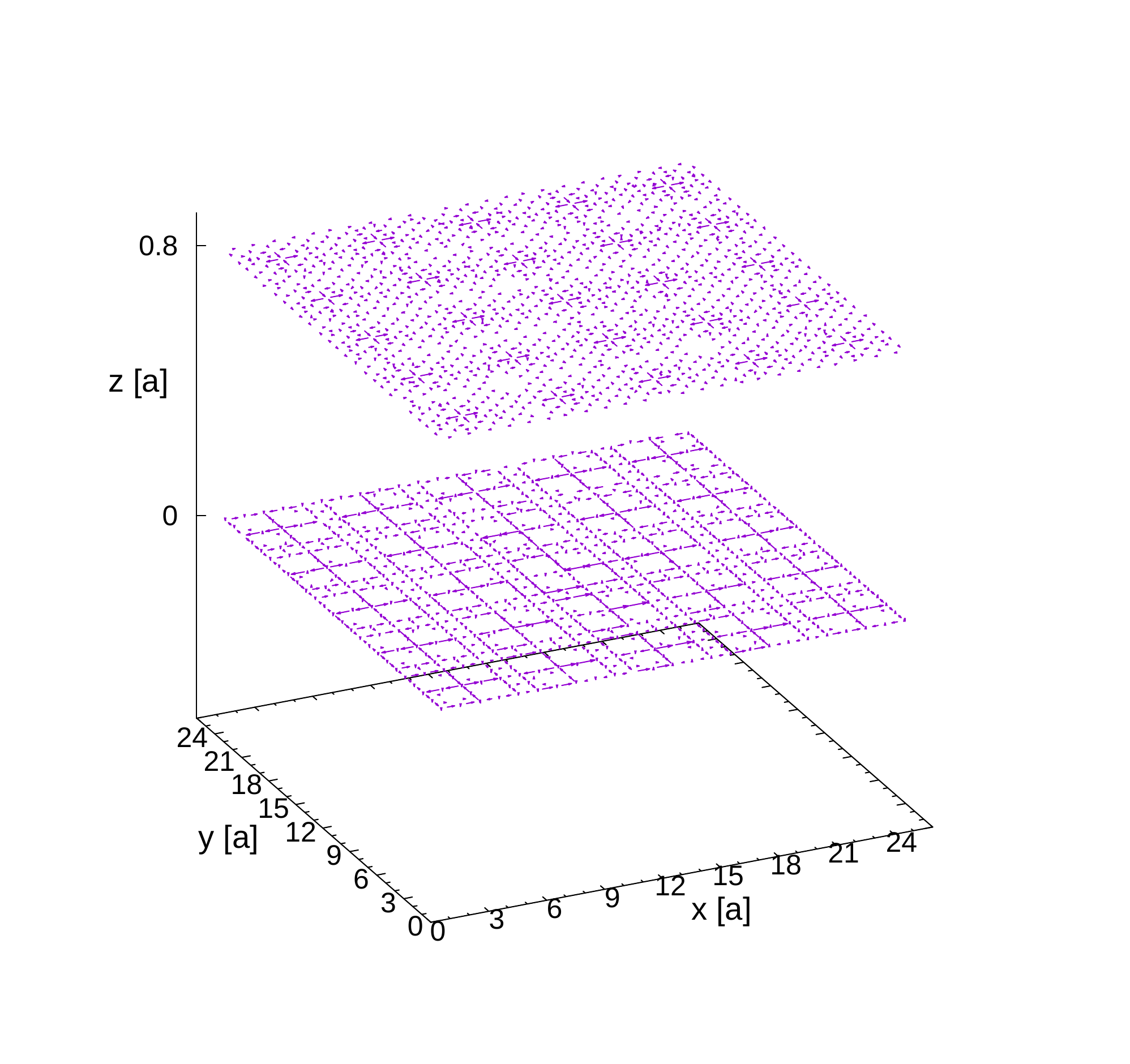}
      \caption{Current distribution.}
      \label{fig:25_current_mmma_all}
    \end{subfigure}
    \hspace*{1mm}
    \begin{subfigure}{0.31\columnwidth}
      \centering
      \includegraphics[width=3cm]{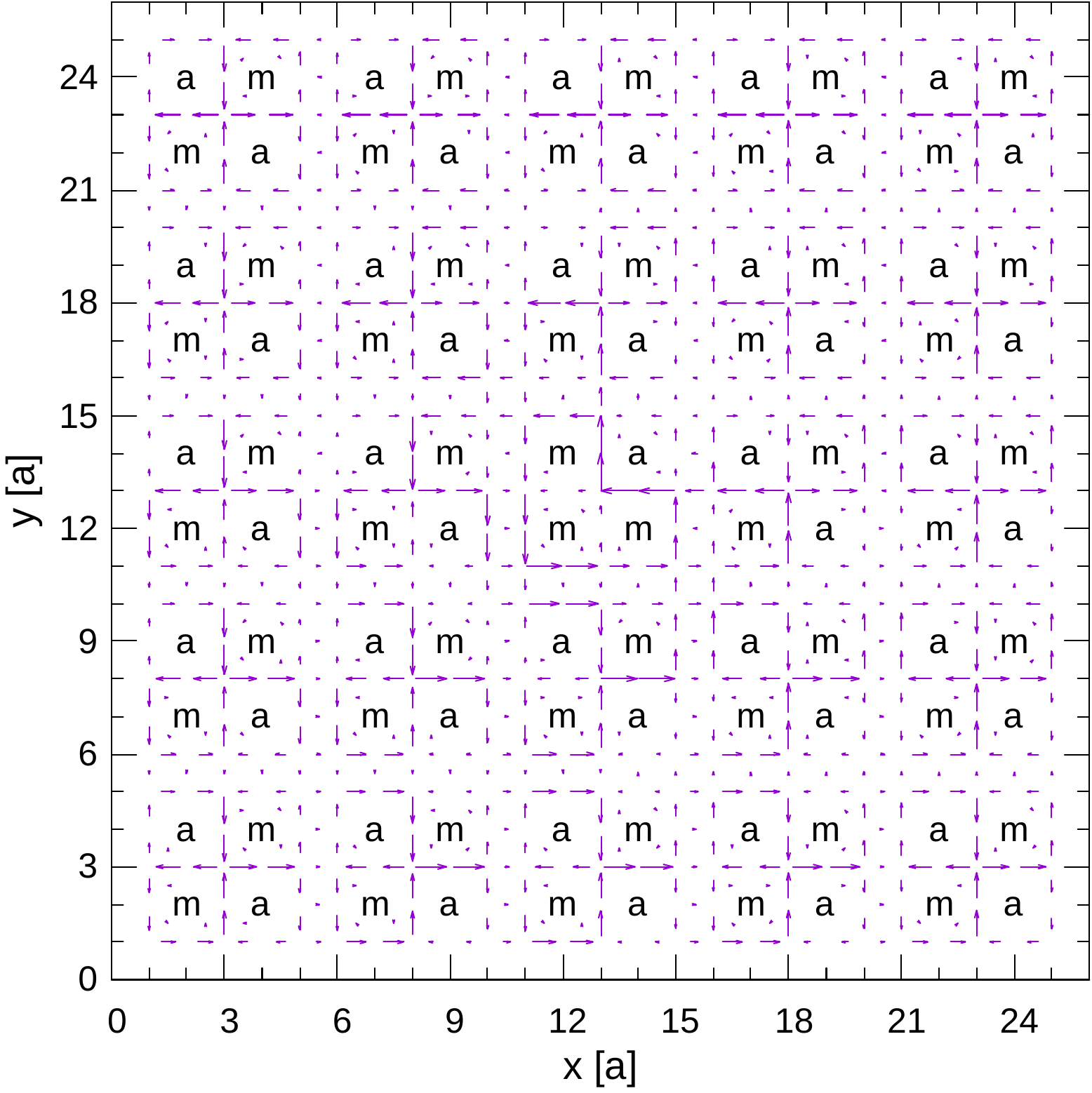}
      \caption{The bulk layer current.}
      \label{fig:25_current_mmma_bulk}
    \end{subfigure}
    \hspace*{1mm}
    \begin{subfigure}{0.31\columnwidth}
      \centering
      \includegraphics[width=3cm]{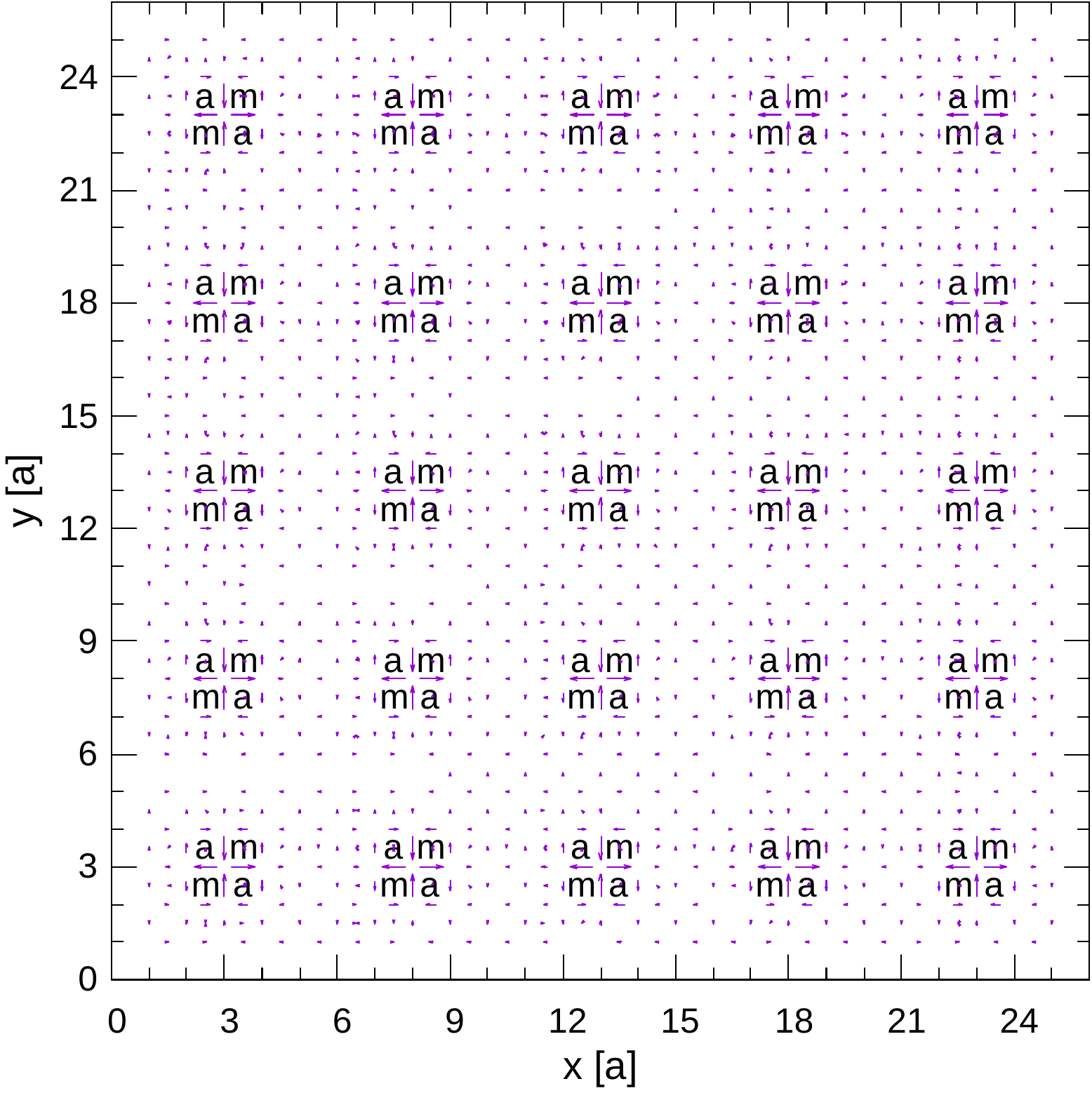}
      \caption{The surface layer current.}
      \label{fig:25_current_mmma_surface}
    \end{subfigure}
    \centering
    \begin{subfigure}{0.31\columnwidth}
      \centering
      \includegraphics[width=3.5cm]{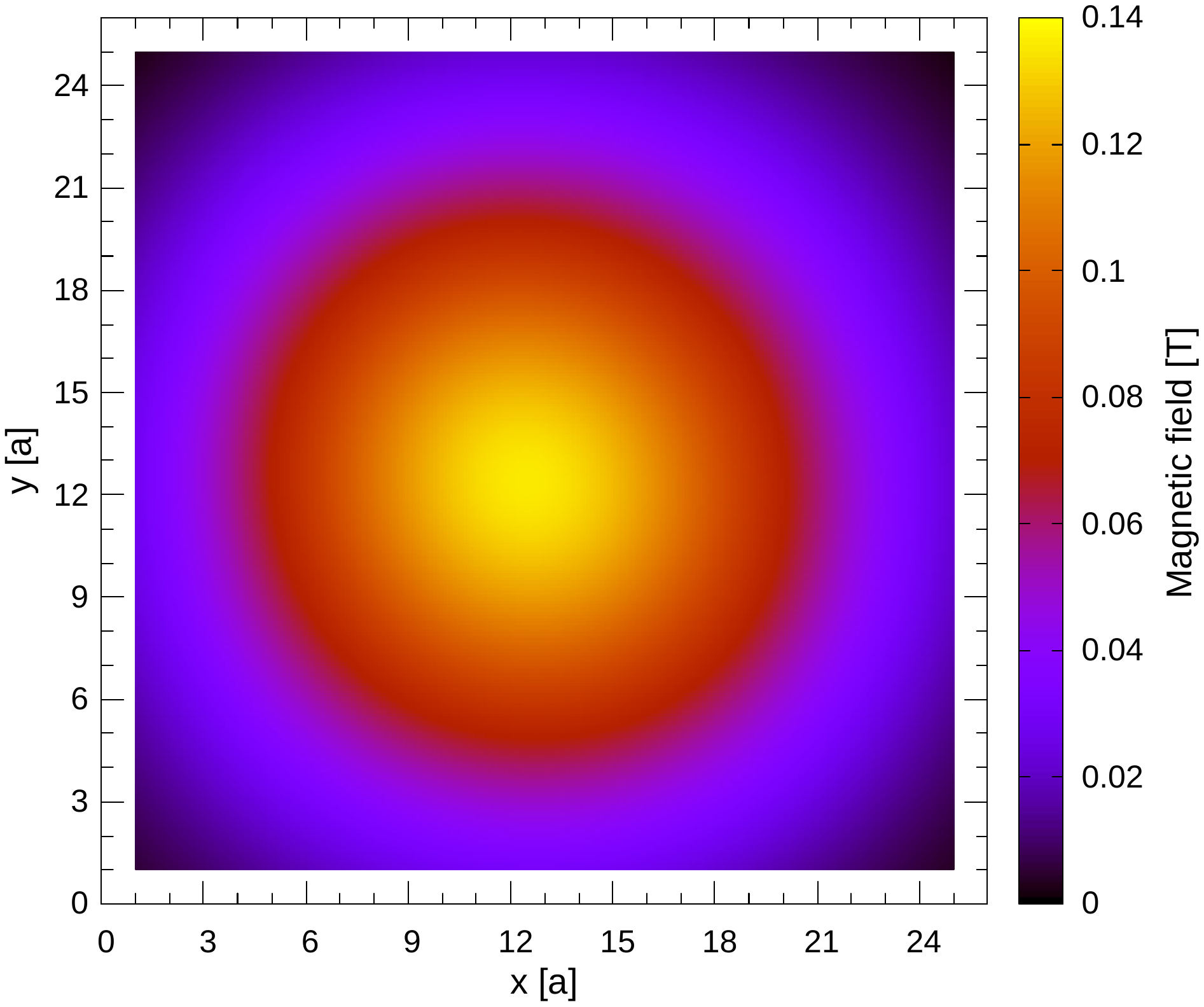}
      \caption{The contour plot of $B_z$ at $z=10$.}
      \label{fig:25_magnetic_mmma_all}
    \end{subfigure}
    \hspace*{1mm}
    \begin{subfigure}{0.31\columnwidth}
      \centering
      \includegraphics[width=3.5cm]{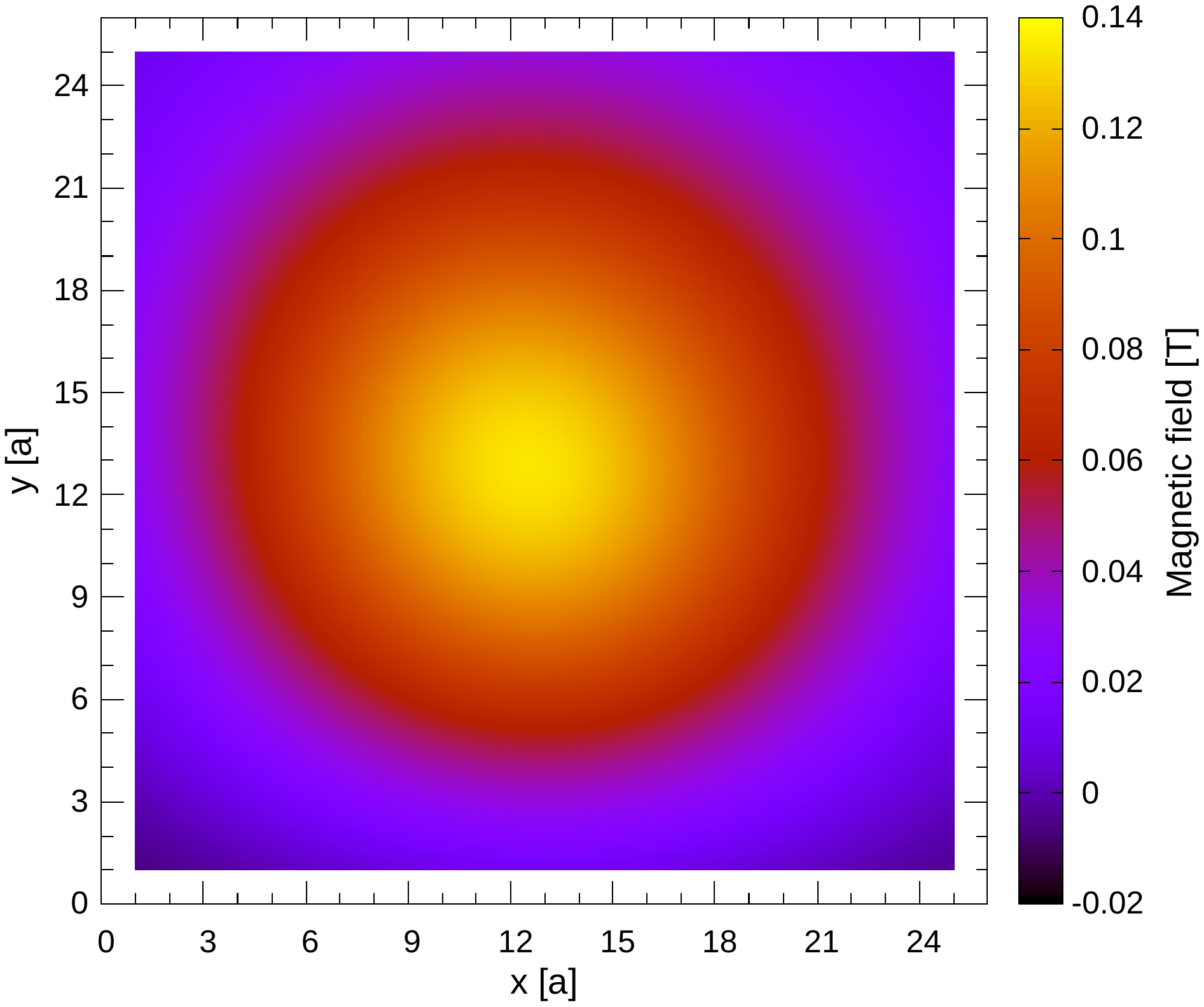}
      \caption{The bulk layer contribution.}
    \label{fig:25_magnetic_mmma_bulk}
    \end{subfigure}
    \hspace*{1mm}
    \begin{subfigure}{0.31\columnwidth}
      \centering
      \includegraphics[width=3.5cm]{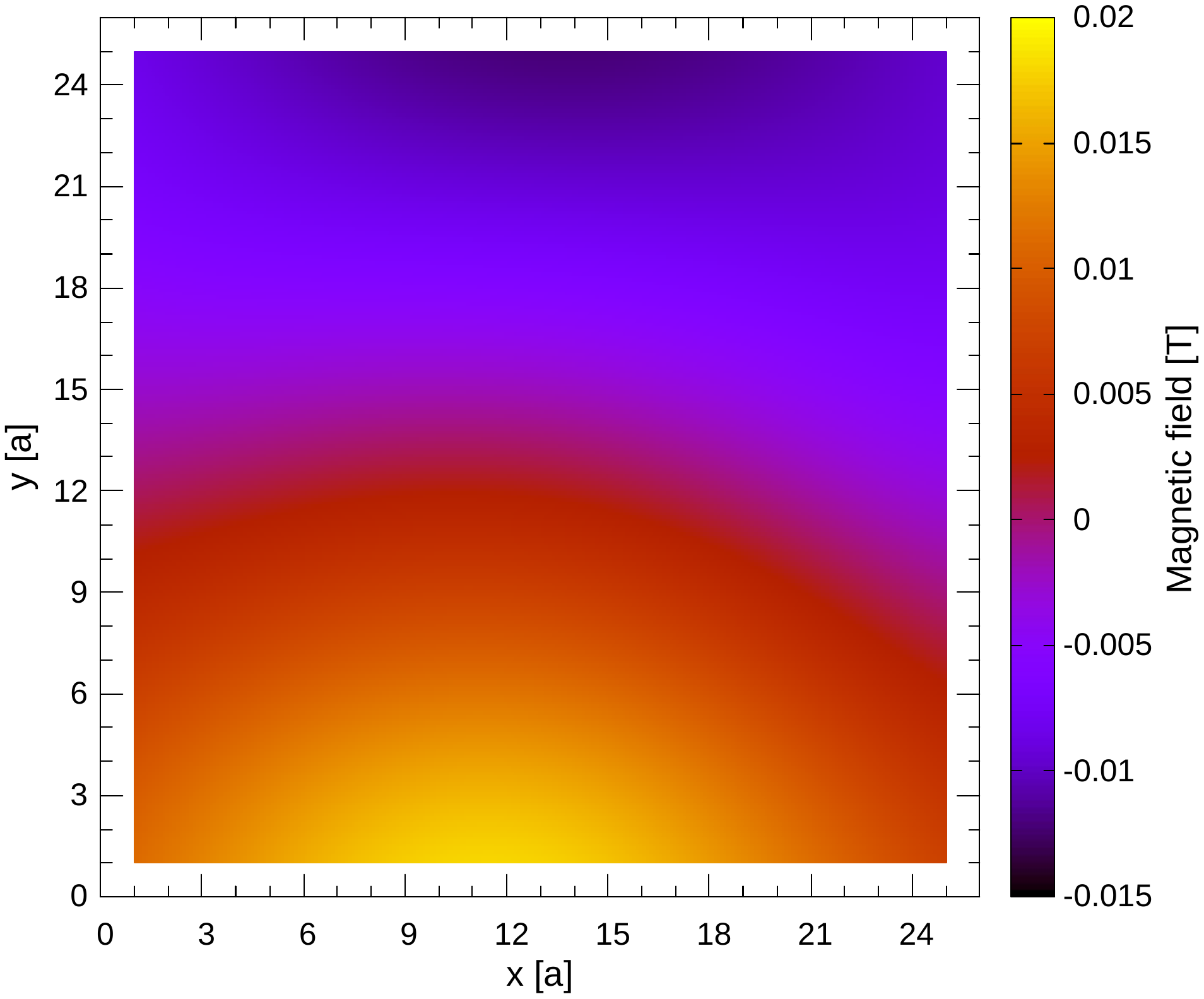}
      \caption{The surface layer contribution.}
      \label{fig:25_magnetic_mmma_surface}
    \end{subfigure}
    \caption{The $25 \times 25$ lattice system with the current pattern around the central SVQ,  `mmma' in Fig.~\ref{fig:result:1NI_16_current}, and the contour plot of $B_z$ at $z=10$ generated by it.}
       \label{fig:25_current_mmma}
  \end{figure}

\begin{figure}[H]
    \centering
    \begin{subfigure}{0.31\columnwidth}
      \centering
      \includegraphics[width=3.5cm]{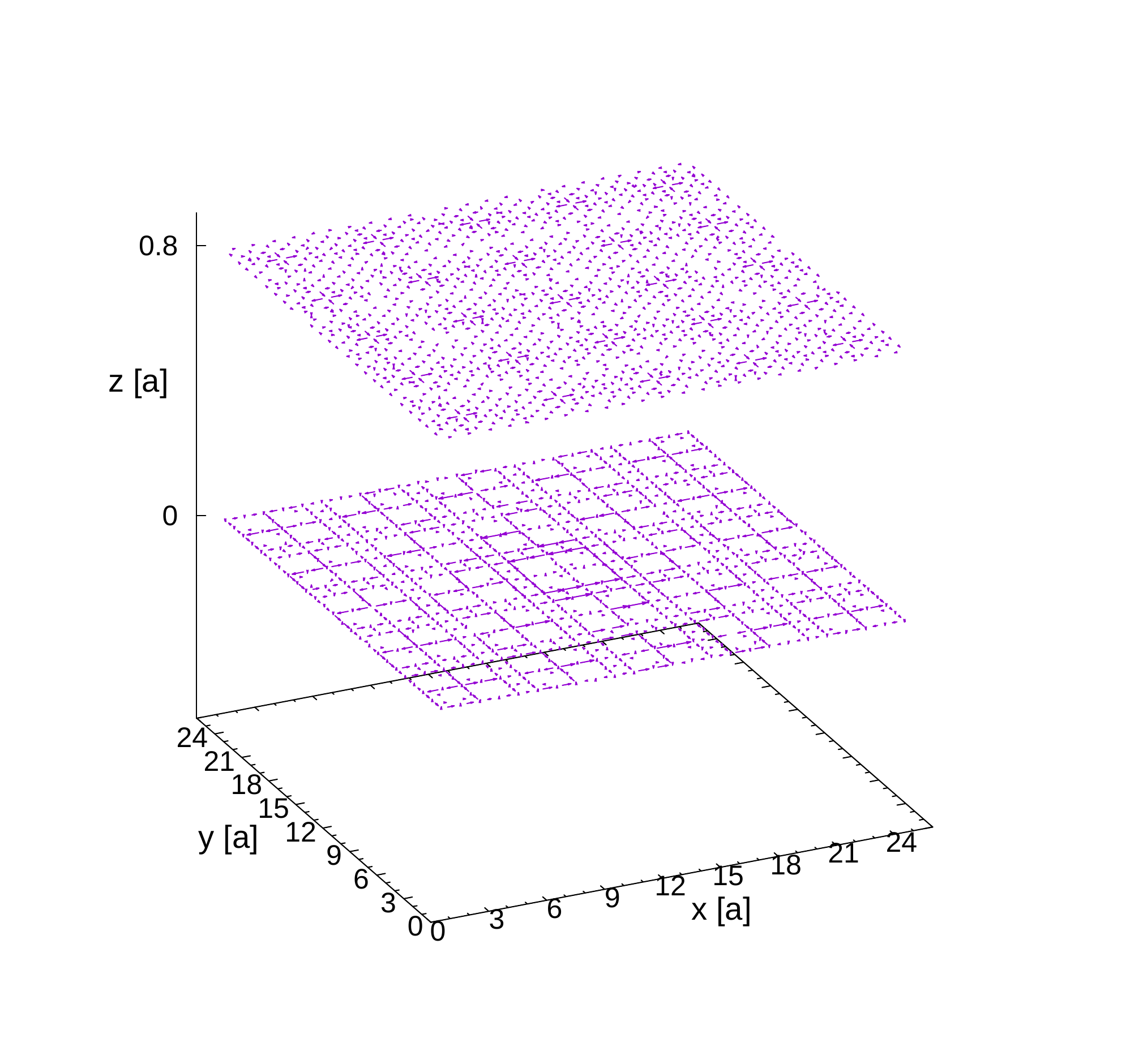}
      \caption{Current distribution.}
      \label{fig:25_current_mmmm_all}
    \end{subfigure}
    \hspace*{1mm}
    \begin{subfigure}{0.31\columnwidth}
      \centering
      \includegraphics[width=3cm]{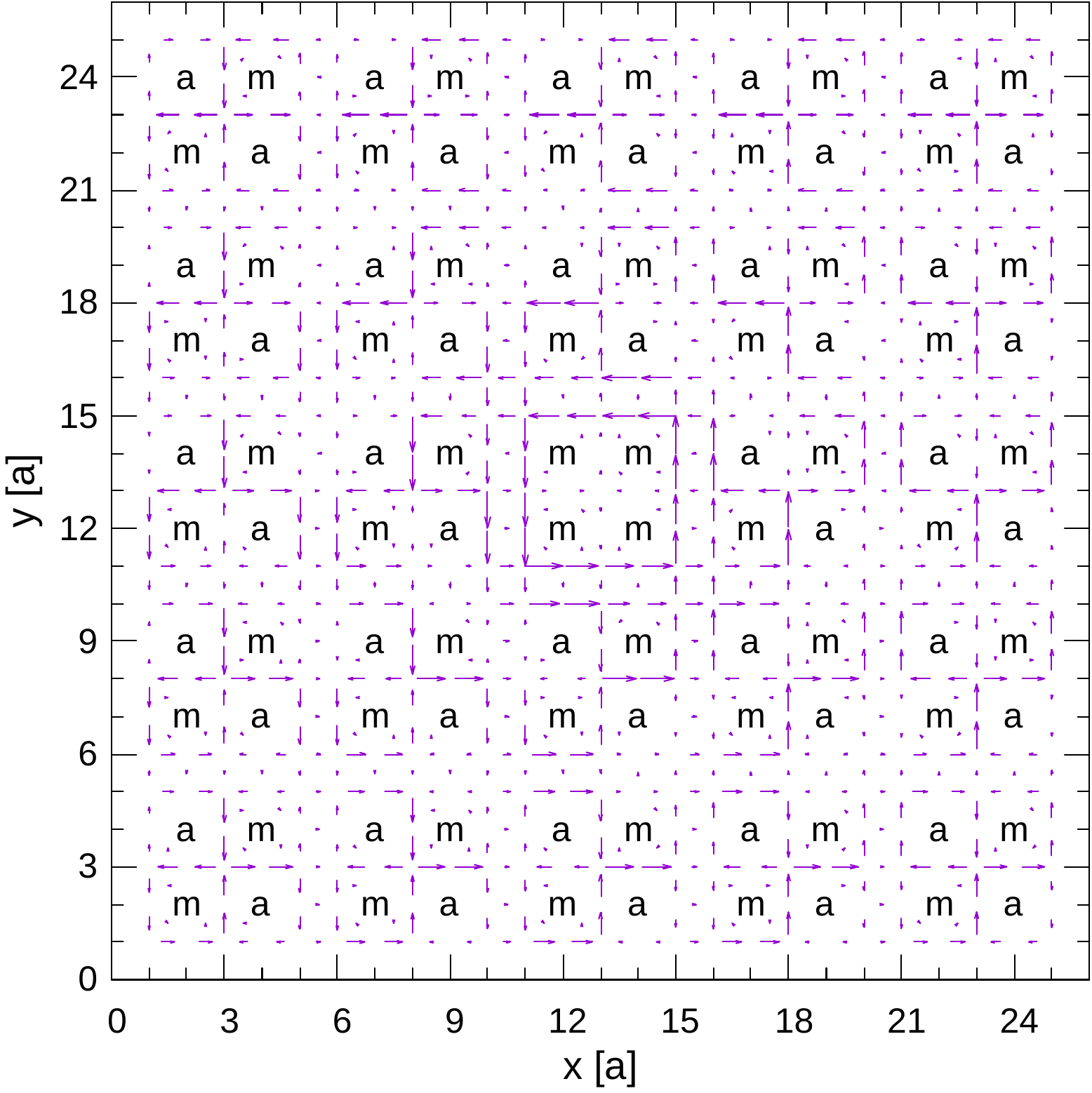}
      \caption{The bulk layer current.}
      \label{fig:25_current_mmmm_bulk}
    \end{subfigure}
    \hspace*{1mm}
    \begin{subfigure}{0.31\columnwidth}
      \centering
      \includegraphics[width=3cm]{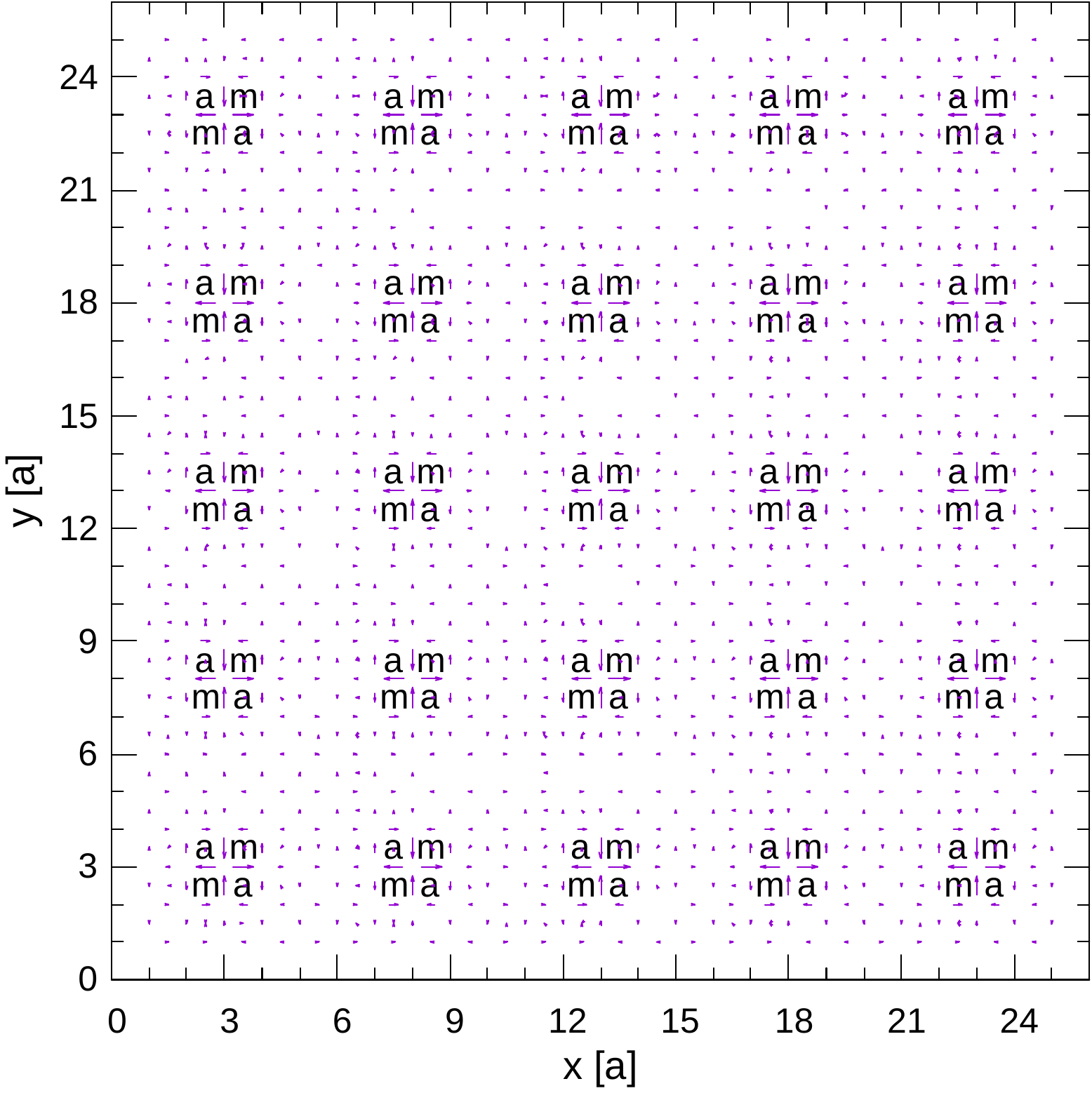}
      \caption{The surface layer current.}
      \label{fig:25_current_mmmm_surface}
    \end{subfigure}
    \begin{subfigure}{0.31\columnwidth}
      \centering
      \includegraphics[width=3.5cm]{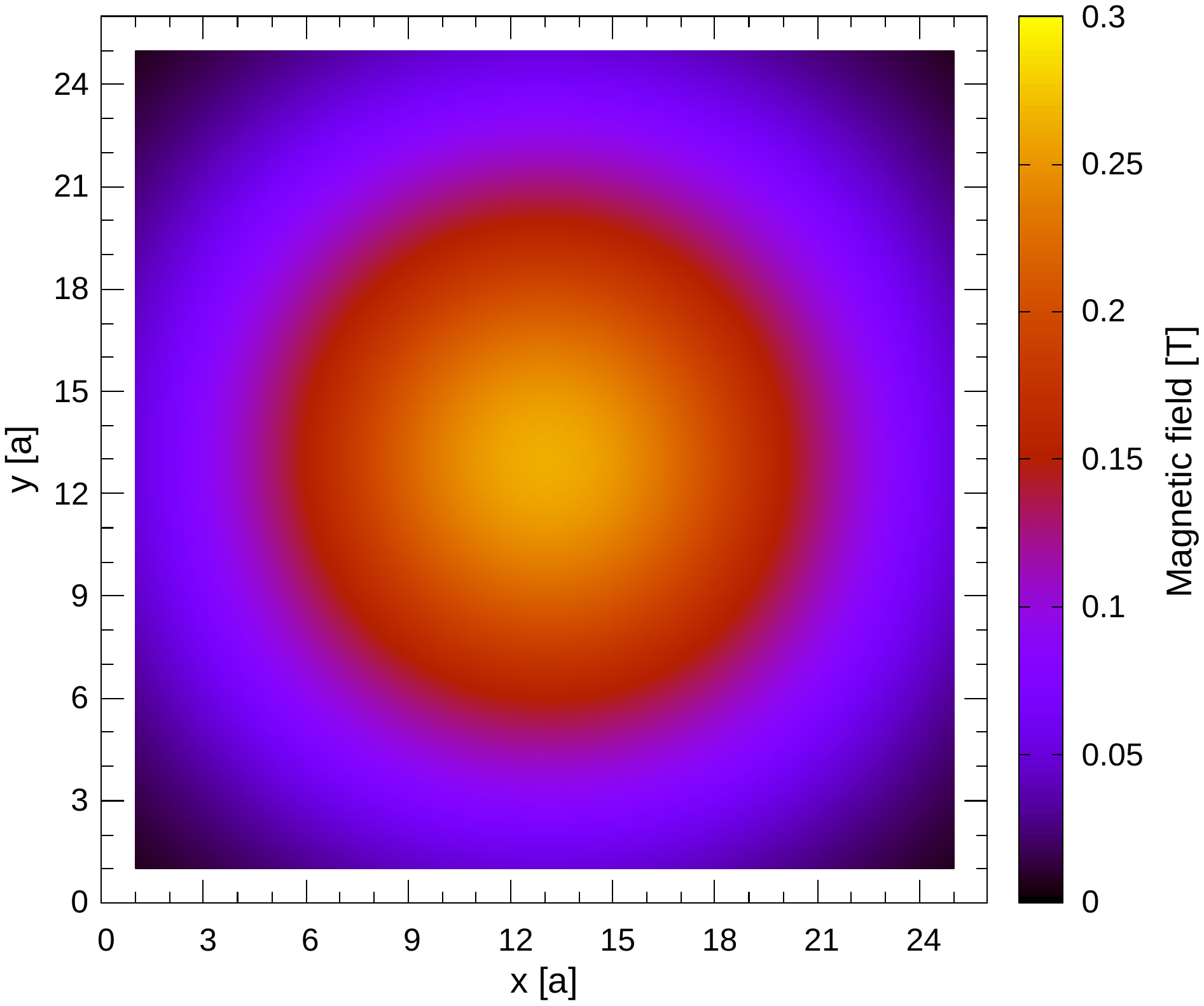}
      \caption{The contour plot of $B_z$ at $z=10$.}
      \label{fig:25_magnetic_mmmm_all}
    \end{subfigure}
    \hspace*{1mm}
    \begin{subfigure}{0.31\columnwidth}
      \centering
      \includegraphics[width=3.5cm]{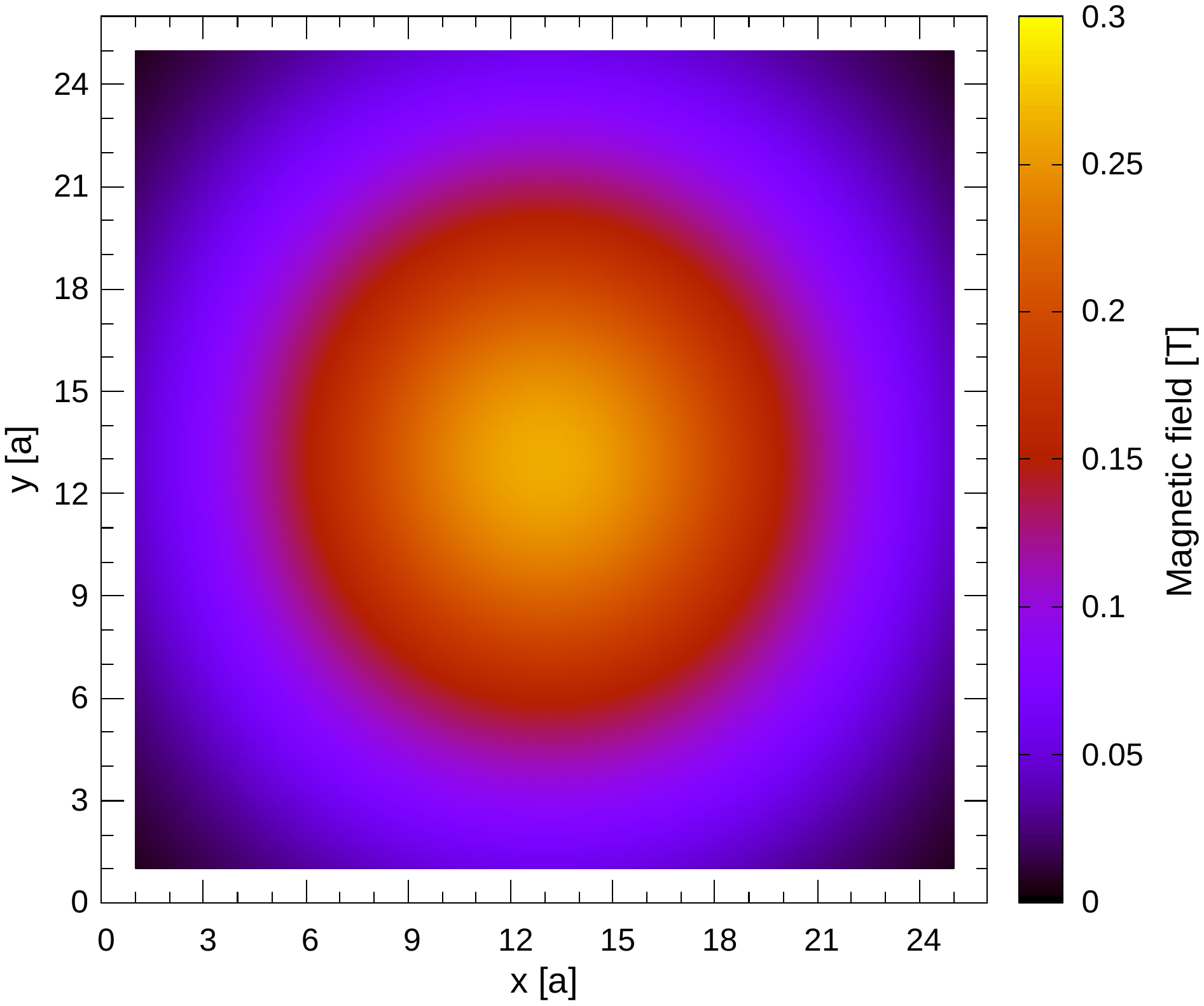}
      \caption{The bulk layer contribution.}
      \label{fig:25_magnetic_mmmm_bulk}
    \end{subfigure}
    \hspace*{1mm}
    \begin{subfigure}{0.31\columnwidth}
      \centering
      \includegraphics[width=3.5cm]{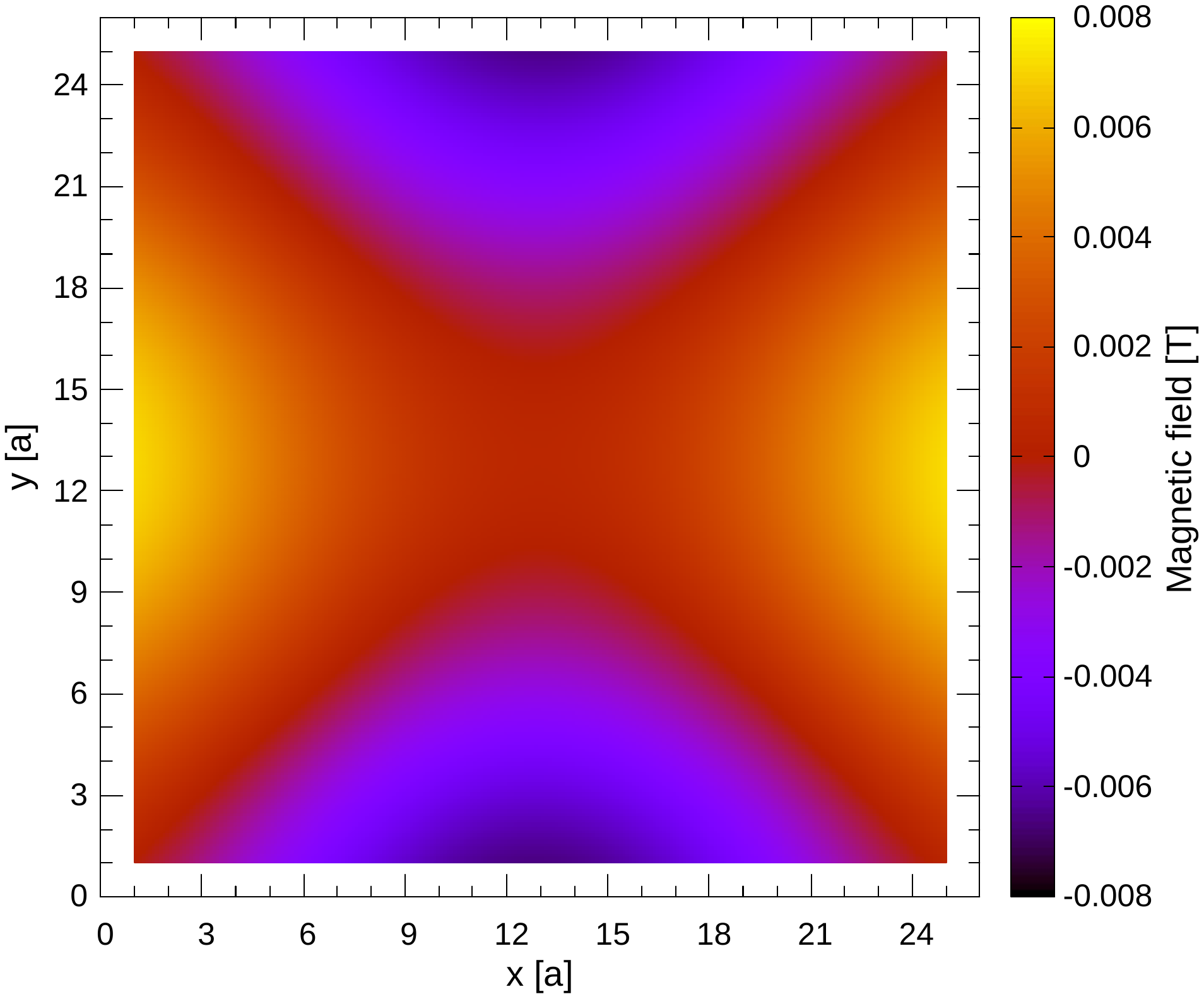}
      \caption{The surface layer contribution.}
      \label{fig:25_magnetic_mmmm_surface}
    \end{subfigure}
    \caption{The $25 \times 25$ lattice system with the current pattern around the central SVQ,  `mmmm' in Fig.~\ref{fig:result:1NI_16_current}, and the contour plot of $B_z$ at $z=10$ generated by it.}
    \label{fig:25_current_mmmm}
  \end{figure}

In order to increase the magnitude of the magnetic field generated by the SVILCs
we modify the current patterns. In Figs.~\ref{fig:25_current_amma}-\ref{fig:25_current_mmmm}, magnetic fields from different current patters are shown. 
In experiments, such different current patterns may be realized by applying a pulsed magnetic field; the electromotive force produced by the pulsed magnetic field 
may lead to the change of the current patters; and the obtained current may persist long enough for allowing the detection.

In Fig.~\ref{fig:25_current_amma}, the case where the SVILCs for the central SVQ is changed from `maam' to `amma' (see Fig.~\ref{fig:result:1NI_16_current} for the meaning of  `maam' and `amma')
is depicted.
The magnitude of the produced magnetic field is in the order of mT, thus, it may be detectable.

In Fig.~\ref{fig:25_current_mama}, the case where the SVILCs  for the central SVQ is changed from `maam' to `mama' is depiced; the magnitude of the produced magnetic field is in the order of 10 mT, thus, it will be detectable.
In Fig.~\ref{fig:25_current_mmma}, the case where the SVILCs for the central SVQ is changed from `maam' to `mmma'  is depicted; the magnitude of the produced magnetic field is in the order of 100 mT, thus, it will be detectable.
Lastly, in Fig.~\ref{fig:25_current_mmmm}, the case where the SVILCs for the central SVQ is changed from `maam' to `mmmm' is depicted; the magnitude of the produced magnetic field is in the order of 300 mT, thus, it will be easily detectable if this current pattern persist long enough for the detection.

We do not know the exact stability of the current patters above; however, they are protected by topological quantum numbers, thus, expected to be fairly stable.
The results here are very encouraging for the experimental confirmation of the existence of SVILCs.

  \section{SVILC qubit: One nano-island system}
  \label{subss:result:qubit_state}
  
     \begin{figure}[H]
    \centering
    \begin{subfigure}{0.48\columnwidth}
      \centering
      \includegraphics[width=4cm]{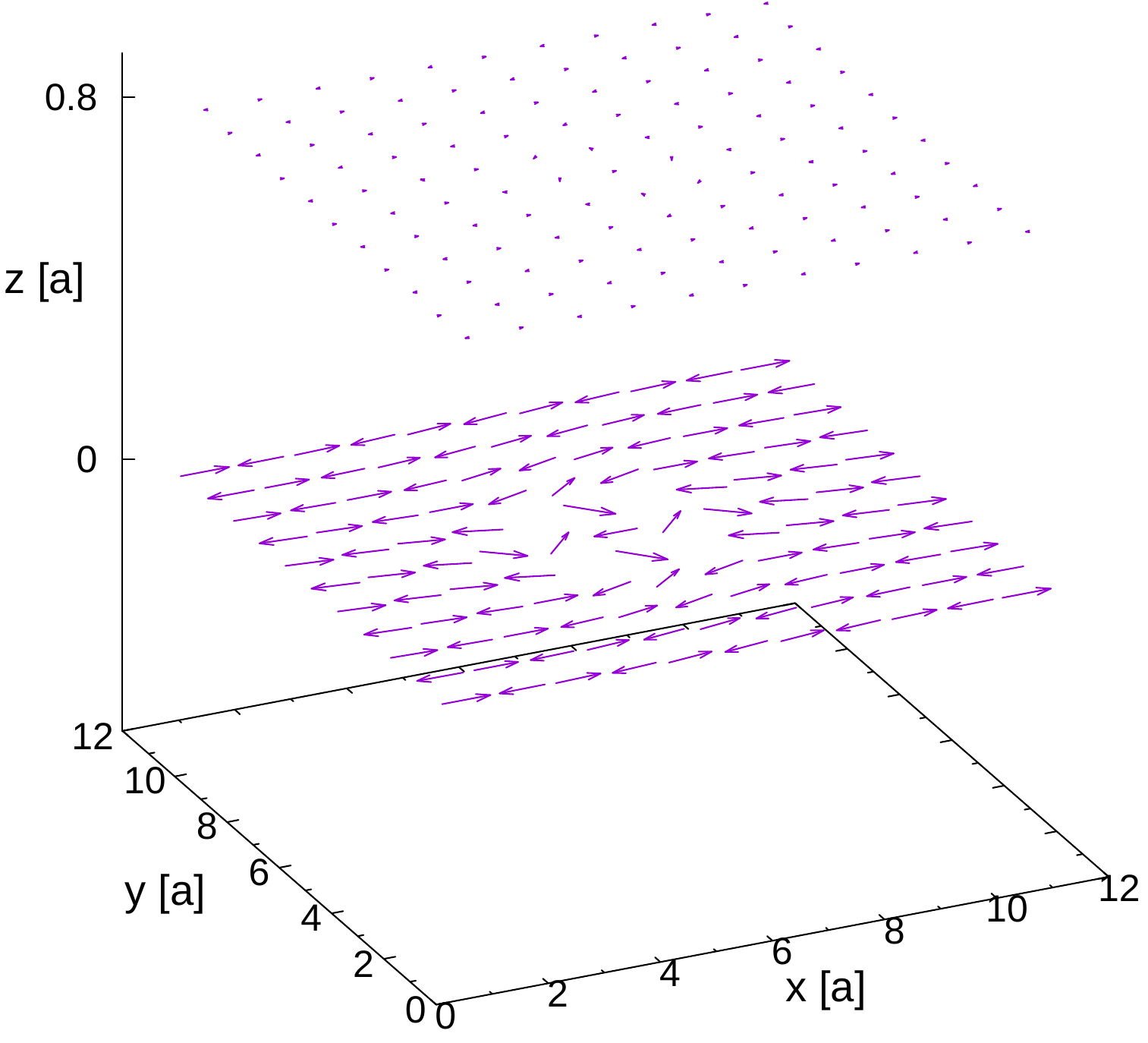}
      \caption{Spin moment distribution.}
      \label{fig:result:1NI_spin}
    \end{subfigure}
    \begin{subfigure}{0.48\columnwidth}
      \centering
      \includegraphics[width=4cm]{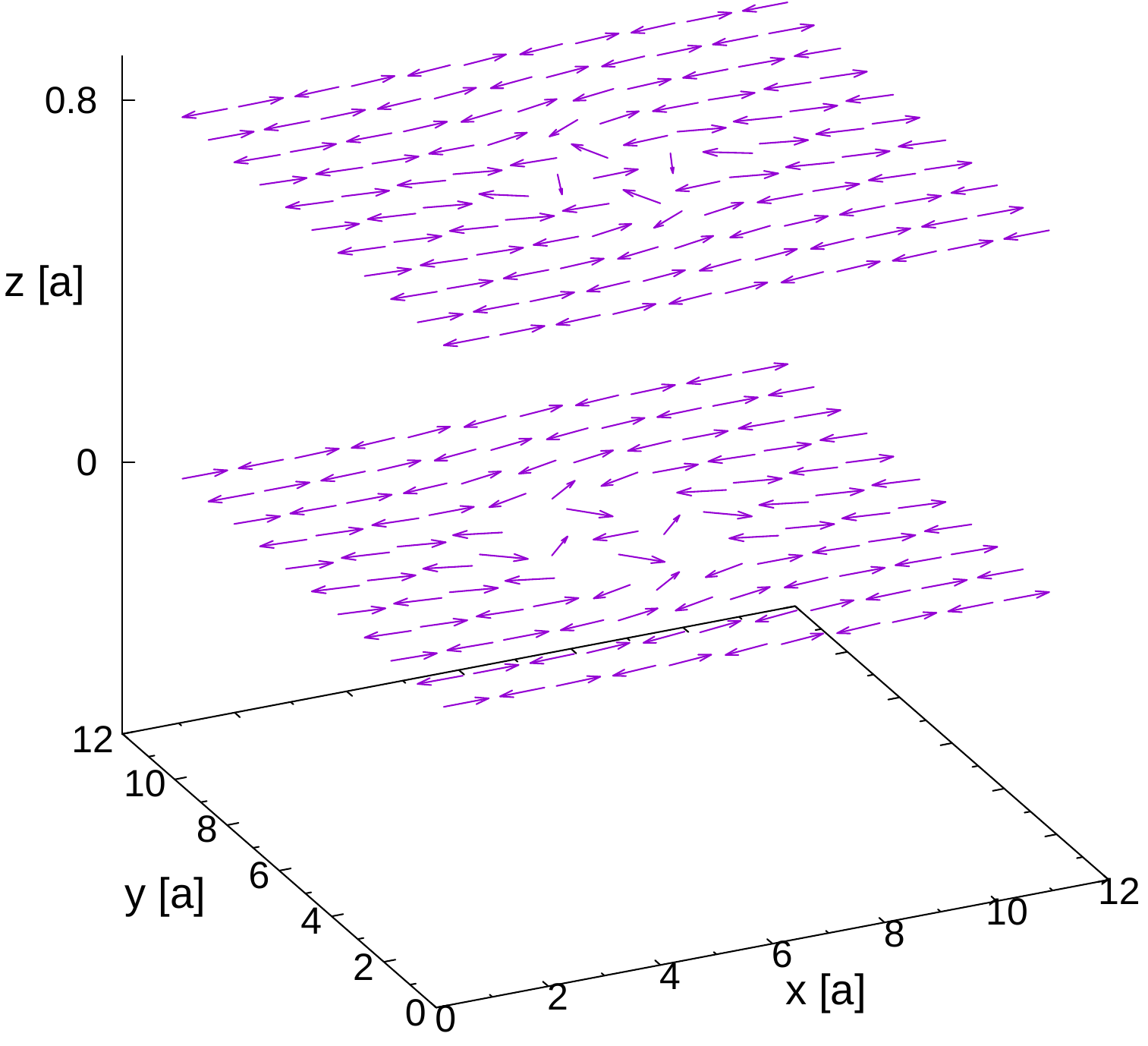}
      \caption{Spin moment distribution normalized for each layer.}
      \label{fig:result:1NI_spin_n}
    \end{subfigure}
    \begin{subfigure}{0.48\columnwidth}
      \centering
      \includegraphics[width=3cm]{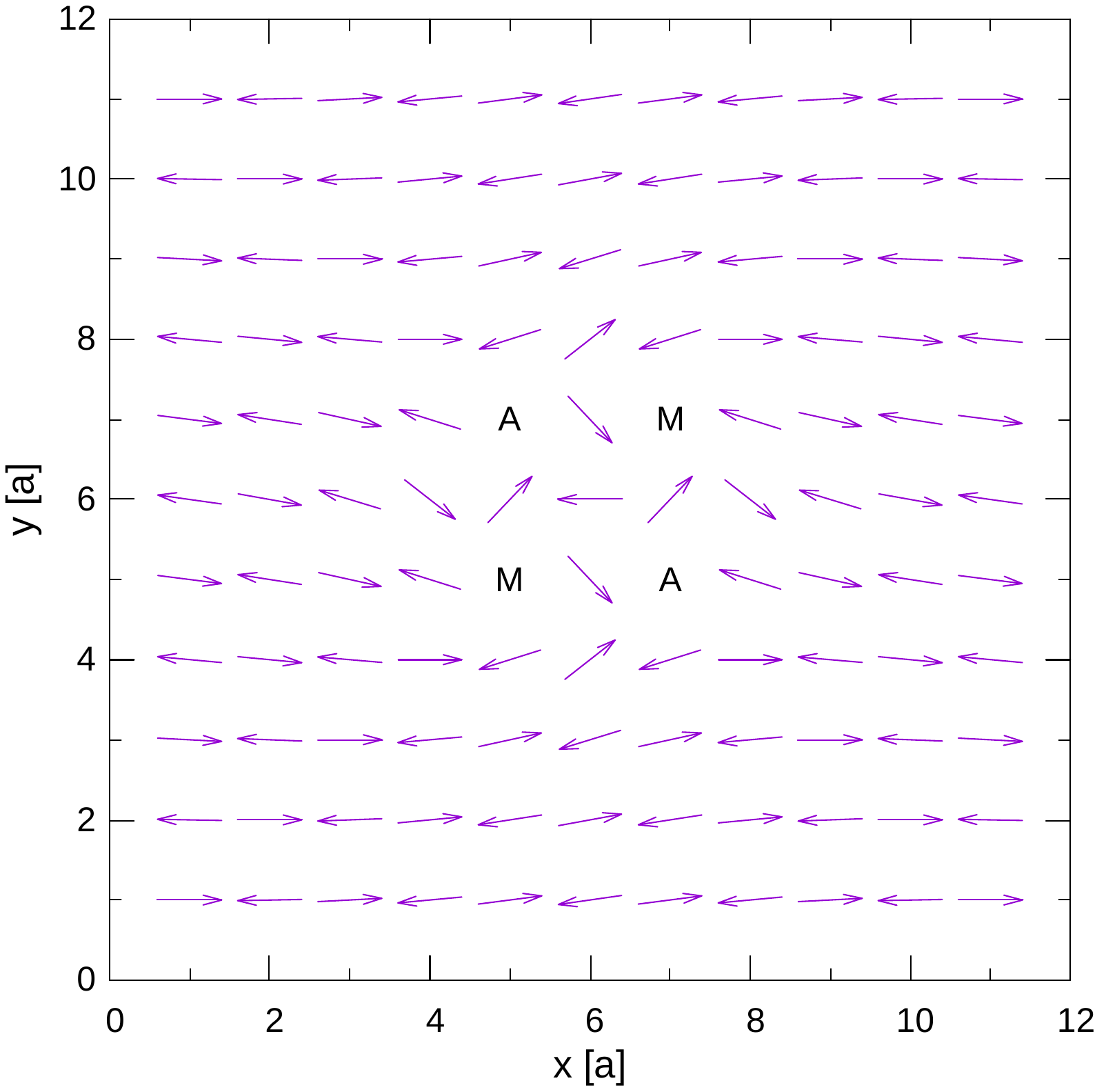}
      \caption{The normalized bulk layer distribution.}
      \label{fig:result:1NI_spin_bulk}
    \end{subfigure}
    \begin{subfigure}{0.48\columnwidth}
      \centering
      \includegraphics[width=3cm]{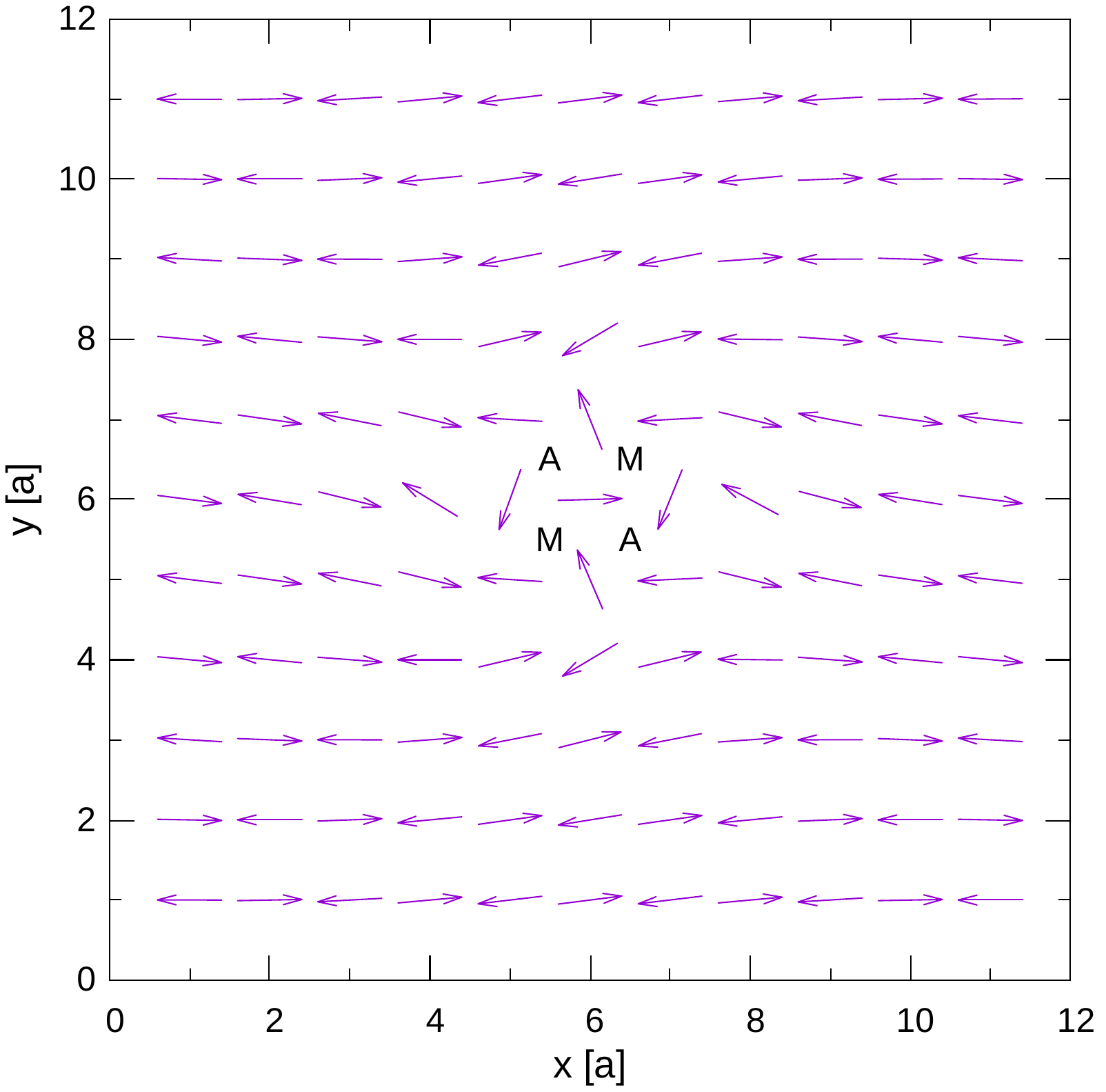}
      \caption{The normalized surface layer distribution.}
      \label{fig:result:1NI_spin_surface}
    \end{subfigure}
    \caption{The spin moment distribution for the one nano-island system.}
    \label{fig:result:1NI_spin_vortex}
  \end{figure}
  
\begin{figure}[H]
    \centering
    \begin{subfigure}{0.23\columnwidth}
      \centering
      \includegraphics[width=\columnwidth]{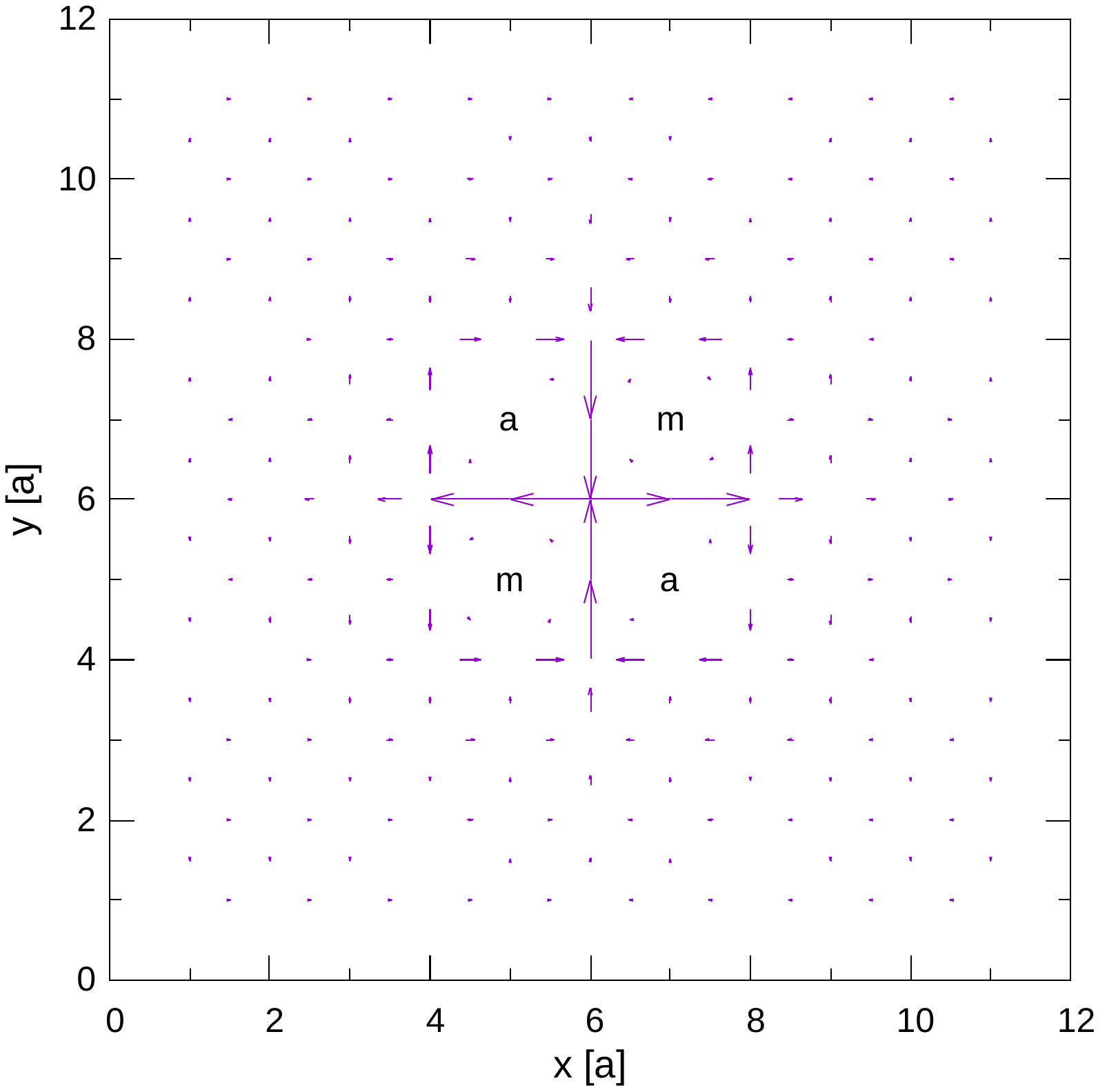}
      \caption{maam}
      \label{fig:result:1NI_maam}
    \end{subfigure}
    \hspace*{1mm}
    \begin{subfigure}{0.23\columnwidth}
      \centering
      \includegraphics[width=\columnwidth]{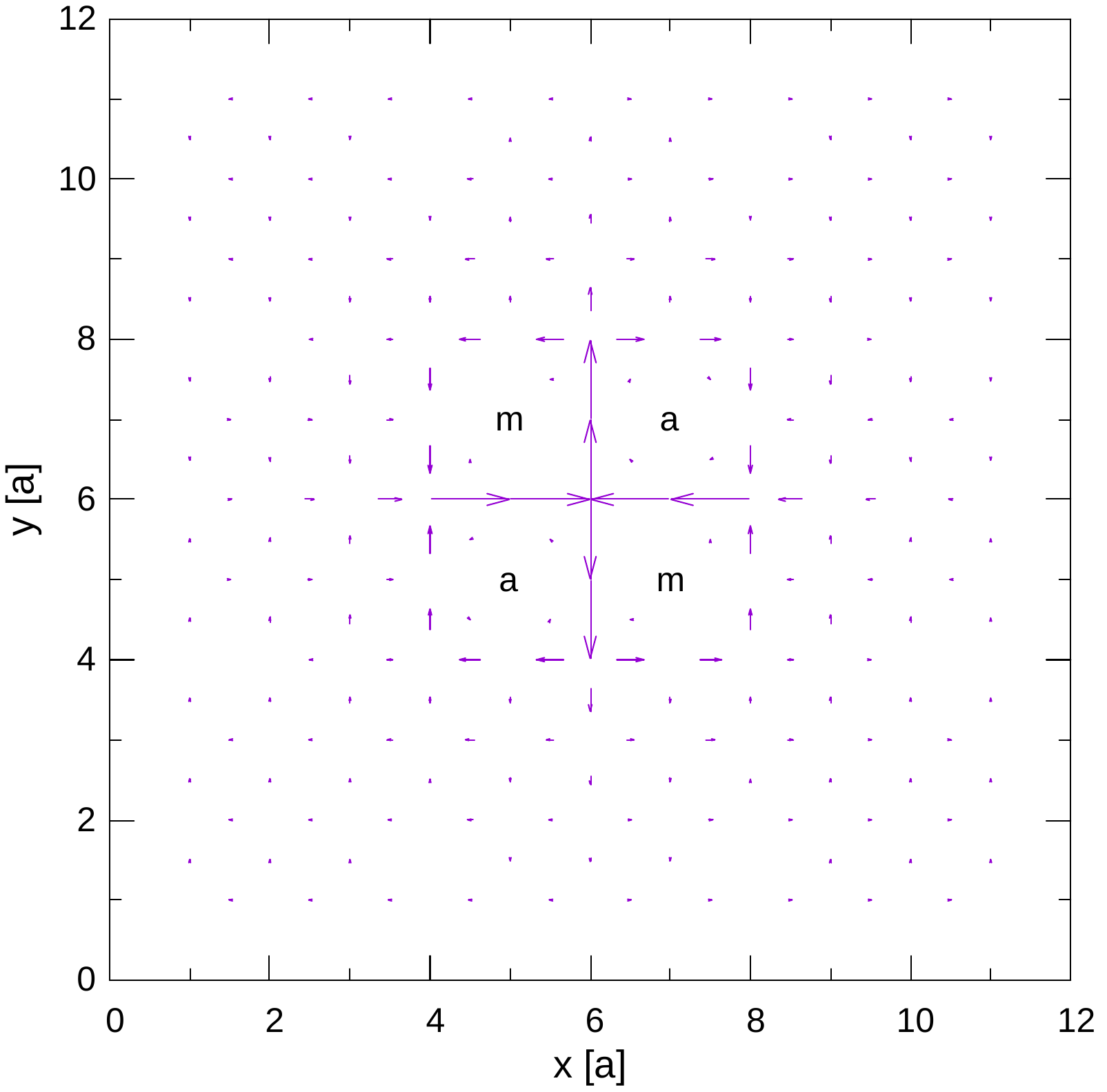}
      \caption{amma}
      \label{fig:result:1NI_amma}
    \end{subfigure}
    \hspace*{1mm}
    \begin{subfigure}{0.23\columnwidth}
      \centering
      \includegraphics[width=\columnwidth]{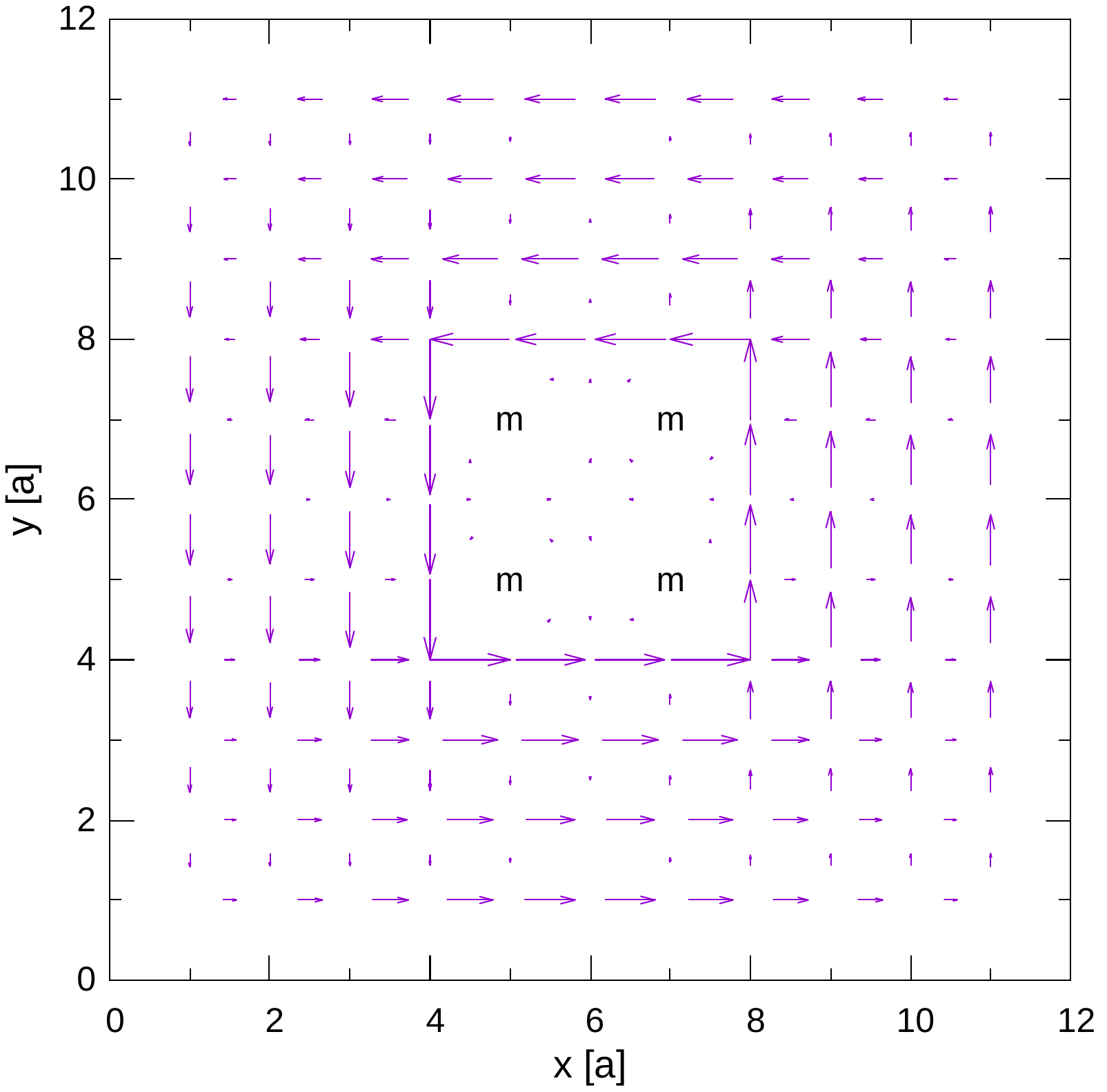}
      \caption{mmmm}
      \label{fig:result:1NI_mmmm}
    \end{subfigure}
    \hspace*{1mm}
    \begin{subfigure}{0.23\columnwidth}
      \centering
      \includegraphics[width=\columnwidth]{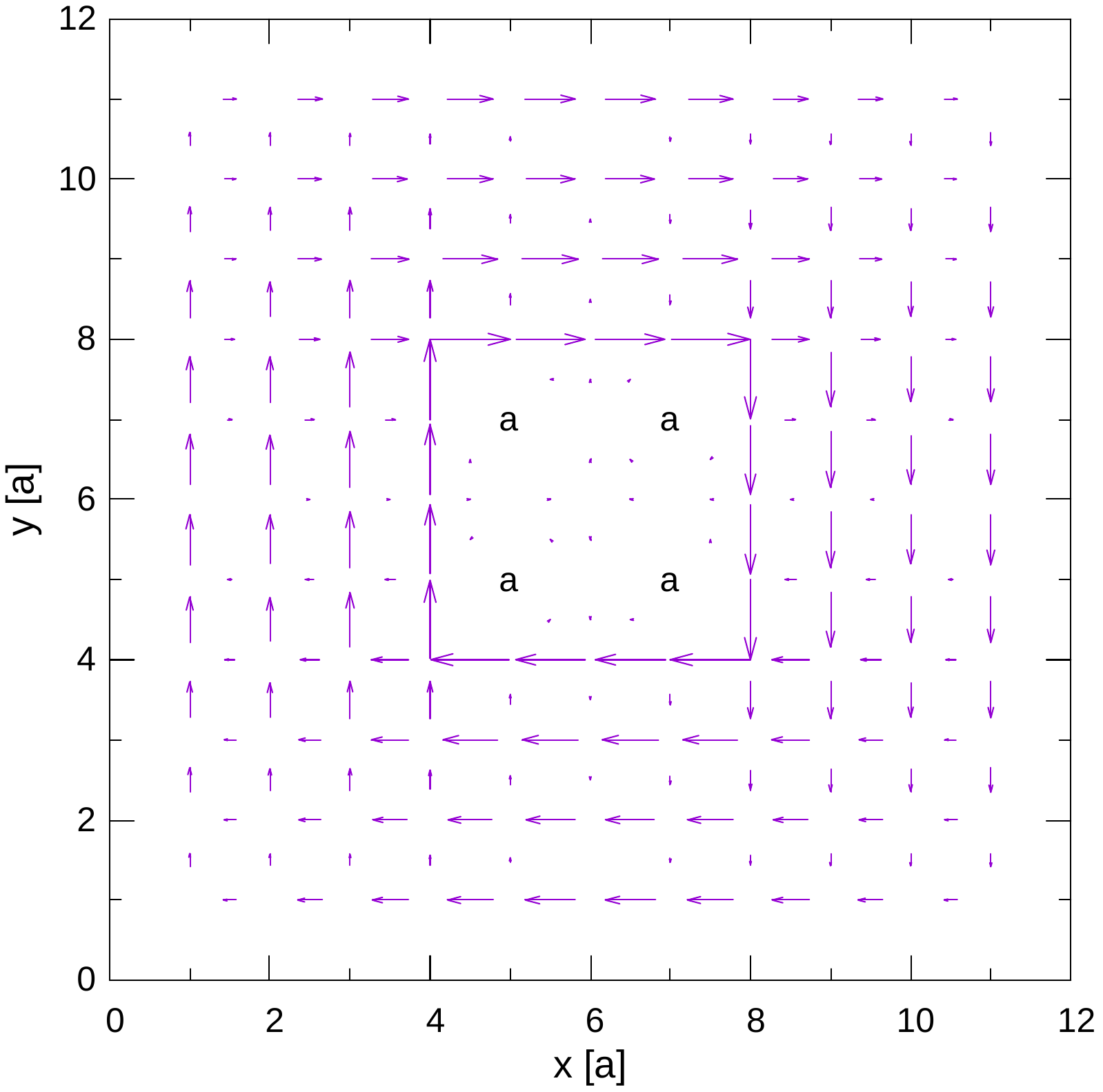}
      \caption{aaaa}
      \label{fig:result:1NI_aaaa}
    \end{subfigure}
    \begin{subfigure}{0.23\columnwidth}
      \centering
      \includegraphics[width=\columnwidth]{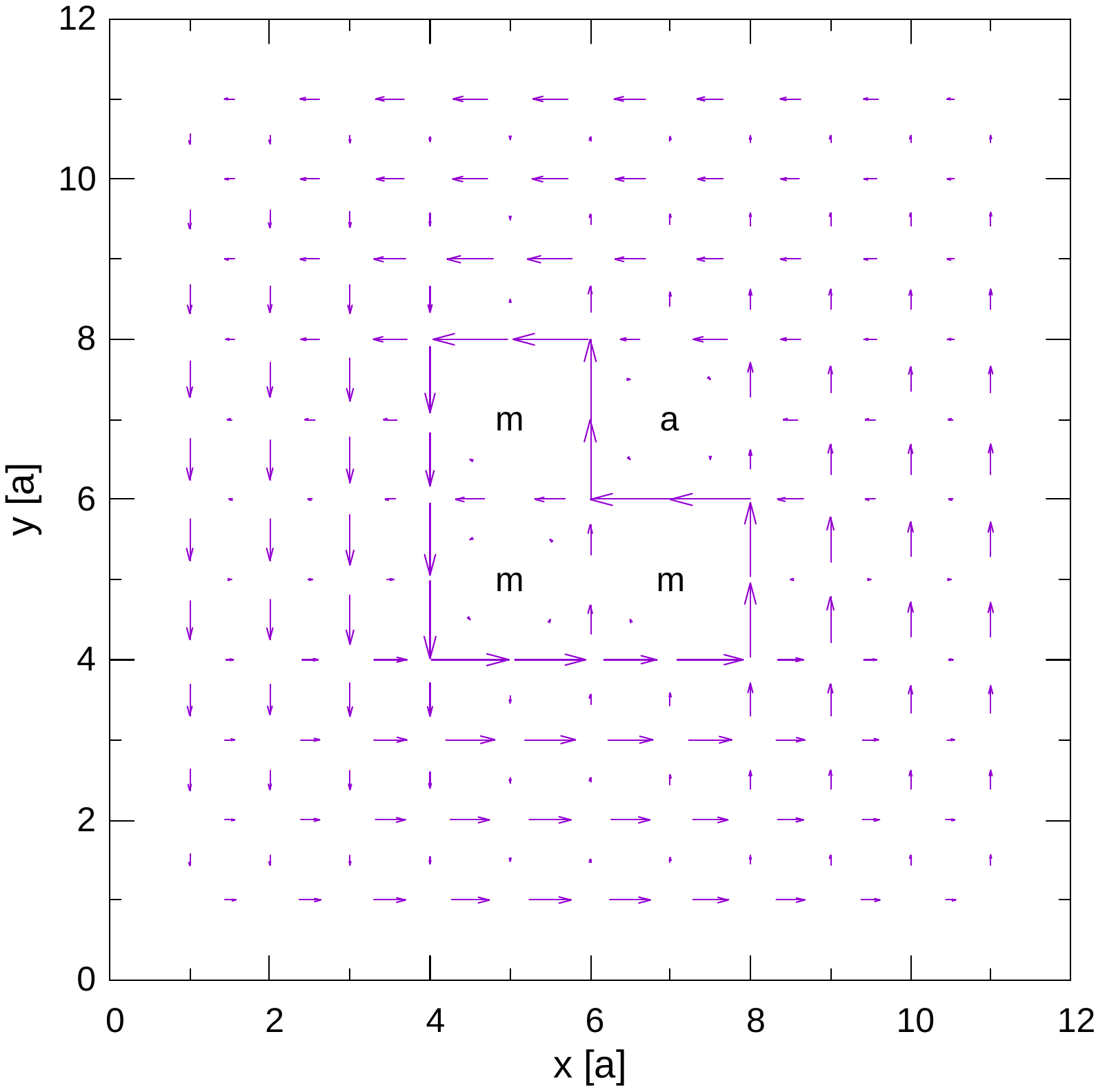}
      \caption{mmma}
      \label{fig:result:1NI_mmma}
    \end{subfigure}
    \hspace*{1mm}
    \begin{subfigure}{0.23\columnwidth}
      \centering
      \includegraphics[width=\columnwidth]{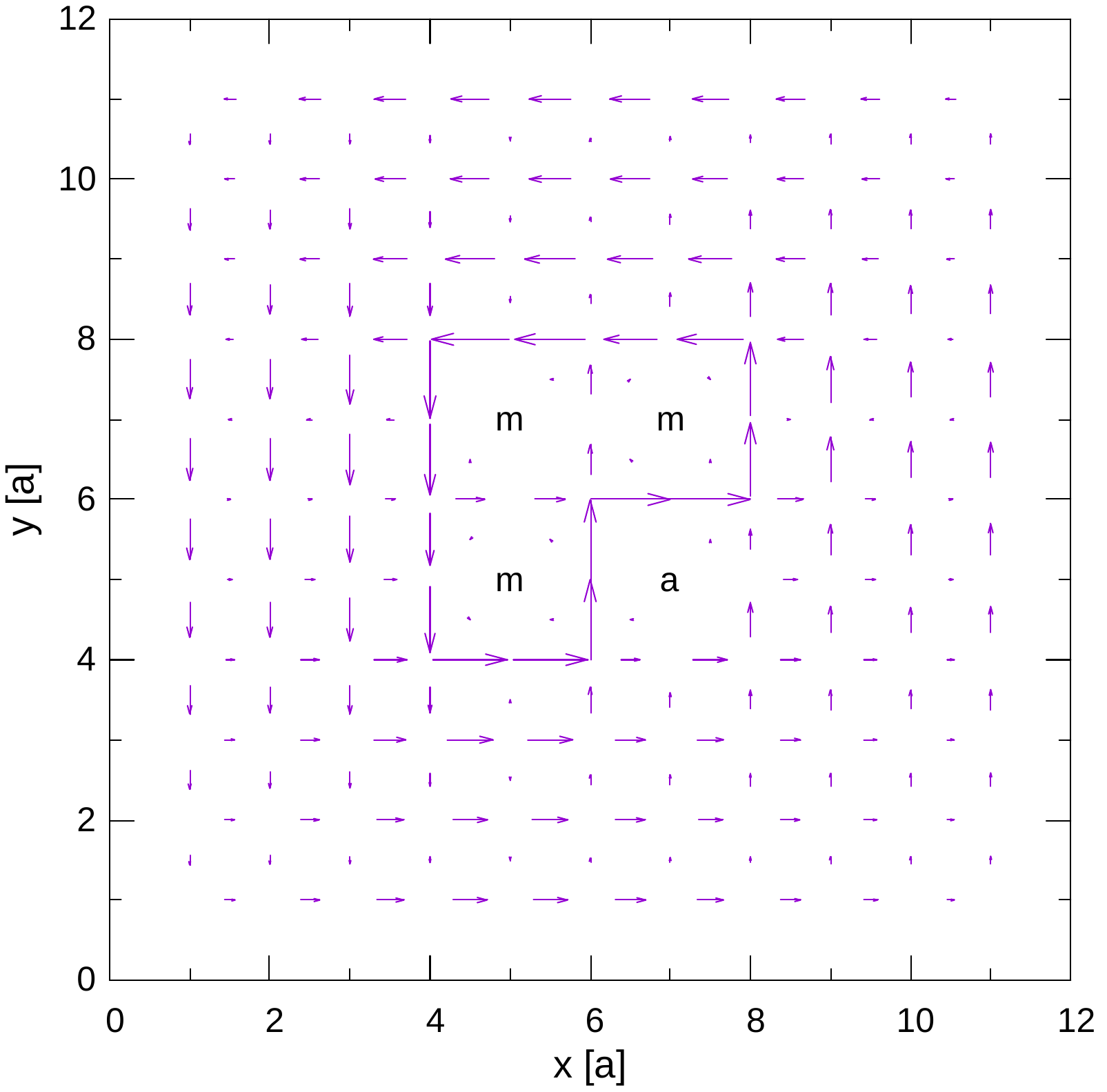}
      \caption{mmam}
      \label{fig:result:1NI_mmam}
    \end{subfigure}
    \hspace*{1mm}
    \begin{subfigure}{0.23\columnwidth}
      \centering
      \includegraphics[width=\columnwidth]{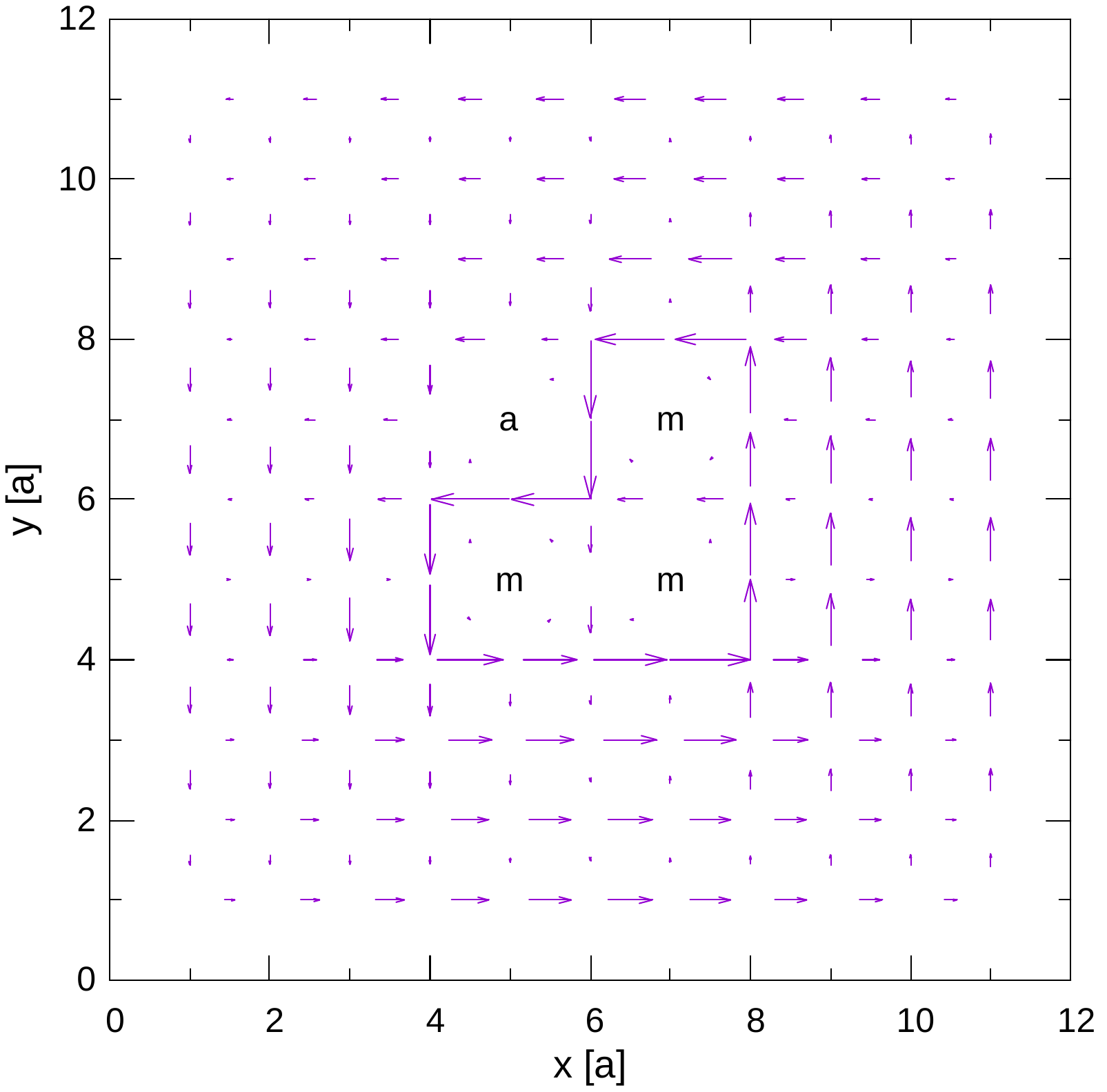}
      \caption{mamm}
      \label{fig:result:1NI_mamm}
    \end{subfigure}
    \hspace*{1mm}
    \begin{subfigure}{0.23\columnwidth}
      \centering
      \includegraphics[width=\columnwidth]{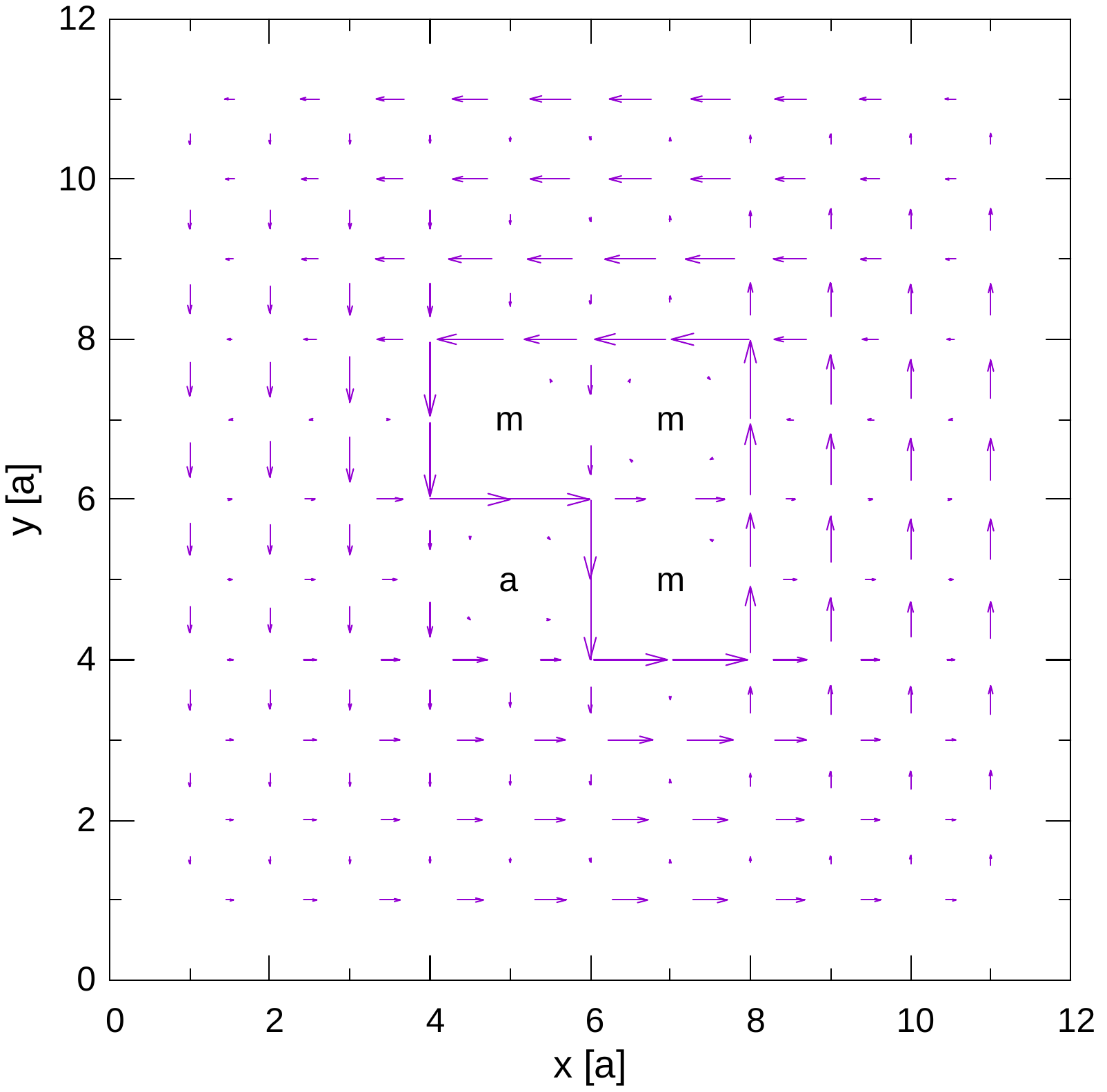}
      \caption{ammm}
      \label{fig:result:1NI_ammm}
    \end{subfigure}
    \begin{subfigure}{0.23\columnwidth}
      \centering
      \includegraphics[width=\columnwidth]{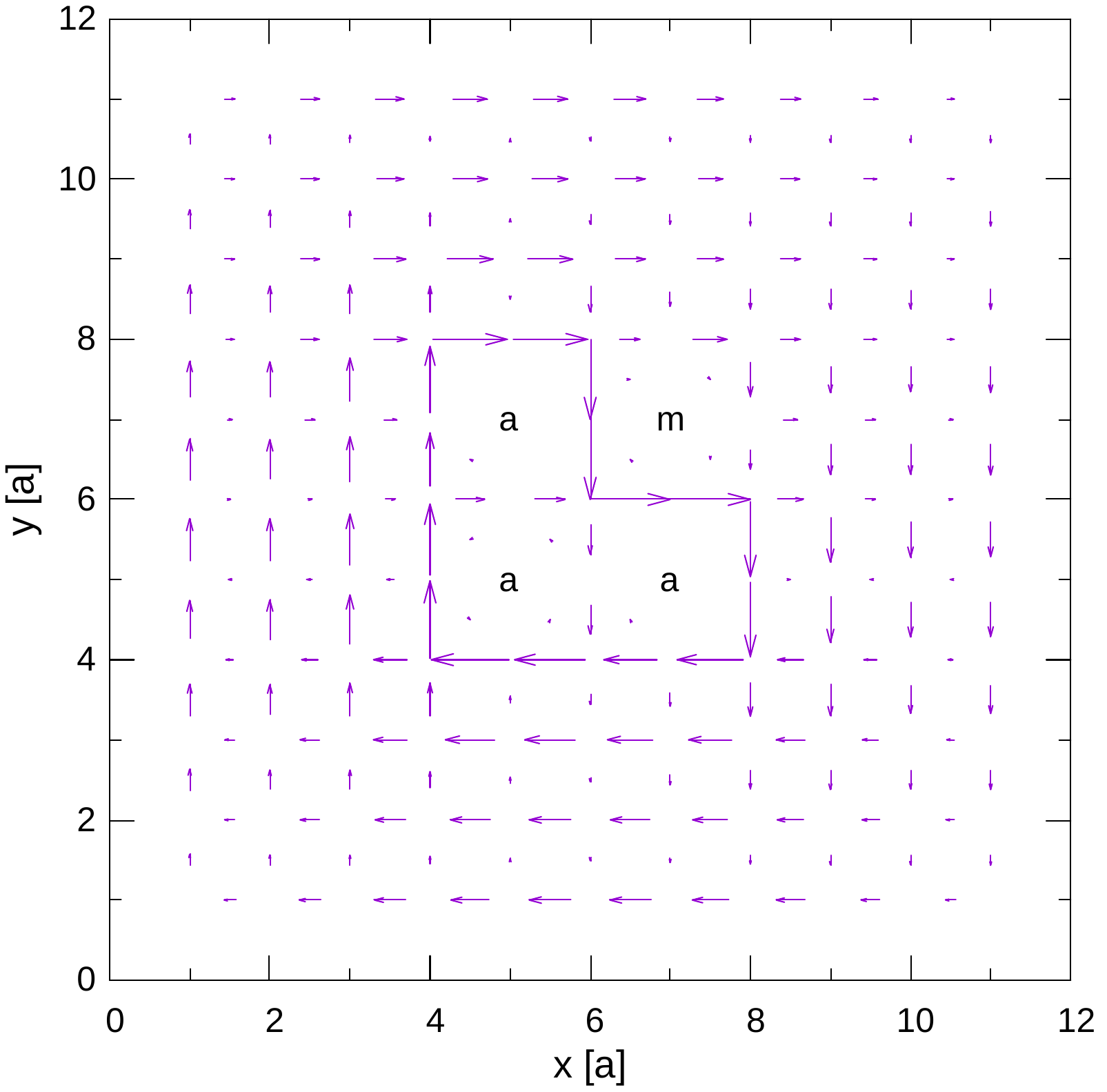}
      \caption{aaam}
      \label{fig:result:1NI_aaam}
    \end{subfigure}
    \hspace*{1mm}
    \begin{subfigure}{0.23\columnwidth}
      \centering
      \includegraphics[width=\columnwidth]{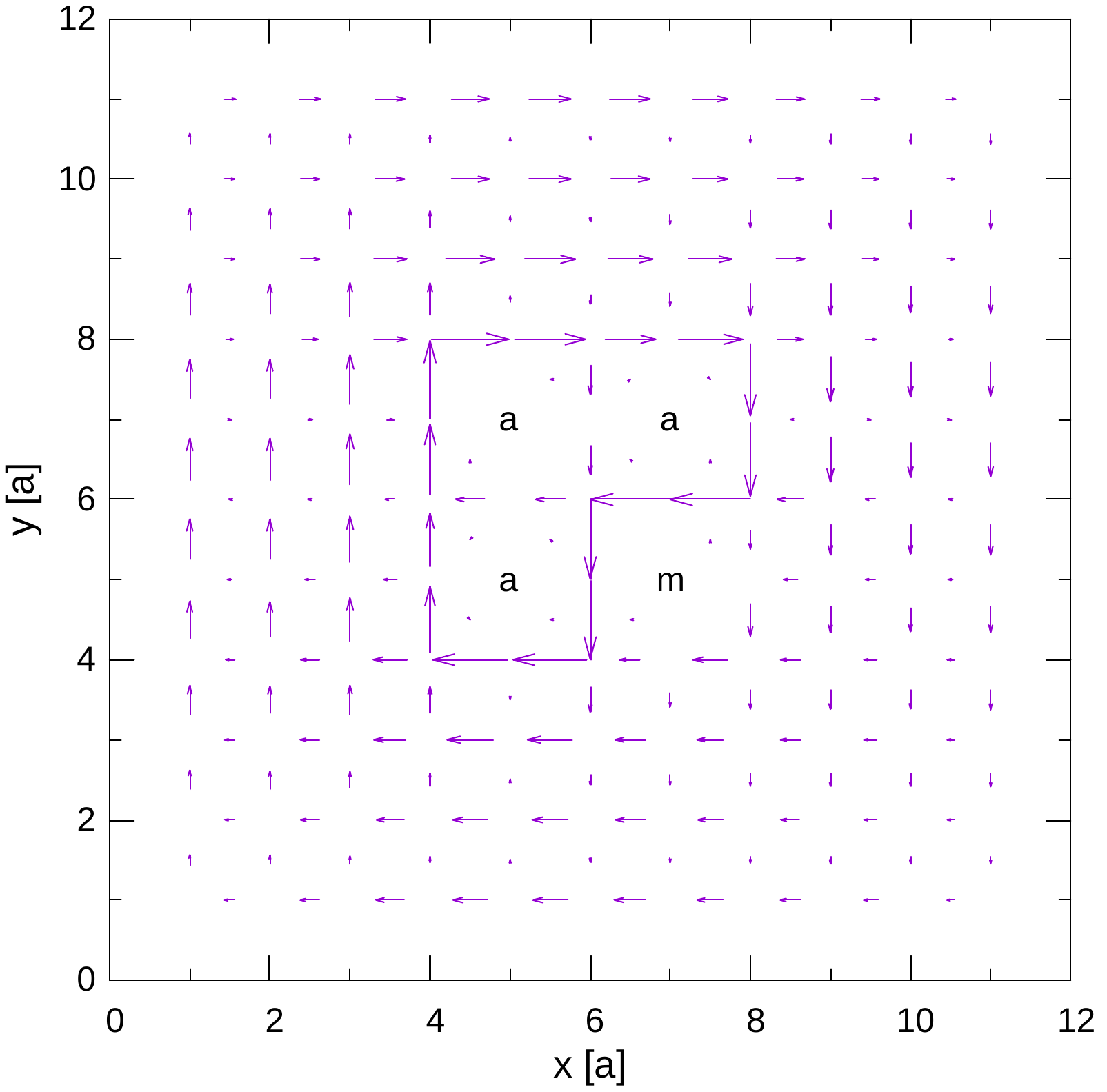}
      \caption{aama}
      \label{fig:result:1NI_aama}
    \end{subfigure}
    \hspace*{1mm}
    \begin{subfigure}{0.23\columnwidth}
      \centering
      \includegraphics[width=\columnwidth]{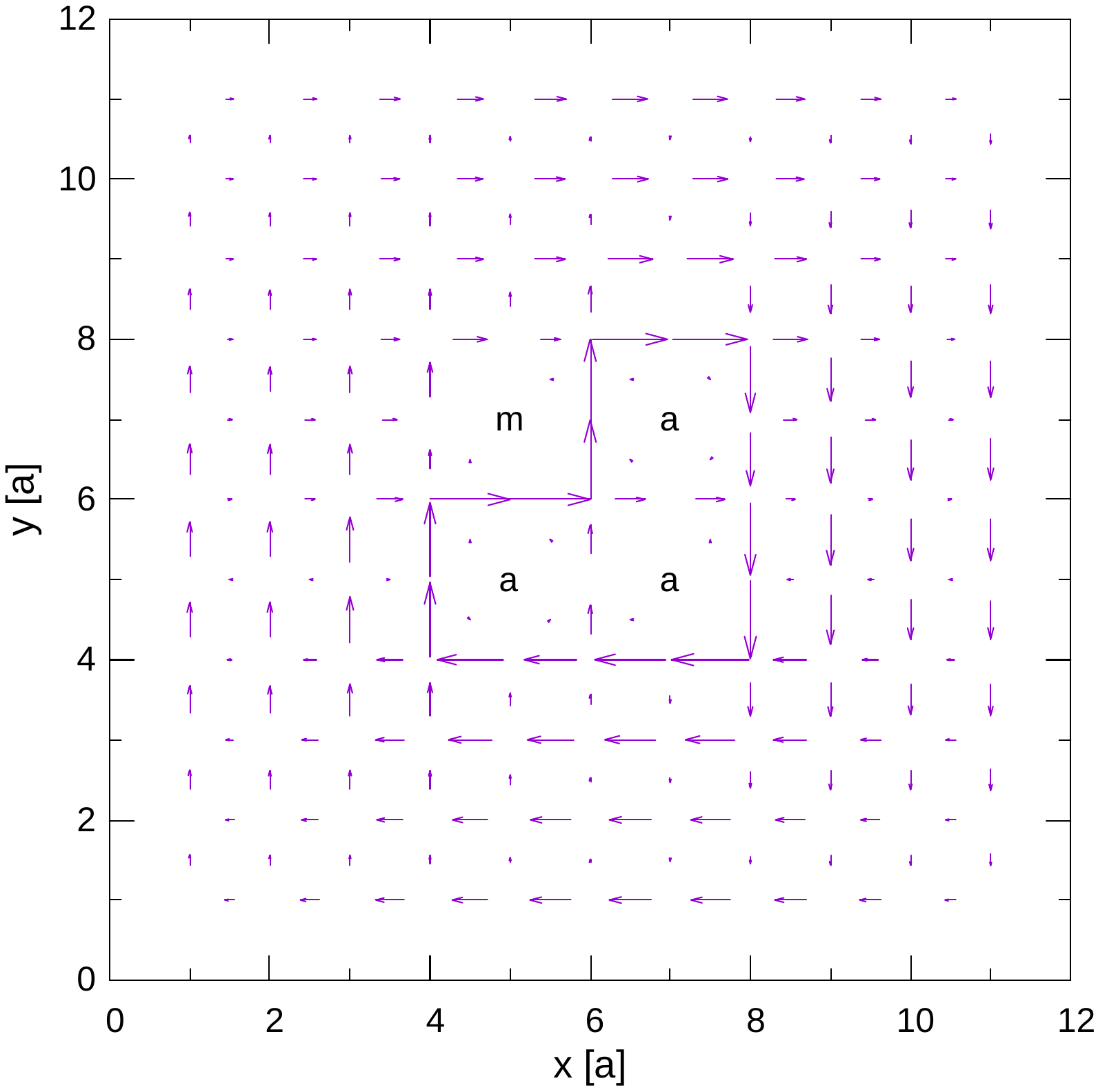}
      \caption{amaa}
      \label{fig:result:1NI_amaa}
    \end{subfigure}
    \hspace*{1mm}
    \begin{subfigure}{0.23\columnwidth}
      \centering
      \includegraphics[width=\columnwidth]{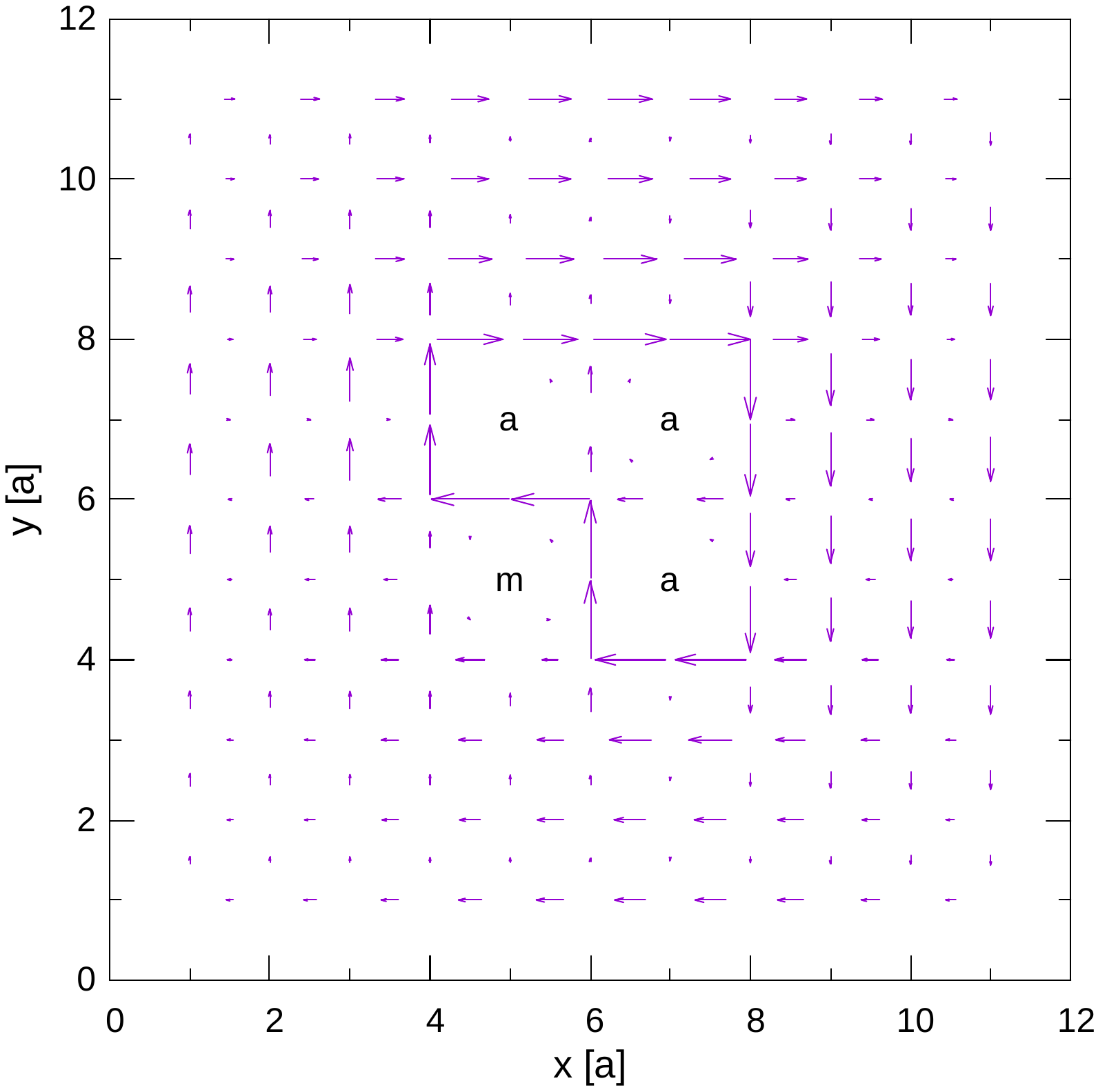}
      \caption{maaa}
      \label{fig:result:1NI_maaa}
    \end{subfigure}
    \begin{subfigure}{0.23\columnwidth}
      \centering
      \includegraphics[width=\columnwidth]{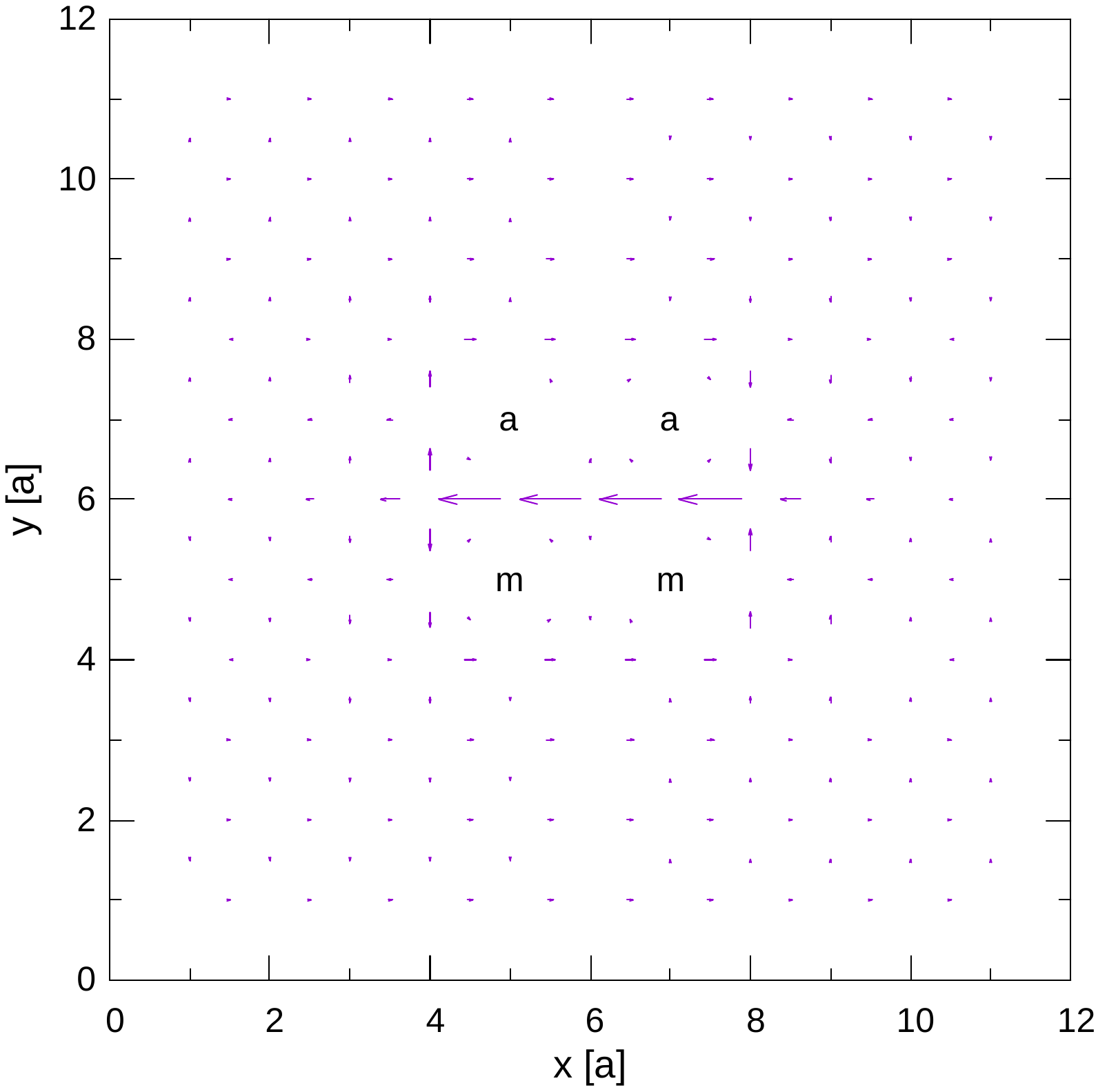}
      \caption{mama (A)}
      \label{fig:result:1NI_mama}
    \end{subfigure}
    \hspace*{1mm}
    \begin{subfigure}{0.23\columnwidth}
      \centering
      \includegraphics[width=\columnwidth]{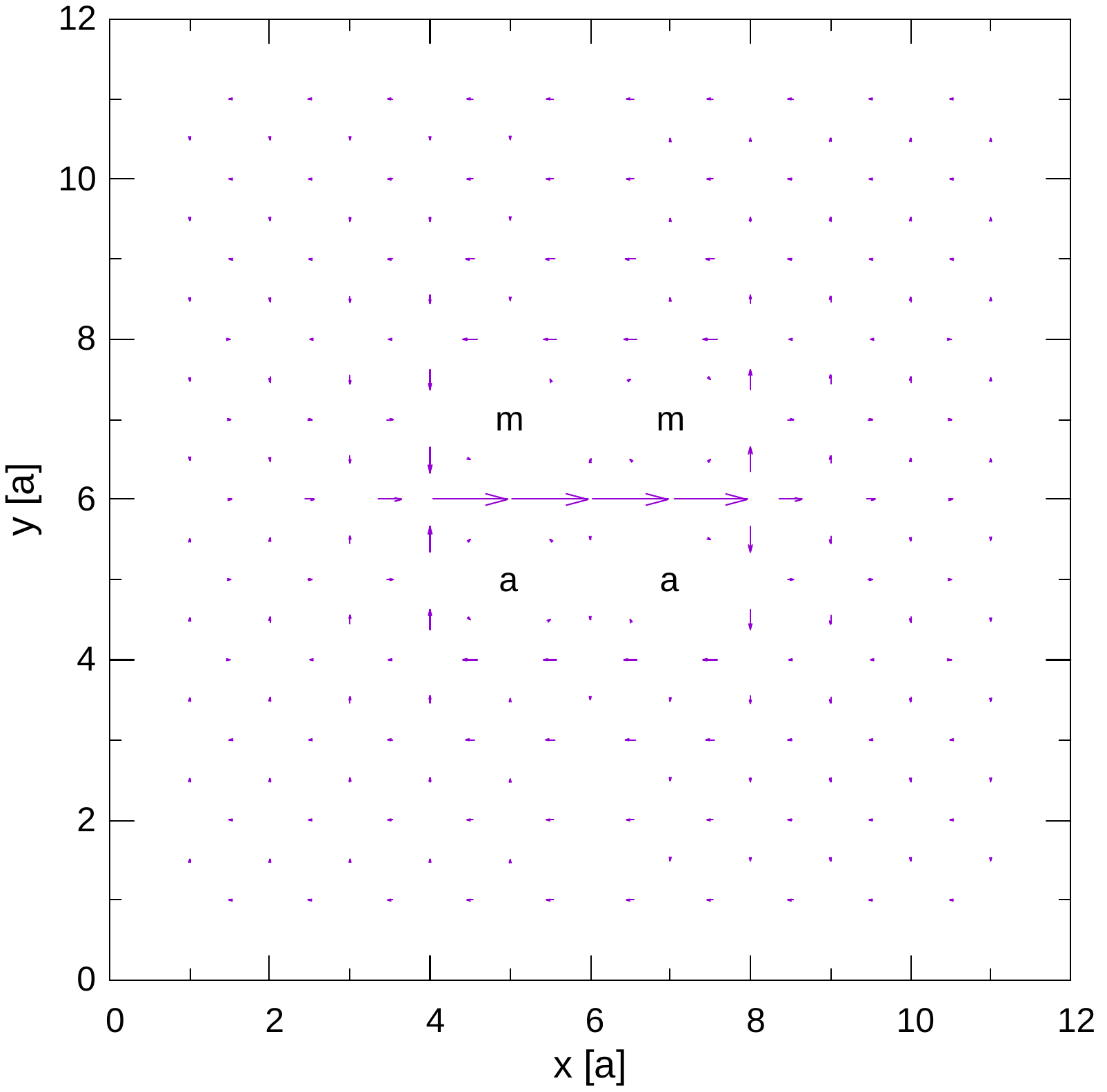}
      \caption{amam (B)}
      \label{fig:result:1NI_amam}
    \end{subfigure}
    \hspace*{1mm}
    \begin{subfigure}{0.23\columnwidth}
      \centering
      \includegraphics[width=\columnwidth]{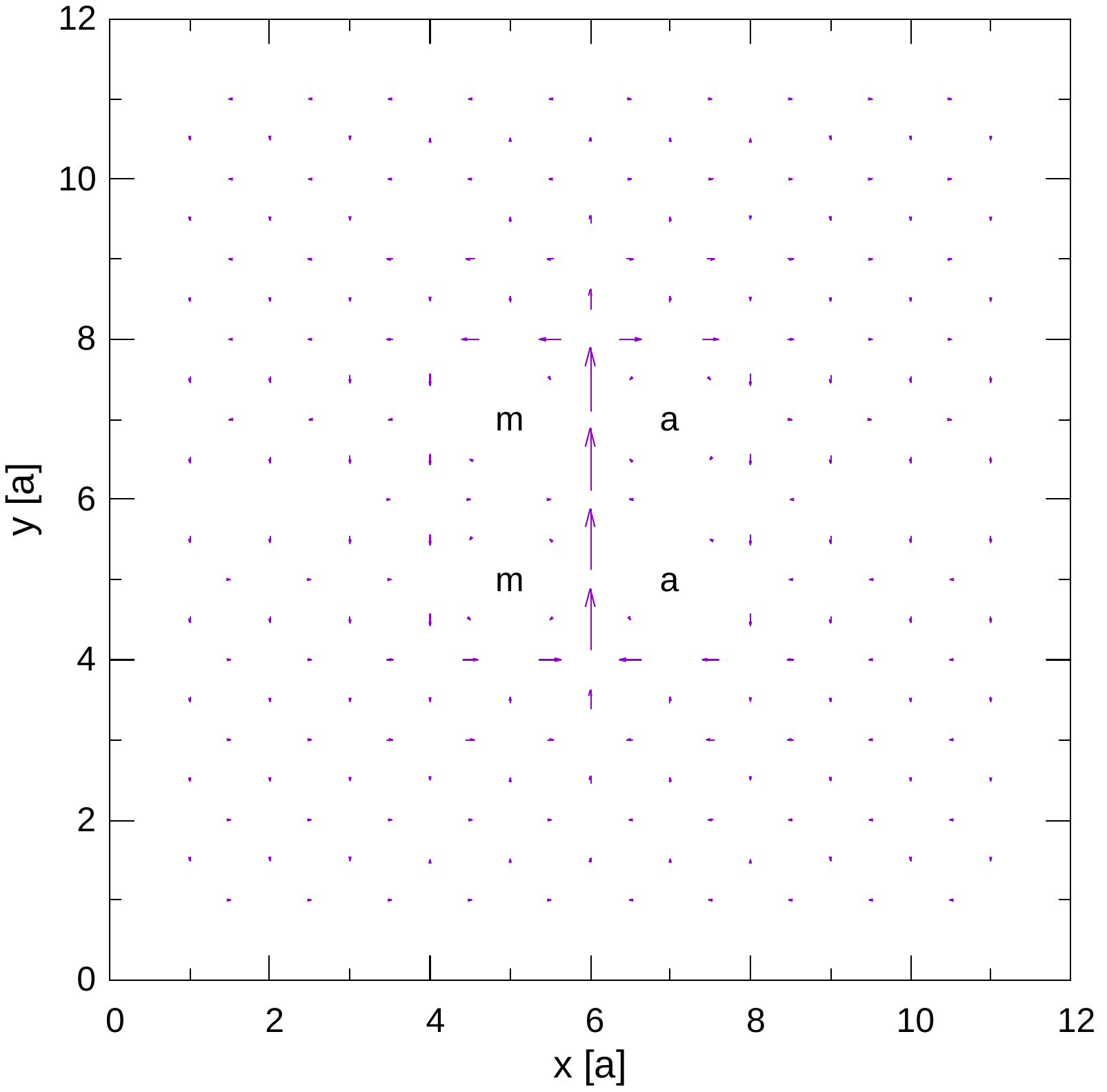}
      \caption{mmaa (C)}
      \label{fig:result:1NI_mmaa}
    \end{subfigure}
    \hspace*{1mm}
    \begin{subfigure}{0.23\columnwidth}
      \centering
      \includegraphics[width=\columnwidth]{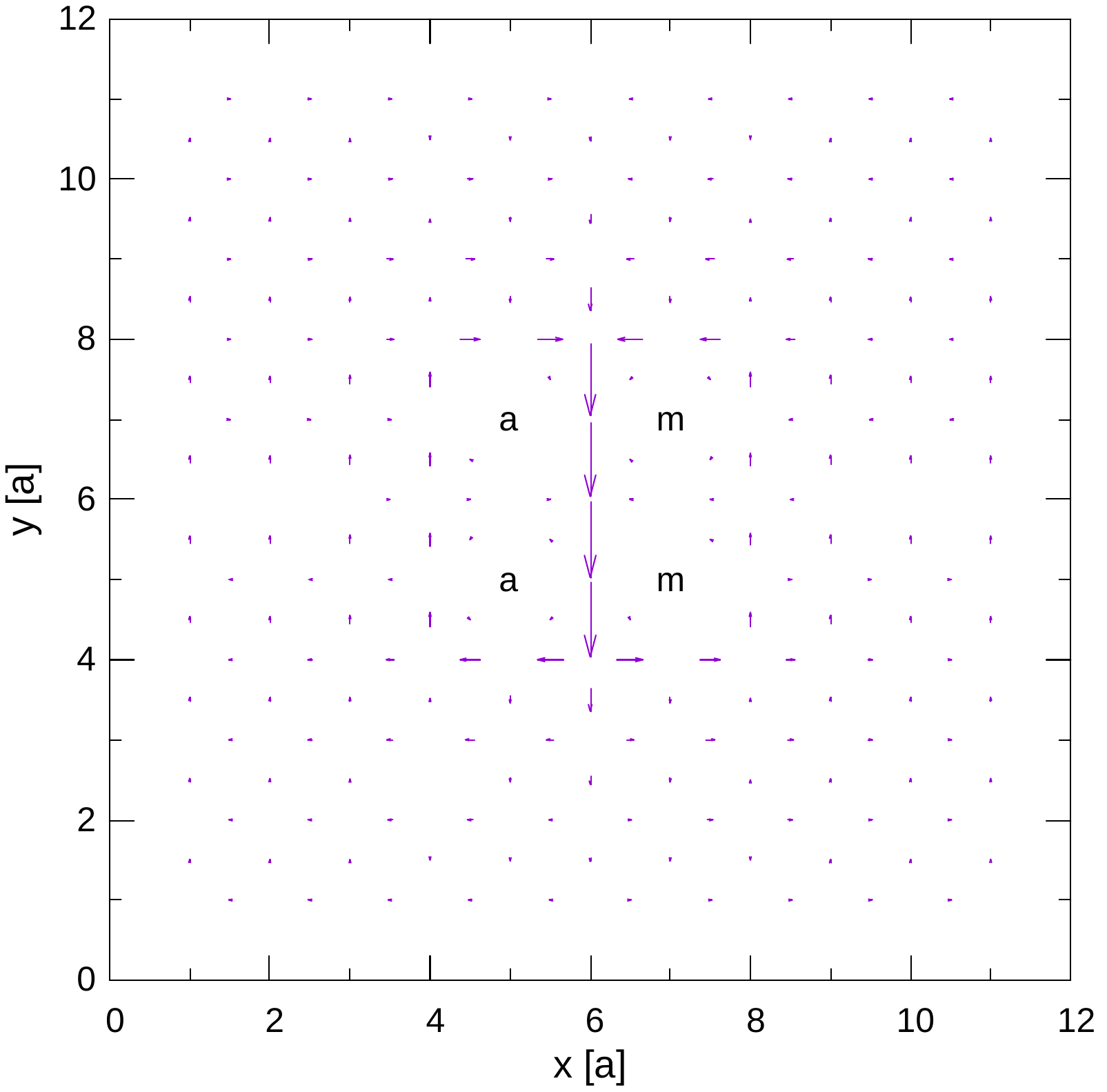}
      \caption{aamm (D)}
      \label{fig:result:1NI_aamm}
    \end{subfigure}
    \caption{Sixteen current patterns for one SVILC qubit system depicted in
     Fig.~\ref{fig:result:1NI_spin_vortex}. Each current pattern label indicates
    the winding numbers for the four loop currents from the left-down $\rightarrow$ left-up
    $\rightarrow$ right-down   $\rightarrow$ right-up; for example `maam' in (a) indicates
     $+1$   $\rightarrow$ $-1$ $\rightarrow$ $-1$   $\rightarrow$ $+1$. 
     (m), (n), (o), and (p) are used as qubit states denoted as A, B, C, and D, respectively. }
    \label{fig:result:1NI_16_current}
  \end{figure}

  Let us consider the qubit application of the SVILCs. We consider the $11 \times 11$ nano-island equipped with one SVQ as shown in Fig.~\ref{fig:result:1NI_spin_vortex}.
  There are 16 current patterns available as depicted in Fig.~\ref{fig:result:1NI_16_current} in this system.

\begin{figure}[H]
    \begin{subfigure}{0.9\columnwidth}
      \centering
   \includegraphics[width=4.0cm]{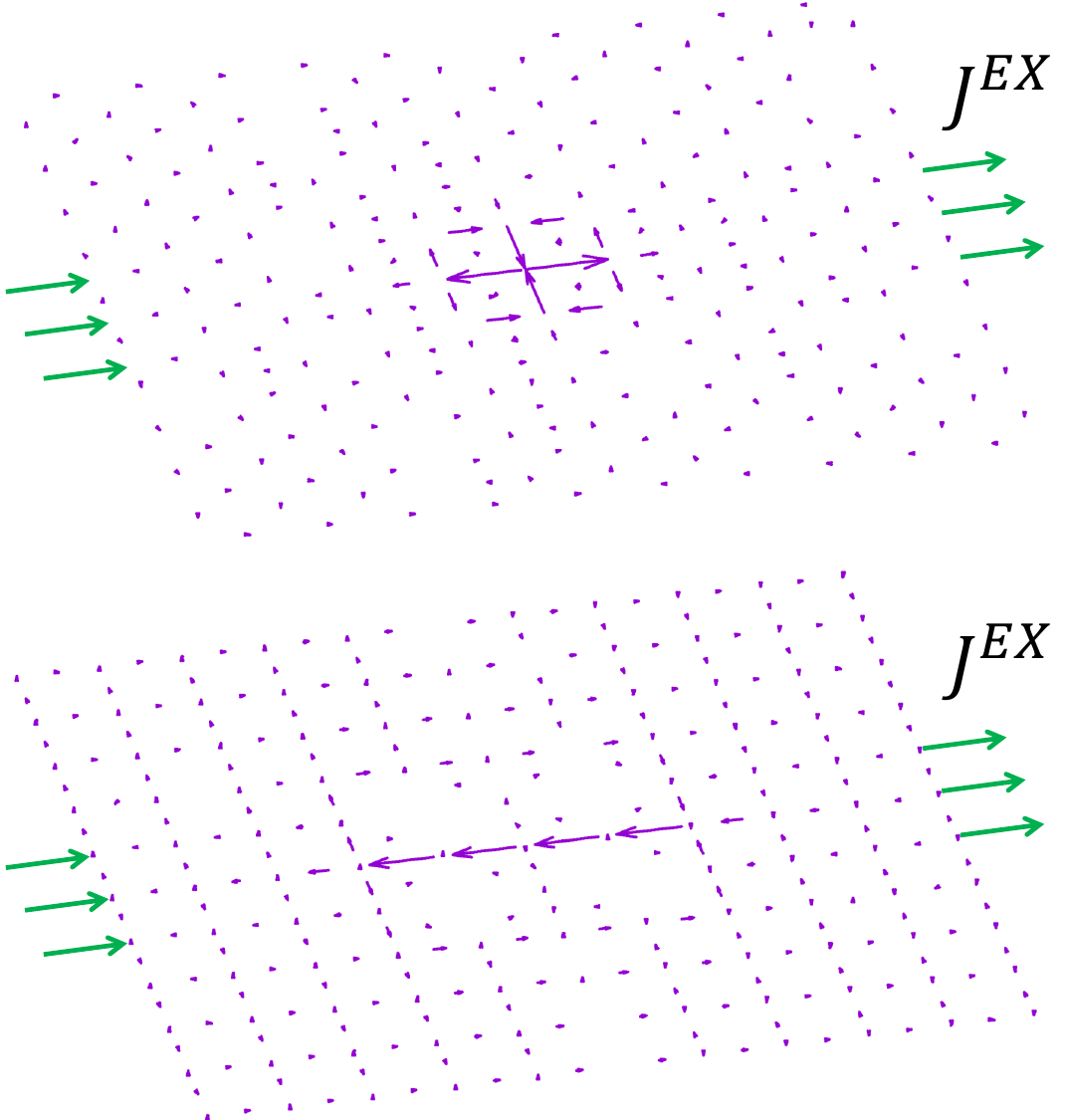}
    \caption{The external feeding current $J^{\rm EX}$.}
    \label{fig:result:1NI_external_current_x}
       \end{subfigure}
         \hspace*{1mm}
    \centering
    \begin{subfigure}{0.48\columnwidth}
      \centering
      \includegraphics[width=4cm]{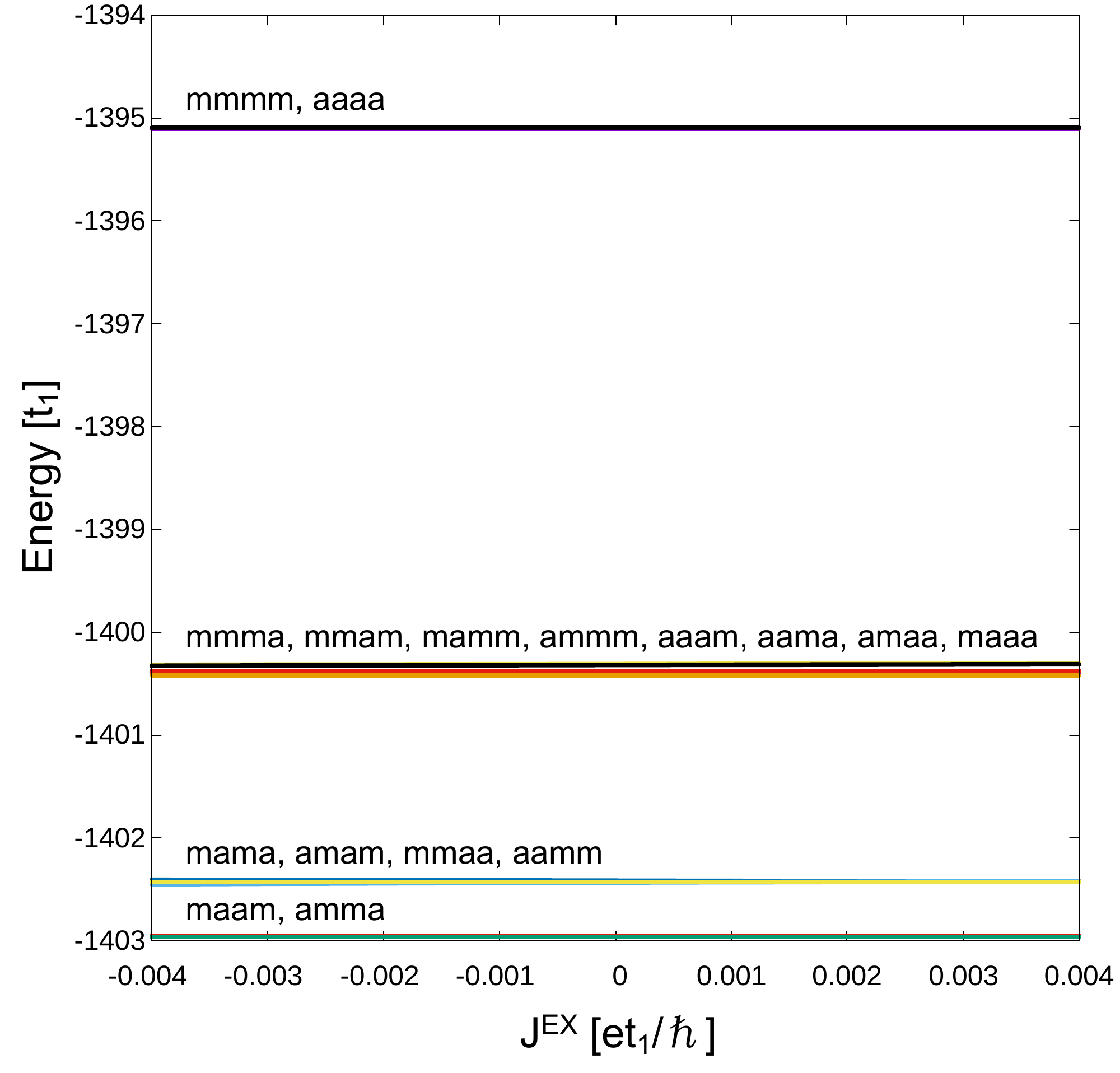}
      \caption{$J^{\rm EX}$ dependence of the energies of the all sixteen states.}
      \label{fig:result:1NI_16}
    \end{subfigure}
    \hspace*{1mm}
    \begin{subfigure}{0.48\columnwidth}
      \centering
      \includegraphics[width=4cm]{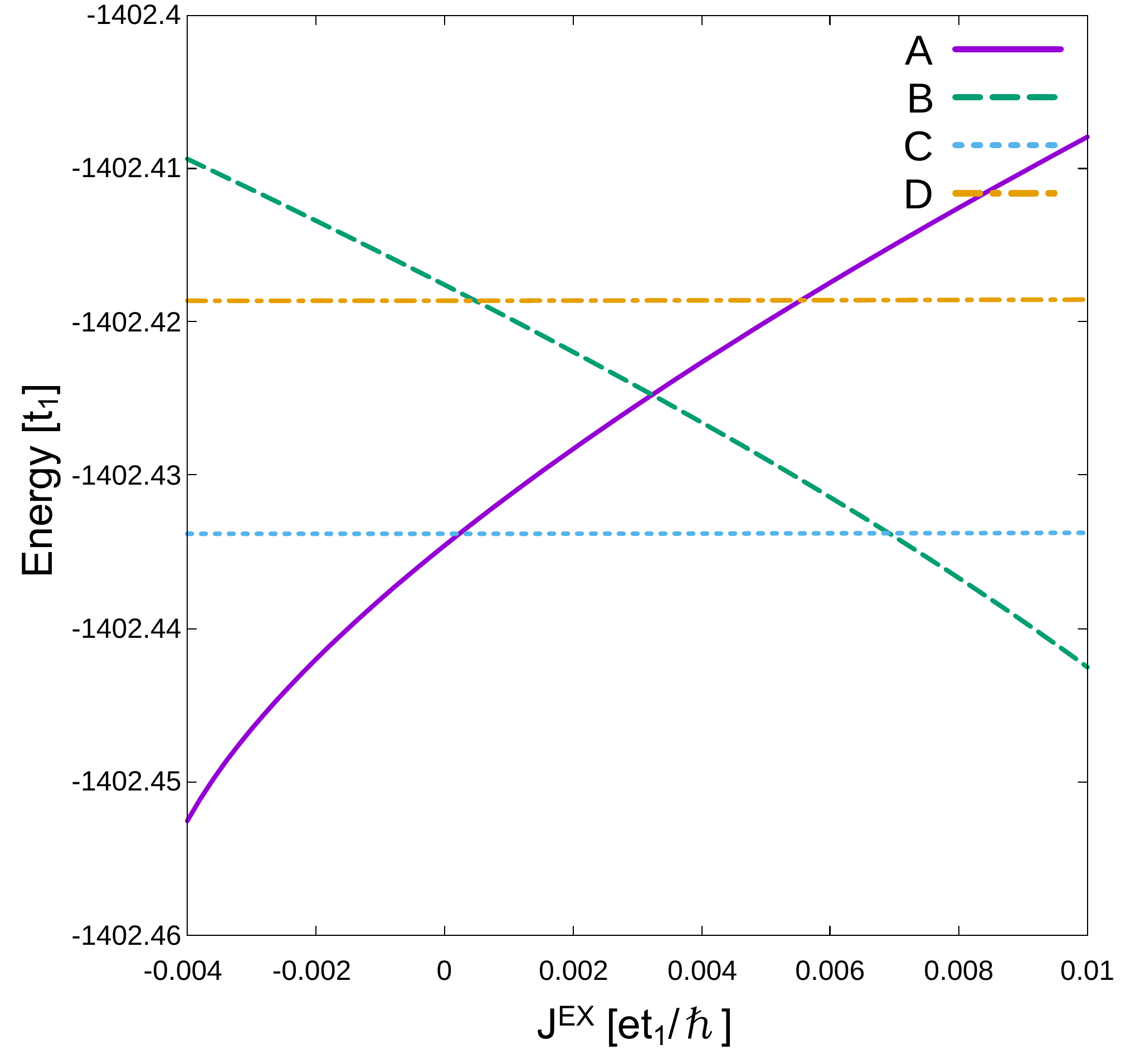}
      \caption{$J^{\rm EX}$ dependence of the energies of A, B, C, and D  states (corresponding to `mama', `amam', `mmaa', and `aamm', respectively).}
      \label{fig:result:1NI_ABCD}
    \end{subfigure}
    \caption{The change of the 16 energy levels by the external current $J^{\rm EX}$.}
    \label{fig:result:1NI_external_current_energy}
  \end{figure}

We consider the control of the qubit states by external current feeding as in our previous work \cite{Koizumi:2022aa}.
We first consider the case where $J^{\rm EX}$ depicted in 
Fig.~\ref{fig:result:1NI_external_current_energy}(a) is used.
The 16 qubit states are split into four groups as shown in Fig.~\ref{fig:result:1NI_external_current_energy}(b).
In the present work, we only consider the group composed of `mama', `amam', `mmaa', and `aamm';
they have linear current flow regions as shown in Fig.~\ref{fig:result:1NI_16_current}(m)-(p).
The external current $J^{\rm EX}$ changes the energies of the energy levels A (`mama') and B (`amam'), but does not change significantly for C (`mmaa') and D (`aamm'). 
The energy level crossing occurs between A and B by varying  $J^{\rm EX}$ as shown 
in Fig.~\ref{fig:result:1NI_external_current_energy}(c).
Then, using the Landau-Zener transition with the transition parameter $J^{\rm EX}$,
the qubit control of $A$ and $B$ states can be achieved \cite{Koizumi:2022aa}.

\begin{figure}[H]
    \begin{subfigure}{0.9\columnwidth}
    \centering
    \includegraphics[width = 4.0cm]{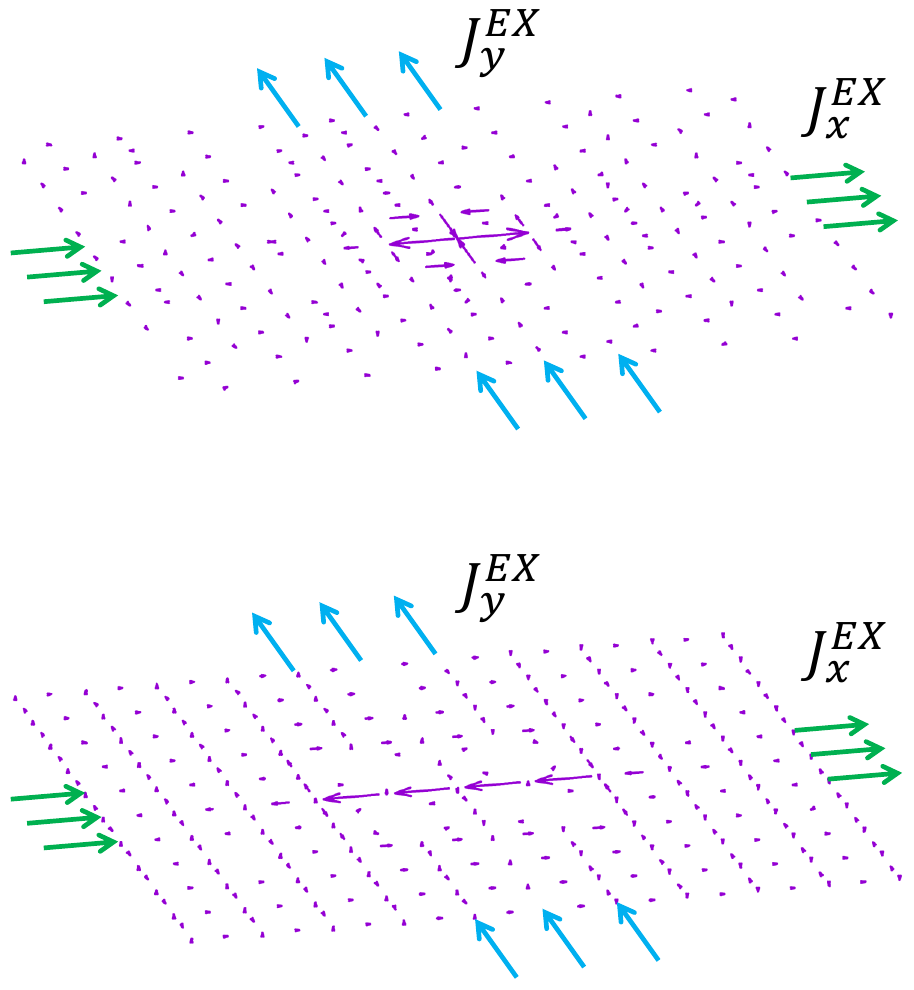}
    \caption{Two external feeding currents $J_x^{\rm EX}$ and  $J_y^{\rm EX}$ for the controlling the qubit states.}
    \label{fig:result:1NI_external_current_xy}
  \end{subfigure}
     \hspace*{1mm}
    \centering
    \begin{subfigure}{0.48\columnwidth}
      \centering
      \includegraphics[width=4cm]{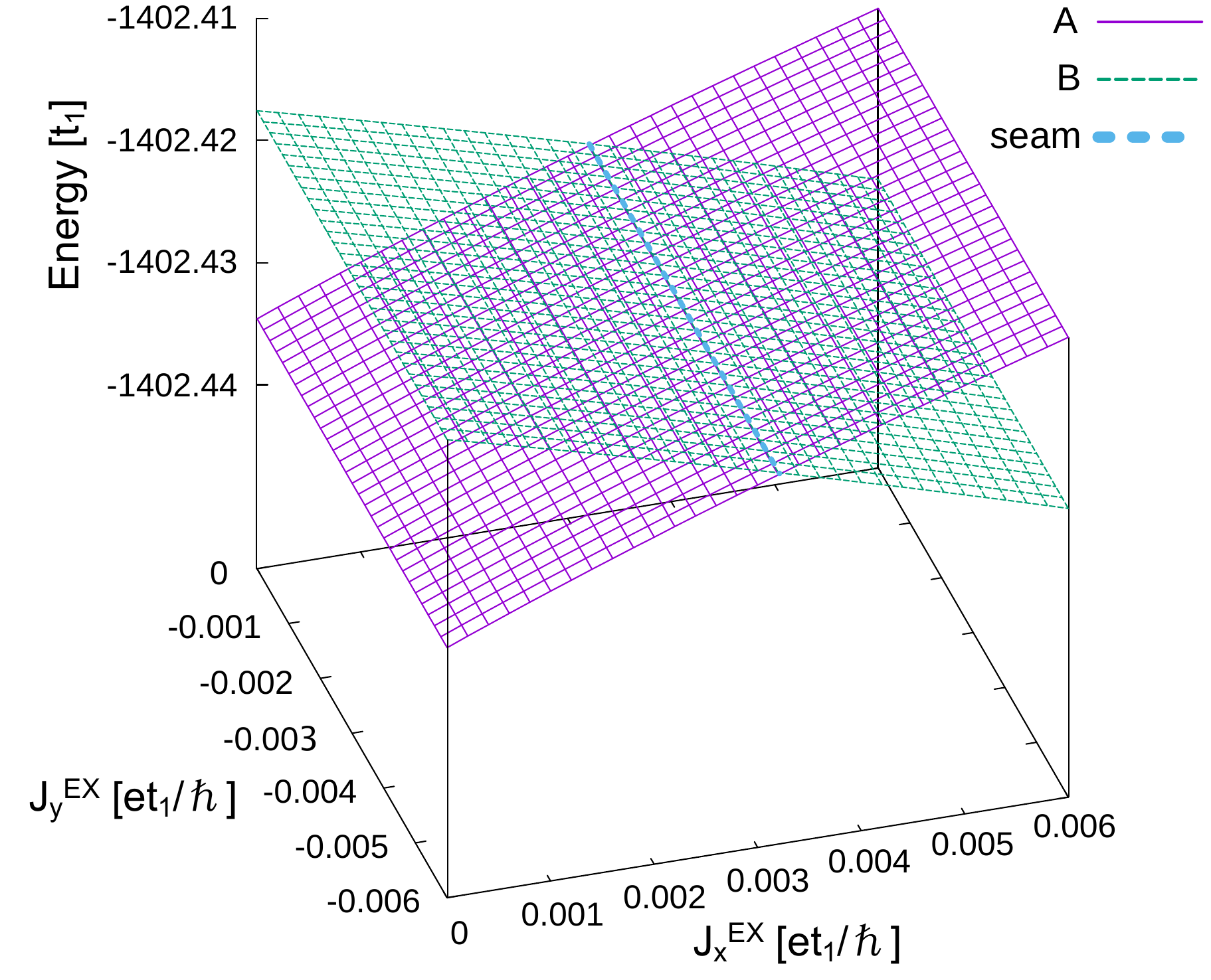}
      \caption{The seam of crossings of A and B states by the external current feeding.}
      \label{fig:result:1NI_AB+}
    \end{subfigure}
    \hspace*{1mm}
    \begin{subfigure}{0.48\columnwidth}
      \centering
      \includegraphics[width=4cm]{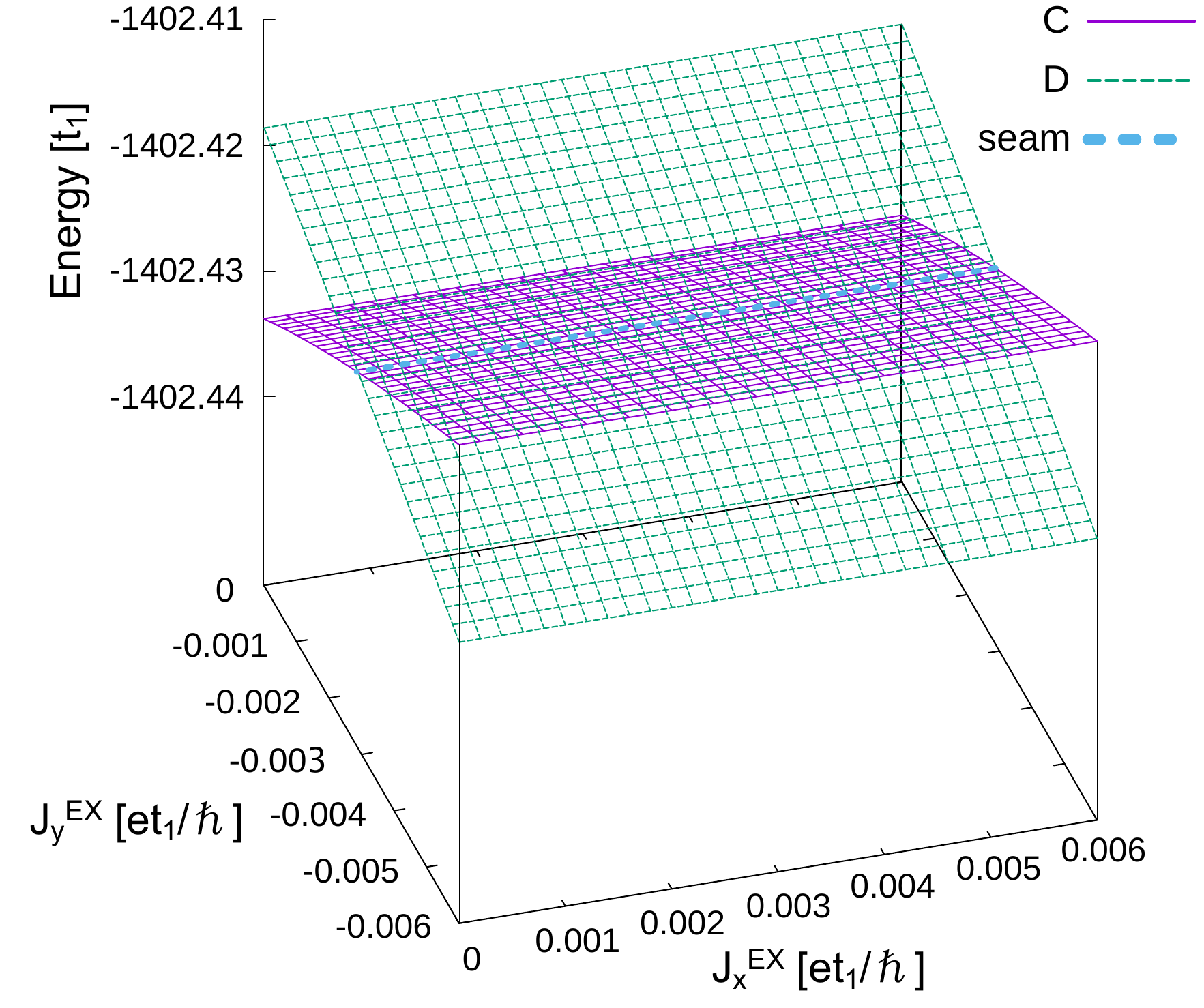}
      \caption{The seam of crossings of C and D states by the external current feeding.}
      \label{fig:result:1NI_CD+}
    \end{subfigure}
       \hspace*{1mm}
      \begin{subfigure}{0.9\columnwidth}
    \centering
    \includegraphics[width = 6.0cm]{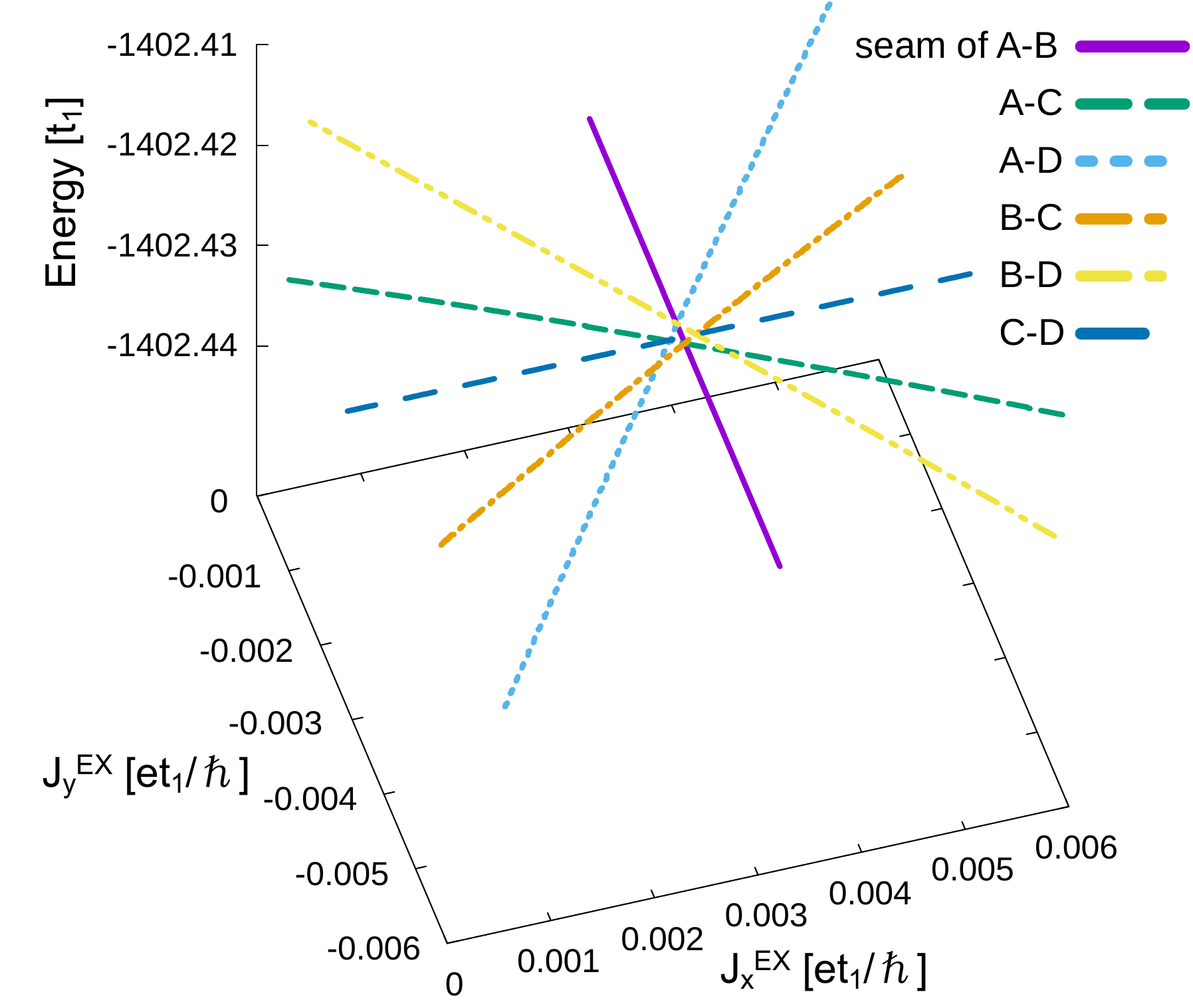}
    \caption{Six seams of crossings arising from the qubit states A, B, C, and D by the external current feeding.}
    \label{fig:result:1NI_external_current_xy_seam}
  \end{subfigure}
      \caption{Crossing of the qubit states by the external current feeding $J_x^{\rm EX}$ and  $J_y^{\rm EX}$.}
    \label{fig:result:1NI_external_current_xy_energy}
  \end{figure}

Now we consider the control of the qubit states using two external currents, $J^{\rm EX}_x$ and 
 $J^{\rm EX}_y$ as depicted in 
Fig.~\ref{fig:result:1NI_external_current_xy_energy}(a).
The current $J^{\rm EX}_x$ is the same as $J^{\rm EX}$ in  Fig.~\ref{fig:result:1NI_external_current_energy}(a).
The energy levels for A  and B changes by $J^{\rm EX}_x$ but do not change significantly by $J^{\rm EX}_y$;
as a result, a seam of energy level crossings along the $J^{\rm EX}_y$ axis as seen in Fig.~\ref{fig:result:1NI_external_current_xy_energy}(b).
For the energy levels  C  and B, their energies change by $J^{\rm EX}_y$ but do not change significantly
by $J^{\rm EX}_x$;
as a result, a seam of energy level crossings along the $J^{\rm EX}_x$ axis as seen in Fig.~\ref{fig:result:1NI_external_current_xy_energy}(c).
The total of six seams of crossings exist in the $J^{\rm EX}_x$-$J^{\rm EX}_y$ parameter plane as shown in 
Fig.~\ref{fig:result:1NI_external_current_energy}(d).
Using those seams of energy crossings, a variety of Landau-Zener transitions will be possible.

\section{Three nano-island SVILC qubit architecture}
  \label{subss:3Nano-Islands}

\begin{figure}[H]
    \centering
      \begin{subfigure}{0.9\columnwidth}
       \centering
    \includegraphics[width=7cm]{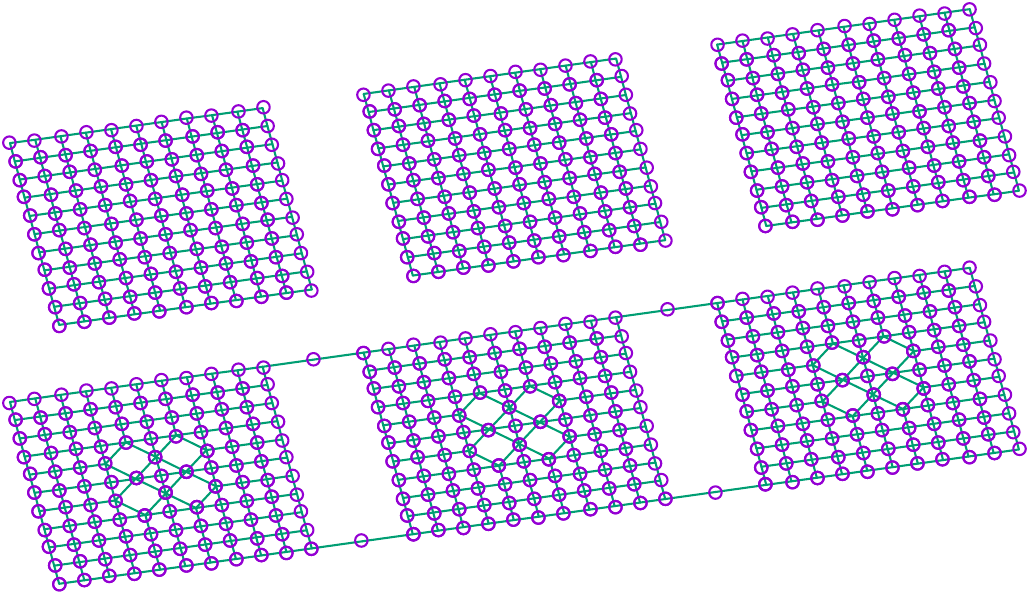}
    \caption{Three nano-islands connected by four atomic-sized quantum dots.}
    \label{fig:result:3NI_model}
  \end{subfigure}
    \centering
    \begin{subfigure}{0.48\columnwidth}
      \centering
      \includegraphics[width=5cm]{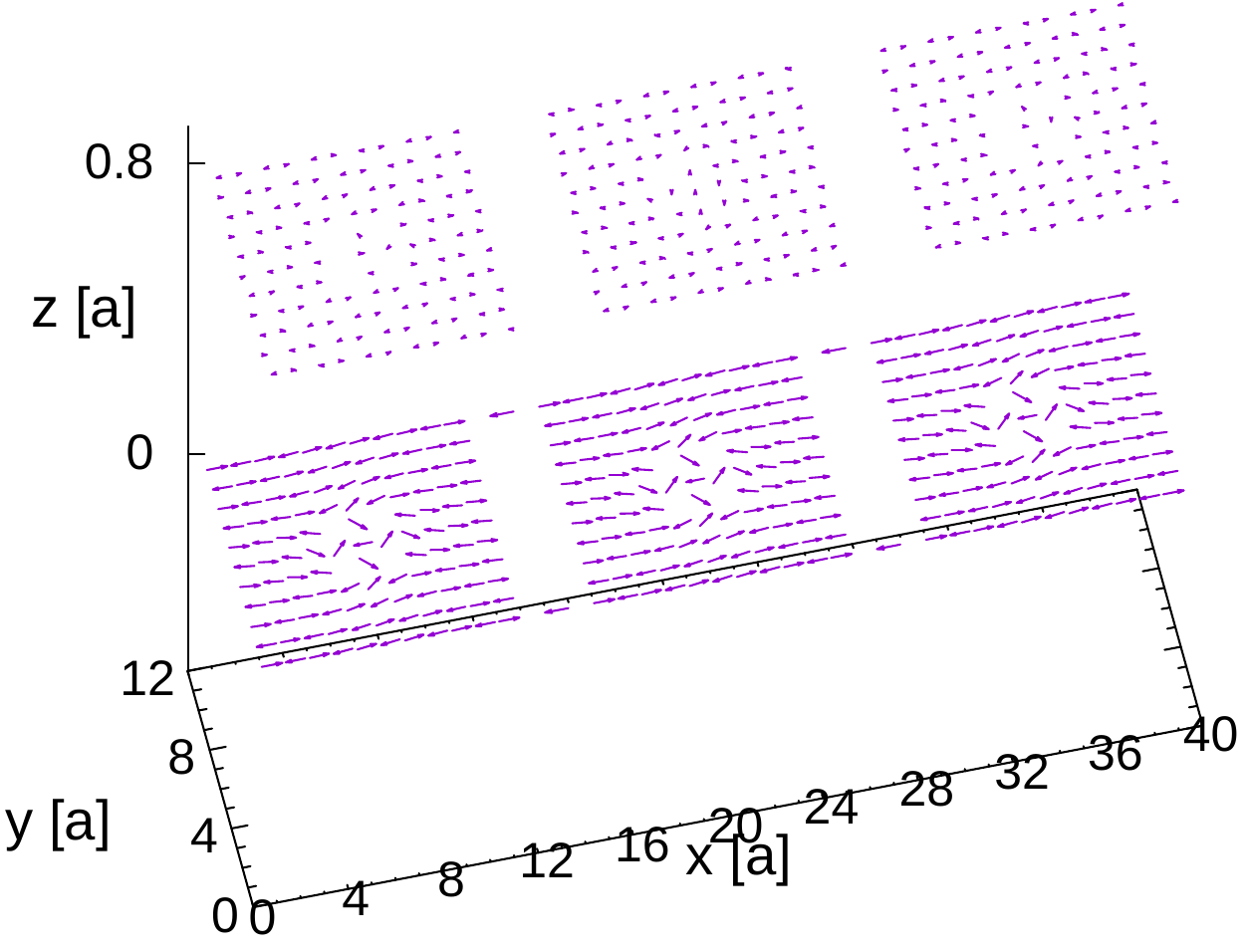}
      \caption{The spin moment distribution.}
      \label{fig:result:3NI_spin}
    \end{subfigure}
    \hspace*{1mm}
    \begin{subfigure}{0.48\columnwidth}
      \centering
      \includegraphics[width=5cm]{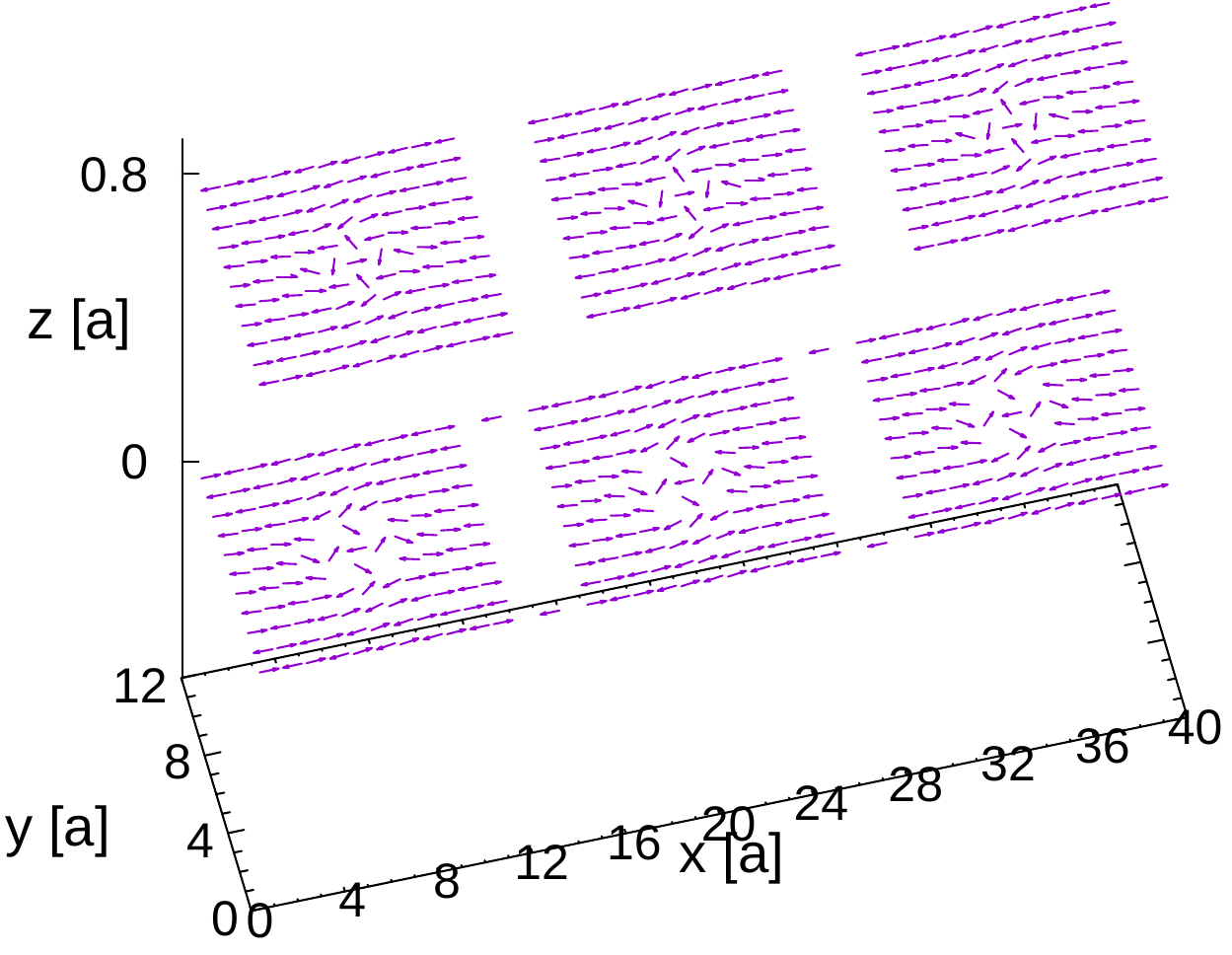}
      \caption{The normalized spin moment distribution.}
      \label{fig:result:3NI_spin_n}
    \end{subfigure}
    \begin{subfigure}{0.48\columnwidth}
      \centering
      \includegraphics[width=4.5cm]{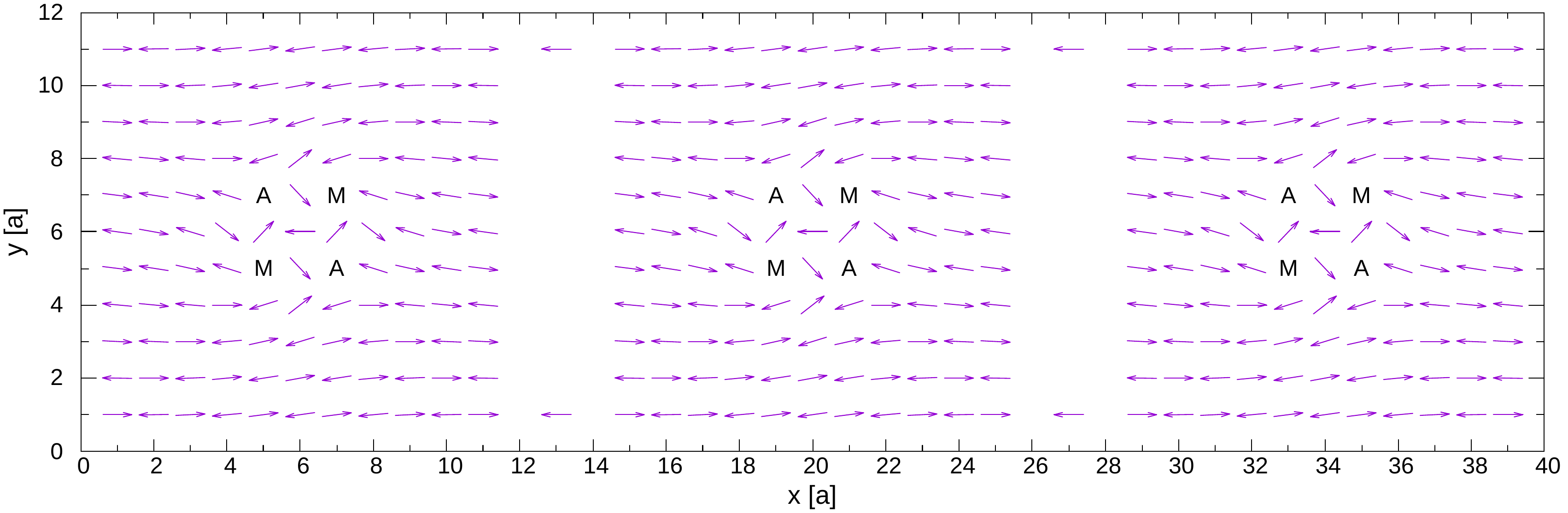}
      \caption{The bulk layer distribution.}
      \label{fig:result:3NI_spin_bulk}
    \end{subfigure}
    \hspace*{1mm}
    \begin{subfigure}{0.48\columnwidth}
      \centering
      \includegraphics[width=4.5cm]{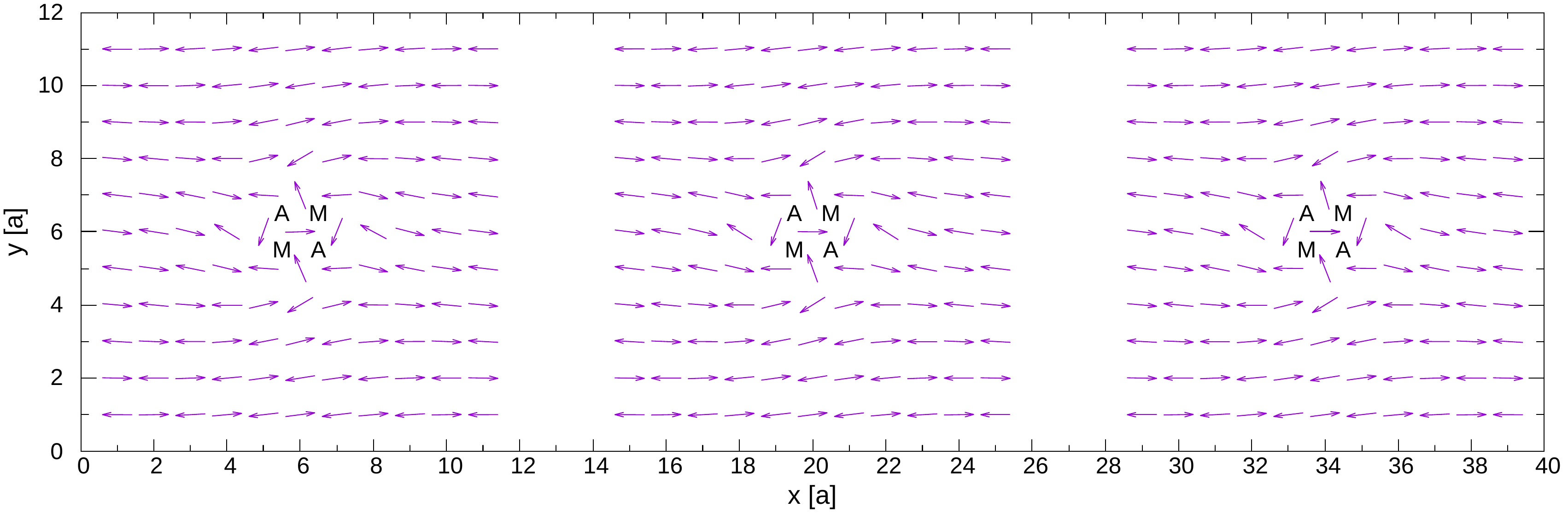}
      \caption{The normalized surface layer distribution.}
      \label{fig:result:3NI_spin_surface}
    \end{subfigure}
    \caption{Three nano-island SVILC qubit system connected by four atomic size quantum dots.}
    \label{fig:result:3NI}
  \end{figure}

\begin{figure}[H]
    \centering
    \begin{subfigure}{0.9\columnwidth}
    \centering
    \includegraphics[width=9cm]{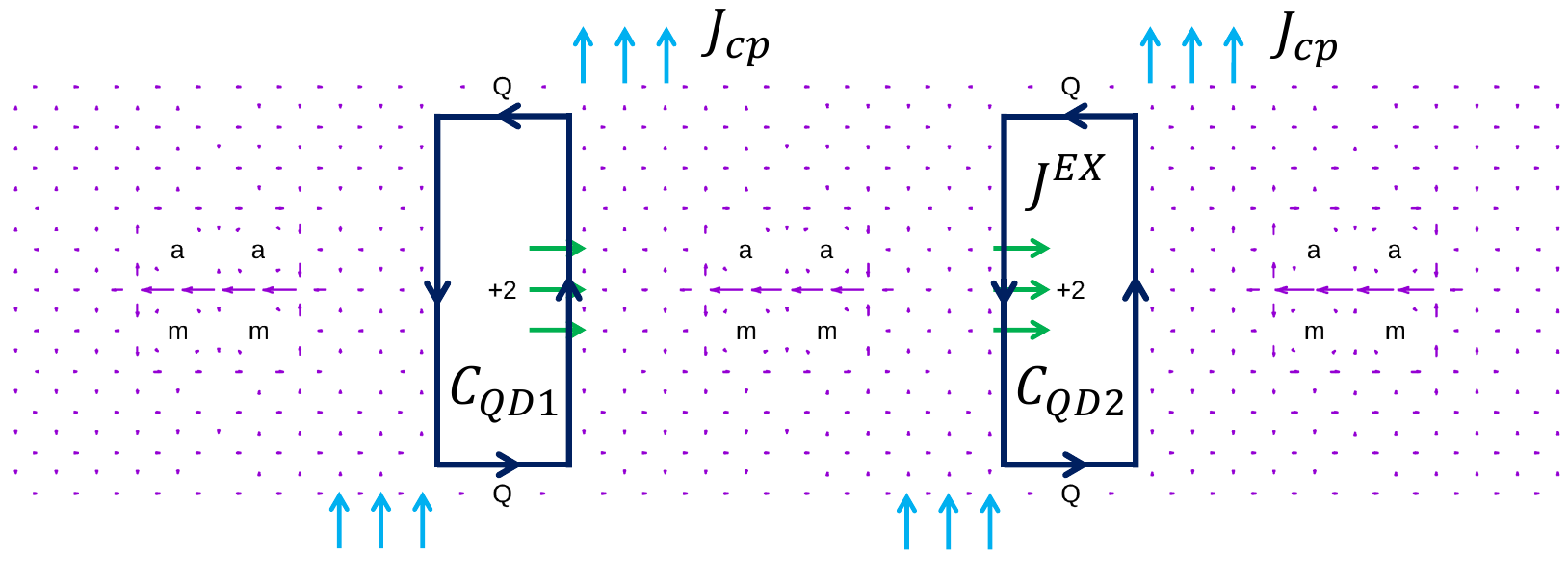}
    \caption{Two loops $C_{QD1}$ and $C_{QD2}$ containing the quantum dots, and the coupling current $J_{\rm cp}$}.
    \label{fig:result:3NI_loop}
  \end{subfigure}
     \begin{subfigure}{0.9\columnwidth}
  \centering
    \includegraphics[width=8cm]{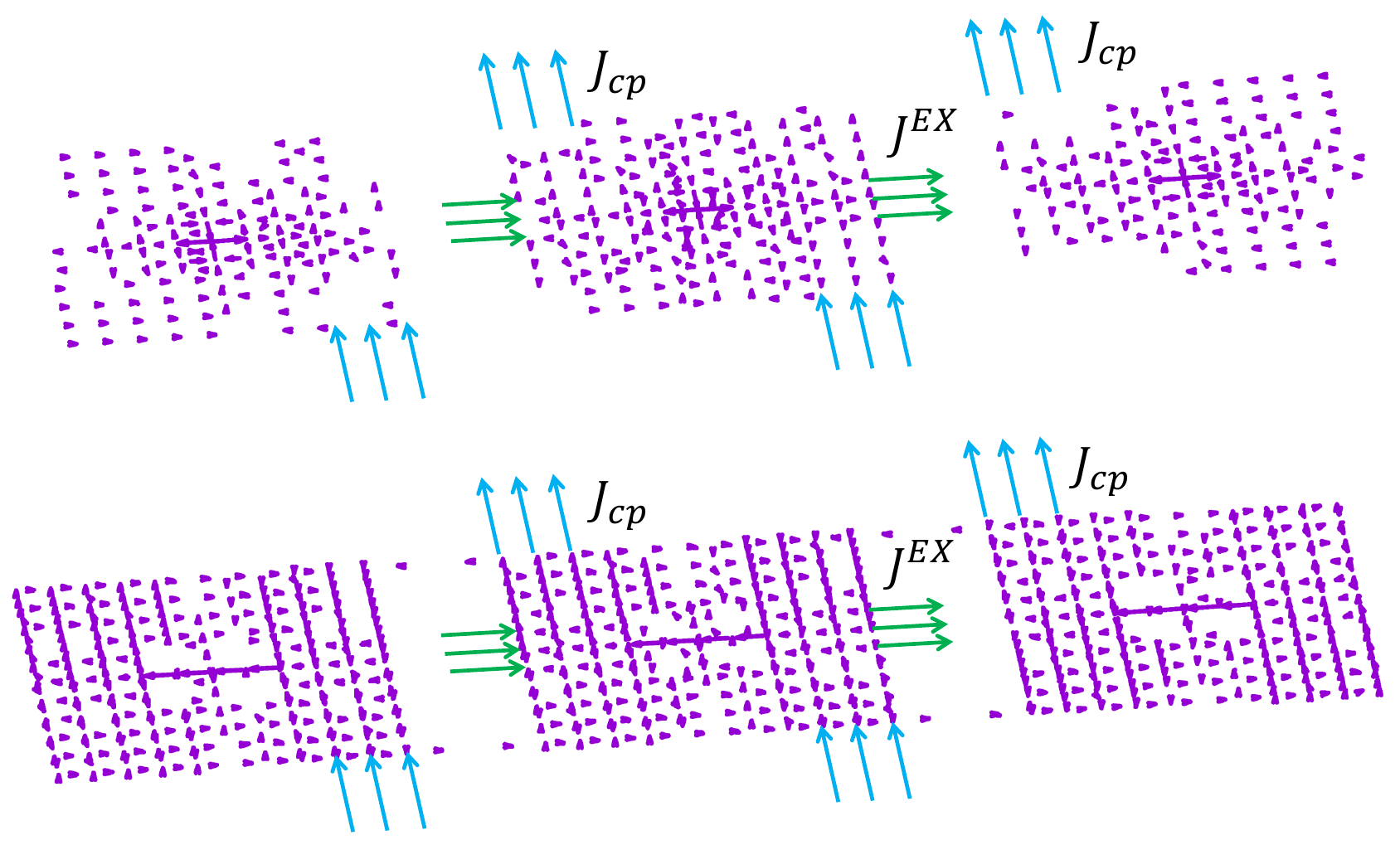}
     \caption{Three nano-island system with external currents $J^{\rm EX}$ and $J_{\rm cp}$.
       $J^{\rm EX}$ here corresponds to $J^{\rm EX}_x$ in 
       Fig.~\ref{fig:result:1NI_external_current_xy_energy}(a). }
       \end{subfigure}
    \caption{Three nano-island SVILC qubit system controlled by the external feeding current $J^{\rm EX}$ and the coupling current $J_{\rm cp}$. The winding numbers for $\chi$ around the loops, $C_{QD1}$ and $C_{QD2}$ are also used.}
    \label{fig:result:3NI_current}
  \end{figure}

 Now we consider the three nano-island SVILC qubit architecture shown in 
  Fig.~\ref{fig:result:3NI_current}. Three islands are connected by four atomic-sized quantum dots
  as shown in Fig.~\ref{fig:result:3NI}(a). In this architecture, the coupling between quits can be controlled by the current that flows across the quantum dots. We call such a current, the {\em coupling current}.
  
  The Hamiltonian for the connection via each quantum dot is given by
  \begin{eqnarray}
    \label{eq:qdot_hamiltonian}
    H_{\textrm{QD}} = - t_D \sum_{\langle j_D, k_I \rangle_1, \sigma} \left(c_{j_D \sigma}^\dag c_{k_I \sigma}
    + c_{k_I \sigma}^\dag c_{j_D \sigma}\right)
    +U_D \sum_{j_D} c^\dag_{j_D \uparrow} c_{j_D \uparrow} c^\dag_{j_D \downarrow} c_{j_D \downarrow}
    - \mu_D \sum_{j_D} c_{j_D \uparrow}^\dag c_{j_D \downarrow}
    \nonumber
    \\
  \end{eqnarray}
  where we use the following parameter values: $t_D=0.15t$, $U_D=8t_1$, and $\mu_D=4t_1$.
  Actually, we employ the mean-field version of $H_{\textrm{QD}}$ given by
  \begin{eqnarray}
    \label{eq:qdot_hamiltonian_HF}
      H_{\textrm{QD}}^{\textrm{HF}} 
      &=&- t_D \sum_{\langle j_D, k_I \rangle_1, \sigma} \left(c_{j_D \sigma}^\dag c_{k_I \sigma}
      + c_{k_I \sigma}^\dag c_{j_D \sigma}\right) - \mu_D \sum_{j_D} c_{j_D \uparrow}^\dag c_{j_D \downarrow}
      \nonumber
       \\ 
      &&+ U_D \sum_{j_D} 
      \left[
      \left(\frac{1}{2} - \frac{2}{3}\langle S^z_{j_D} \rangle \right)c_{j_D \uparrow}^\dag c_{j_D \uparrow}
      + \left(\frac{1}{2} + \frac{2}{3}\langle S^z_{j_D} \rangle \right) c_{j_D \downarrow}^\dag c_{j_D \downarrow}- \frac{2}{3}\langle {\bf S}_{j_D} \rangle ^2 
      \right.
      \nonumber
      \\ &&
      \left.
      - \frac{2}{3}\left(\langle S^x_{j_D} \rangle - i\langle S^y_{j_D} \rangle \right)c_{j_D \uparrow}^\dag c_{j_D \downarrow}
      - \frac{2}{3}\left(\langle S^x_{j_D} \rangle + i\langle S^y_{j_D} \rangle \right)c_{j_D \downarrow}^\dag c_{j_D \uparrow}
      \right] 
  \end{eqnarray}
  
The control of the qubit states is achieved using the feeding current $J^{\rm EX}$ and the coupling current $J_{\rm cp}$. We also use the winding numbers for $\chi$ around the loops, $C_{QD1}$ and $C_{QD2}$ depicted in Figs.~\ref{fig:result:3NI_current}(a) and (b) as controlling parameters. Actually, these winding numbers are crucially important as will be shown below.
The qubit states are denoted by S$_1$-S$_2$-S$_3$, where S$_i, \ i=1,2,3$ indicates each island state specified by the current pattern.

\begin{figure}[H]
   \centering
    \begin{subfigure}{0.46\columnwidth}
     \centering
      \includegraphics[width=\columnwidth]{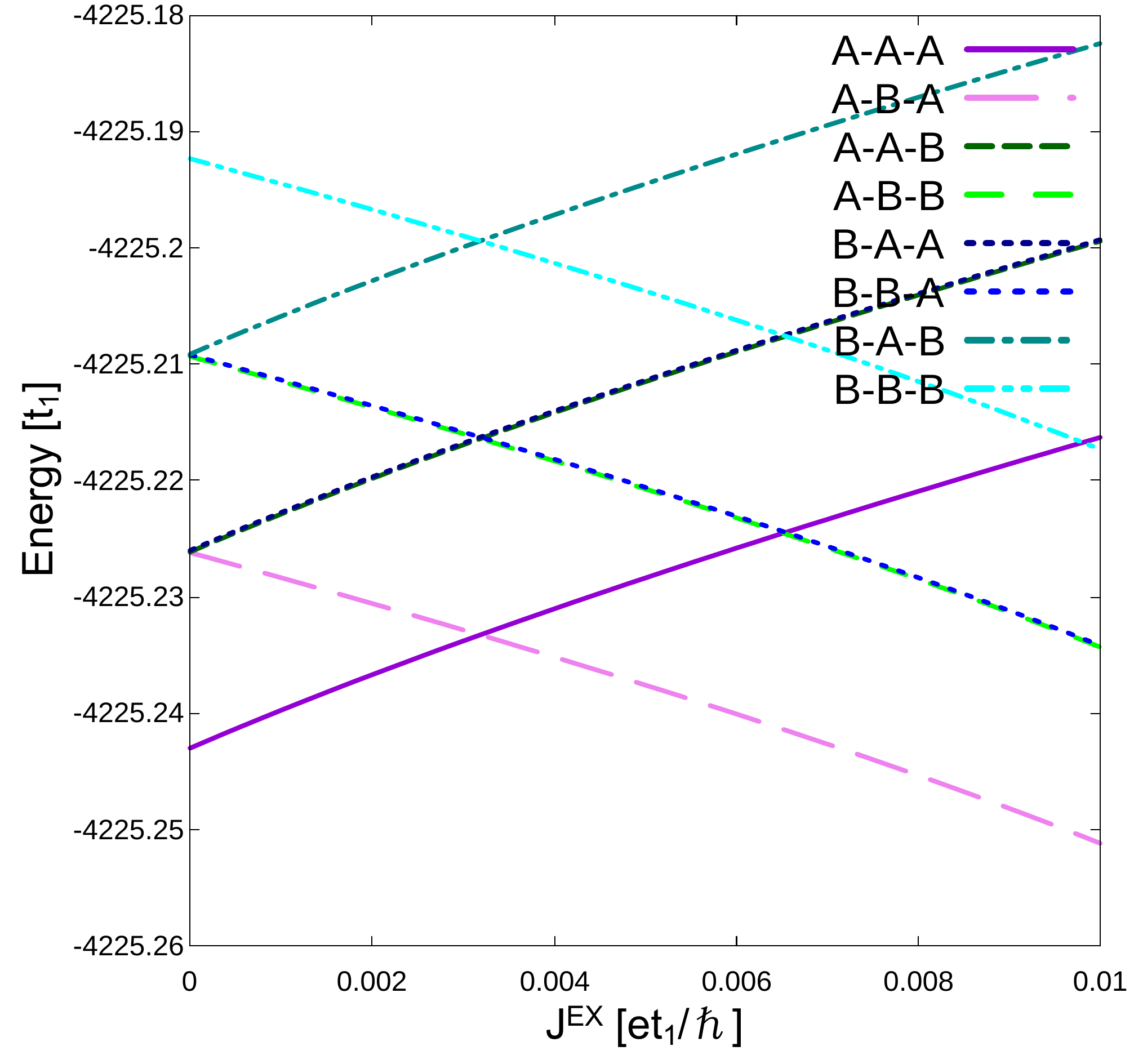}
      \caption{$J_{\rm cp}=0.0$, $w_1=0$, $w_2=0$}
    \end{subfigure}
    \hspace*{0.05\columnwidth}
    \begin{subfigure}{0.46\columnwidth}
      \centering
      \includegraphics[width=\columnwidth]{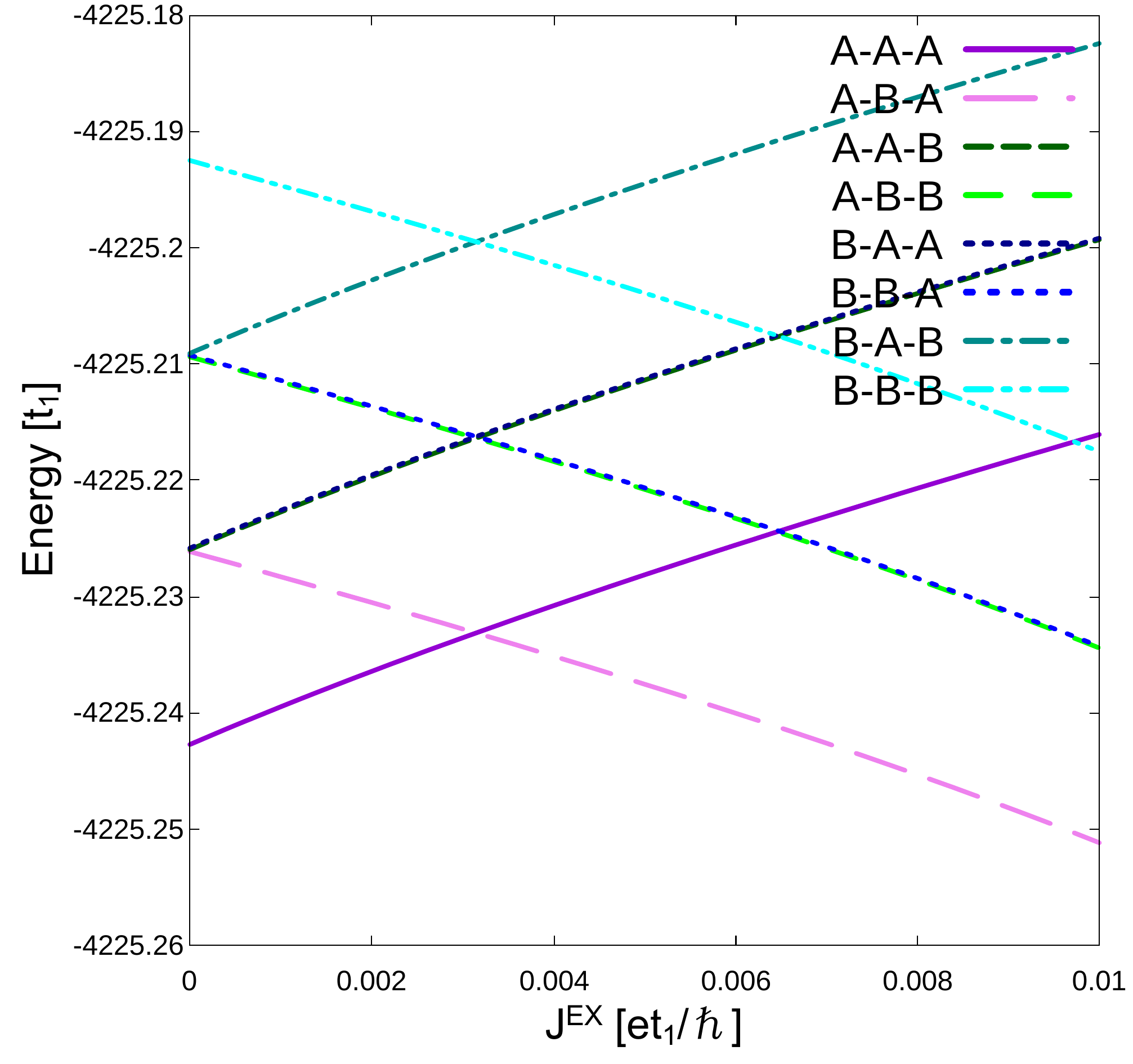}
      \caption{$J_{\rm cp}=0.00008$, $w_1=0$, $w_2=0$}
    \end{subfigure}
    \begin{subfigure}{0.46\columnwidth}
      \centering
      \includegraphics[width=\columnwidth]{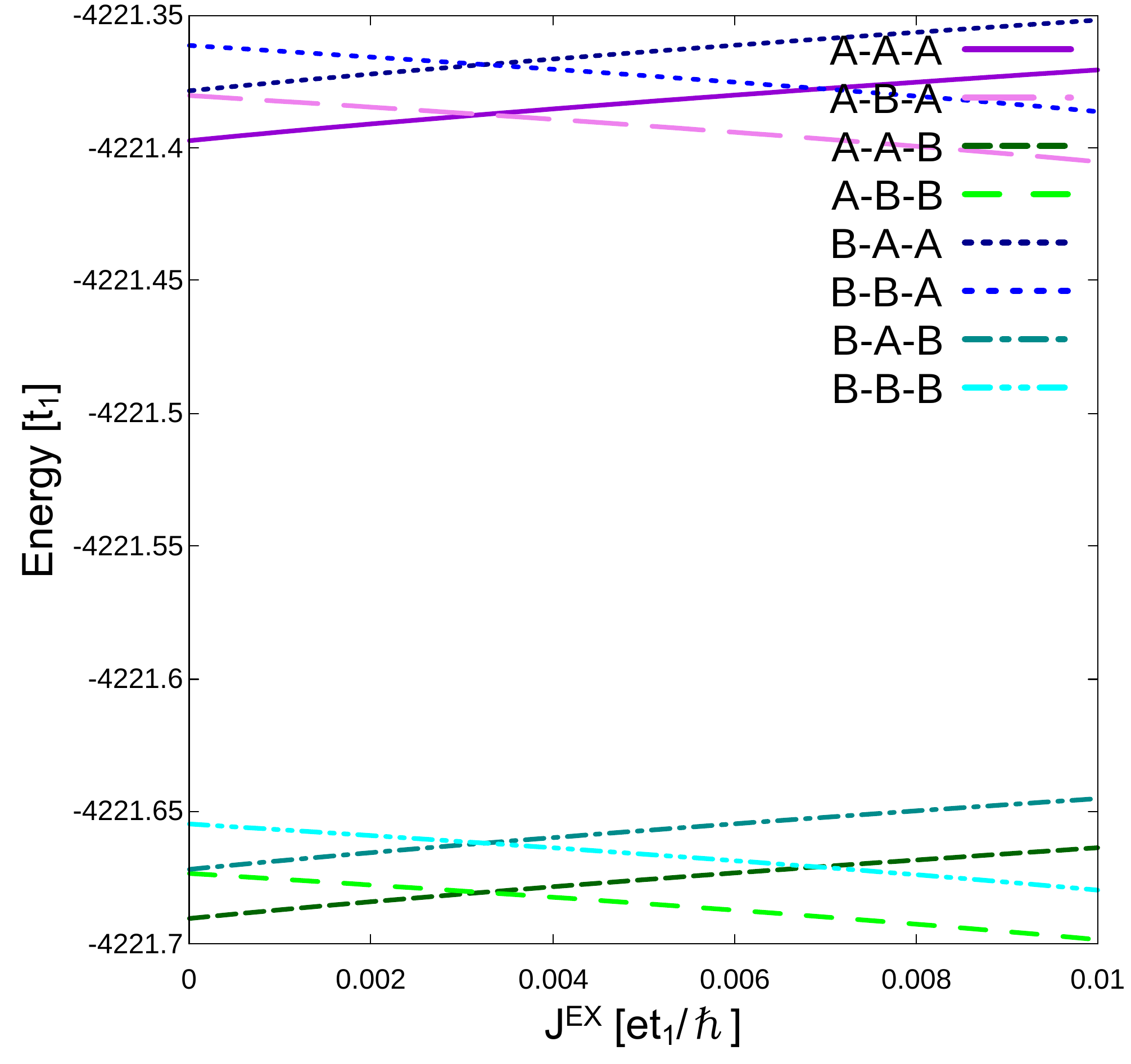}
      \caption{$J_{\rm cp}=0.0$, $w_1=2$, $w_2=2$}
    \end{subfigure}
    \hspace*{0.05\columnwidth}
    \begin{subfigure}{0.46\columnwidth}
      \centering
      \includegraphics[width=\columnwidth]{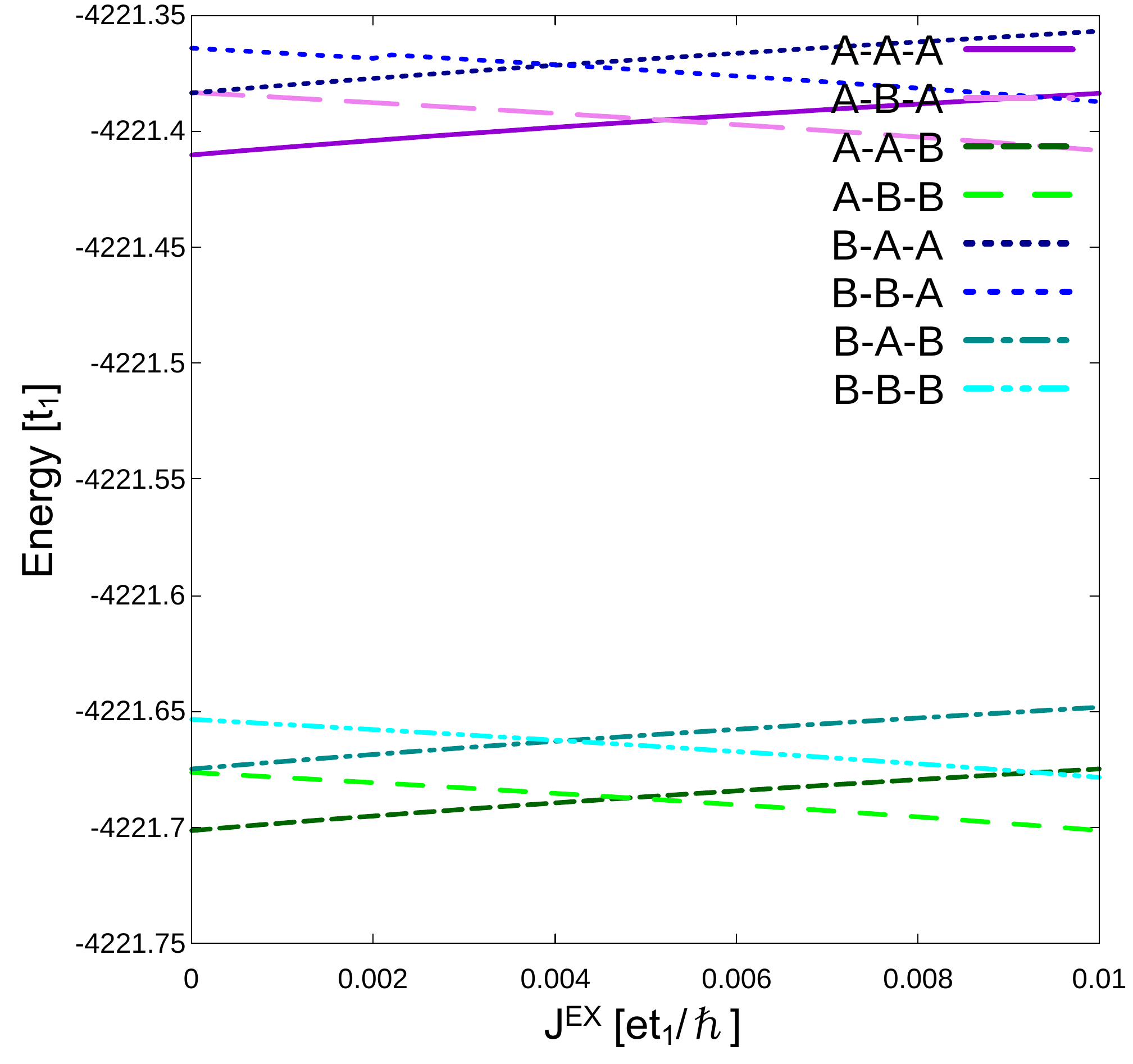}
       \caption{$J_{\rm cp}=0.00008$, $w_1=2$, $w_2=2$}
    \end{subfigure}
    \begin{subfigure}{0.46\columnwidth}
      \centering
      \includegraphics[width=\columnwidth]{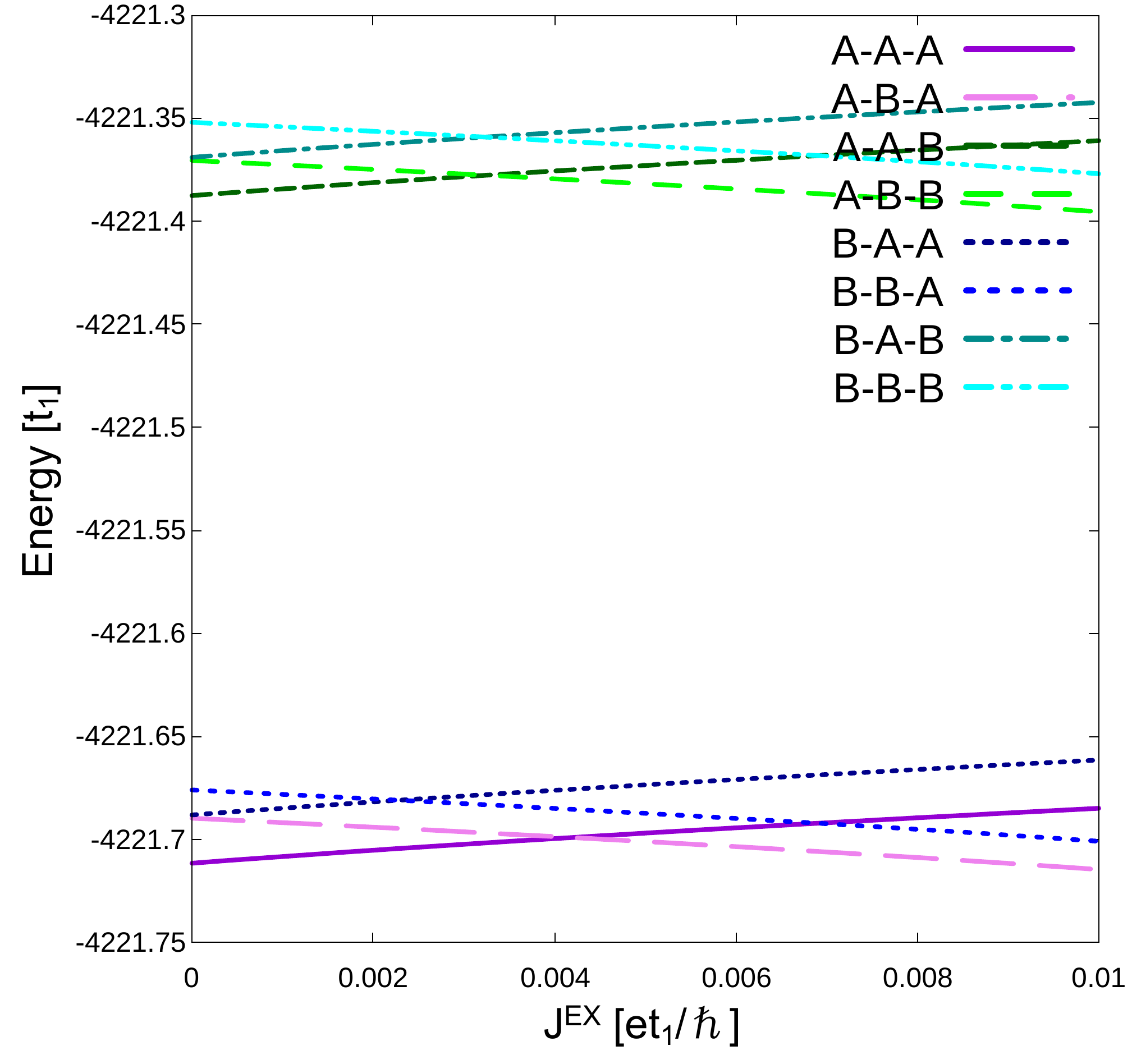}
       \caption{$J_{\rm cp}=0.0$, $w_1=-2$, $w_2=-2$}
    \end{subfigure}
    \hspace*{0.05\columnwidth}
    \begin{subfigure}{0.46\columnwidth}
      \centering
      \includegraphics[width=\columnwidth]{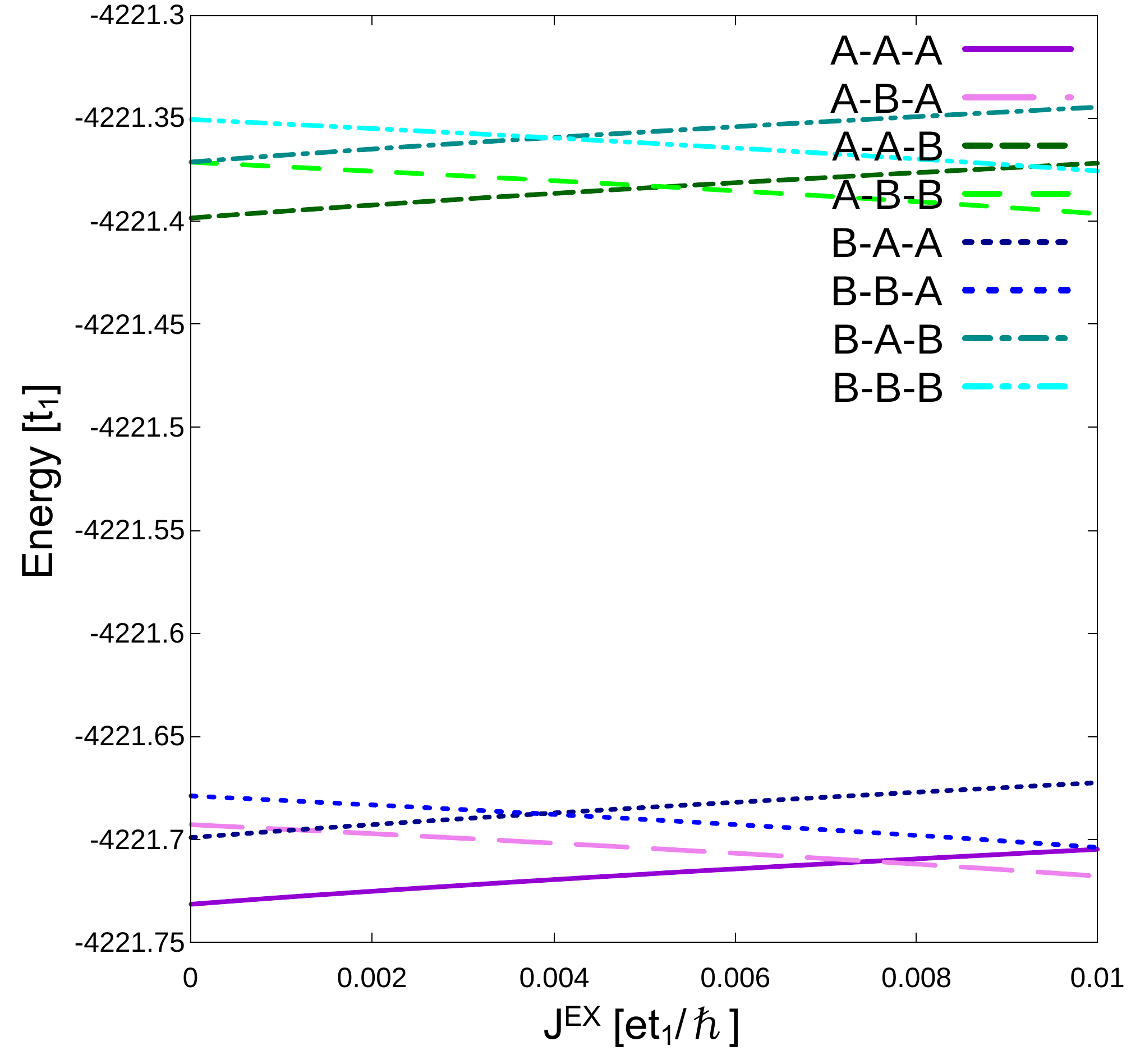}
       \caption{$J_{\rm cp}=0.00008$, $w_1=-2$, $w_2=-2$}
    \end{subfigure}
    \caption{Crossing of SVILC qubit states of the three nano-islands architecture
    by changing currents $J^{\rm  EX}$ and $J_{\rm cp}$, and the winding numbers $\chi$ along $C_{QD1}$ (denoted by $w_1$) and $C_{QD2}$ (denoted by $w_2$).}
    \label{fig:result:3NI_energy}
  \end{figure}

The results are shown in Fig.~\ref{fig:result:3NI_energy}.
The energies of the qubit states vary with $J^{\rm EX}$.
When the winding number for  $\chi$ along $C_{QD1}$ (denoted by $w_1$)
and that along $C_{QD}$ (denoted by $w_2$) are zero, the coupling between qubit states in 
different islands is absent as shown in Figs.~\ref{fig:result:3NI_energy}(a) and (b);
and the coupling is realized when the winding numbers are nonzero (see Figs.~\ref{fig:result:3NI_energy}(c)-(f)). Note that the condition in Eq.~(\ref{wcomb-eq}) requires $w_1$ and $w_2$ to be even.
This indicates that we can turn-on and off the coupling
by changing the winding numbers.
The coupling can be enhanced by feeding the coupling current $J_{\rm cp}$, as seen in
Figs.~\ref{fig:result:3NI_energy}(d) and (f).

  \begin{figure}[H]
      \begin{subfigure}{0.9\columnwidth}
    \centering
    \includegraphics[width=7cm]{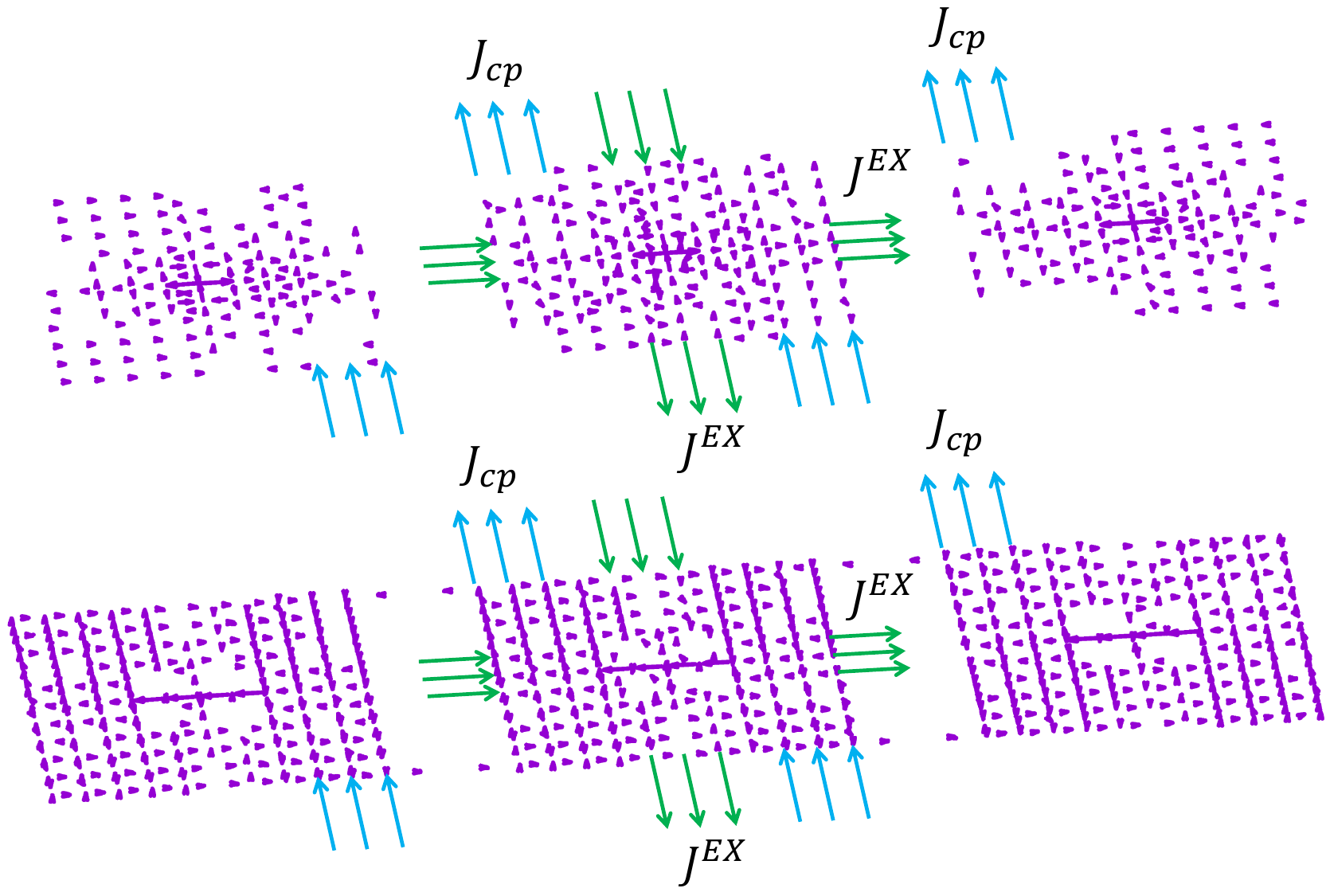}
    \caption{Three nano-islands architecture with external currents $J^{\rm EX}$ and $J_{\rm cp}$.
     $J^{\rm EX}$ here corresponds to applying $J^{\rm EX}_x$ and  $-J^{\rm EX}_y$ in 
     Fig.~\ref{fig:result:1NI_external_current_xy_energy}(a), simultaneously. }
    \label{fig:result:3NI_current_xy}
    \end{subfigure}
    \begin{subfigure}{0.48\columnwidth}
      \centering
      \includegraphics[width=4.5cm]{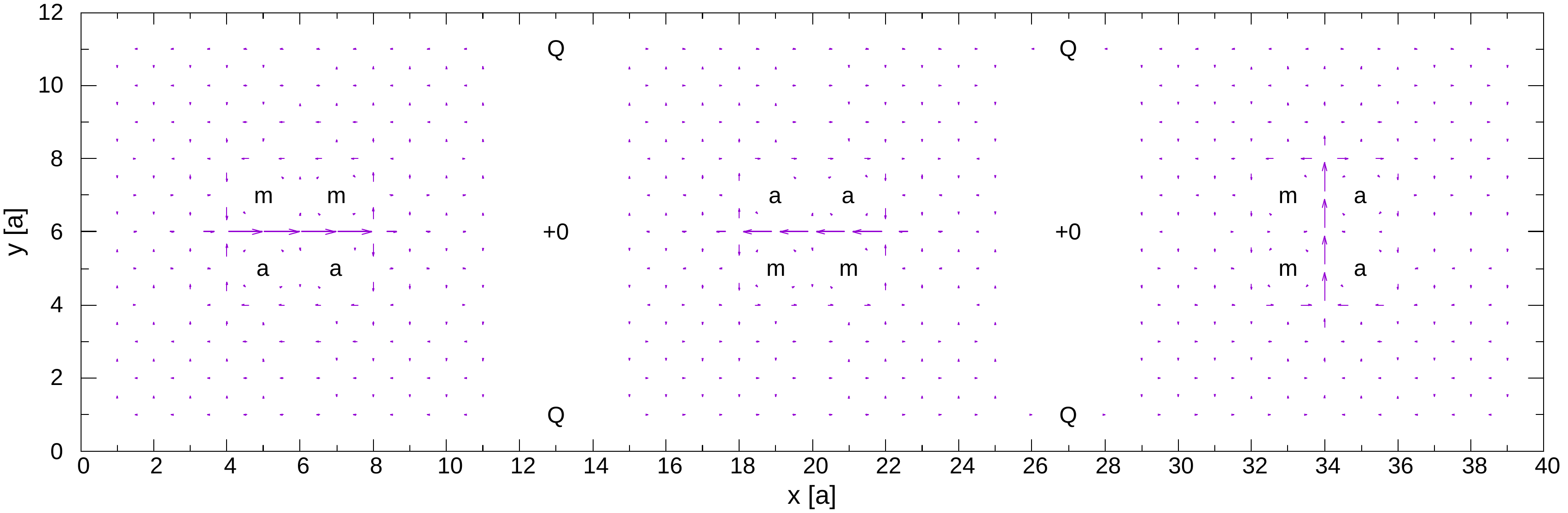}
      \caption{
      $J^{\rm EX}=0.0$, $J_{\rm cp}=0.0$, $w_1=0$, $w_2=0$}
    \end{subfigure}
    \begin{subfigure}{0.48\columnwidth}
      \centering
      \includegraphics[width=4.5 cm]{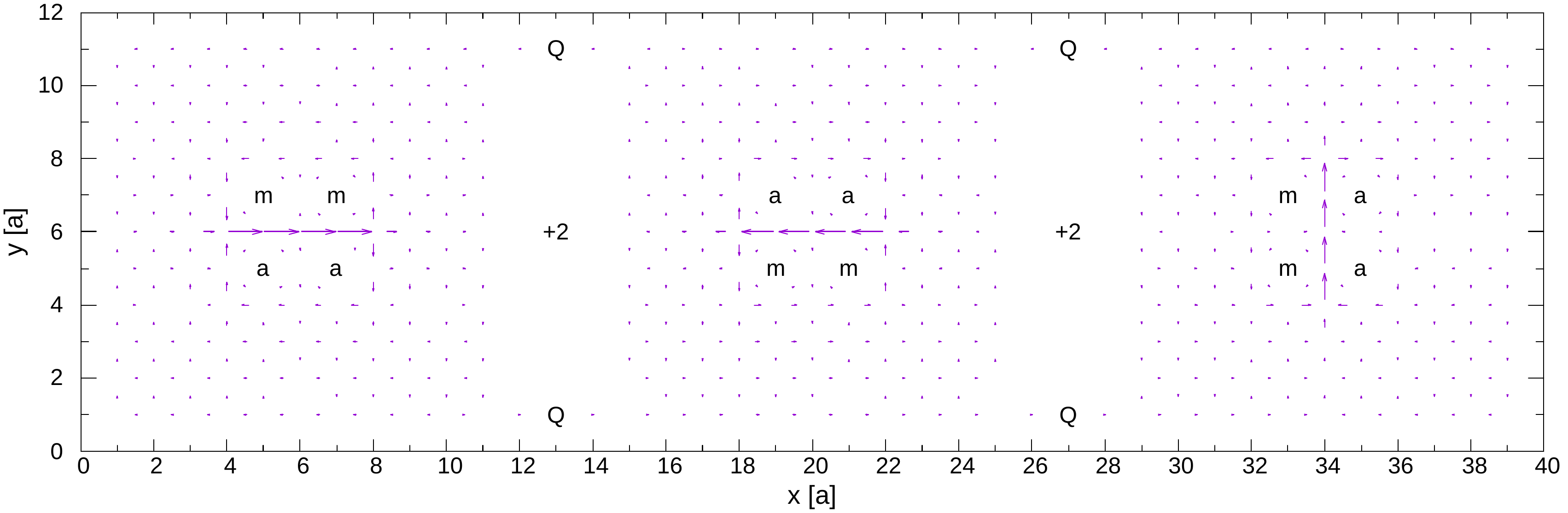}
      \caption{$J^{\rm EX}\!=\!0.005$, $J_{\rm cp}\!=\!0.00008$, 
      $w_2\!=\!2$, $w_2\!=\!2$}
    \end{subfigure}
    \begin{subfigure}{0.48\columnwidth}
      \centering
      \includegraphics[width=4.5 cm]{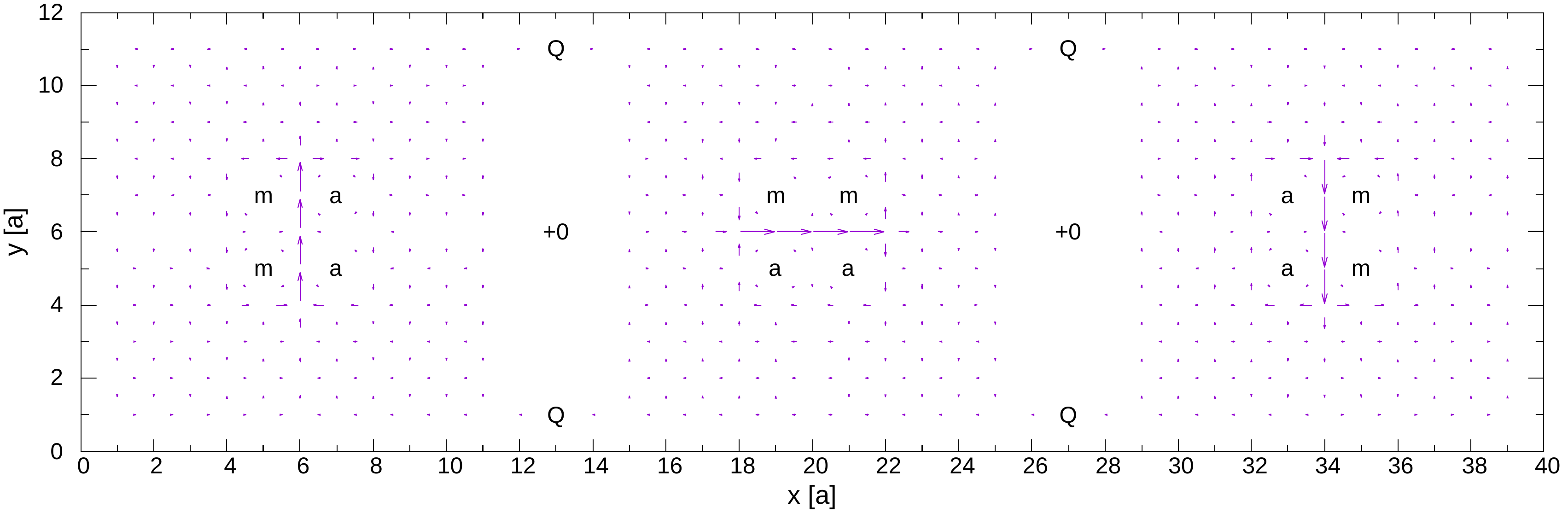}
      \caption{$J^{\rm EX}=0$, $J_{\rm cp}=0.0$, $w_1=0$, $w_2=0$}
    \end{subfigure}
    \begin{subfigure}{0.48\columnwidth}
      \centering
      \includegraphics[width=4.5 cm]{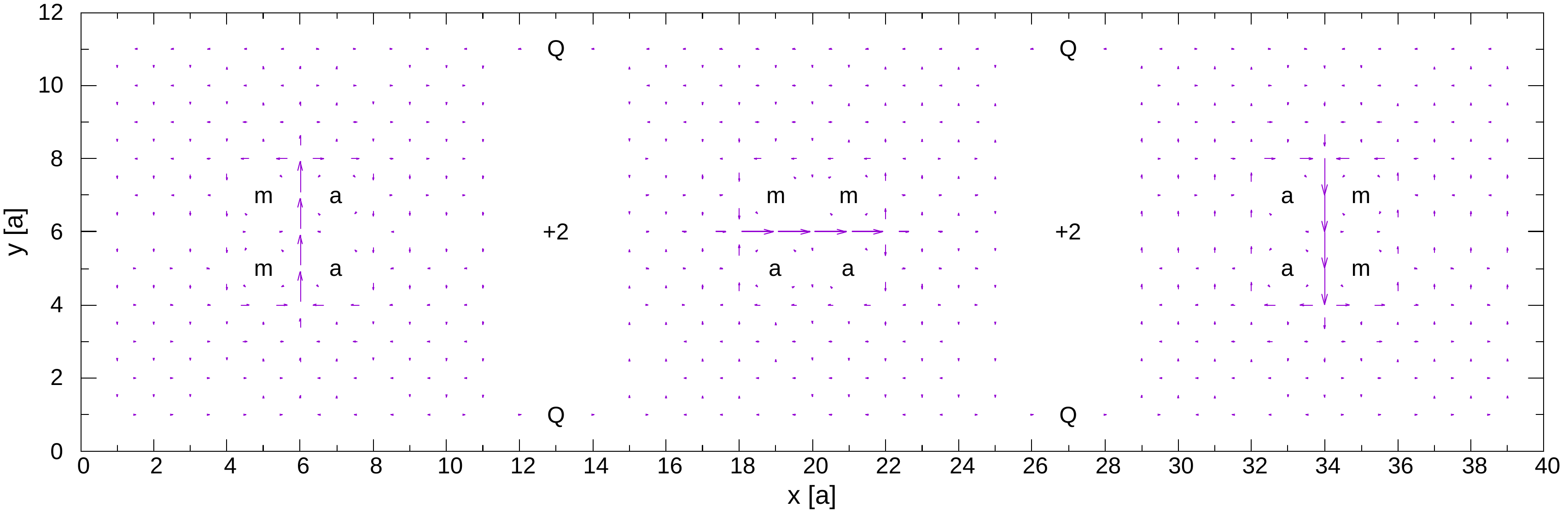}
      \caption{$J^{\rm EX}\!=\!0.005$, $J_{\rm cp}\!=\!0.00008$, 
      $w_2\!=\!2$, $w_2\!=\!2$}
    \end{subfigure}
    \begin{subfigure}{0.48\columnwidth}
      \centering
      \includegraphics[width=4.5 cm]{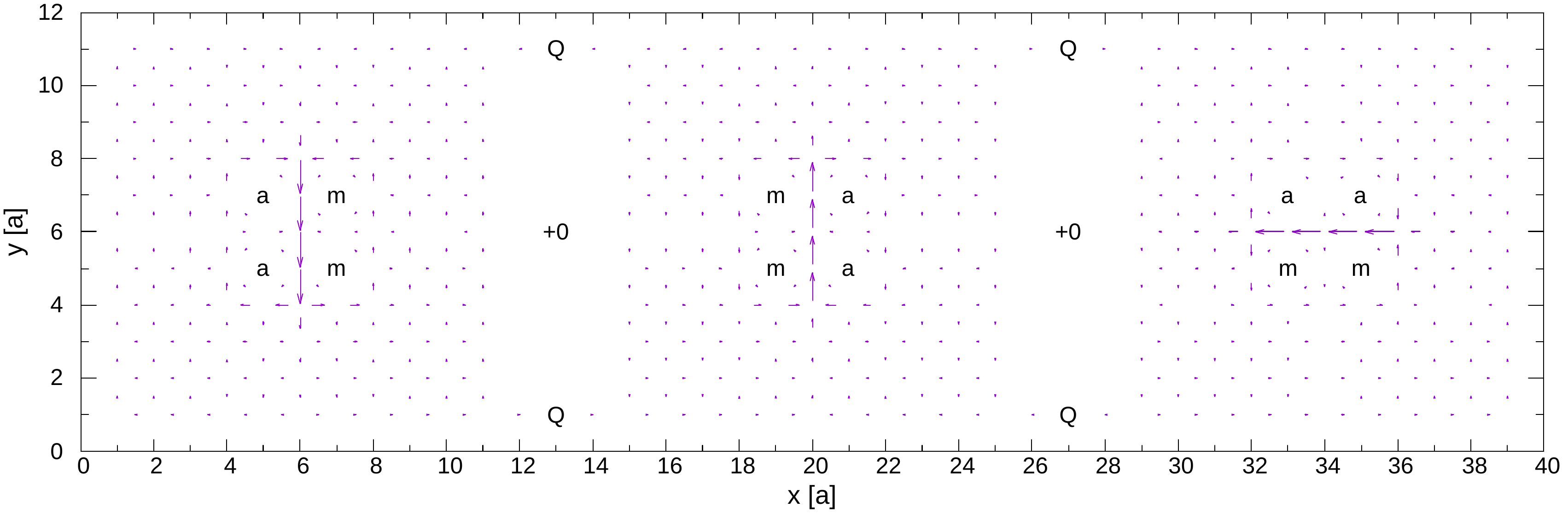}
      \caption{$J^{\rm EX}=0.0 $, $J_{\rm cp}=0.0$, $w_1=0$, $w_2=0$}
    \end{subfigure}
    \begin{subfigure}{0.48\columnwidth}
      \centering
      \includegraphics[width=4.5 cm]{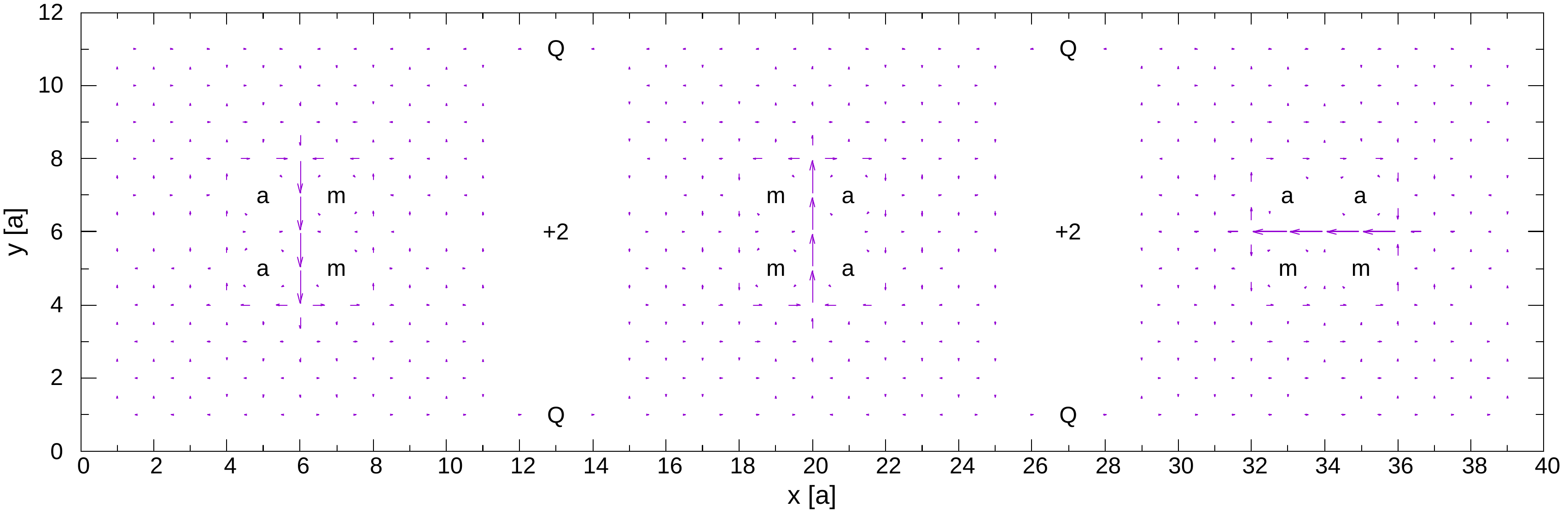}
      \caption{$J^{\rm EX}\!=\!0.005$, $J_{\rm cp}\!=\!0.00008$, 
      $w_2\!=\!2$, $w_2\!=\!2$}
    \end{subfigure}
    \begin{subfigure}{0.48\columnwidth}
      \centering
      \includegraphics[width=4.5 cm]{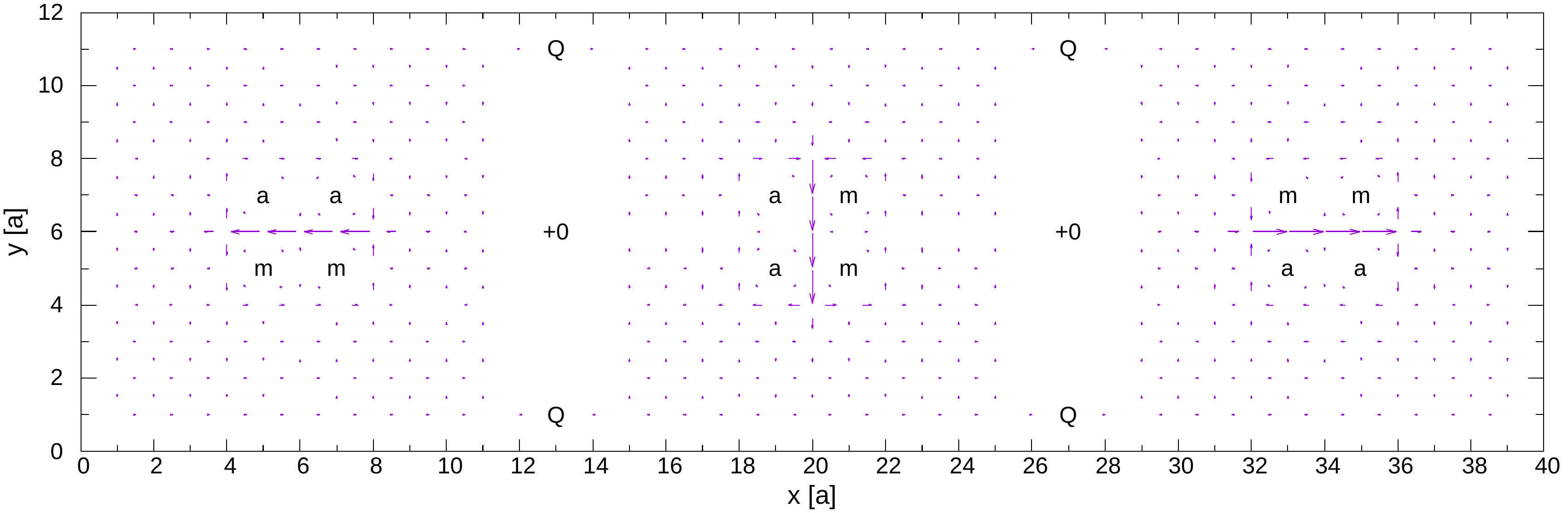}
      \caption{$J^{\rm EX}=0.0$, $J_{\rm cp}=0.0$, $w_1=0$, $w_2=0$}
          \end{subfigure}
    \begin{subfigure}{0.48\columnwidth}
      \centering
      \includegraphics[width=4.5 cm]{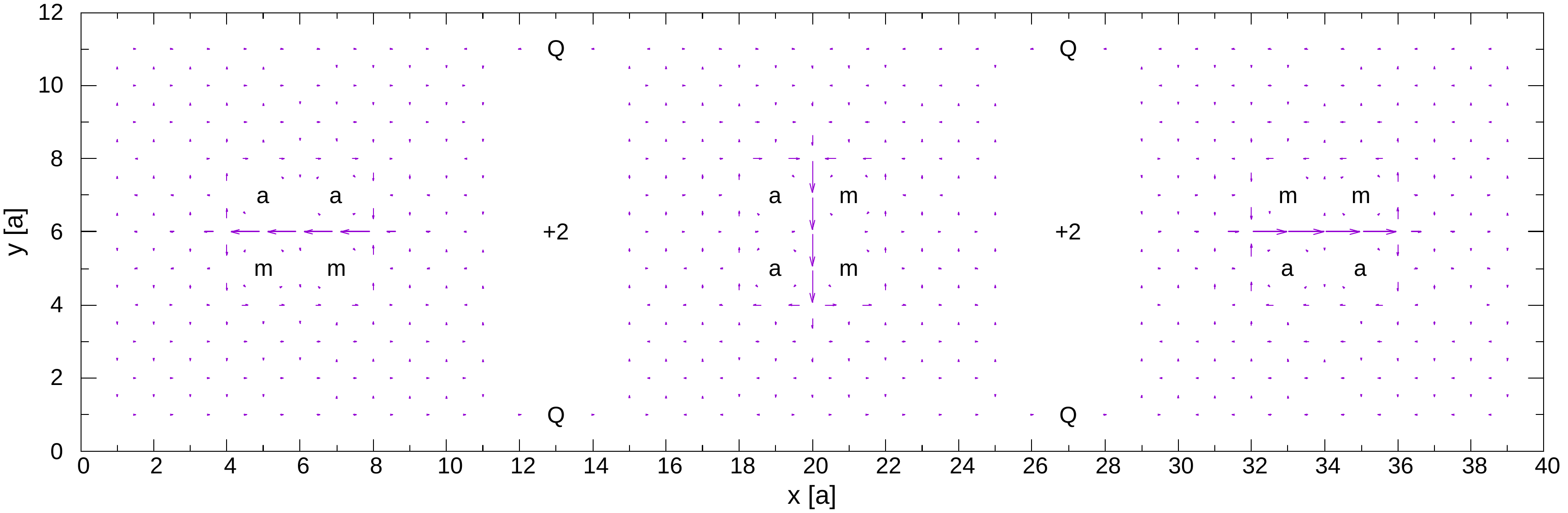}
       \caption{$J^{\rm EX}\!=\!0.005$,$J_{\rm cp}\!=\!0.00008$,
      $w_2\!=\!2$,$w_2\!=\!2$}
    \end{subfigure}
    \label{fig:result:3NI_current_readout}
    \centering
    \begin{subfigure}{0.48\columnwidth}
      \centering
      \includegraphics[width=5.5cm]{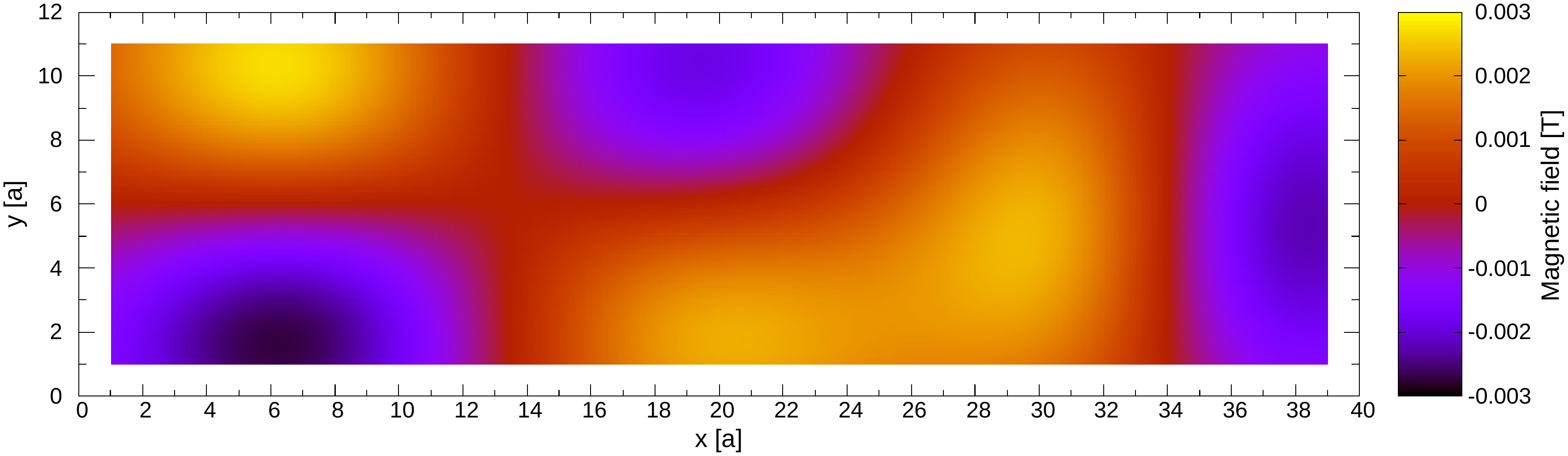}
      \caption{$B_z$ at $z=10$ for (b)}
    \end{subfigure}
    \begin{subfigure}{0.48\columnwidth}
      \centering
      \includegraphics[width=5.5cm]{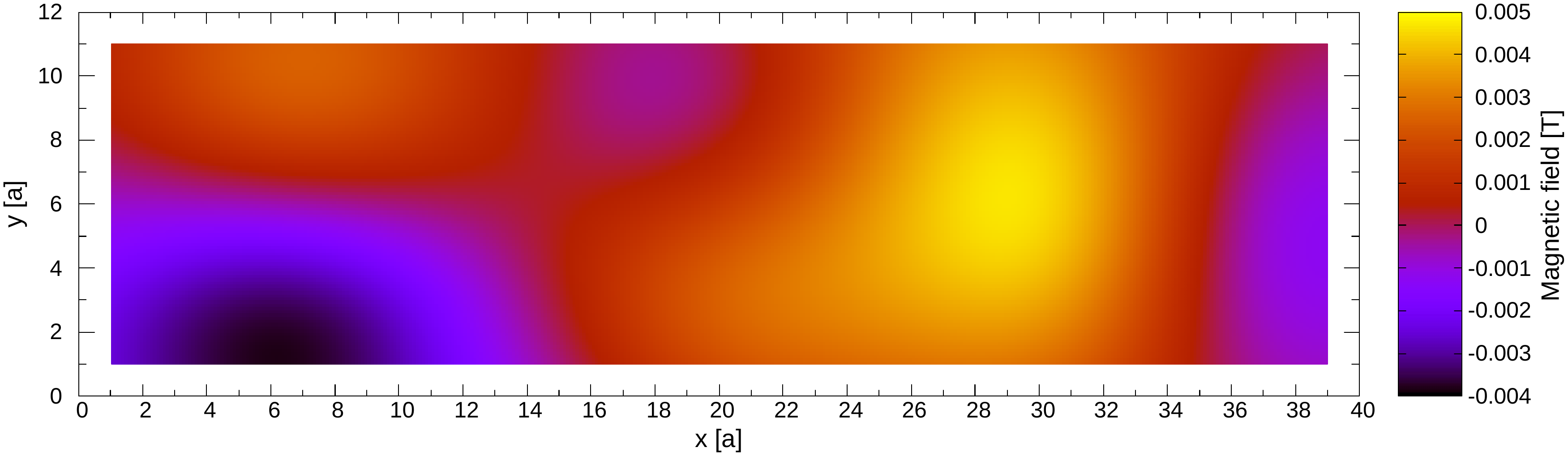}
       \caption{$B_z$ at $z=10$ for (c)}
          \end{subfigure}
    \begin{subfigure}{0.48\columnwidth}
      \centering
      \includegraphics[width=5.5cm]{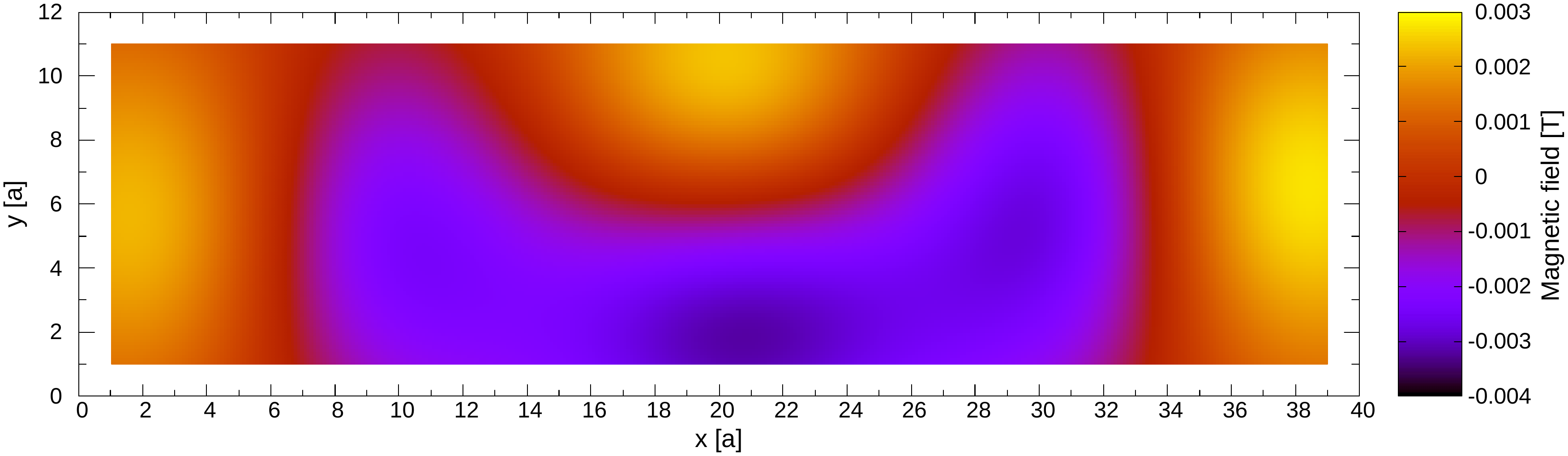}
     \caption{$B_z$ at $z=10$ for (c)}
    \end{subfigure}
    \begin{subfigure}{0.48\columnwidth}
      \centering
      \includegraphics[width=5.5cm]{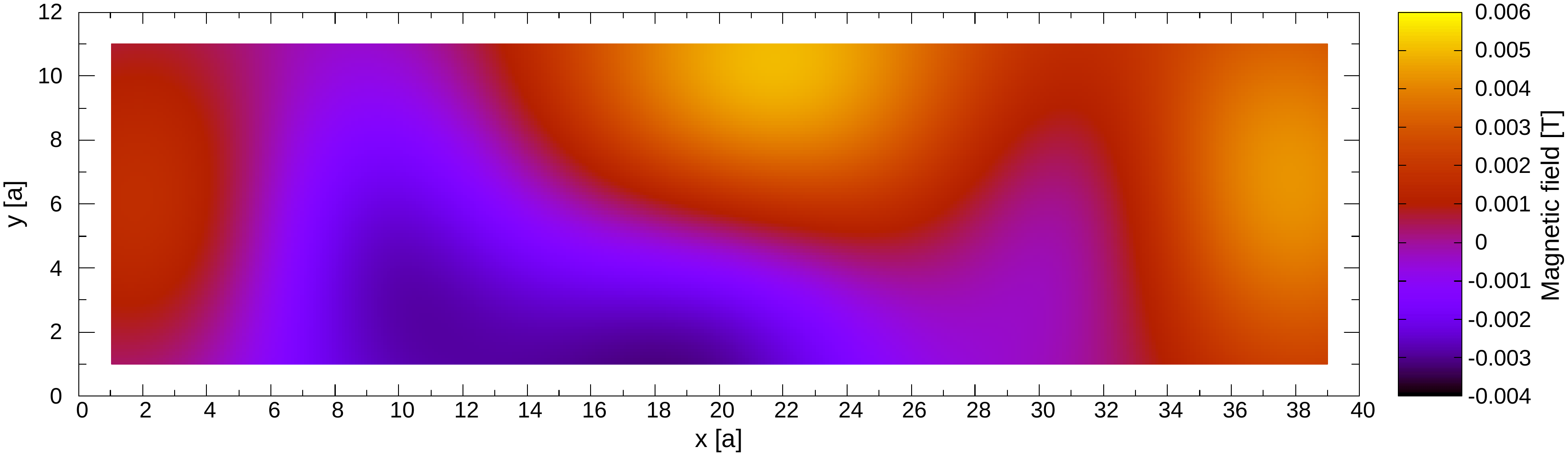}
    \caption{$B_z$ at $z=10$ for (e)}
    \end{subfigure}
    \begin{subfigure}{0.48\columnwidth}
      \centering
      \includegraphics[width=5.5cm]{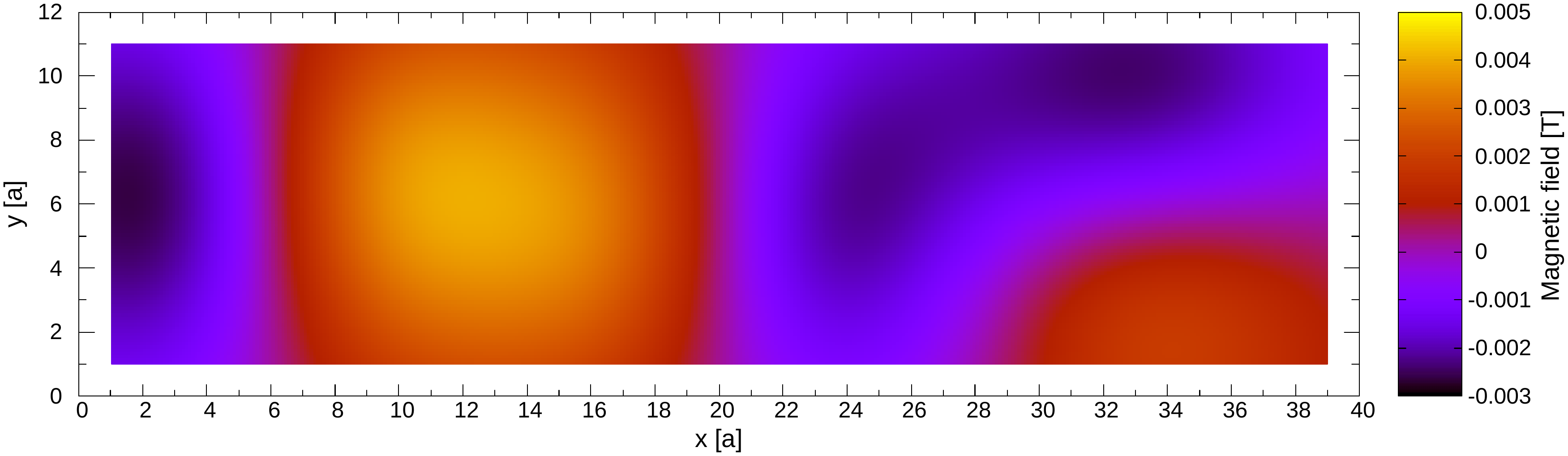}
     \caption{$B_z$ at $z=10$ for (f)}
    \end{subfigure}
    \begin{subfigure}{0.48\columnwidth}
      \centering
      \includegraphics[width=5.5cm]{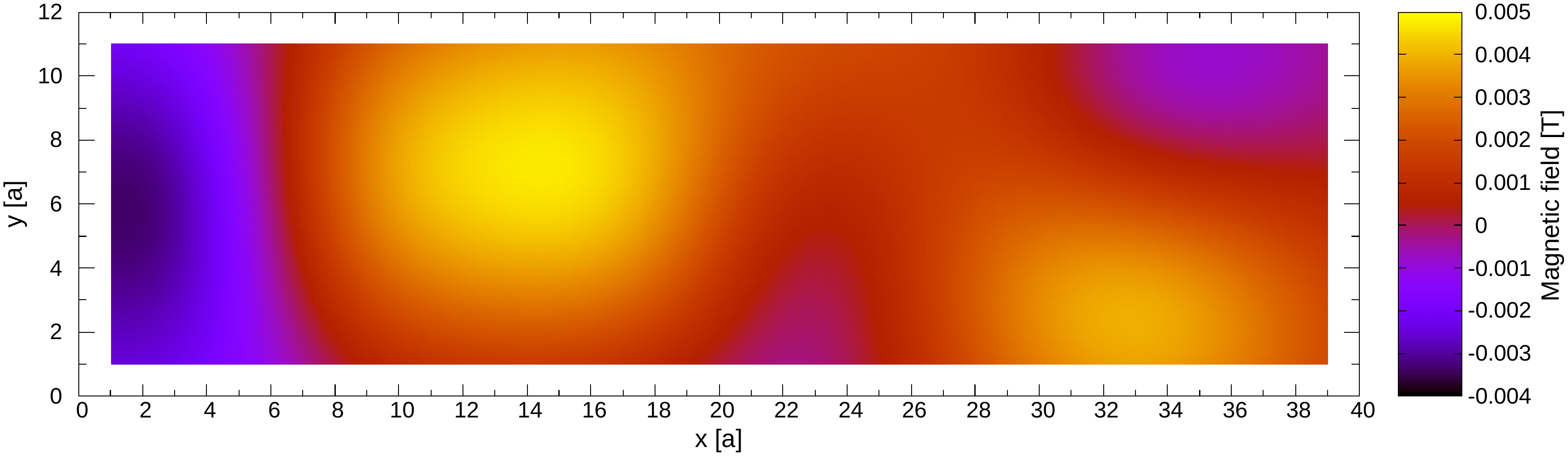}
     \caption{$B_z$ at $z=10$ for (g)}
    \end{subfigure}
    \begin{subfigure}{0.48\columnwidth}
      \centering
      \includegraphics[width=5.5cm]{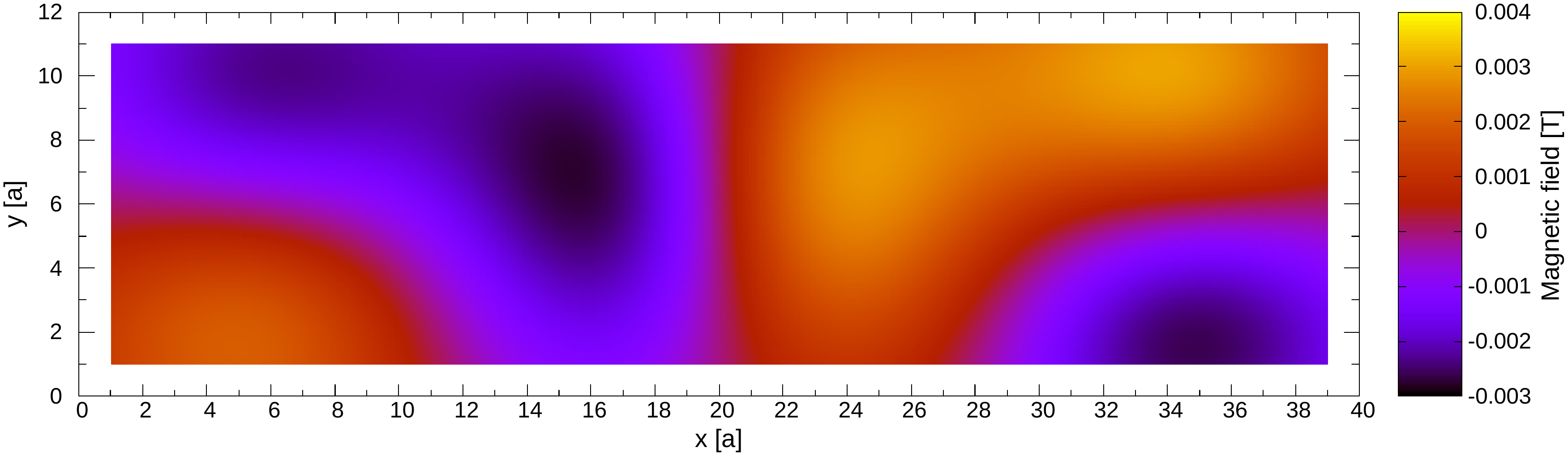}
     \caption{$B_z$ at $z=10$ for (h)}
          \end{subfigure}
    \begin{subfigure}{0.48\columnwidth}
      \centering
      \includegraphics[width=5.5cm]{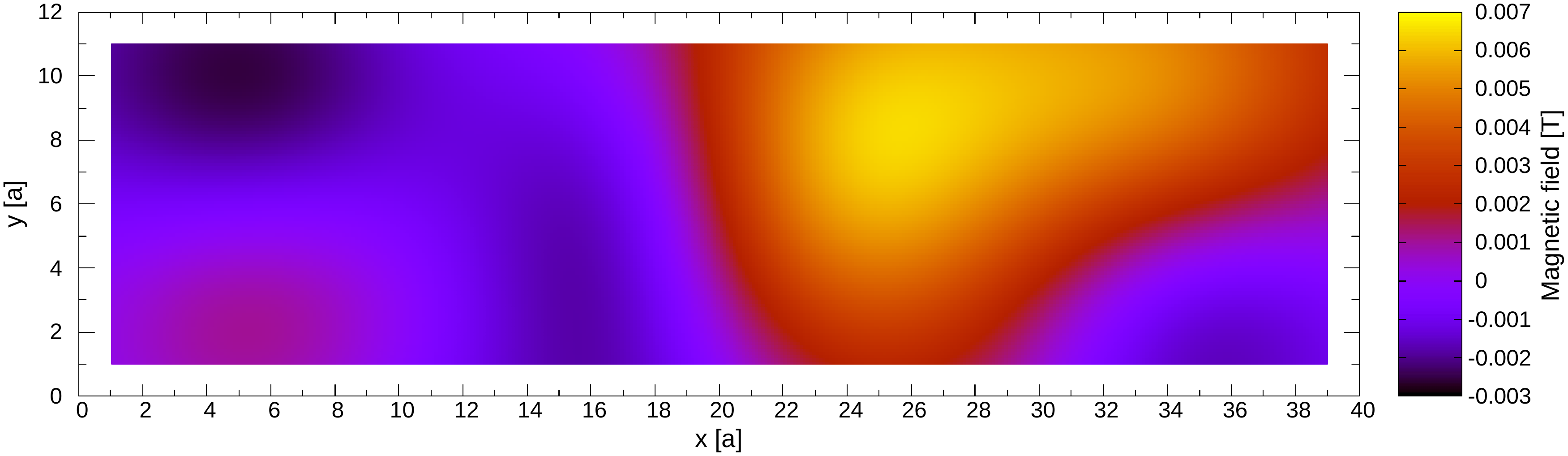}
     \caption{$B_z$ at $z=10$ for (i)}
    \end{subfigure}
    \caption{SVILC qubit currents and magnetic fields generated by them.}
    \label{fig:result:3NI_magnetic}
  \end{figure}
  
In Fig.~\ref{fig:result:3NI_magnetic}, the four qubit states, B-A-C, C-B-D, D-C-A, and A-D-B,  and the generated magnetic fields are depicted.
These results indicate current patterns are detectable from the magnetic field distributions.
The magnetic field is somewhat enhanced when the coupling is introduced.
These results indicate the observation of the magnetic field may be used for the readout of the qubit states.

\section{Concluding remarks}
\label{sec6}
 
The existence of nano-sized loop currents in the cuprate is supported by many experiments.
The existence of spin-vortices is confirmed by experiments although their size is much larger than the one considered in this work \cite{Wang:2023aa}.
It is important to check if the nano-sized loop currents in the cuprate are whether the SVILCs or not. The present work indicates that if they are the SVILCs they will produce detectable magnetic fields. 
Since the spin moments of spin-vortices are lying in the CuO$_2$ planes, their magnetic field does not have the component perpendicular to the CuO$_2$ planes.
On the other hand, the magnetic field produced by the SVILCs has the perpendicular component,
thus, both the spin-vortices and SVILCs can be separately detected.
If their existence is confirmed, it will lead to the elucidation of the mechanism of the cuprate superconductivity. Further, it may provide new qubits as suggested in the present work. The SVILC qubits may realize practical quantum computers with more than 100 logical qubits.

\appendix

\section{The particle number conserving Bogoliubov-de Gennes formalism}
\label{ Bogoliubov}

We succinctly explain the particle number conserving 
Bogoliubov-de Gennes (PNC-BdG) formalism \cite{koizumi2023,koizumi2021}
in this Appendix.

The BdG formalism \cite{Zhu2016,deGennes} starts with the observation that the BCS superconducting state given by
\begin{eqnarray}
|{\rm BCS} (\theta) \rangle=\prod_{\bf k}\left(u_{\bf k}+v_{\bf k}c^{\dagger}_{{\bf k} \uparrow}c^{\dagger}_{-{\bf k} \downarrow}
e^{ {i}{\theta}} \right)|{\rm vac} \rangle
\label{BCS}
\end{eqnarray}
where real parameters $u_{\bf k}$ and $v_{\bf k}$ satisfy $u_{\bf k}^2+v_{\bf k}^2=1$,
can be viewed as the vacuum of the 
Bogoliubov quasiparticles, whose annihilation operators are given by
\begin{eqnarray}
 \gamma^{\rm BCS}_{{\bf k} \uparrow }&=& u_{k} e^{-{i \over 2}\theta} c_{{\bf k} \uparrow} -  v_{ k} e^{{i \over 2}\theta} c^{\dagger}_{-{\bf k} \downarrow}
\nonumber
\\
\gamma^{\rm BCS}_{-{\bf k} \downarrow }&=& u_{k} e^{-{i \over 2}\theta} c_{-{\bf k} \downarrow} +  v_{ k} e^{{i \over 2}\theta} c^{\dagger}_{{\bf k} \uparrow}
\end{eqnarray}
and satisfy 
\begin{eqnarray}
 \gamma^{\rm BCS}_{{\bf k} \uparrow }|{\rm BCS} \rangle=0, \quad \gamma^{\rm BCS}_{-{\bf k} \downarrow }|{\rm BCS} \rangle=0
 \end{eqnarray}
 
The above change of the vacuum can be implemented by changing the electron field operators
using the the Bogoliubov operators 
\begin{eqnarray}
\hat{\Psi}^{\rm BCS}_{\uparrow}({\bf r})&=& \sum_{\bf k} e^{{i \over 2}\theta} \left[ \gamma^{\rm BCS}_{{\bf k} \uparrow } u_{\bf k}({\bf r}) -  (\gamma^{\rm BCS})^{\dagger}_{-{\bf k} \downarrow } v_{\bf k}({\bf r}) \right]
\nonumber
\\
\hat{\Psi}^{\rm BCS}_{\downarrow}({\bf r})&=& \sum_{\bf k} e^{{i \over 2}\theta}\left[ \gamma^{\rm BCS}_{{\bf k} \downarrow } u_{\bf k}({\bf r}) + (\gamma^{\rm BCS})^{\dagger}_{-{\bf k} \uparrow } v_{\bf k}({\bf r}) \right]
\label{fieldOp}
\end{eqnarray}
where the following coordinate dependent functions are introduced
\begin{eqnarray}
u_{\bf k}({\bf r})={1 \over \sqrt{V}}e^{ i {\bf k} \cdot {\bf r}}u_k, \quad
v_{\bf k}({\bf r})={1 \over \sqrt{V}}e^{i {\bf k} \cdot {\bf r}}v_k
\end{eqnarray}

We further allow the coordinate dependent functions that are different from plane waves;
and also allows the coordinate dependence in the phase factor $e^{{i \over 2}\theta}$ as $e^{{i \over 2}\theta({\bf r})}$. Using the label $n$ in place of the wave number ${\bf k}$,
the field operators become
 \begin{eqnarray}
\hat{\Psi}^{\rm BCS}_{\uparrow}({\bf r})&=&\sum_{n} e^{{i \over 2}\theta({\bf r}) }\left[ \gamma^{\rm BCS}_{{n} \uparrow } u_{n}({\bf r})  -(\gamma^{\rm BCS})^{\dagger}_{{n} \downarrow } v^{\ast}_{n}({\bf r}) \right]
\nonumber
\\
\hat{\Psi}^{\rm BCS}_{\downarrow}({\bf r})&=&
\sum_{n} e^{{i \over 2}\theta({\bf r})} \left[ \gamma^{\rm BCS}_{{n} \downarrow } u_{n}({\bf r}) + (\gamma^{\rm BCS})^{\dagger}_{{n} \uparrow } v^{\ast}_{n}({\bf r}) \right]
\label{field2p}
\end{eqnarray}
By solving the Bogoliubov-de Gennes equations, $e^{{i \over 2}\theta({\bf r})} u_{n}({\bf r})$ and $e^{-{i \over 2}\theta({\bf r})} v_{n}({\bf r})$ are obtained.

In the particle-number conserving Bogoliubov-de Gennes formalism, the following replacement occurrs
\begin{eqnarray}
\theta \rightarrow -\hat{\chi}
\label{eqA9}
\end{eqnarray}
where $\hat{\chi}$ is the operator version of $\chi$,
with $e^{-{i \over 2}\chi({\bf r})}$ describing the collective flow motion of electrons (we call it , the '$\chi$ mode').

The quantization condition for $\hat{\chi}$ is given by
\begin{eqnarray}
[\hat{\chi}({\bf r}, t), \hat{\rho}_\chi({\bf r}',t)] = 2i\delta({\bf r}- {\bf r}')
\label{Canonical}
\end{eqnarray}
where ${\rho}_\chi$ is the operator for the number density of electrons that participate in the $\chi$ mode. The appearance of $\chi$ and its quantization are explained in Appendix B.
Using the commutation relation in Eq.~(\ref{Canonical}), it can be shown that 
$e^{ {i \over 2}\hat{\chi}}$ is the number changing operators that increases the number of electrons participating the collective motion by one, and $e^{ -{i \over 2}\hat{\chi}}$ decreases the number by one.

By replacing $\theta$ by $-\hat{\chi}$ in Eq.~(\ref{field2p}), the field operators for the new theory are given by
\begin{eqnarray}
\hat{\Psi}_{\uparrow}({\bf r})&=&\sum_{n} e^{-{i \over 2}\hat{\chi} ({\bf r})}\left[ \gamma_{{n} \uparrow } u_{n}({\bf r})  -\gamma^{\dagger}_{{n} \downarrow } v^{\ast}_{n}({\bf r}) \right]
\nonumber
\\
\hat{\Psi}_{\downarrow}({\bf r})&=&
\sum_{n} e^{-{i \over 2}\hat{\chi} ({\bf r})} \left[ \gamma_{{n} \downarrow } u_{n}({\bf r}) +\gamma^{\dagger}_{{n} \uparrow } v^{\ast}_{n}({\bf r}) \right]
\label{new-fieldO}
\end{eqnarray}
where $\gamma_{n \sigma}, \gamma^\dagger_{n \sigma}$ are the Bogoliubov operators of the new theory. 
The important point here is that the above Bogoliubov operators conserve particle numbers,
thus, we call them the particle number conserving Bogoliubov operators (PNC-BOs). The PNC-BOs cause the fluctuation of the number of electrons in the collective $\chi$ mode. 

The PNC-BOs satisfy
\begin{eqnarray}
\gamma_{n \sigma}|{\rm Gnd}(N) \rangle=0, \quad  \langle {\rm Gnd}(N)| \gamma^{\dagger}_{n \sigma}=0
\end{eqnarray}
where $N$ is the total number of particles.
They contain an additional operator $e^{{i \over 2} \hat{\chi}({\bf r})}$.
We take the ground state to be the eigenstate of it that satisfies 
\begin{eqnarray}
e^{{i \over 2} \hat{\chi}({\bf r})}|{\rm Gnd}(N) \rangle= e^{{i \over 2} {\chi}({\bf r})}|{\rm Gnd}(N+1) \rangle
\label{eqBC}
\end{eqnarray}

Using the PNC-BO, we can construct the particle number conserving Bogoliubov-de Gennes equations (PNC-BdG equations). 
Let us consider the following electronic Hamiltonian
\begin{eqnarray}
H_e=\sum_{\sigma} \int d^3 r \hat{\Psi}^{\dagger}_{\sigma}({\bf r}) h({\bf r}) \hat{\Psi}_{\sigma}({\bf r}) 
-{1 \over 2} \sum_{\sigma, \sigma'}\int d^3 r d^3 r' V_{\rm eff}({\bf r}, {\bf r}') \hat{\Psi}^{\dagger}_{\sigma}({\bf r}) \hat{\Psi}^{\dagger}_{\sigma'}({\bf r}') \hat{\Psi}_{\sigma'}({\bf r}') \hat{\Psi}_{\sigma}({\bf r}) 
\nonumber
\\
\end{eqnarray}
where $h({\bf r})$ is the single-particle Hamiltonian given by
\begin{eqnarray}
h({\bf r})={ 1 \over {2m_e}} \left( { \hbar \over i} \nabla +{e \over c} {\bf A} \right)^2+U({\bf r})-\mu 
\end{eqnarray}
and $-V_{\rm eff}$ is the effective interaction between electrons.

We perform the mean field approximation on $H_e$. The result is 
\begin{eqnarray}
H_e^{\rm MF}&=&\sum_{\sigma} \int d^3 r \hat{\Psi}^{\dagger}_{\sigma}({\bf r}) h({\bf r}) \hat{\Psi}_{\sigma}({\bf r}) 
+\int d^3 r d^3 r' 
\left[ \Delta({\bf r}, {\bf r}')\hat{\Psi}^{\dagger}_{\uparrow}({\bf r}) \hat{\Psi}^{\dagger}_{\downarrow}({\bf r}') e^{-{i \over 2}(\hat{\chi}({\bf r}) +\hat{\chi}({\bf r}')) }
+{\rm H. c.} \right]
\nonumber
\\
&+&\int d^3 r d^3 r' 
{ {|\Delta({\bf r}, {\bf r}')|^2} \over {V_{\rm eff}({\bf r}, {\bf r}') }}
\end{eqnarray}
where the gap function $\Delta({\bf r}, {\bf r}')$ is defined as 
\begin{eqnarray}
 \Delta({\bf r}, {\bf r}')= V_{\rm eff}({\bf r}, {\bf r}')\langle e^{{i \over 2}(\hat{\chi}({\bf r}) +\hat{\chi}({\bf r}')) }
\hat{\Psi}_{\uparrow}({\bf r}) \hat{\Psi}_{\downarrow} ({\bf r'}) \rangle
\end{eqnarray}
Due to the factor $ e^{{i \over 2}(\hat{\chi}({\bf r}) +\hat{\chi}({\bf r}'))}$ the expectation value in $\Delta$ can be calculated using the particle number fixed state.

Using commutation relations for $\hat{\Psi}^{\dagger}_{\sigma }({\bf r})$ and $\hat{\Psi}_{\sigma }({\bf r})$,
\begin{eqnarray}
&&\{ \hat{\Psi}_{\sigma }({\bf r}),\hat{\Psi}^{\dagger}_{\sigma' }({\bf r}') \}=\delta_{\sigma \sigma'}\delta({\bf r} -{\bf r}')
\nonumber
\\
&&\{ \hat{\Psi}_{\sigma }({\bf r}),\hat{\Psi}_{\sigma' }({\bf r}') \}=0
\nonumber
\\
&&\{ \hat{\Psi}^{\dagger}_{\sigma }({\bf r}),\hat{\Psi}^{\dagger}_{\sigma' }({\bf r}') \}=0
 \end{eqnarray}
the following relations are obtained
\begin{eqnarray}
\left[\hat{\Psi}_{\uparrow }({\bf r}) , {H}_{\rm MF} \right]&=&
{h}({\bf r})\hat{\Psi}_{\uparrow }({\bf r})+\int d^3 r' \Delta({\bf r},{\bf r}')\hat{\Psi}^{\dagger}_{\downarrow }({\bf r}')e^{-{i \over 2}(\hat{\chi}({\bf r}) +\hat{\chi}({\bf r}')) }
\nonumber
\\
\left[\hat{\Psi}_{\downarrow }({\bf r}) , {H}_{\rm MF} \right] &=&{h}({\bf r})\hat{\Psi}_{\downarrow }({\bf r})-\int d^3 r' \Delta({\bf r},{\bf r}')\hat{\Psi}^{\dagger}_{\uparrow }({\bf r}')e^{-{i \over 2}(\hat{\chi}({\bf r}) +\hat{\chi}({\bf r}')) }
\nonumber
\\
\label{deG1}
\end{eqnarray}

The PNC-BOs $\gamma_{n \sigma}$ and $\gamma^{\dagger}_{n \sigma}$ obey fermion commutation relations. They are chosen to satisfy
\begin{eqnarray}
\left[ {H}_{\rm MF}, \gamma_{n \sigma } \right]=-\epsilon_n \gamma_{n \sigma}, \quad \left[{H}_{\rm MF}, \gamma^{\dagger}_{n \sigma } \right] =\epsilon_n \gamma^{\dagger}_{n \sigma}
\label{deG2}
\end{eqnarray}
with $\epsilon_n \geq 0$. Then, ${H}_{\rm MF}$ is diagonalized as
\begin{eqnarray}
{H}_{\rm MF}=E_g + {\sum_{n, \sigma}}' \epsilon_n \gamma^{\dagger}_{n \sigma}\gamma_{n \sigma}
\label{deG3}
\end{eqnarray}
where $E_g$ is the ground state energy, and `$\sum' $' indicates the sum is taken over $\epsilon_n \geq 0$ states.

From Eqs.~(\ref{deG1}), (\ref{deG2}), and (\ref{deG3}), we obtain the following system of equations
\begin{eqnarray}
\epsilon_n u_n({\bf r})&=&
e^{{i \over 2} \hat{\chi}({\bf r})}h({\bf r}) e^{-{i \over 2}\hat{\chi}({\bf r})}u_n({\bf r})+e^{{i \over 2} \hat{\chi}({\bf r})}\int d^3 r' \Delta ({\bf r},{\bf r}')e^{-{i \over 2}\hat{\chi}({\bf r})}v_n({\bf r}')
\nonumber
\\
\epsilon_n v^{\ast}_n({\bf r})&=&-e^{{i \over 2} \hat{\chi}({\bf r})}
 h({\bf r}) e^{-{i \over 2} \hat{\chi}({\bf r})}v^{\ast}_n({\bf r})+e^{{i \over 2} \hat{\chi}({\bf r})}\int d^3r' \Delta ({\bf r},{\bf r}')e^{-{i \over 2}\hat{\chi}({\bf r})}u^{\ast}_n({\bf r}')
\end{eqnarray}

We take the expectation value of the above equations for the state $|{\rm Gnd}(N) \rangle$. Using the relation in Eq.~(\ref{eqBC}),
the above are cast into the following,
\begin{eqnarray}
\epsilon_n u_n({\bf r})&=&
\bar{h}({\bf r}) u_n({\bf r})+\int d^3 r'\Delta ({\bf r},{\bf r}')v_n({\bf r}')
\nonumber
\\
\epsilon_n v_n({\bf r})&=&-
 \bar{h}^{\ast}({\bf r}) v_n({\bf r})+\int d^3 r'\Delta^{\ast}({\bf r},{\bf r}')u_n({\bf r}')
 \label{e1}
\end{eqnarray}
where the single particle Hamiltonian $\bar{h}$ is
\begin{eqnarray}
\bar{h}({\bf r})={ 1 \over {2m_e}} \left( { \hbar \over i} \nabla +{e \over c} {\bf A}-{ \hbar \over 2} \nabla \chi \right)^2+U({\bf r})-\mu 
 \label{e2}
\end{eqnarray}
the pair potential $\Delta({\bf r}, {\bf r}')$ is
\begin{eqnarray}
\Delta({\bf r}, {\bf r}')=V_{\rm eff}{\sum_n}' \left[ u_n({\bf r}) v^{\ast}_n({\bf r}')(1- f(\epsilon_n))-u_n({\bf r}') v^{\ast}_n({\bf r})f(\epsilon_n) \right]
 \label{e3}
\end{eqnarray}
and $f(\epsilon_n)$ is the Fermi function ($T \rightarrow 0$ limit should be considered for the ground state). The number density is given by
\begin{eqnarray}
\rho({\bf r})={\sum_n}' \left[ |u_n({\bf r})|^2f(\epsilon_n)+|v_n({\bf r})|^2({\bf r})(1- f(\epsilon_n)) \right]
 \label{e4}
\end{eqnarray}
 They are the PNC-BdG equations \cite{koizumi2019}.
 
 The important point in  Eq.~(\ref{e2}) is that instead of the usual momentum operator ${\bf p}={\hbar \over i}\nabla$,  ${\bf p}={\hbar \over i}\nabla\!-\!{\hbar \over 2} \nabla \chi$ appears.
 
 %
 
 \section{The $\chi$ mode from the neglected $U(1)$ phase by Dirac in the Schr\"{o}dinger wave mechanics formalism of quantum mechanics, and its quantization }
  \label{chi-mode}
  
  In this Appendix, the appearance of $\chi$ and its quantization are explained.
From a many-electron wave function $\Psi$, we can always obtain a `Berry connection' \cite{Berry} defined by
\begin{eqnarray}
\!{\bf A}^{\rm MB}_{\Psi}({\bf r},t)\!=\!
{{{\rm Re} \left\{
 \int d\sigma_1  d{\bf x}_{2}  \cdots d{\bf x}_{N}
 \Psi^{\ast}({\bf r}, \sigma_1, \cdots, {\bf x}_{N},t)
  (-i \hbar \nabla )
\Psi({\bf r}, \sigma_1, \cdots, {\bf x}_{N},t) \right\}
 }
 \over {\hbar N^{-1}\rho({\bf r},t)}} 
 \nonumber
 \\
 \label{Berry}
\end{eqnarray}
This is the Berry connection from many-body wave functions, and it contains the information about many-body effects.
Here,
`$\rm{Re}$' denotes the real part, $\Psi$ is the total wave function, ${\bf x}_i$ collectively stands for the coordinate ${\bf r}_i$ and the spin $\sigma_i$ of the $i$th electron, $N$ is the total number of electrons, and $\rho$ is the number density calculated from $\Psi$. 

The emergence of nontrivial ${\bf A}_{\Psi}^{\rm MB}$ contradicts the following assertion made by Dirac \cite{DiracSec22}:
In the Schr\"{o}dinger's representation, 
the momenta $p_r$'s conjugate to canonical coordinates $q_r$'s given 
\begin{eqnarray}
p_r=-i \hbar {{\partial} \over {\partial q_r}}, \quad r=1, \cdots, n
\label{pr1}
\end{eqnarray}
where $n$ is the degrees of freedom in the many-body system. In the above case $n=3N$ and canonical coordinates are ${\bf r}_1 \cdots {\bf r}_N$. Dirac argued that 
according to the Heisenberg formulation of quantum theory \cite{Born-Jordan}, commutation relations
\begin{eqnarray}
[q_r, q_s]=0, \quad [p_r, p_s]=0, \quad  [q_r, p_s]=i \hbar \delta_{rs}
\end{eqnarray}
are more fundamental than the derivative representation of the momenta; thus, the following  $p_r$'s are also legitimate
\begin{eqnarray}
p_r=-i \hbar {{\partial} \over {\partial q_r}}+{{\partial F} \over {\partial q_r}}, \quad r=1, \cdots, n
\label{pr2}
\end{eqnarray}
where $F$ is a function of $q_1, \cdots, q_n$, and also of time $t$. 
However the above $F$ can be removed by
the following transformation of the wave function
\begin{eqnarray}
\psi(q_1, \cdots, q_n) \rightarrow e^{i \gamma} \psi(q_1, \cdots, q_n) 
\label{gamma}
\end{eqnarray}
where $\gamma$ is related to $F$ by
$
F=-\hbar \gamma + \mbox{constant}
$.
Therefore, we can always use $p_r$'s in Eq.~(\ref{pr1}). In the present case $e^{i \gamma} \psi(q_1, \cdots, q_n)$ corresponds to $\Psi$. By following this assertion, the standard calculation obtains $e^{i \gamma} \psi(q_1, \cdots, q_n)$ as a whole by employing a finite number of basis functions.
However, the appearance of the non-trivial $\chi$ violates this assertion; in this case, $e^{i \gamma}$ needs to be considered as an extra degree-of-freedom \cite{koizumi2022,koizumi2022b}.

In general, a many-body wave function is expressed using the Berry connection in Eq.~(\ref{Berry}) as 
\begin{eqnarray}
\Psi({\bf x}_1, \cdots, {\bf x}_{N},t)=\exp\left( i \sum_{j=1}^{N} \int_{0}^{{\bf r}_j} {\bf A}_{\Psi}^{\rm MB}({\bf r}',t) \cdot d{\bf r}' \right)\Psi_0({\bf x}_1, \cdots, {\bf x}_{N},t)
\label{eqPsi0}
\end{eqnarray} 
where $\Psi_0$ is given by
\begin{eqnarray}
\Psi_0 ({\bf x}_1, \cdots, {\bf x}_{N},t)=\Psi ({\bf x}_1, \cdots, {\bf x}_{N},t)\exp\left(- i \sum_{j=1}^{N} \int_{0}^{{\bf r}_j} {\bf A}_{\Psi}^{\rm MB}({\bf r}',t) \cdot d{\bf r}' \right)
\label{wavef0}
\nonumber
\\
\end{eqnarray}
Note that $\Psi_0$ is currentless since the current density calculated by $\Psi$ is canceled by the current density from 
$\exp\left(- i \sum_{j=1}^{N} \int_{0}^{{\bf r}_j} {\bf A}_{\Psi}^{\rm MB}({\bf r}',t) \cdot d{\bf r}' \right)$.
The velocity field ${\bf v}$ in Eq.~(\ref{eqvelo}) is obtained using $\Psi$ and the velocity operator for many-electron systems with including the electromagnetic vector potential ${\bf A}$
\cite{Koizumi2024Lorentz}.

As will be shown later, ${\bf A}_{\Psi}^{\rm MB}$ in the present model arises
from spin-twisting itinerant motion of conduction electrons, and can be expressed as
 \begin{eqnarray}
{\bf A}_{\Psi}^{\rm MB}=-{1 \over 2} \nabla \chi
\label{eq-chi}
\end{eqnarray}
where $\chi$ is an angular variable with period $2\pi$.
Then, $\gamma$ in Eq.~(\ref{gamma}) becomes
\begin{eqnarray}
\gamma= -{1 \over 2}\sum_{j=1}^{N} \chi({\bf r}_j,t)
\label{gamma2}
\end{eqnarray}

Due to the presence of the phase factor, the total energy calculated for the $N$-electron  Hamiltonian
\begin{eqnarray}
H=\sum_{j=1}^N \left[-{ \hbar^2 \over {2 m_e}}\nabla_j^2 +U({\bf r}_j)\right]+{ 1 \over 2}\sum_{i\neq j} V({\bf r}_i, {\bf r}_j),
\label{bloch-15}
\end{eqnarray}
with  $U({\bf r}_j)$ being a single-particle potential energy, and $V({\bf r}_i, {\bf r}_j)$ a two-particle interaction energy,
is given by
\begin{eqnarray}
E[\chi]-E_0={\hbar^2 \over {8 m_e}}\int d^3r \rho({\bf r}, t) \left[\nabla \chi ({\bf r},t)\right]^2
\label{eqEchi}
\end{eqnarray}
where $E[\chi]$ is the total energy for $\Psi$ which is a functional of $\chi$, and $E_0$ is the total energy for $\Psi_0$.
This formula indicates that $\nabla \chi$ gives rise to a velocity field, and $\rho$ is the electron density that moves with this velocity field. Actually, $\chi$ and $\rho$ are canonical conjugate fields; then it is possible to quantize $\rho$ and $\chi$ as field operators $\hat{\rho}$ and $\hat{\chi}$, respectively. 
By treating $\rho$ as the number density of electrons participating in the $\chi$ mode,
the quantization allows the change of number of electrons participating in this $\chi$ mode.
 
 Let us examine the quantization of $\chi$ and its conjugate variable. In order to find the conjugate variable 
we use the following Lagrangian 
\begin{eqnarray}
L&=&\langle \Psi | i\hbar {\partial \over {\partial t}} -H |\Psi \rangle
\nonumber
\\
&=&\int d^3 r \hbar { \dot{\chi} \over 2}\rho+ i\hbar \langle \Psi_0 |{\partial \over {\partial t}}|\Psi_0 \rangle-\langle \Psi | H |\Psi \rangle
\end{eqnarray}
The above indicates that the conjugate field of $\chi$, $\pi_{\chi}$, is
\begin{eqnarray}
\pi_{\chi}={{\delta L} \over {\delta \dot{\chi} }}={ \hbar \over 2} \rho
\end{eqnarray}
Then, the canonical quantization condition 
\begin{eqnarray}
[\hat{\chi}({\bf r},t),\hat{\pi}_{\chi} ({\bf r}',t)]=i\hbar \delta({\bf r}-{\bf r}')
\end{eqnarray}
yields $
[\hat{\chi}({\bf r},t),\hat{\rho}({\bf r}',t)]=2i\delta({\bf r}-{\bf r}')$
for the quantization condition for $\hat{\chi}$.
At this point, we make a departure from the wave function description of the physical state;
the use of $\hat{\chi}$ allows the possibility that the number of electrons participating in the $\chi$ mode can be less than the total number. Namely, we use $\rho_\chi$ instead of $\rho$  as the conjugate variable, where the total number of electrons calculated by $\rho_\chi$ is less than $N$. Since this $\chi$ mode gives rise to the velocity field ${\bf v}$, $\rho_\chi$ is identified as the superconducting electron density.


The quantization condition for $\hat{\chi}$ is given in Eq.~(\ref{Canonical}) with
 ${\rho}_\chi$ being the operator for the number density of electrons that participate in the $\chi$ mode. 
Using the commutation relation in Eq.~(\ref{Canonical}), it can be shown that 
$e^{ {i \over 2}\hat{\chi}}$ is the number changing operators that increases the number of electrons participating the collective motion by one, and $e^{ -{i \over 2}\hat{\chi}}$ decreases the number by one.

\bibliographystyle{spphys}

 
\end{document}